\documentclass[a4paper,12pt]{article}
\usepackage[margin=24mm]{geometry}
\makeatletter
 
\@addtoreset{section}{part}
\def\@part[#1]#2{%
	\ifnum \c@secnumdepth >\m@ne
	\refstepcounter{part}%
	\addcontentsline{toc}{part}{\thepart\hspace{1em}#1}%
	\else
	\addcontentsline{toc}{part}{#1}%
	\fi
	{\parindent \z@ \raggedright
		\interlinepenalty \@M
		\normalfont\centering
		\ifnum \c@secnumdepth >\m@ne
		\Large\bfseries \partname\nobreakspace\thepart
		\par\nobreak
		\fi
		\huge \bfseries #2%
		\markboth{}{}\par}%
	\nobreak
	\vskip 3ex
	\@afterheading}
\renewcommand\partname{ }
\makeatother
\RequirePackage{amsthm,amsmath,amsfonts,amssymb}
\RequirePackage[authoryear]{natbib}
\RequirePackage[colorlinks,citecolor=blue,urlcolor=blue]{hyperref}
\RequirePackage{graphicx}

\usepackage{setspace}
\usepackage{mathrsfs} 
\usepackage{dsfont}
\usepackage{wrapfig}

\usepackage{tikz} 
\usetikzlibrary{positioning}
\usetikzlibrary{calc}

\usepackage{fancyhdr}
\usepackage{float}
\usepackage{bigints} 

\usepackage{marginnote,bm}
\usepackage{xcolor,colortbl} 

\newtheorem{theorem}{Theorem}[section]

\theoremstyle{remark}
\newtheorem{definition}[theorem]{Definition}

\newtheorem{proposition}[theorem]{Proposition}
\usepackage{url} 
\usepackage{subcaption}

\definecolor{ashgrey}{rgb}{0.7, 0.75, 0.71}
\definecolor{darkgray}{rgb}{0.66, 0.66, 0.66}
\definecolor{cadetgrey}{rgb}{0.57, 0.64, 0.69}
\definecolor{aoe}{rgb}{0.0, 0.5, 0.0}
\definecolor{burntorange}{rgb}{0.8, 0.33, 0.0}
\definecolor{bostonuniversityred}{rgb}{0.8, 0.0, 0.0}
\definecolor{burgundy}{rgb}{0.0, 0.0, 1}
\colorlet{shadecolor}{gray!20}

\newcommand{\bmath}{\boldsymbol}

\newcommand{\R}{\mathds{R}}

\newcommand{\ddr}{\mathrm{d}}

\newcommand{\ep}{\varepsilon}

\def\simind{\stackrel{\mbox{\scriptsize{\rm ind}}}{\sim}}
\def\simiid{\stackrel{\mbox{\scriptsize{\rm iid}}}{\sim}}
\def\eqd{\stackrel{\mbox{\scriptsize{d}}}{=}}
\usepackage{accents}
\newcommand{\ddtilde}[1]{\accentset{\approx}{#1}}

\providecommand{\keywords}[1]
{
	\small	
	\textbf{\textit{Keywords---}} #1
}

\title{Model Selection for  Maternal Hypertensive Disorders with Symmetric Hierarchical Dirichlet Processes}

\author{Beatrice Franzolini$^1$\and Antonio Lijoi$^2$ \and Igor Pr\"unster$^2$}

\date{
\small
$^1$ Singapore Institute for Clinical Sciences (SICS),\\
Agency for Science, Technology and Research (A*STAR), Singapore\\
	$^2$ Department of Decision Sciences and BIDSA, Bocconi University Milan, Italy
}

\begin{document}
\maketitle

	\small 
{\setstretch{1.0}
\begin{abstract}
Hypertensive disorders of pregnancy occur in about 10\% of pregnant women around the world. 
Though there is evidence that hypertension impacts maternal cardiac functions, the relation between hypertension and cardiac dysfunctions is only partially understood. 
The study of this relationship can be framed as a joint inferential problem on multiple populations, each corresponding to a different hypertensive disorder diagnosis, that combines multivariate information provided by a collection of cardiac function indexes.
A Bayesian nonparametric approach seems particularly suited for  this setup and we demonstrate it on a dataset consisting of transthoracic echocardiography results of a cohort of Indian pregnant women. We are able to perform model selection, provide density estimates of cardiac function indexes and a latent clustering of patients: these readily interpretable inferential outputs allow to single out modified cardiac functions in hypertensive patients compared to healthy subjects and progressively increased alterations with the severity of the disorder.
The analysis is based on a Bayesian nonparametric model that relies on a novel hierarchical structure, called symmetric hierarchical Dirichlet process. This is suitably designed so that
the mean parameters are identified and used for model selection across populations, a penalization for multiplicity is enforced, and the presence of unobserved relevant factors is investigated through a latent clustering of subjects. Posterior inference relies on a suitable Markov Chain Monte Carlo algorithm and the model behaviour is also showcased on simulated data.
\end{abstract}}


\normalsize
\section{Introduction}
\label{sec:intro}

Hypertensive disorders of pregnancy are a class of high blood pressure disorders that occur during the second half of pregnancy, which include gestational hypertension, preeclampsia and severe preeclampsia. They are characterized by a diastolic blood pressure higher than 90 mm Hg and/or a systolic blood pressure higher than 140 mm Hg and they are often accompanied by proteinuria. These disorders affect about 10\% of pregnant women around the world, with preeclampsia occurring in 2--8\% of all pregnancies \citep{timokhina2019}. These disorders represent one of the leading causes 
of maternal and fetal morbidity and mortality, contributing to 7--8\% of maternal death worldwide \citep{dolea2003,shah2009,mcclure2009}. The World 
Health Organization estimates that the incidence of preeclampsia is seven 
times higher in developing countries than in developed countries. However, the occurrence  of these diseases appears under-reported in low and middle income countries, implying that the true incidence is unknown \citep{igberase2006,malik2018}. While there is evidence that hypertensive disorders of pregnancy are related with the development of cardiac dysfunctions both in the mother and in the child \citep{bellamy2007,davis2012, ambrovzic2020dynamic, garcia2020maternal, aksu2021cardiac, demartelly2021long}, there is no common agreement on the relation between the severity of hypertension and cardiac dysfunction \citep{tatapudi2017} and echocardiography is not included in baseline evaluation of hypertensive disorders of pregnancy. Further investigations on these disorders are needed, especially for developing countries, where women often give birth at a younger age with respect to developed countries. 

The goal of this work is to detect which cardiac function is altered and 
under which hypertensive disorders by relying on a principled Bayesian nonparametric approach. An interesting case-control study to explore the relation between cardiac dysfunction and hypertensive disorders is provided by \citet{data}, where the measures of ten different cardiac function indexes were recorded in four groups of pregnant women in India. Groups of women are characterized by different hypertensive disorder diagnoses, that are naturally ordered based on the severity of the diagnosed disorder: healthy (C), gestational hypertension (G), mild preeclampsia (M) and severe preeclampsia (S). 
Hypertensive diagnoses are used as identifiers for what we call populations of patients and we refer to cardiac function indexes also with the term response variables. For each response variable we want to determine a partition of the four populations of patients. This amounts to identifying similarities between different hypertensive disorders, with respect to each cardiac index. Supposing, for instance, that the selected partition assigns all the populations to the same cluster, one can conclude that no 
alteration is shown for the corresponding  cardiac index across different hypertensive diseases.

Our goal of identifying a partition of the four patients' populations for each of the ten responses can be rephrased as a problem of multiple model selection: we want to select the most plausible partition for each cardiac index. 
Frequentist hypothesis testing does not allow to deal with more than two 
populations in a straightforward way
and pairwise comparisons may lead 
to conflicting conclusions. Conversely, a Bayesian approach yields the posterior distribution on the space of partitions, which can be used for simultaneous comparisons. 
Moreover, the presence of $M=10$ jointly tested cardiac indexes requires to perform model selection repeatedly ten times. Once again, a Bayesian approach seems to be preferred, because, as observed for instance by \citet{scott2006}, it does not require the introduction of a penalty term for multiple comparison, thanks to the prior distribution build-in penalty. 

Here we design
a Bayesian nonparametric model, that is tailored to deal with both a collection of ordered populations and the multivariate information of the response variables, while preserving the typical flexibility of nonparametric models and producing easily interpretable results.
When applied to the dataset on transthoracic echocardiography results for a cohort of Indian pregnant women in Section \ref{s:results}, our model effectively identifies 
modified cardiac functions in hypertensive patients compared to healthy subjects and progressively increased alterations with the severity of the disorder, in addition to other more subtle findings. The observed data $X_{i,j,m}$ represent the measurement of the $m$-th response variable (cardiac index) on the $i$-th individual (pregnant woman) in the $j$-th population (hypertensive disorder) 
and, as in standard univariate ANOVA models, they are assumed to be partially exchangeable across disorders.
This means that for every $m\in\{1,\ldots,M\}$,  
the law of $(\,(X_{i,1,m})_{i\ge 1},\ldots,(X_{i,J,m})_{i\ge 1})$ is invariant with respect to permutations within each sequence of random variables, namely for any positive integers $n_1,\ldots,n_J$ 
\[
(\,(X_{i,1,m})_{i=1}^{n_1},\ldots,(X_{i,J,m})_{i=1}^{n_J})\eqd 
(\,(X_{{\sigma_1(i)},1,m})_{i=1}^{n_1},\ldots,(X_{{\sigma_J(i)},J,m})_{i=1}^{n_J})
\]
for all permutations $\sigma_j$ of $(1,\ldots,n_j)$, with $j=1,\ldots,J$. This is a natural generalization of exchangeability to tackle heterogeneous data and, by de Finetti's representation theorem, it amounts to assuming the existence of a collection of (possibly dependent) random probability measures $\{\pi_{j,m}:\:\: j=1,\ldots,J\;\: m=1,\ldots,M\}$ such that
\[
X_{i,j,m} \mid \pi_{j,m} \simiid \pi_{j,m} \qquad i=1,\ldots,n_j
\]  
Hence, for any two populations $j\ne j'$, homogeneity corresponds to $\pi_{j,m}=\pi_{j',m}$ (almost surely). However, a reliable assessment of this type of homogeneity is troublesome when having just few patients per 
diagnosis, as it happens in the mild preeclampsia subsample.  
Without relying on simplifying parametric assumptions, a small sub-sample size may not be sufficiently informative to infer equality of entire unknown distributions. 
To overcome this issue, without introducing parametric assumptions, we resort to an alternative weaker notion of homogeneity between populations $j$ and $j'$: we only require the conditional means of the two populations to (almost surely) coincide
\begin{equation}
	\label{eq:equality_means}
	\mathbb{E}(X_{i,j,m}\mid\pi_{j,m})=\mathbb{E}(X_{i,j',m}\mid\pi_{j',m}).
\end{equation}
According to this definition, the detection of heterogeneities in cardiac function indexes amounts to inferring which cardiac indexes have means that differ 
across diagnoses, as it is done in standard parametric ANOVA models.
Besides clustering populations according to \eqref{eq:equality_means}, it is also of interest to cluster patients, both within and across different groups, once the effect of the specific hypertensive disorder is taken 
into account. This task may be achieved by assuming a model that decomposes the observations as 
\begin{equation}
	\label{eq:decomp_x}
	X_{i,j,m}=\theta_{j,m}+\ep_{i,j,m}\qquad
	\ep_{i,j,m}|(\xi_{i,j,m},\sigma_{i,j,m}^2)\simind \mbox{N}(\xi_{i,j,m},\sigma_{i,j,m}^2)
\end{equation} 
and the $\xi_{i,j,m}$ have a symmetric distribution around the origin, in order to ensure $E(\xi_{i,j,m})=0$. In view of this decomposition, we 
will let $\theta_{j,m}$ govern the clustering of populations while the  $(\xi_{i,j,m},\sigma^2_{i,j,m})$'s determine the clustering of individuals, namely patients, 
after removing the effect of the specific hypertensive disorder. In order to pursue this, for each cardiac index $m$, we will specify a
hierarchical process prior for $(\xi_{i,j,m},\sigma^2_{i,j,m})$ that is suited to infer the clustering structure both within and across different 
hypertensive disorders for a specific cardiac index. In particular, we will deploy a novel instance of hierarchical Dirichlet process, introduced in \cite{teh2006}, that we name \textit{symmetric}, to highlight its centering in $0$.

Early examples of Bayesian nonparametric models for ANOVA can be found in \cite{cifreg78} and \cite{mulierepetrone}, while the first popular proposal, due to \citet{de2004}, uses the dependent Dirichlet process (DDP) \citep{maceachern2000dependent} and is therefore termed ANOVA-DDP. This model is mainly tailored to estimate populations' probability distributions, while we draw inferences over clusters of populations' means and obtain estimates of the unknown distributions as a by-product. Moreover, the ANOVA-DDP of \citet{de2004} was not introduced as a model selection procedure. A popular Bayesian nonparametric model, that does cluster 
populations and can be used for model selection, is the nested Dirichlet process of \citet*{rodriguez2008}. As shown in \citet{camerlenghi2019}, such a prior is biased towards homogeneity, in the sense that even a single 
tie between populations $j$ and $j'$, namely $X_{i,j,m}=X_{i',j',m}$ for some $i$ and $i'$, entails $\pi_{j,m}=\pi_{j'm}$ (almost surely). In order to overcome such a drawback, a novel class of nested, and more flexible, priors has been proposed in   
\citet{camerlenghi2019}. See also \cite{soriano_ma} for related work. Interesting alternatives that extend the analysis to more than two populations can be found in \cite{ChristMa}, \cite*{Rebaudo} and in \cite*{Beraha}. Another similar proposal is the one by \citet{gutierrez2019bayesian}, whose model identifies differences over cases' distributions and the control group. These models imply that two populations belong to the same cluster if they share the entire distribution. However, as already mentioned, distribution-based clustering is not ideal when dealing with scenarios as the one of hypertensive dataset. Further evidence will be provided in Section~\ref{ss:sim}, through simulation 
studies. In addition, note that all these contributions deal with only one response variable and would need to be suitably generalized to fit 
the setup of this paper. As far as the contributions treating multiple response variables are concerned, uses of nonparametric priors for multiple 
testing can be found, for instance, in \citet{gopalan1998}, \citet*{do2005bayesian}, \citet{dahl2007multiple}, \citet*{guindani2009}, \citet{martin2012} and more recently in \citet*{cipolli2016bayesian}, who propose an approximate finite P\'olya tree multiple testing procedure to compare two-samples' locations, and in \citet{dentitwo}. However, in all these contributions, models are developed directly over summaries of the original data (e.g. averages, z-scores) and, as such, do not allow to draw any inference on the entire distributions and clusters of subjects. 

The outline of the paper is as follows. In Section~\ref{s:model} we introduce the model, which makes use of an original hierarchical prior structure for symmetric distributions (Section~\ref{ss:s-HDP}). In Section~\ref{s:partition} we derive the prior law of the random partitions induced by 
the model, key ingredient for the Gibbs sampling scheme devised in Section~\ref{s:inference}. In Section~\ref{s:results}, we first present a series of simulation studies that highlight the behaviour of the model before applying it to obtain our results on cardiac dysfunction in hypertensive disorders. Section~\ref{s:conclusion} contains some concluding remarks. As Supplementary Material we provide the datasets and Python codes, some further background material and details about the derivation of 
the posterior sampling scheme as well as additional simulation studies and results on the application, including an analysis of prior sensitivity.

\section{The Bayesian nonparametric model}
\label{s:model}

The use of discrete nonparametric priors for Bayesian model-based clustering has become standard practice. The Dirichlet process (DP) \citep{ferguson1973} is the most popular instance, and clustering is typically addressed by resorting to a mixture model, which with our data structure amounts to
\begin{equation*}
	\label{eq:mixture_kernel}
	X_{i,j,m}|\psi_{i,j,m}\simind k(X_{i,j,m};\psi_{i,j,m}),\qquad
	\psi_{i,j,m}|\tilde p_{j,m}\simind \tilde p_{j,m} 
\end{equation*}
for $m=1,\ldots,M$, $j=1,\ldots,J$ and $i=1,\ldots,n_j$. Here $k(\,\cdot\,;\,\cdot\,)$ is some kernel and the $\tilde p_{j,m}$'s are discrete random probability measures. Hence, the $\psi_{i,j,m}$'s may exhibit ties. The model specification for $\tilde p_{j,m}$ will be tailored to address the following goals: (i) cluster the $J$ probability distributions based on their means; (ii) cluster the observations $X_{i,j,m}$ according to the ties induced on the $\psi_{i,j,m}$'s by the $\tilde p_{j,m}$'s for 
a given fixed $j$ and across different $j$'s. These two issues will be targeted separately: we first design a clustering scheme for the populations, through the specification of a prior on the means of the $X_{i,j,m}$'s and, then, we cluster the data using a hierarchical DP having a specific invariance structure that is ideally suited to the application at hand. 

\subsection{The prior on disease-specific locations}
\label{ss:like}
As a model for the observations we consider a nonparametric mixture of Gaussian distributions specified as
\begin{equation} 
	\label{eq:model}
	X_{i,j,m}\,|\,(\bm{\theta}_m,\bm{\xi}_m,\bm{\sigma}^2_m)
	\simind \mathcal{N}(\theta_{j,m}+\xi_{i,j,m},\sigma_{i,j,m}^2)
\end{equation}
where $\bm{\theta}_m=(\theta_{1,m},\ldots,\theta_{J,m})$, $\bm{\xi}_m=(\xi_{1,1,m},\ldots,\xi_{1,n_1,m},\xi_{2,1,m},\ldots,
\xi_{n_J,J,m})$, with a similar definition for the vector $\bm{\sigma}_m^2$, and $\mathcal{N}(\,\mu,\,\sigma^2\,)$ denotes a normal distribution with mean $\mu$ and variance $\sigma^2$. The assumption in \eqref{eq:model} clearly reflects \eqref{eq:decomp_x}. Moreover, in order to account for the two levels of clustering we are interested in, we will assume that 
\begin{equation}
	\label{eq:model_parameters}(\bm{\theta}_{1},\ldots,\bm{\theta}_{M})\sim P,\qquad
	(\xi_{i,j,m},\sigma_{i,j,m}^2)\,|\,\tilde{q}_{j,m}\simiid \tilde q_{j,m}\quad (i=1,\ldots,n_j)
\end{equation}
where $\tilde q_{1,m},\ldots,\tilde q_{J,m}$ are discrete random probability measures independent from  $(\bm{\theta}_{1},\ldots,\bm{\theta}_{M})$. Thus, the likelihood corresponds to
\begin{equation}\label{eq1}
	\prod_{m=1}^M\,\prod_{j=1}^J \,
	\prod_{i=1}^{n_j}
	\frac{1}{\sigma_{i,j,m}}\:
	\varphi\Big(\frac{x_{i,j,m}-\theta_{j,m}-\xi_{i,j,m}}{\sigma_{i,j,m}}\Big)
	\:\tilde{q}_{j,m}(\ddr\xi_{i,j,m},
	\ddr\sigma_{i,j,m})
\end{equation}
with $\varphi$ denoting the standard Gaussian density. Relevant inferences can be carried out if one is able to marginalize this expression with respect to both $(\bm{\theta}_{1},\ldots,\bm{\theta}_{M})$ and $(\tilde q_{1,m},\ldots,\tilde{q}_{J,m})$ for each 
$m=1,\ldots,M$. 

This specification allows to address the model selection problem in the following way. If $\mathcal{M}^m$ stands for the space of all partitions of the $J$ populations for the $m$-th cardiac function index, then $\mathcal{M}^m=\{M_b^m \,:\, b =1,\ldots,\mbox{card}(\mathcal{P}_J)\}$ where $\mathcal{P}_J$ is the collection of all possible partitions of $[J]=\{1,\ldots,J\}$. In our specific case, $J=4$ and card$(\mathcal{P}_J)=15$, thus we have 15 competing models per cardiac index. Each competing model corresponds to a specific partition in $\mathcal{M}^m$. In particular, the partition arises from ties between the population specific means in $\bm{\theta}_m$ and, hence, the distribution $P$ in \eqref{eq:model_parameters} needs to associate positive probabilities to ties between the parameters within the vector $\bm{\theta}_m$, for each $m=1,\ldots,M$.

Let us start considering as distribution $P$ a well-known effective clustering prior, i.e. a mixture of DPs in the spirit of \cite{antoniak1974mixtures}, namely
\begin{equation}
	\label{eq2}
	\begin{aligned}
		\theta_{j,m}\mid \tilde p_m \enskip&\simiid\enskip \tilde p_m \qquad &j=1,\ldots,J \\
		\tilde p_m\mid\omega\enskip &\simiid \enskip \mbox{DP}(\omega,G_{m})\qquad& m=1,\ldots,M \\
		\omega \enskip &\sim \enskip p_\omega&
	\end{aligned}
\end{equation}
where DP$(\omega,G_m)$ denotes the DP with concentration parameter $\omega$ and non-atomic baseline probability measure $G_m$ and $p_{\omega}$ is a probability measure on $\R ^+$. The discreteness of the DP implies the presence (with positive probability) of ties within the vector of locations $\bm{\theta}_m$ associated to a certain cardiac index $m$, as desired. The ties give rise to a random partition: as shown in \cite{antoniak1974mixtures}, the probability of observing a specific partition of the elements in $\bm{\theta}_m$ consisting of $k\le J$ distinct values with respective frequencies $n_1,\ldots,n_k$ coincides with
\begin{equation}
	\label{eq:eppf_dir}
	\Pi_k^{(J)}(n_1,\ldots,n_k)=\frac{\omega^k}{(\omega)_J}\,\prod_{i=1}^k (n_i-1)!
\end{equation}
where $(\omega)_J=\Gamma(\omega+J)/\Gamma(\omega)$. The use of a shared concentration parameter over \eqref{eq:eppf_dir} to address multiple model selection has been already successfully employed in \cite{moser2021multiple}, where they cluster parameters in a probit model.  When there is no pre-experimental information available on competing partitions, the use of \eqref{eq:eppf_dir} as a prior for model selection has some relevant benefits. Indeed, it induces borrowing of strength across diagnoses and, being $\omega$ random, it generates borrowing of information also across cardiac indexes, thus improving the Bayesian learning mechanism. These two features can also be given a frequentist interpretation in terms of desirable penalties.
As a matter of fact, the procedure penalizes for the multiplicity of the model selections that are performed. The penalty has to be meant 
in the following way: while $J$ and/or $M$ increase, the prior odds change in favor of less complex models. For more details on this, see \cite{scott2010bayes}. Summing up, the mixture of DPs automatically induces a prior distribution on $\{\mathcal{M}^m:m=1,\ldots,M\}$, that arises from \eqref{eq:eppf_dir} combined with the prior $p_\omega$ on $\omega$, and it presents desirable properties for model selection 
that can be interpreted either in terms of borrowing of information or in terms of penalties. 

However, in the analysis of hypertensive disorders, some prior information on competing models is available, and this is not yet incorporated in \eqref{eq:eppf_dir}. In fact, as already mentioned, there is a natural order of the diagnoses, which is given by the severity of the disorders, i.e.\ C, G, M, S. Partitions that do not comply with this ordering, e.g. $\{\{C, S\}\{G\},\{M\}\}$, should be excluded from the support of the prior. Thus, we consider a prior over $\mathcal{M}^m$ that associates zero probability to partitions that do not respect the natural order of the diagnoses and a probability proportional to that in \eqref{eq:eppf_dir} for the remaining partitions, i.e.\
\begin{equation}\label{eq:prior2}
	\mathbb{P}(M_b^m\mid \omega) \propto \begin{cases}
	\Pi_k^{(J)}(n_1,\ldots,n_k)&\text{if $M_b^m$ is compatible with the natural order}\\
0 & \text{otherwise}\end{cases}
\end{equation} 
This amounts to a distribution $P$ for $(\bm{\theta}_{1},\ldots,\bm{\theta}_{M})$ given by 
\begin{equation}
	\label{eqnew}
	\begin{aligned}
		(\theta_{1,m},\ldots,\theta_{J,m})\mid \omega \enskip&\simind\enskip P_{\omega, G_m} \qquad m=1,\ldots,M \\
		\omega \enskip &\sim \enskip p_\omega&
	\end{aligned}
\end{equation}
where $P_{\omega, G_m}$ is the distribution obtained sampling a partition according to \eqref{eq:prior2} and associating to each cluster a unique value sampled from $G_m$. 
Using \eqref{eqnew} as a prior for the disease-specific locations, we preserve the desirable properties of the mixture of DPs mentioned before, while incorporating prior information on the severity of the diseases.

As detailed in the next section, we further define random probability measures $\tilde q_{j,m}$ that satisfy the symmetry condition
\begin{equation}
	\label{eq:symmetry}
	\tilde q_{j,m}(A\times B ) 
	= \tilde 
	q_{j,m}((-A)\times B )\qquad\qquad  \text{a.s.}
\end{equation}
for any $A$ and $B$. This condition ensures that the parameters $\theta_{j,m}$, for $j=1,\ldots,J$ and $m=1,\ldots,M$, in \eqref{eq:model} are identified, namely $\mathbb{E}(X_{i,j,m}\mid \bmath{\theta}_{m}, \, \tilde q_{j,m} ) = \theta_{j,m}$ with probability one. This identifiability property is crucial to make inference over the location parameters $\bm{\theta}_m$'s. Similar model specifications for discrete 
exchangeable data have been proposed and studied in \cite{dalal1979nonparametric},  \cite{doss1984}, \cite{diaconis1986} and \cite*{ghosal1999}, of which \eqref{eq1} represents a generalization to 
density functions and partially exchangeable data. 

\subsection{The prior for the error terms}
\label{ss:s-HDP}
While the clustering of populations is governed by \eqref{eq:prior2}, we use a mixture of hierarchical discrete processes for the error terms. This has the advantage of modeling the clustering of the observations, both within and across different samples, once the disease-specific effects are account for. This clustering structure allows to model heterogeneity across patients in a much more realistic way with respect to standard ANOVA models based on assumption of normality. Cardiac indexes may be influenced by a number of factors that are not directly observed in the study, such as pre-existing conditions \citep{hall2011heart} and psychosocial factors \citep{pedersen2017psychosocial}. These unobserved relevant factors may be shared across patients with the same or a different diagnosis and may also result in outliers. To take into account this latent heterogeneity of the data, we introduce the hierarchical
symmetric DP that satisfies the symmetry condition in \eqref{eq:symmetry} and, moreover, allows to model heterogeneous data similarly to the 
hugely popular hierarchical DP \citep{teh2006}.

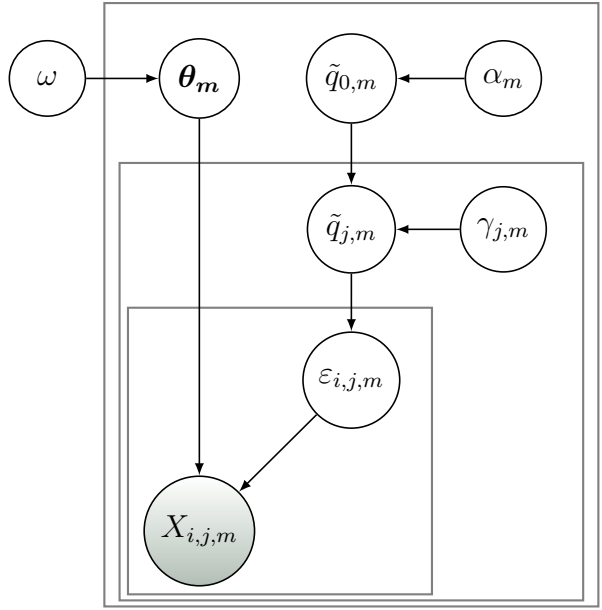
\begin{figure}
	\begin{center}
		\begin{tikzpicture}[-latex ,auto ,node distance =2 cm and 2cm ,on grid ,
			semithick ,
			state/.style ={ circle ,top color =white , bottom color = white ,
				draw,black , text=black , minimum width =1 cm},
			transition/.style = {rectangle, draw=black!50, thick, minimum width=6.1cm, minimum height = 5.8cm},
			transition3/.style = {rectangle, draw=black!50, thick, minimum width=4cm, minimum height = 3.8cm}, 
			transition2/.style = {rectangle, draw=black!50, thick, minimum width=6.5cm, minimum height = 8cm}, scale=0.8]
			\node[state] (C){$\boldsymbol{\theta_{m}}$};
			\node[draw=none] (A) [below =of C] { };
			\node[draw=none] (B) [below =of A] {};
			\node[circle ,top color =white , bottom color = white ,
			draw,black , text=black , minimum width =1 cm]  (D) [left =2 cm of C] {${\omega}$};
			\node[circle ,top color =white , bottom color = ashgrey ,
			draw,black , text=black , minimum width =1 cm] (E) [below = of B] {${X_{i,j,m}}$};
			\node[state] (F) [ right =of B] {$\ep_{i,j,m}$};
			\node[state] (G) [above =of F] {${\tilde q_{j,m}}$};
			\node[state] (Z) [above =of G] {${\tilde q_{0,m}}$};
			\node[state] (X) [right =of Z] {${\alpha_{m}}$};
			\node[state] (Y) [right =of G] {${\gamma_{j,m}}$};
			\node[transition] (H)[ below =0.02cm of F] { };
			\node[transition3] (J)[  above  right =1.5cm of E] { };
			\node[transition2] (L)[ below =1cm of G] { };
			\path (D) edge node[right] {} (C);
			\path (C) edge node[below] {} (E);
			\path (G) edge node[below] {} (F);
			\path (F) edge node[below] {} (E);
			\path (X) edge node[below] {} (Z);
			\path (Y) edge node[below] {} (G);
			\path (Z) edge node[below] {} (G);
		\end{tikzpicture}
		\caption{\small Graphical representation of the model. Each node represents a random variable and each rectangle denotes conditional i.i.d. replications of the model within the rectangle.}
		\label{fig:figure1}
	\end{center}
	\vspace{-0.5 cm}
\end{figure}

The basic building block of the proposed prior is the invariant Dirichlet 
process, which was introduced for a single population ($J=1$) in an exchangeable framework by \cite{dalal1979}. Such a modification of the DP satisfies a symmetry condition, in the sense that it is a random probability measure that is invariant with respect to a chosen group of transformations $\mathcal{G}$. A more formal definition and detailed description of the invariant DP can be found in Section A of the Supplement. For our purposes it is enough to consider the specific case of the symmetric Dirichlet process, 
which can be constructed through a symmetrization of a Dirichlet process. Consider a non-atomic probability measure $P_0$ on $\mathbb{R}$
and let $\tilde Q_0 \sim \text{DP}(\alpha,P_0)$. 
If 
\begin{equation}
	\label{eq:inv_DP}
	\tilde Q(A) = \frac{\tilde Q_0(A) + \tilde Q_0(-A)}{2}\quad \quad \forall A \in \mathcal{B}(\mathbb{R})
\end{equation}
where $-A=\{x\in \mathbb{R}: -x \in A\}$, then $\tilde Q$ is symmetric about 0 (almost surely) and termed symmetric DP, in symbols $\tilde Q\sim\text{s-DP}(\alpha\,,\,P_0)$. For convenience and without loss of generality, we assume that $P_0$ is symmetric: this implies that $P_0$ is the expected value of $\tilde Q$ making it an interpretable parameter.
The random probability measure $\tilde Q$ is the basic building block of the hierarchical process that we use to model the heterogeneity of the error terms across different populations, $j=1,\ldots,J$, in such a way that clusters identified by the unique values can be shared within and across populations. 
This prior is termed \textit{symmetric hierarchical Dirichlet process} (s-HDP) and described as
\begin{equation}
	\label{eq:s-hdp_def}
	\begin{split}
		\tilde q_{j,m}\mid\gamma_{j,m}, \, 
		\tilde 
		q_{0,m}\enskip\simind\enskip\text{s-DP}(\gamma_{j,m},\tilde q_{0,m})\\
		\tilde q_{0,m}\mid\alpha_{m}\enskip\simind\enskip\text{s-DP}(\alpha_{m},P_{0,m})
	\end{split}
\end{equation}
where $\gamma_{j,m}$ and $\alpha_{m}$ are positive parameters and $P_{0,m}$ is a non-atomic probability distribution symmetric about 0. We use the 
notation $(\tilde q_{1,m},\ldots,\tilde q_{J,m})\sim \mbox{s-HDP}(\bm{\gamma}_m,\alpha_m,P_{0,m})$, where $\bm{\gamma}_m=(\gamma_{1,m},\ldots,\gamma_{j,m})$. This definition clearly ensures the validity of \eqref{eq:symmetry}. 
A graphical model representation of the over-all proposed model is displayed in Figure~\ref{fig:figure1}. 

Still referring to the decomposition of the observations into disease-specific locations and an error term, i.e. $X_{i,j,m}=\theta_{j,m}+\ep_{i,j,m}$, it turns out that the $\ep_{i,j,m}$'s are from a symmetric hierarchical DP mixture (s-HDP mixture) with a normal kernel. Hence, the patients' clusters are identified through the $\ep_{i,j,m}$, which, according to \eqref{eq:model}, are conditionally independent from a $\mathcal{N}(\xi_{i,j,m},\sigma_{i,j,m}^2)$ given $(\xi_{i,j,m},\sigma_{i,j,m}^2)$. The choice of the specific invariant DP is aimed at ensuring that $\mathbb{E}(\ep_{i,j,m}|\tilde 
q_{j,m})=0$. 
The clusters identified by the s-HDP mixture can be interpreted as representing common unobserved factors across patients, once the disease-specific locations have been accounted for. Indeed, for any pair of patients, we may consider the decomposition $X_{i,j,m}-X_{i',j',m}=\,\Delta^{(m)}_{\theta} + \Delta^{(m)}_{\xi} + (e_{i,j,m}-e_{i',j',m})$ where $\Delta^{(m)}_{\theta}=\theta_{j,m} - \theta_{j',m}$, $\Delta^{(m)}_{\xi}=\xi_{i,j,m} - \xi_{i',j',m}$ and $e_{i,j,m}$ and $e_{i',j',m}$ are independent and normally distributed random variables with zero mean and variances $\sigma_{i,j,m}^2$ and $\sigma_{i',j',m}^2$, respectively. 

Hence, patients' clustering reflects the residual heterogeneity that is not captured by the disease-specific component $\Delta^{(m)}_{\theta}$ and are related to the subject-specific locations $\Delta^{(m)}_{\xi}$ and to the zero-mean error component $(e_{i,j,m}-e_{i',j',m})$. 
In view of this interpretation, using a s-HDP mixture over error terms offers a three-fold advantage. Firstly, the presence of clearly separated clusters of patients within and across populations will indicate the presence of unobserved relevant factors which affect the cardiac response variables. Secondly, single patients with very low probabilities of co-clustering 
with all other subjects will have to be interpreted as outliers. 
Finally, the estimated clustering structure can also be used to check whether the relative effect of a certain disease (with respect to another) is fully explained by the corresponding $\Delta^{(m)}_{\theta}$. To clarify this last point consider two diseases: if the posterior co-clustering probabilities among patients sharing the same disease are different between the two populations, this will indicate that different diagnoses not only have an influence on disease-specific locations (which is measured by $\Delta^{(m)}_{\theta}$), but they also have 
an impact on the shape of the distribution of the corresponding cardiac index. More details on this can be found in Section D of the Supplement.

\section{Marginal distributions and random partitions}
\label{s:partition}
As emphasized in the previous sections, ties among the $\theta_{j,m}$'s and the $(\xi_{i,j,m},\sigma_{i,j,m}^2)$'s are relevant for inferring the clustering structure both among the populations (hypertensive diseases) and among the individual units (patients). Indeed, for each $m$ (cardiac index) they induce a random partition that emerges as a composition of two partitions generated respectively by the prior in \eqref{eqnew} and the s-HDP. The laws of these random partitions are not only crucial to understand the clustering mechanism, but also necessary in order to derive posterior sampling schemes. In this section such a law is derived and used to compute the predictive distributions that, jointly with the likelihood, determine the full conditionals of the Gibbs sampler devised in Section~\ref{s:inference}. 
To reduce the notational burden, in this and the following section, we remove the dependence of observations and parameters on the specific response variable $m$ and denote with $\phi_{i,j}$ the pair $(\xi_{i,j},\sigma_{i,j}^2)$ and with $\bmath{\phi}$ the collection $(\phi_{1,1},\ldots,\phi_{1,n_1},\phi_{2,1},\ldots \phi_{n_J,J})$.  
 
Conditionally on $\omega$, the law of the partition in \eqref{eq:prior2} leads to the following predictive distribution for the disease-specific locations
\[
\theta_j\,|\,\omega,\theta_1,\ldots,\theta_{j-1}\:\sim\:
a_j(\omega,\theta_1,\ldots,\theta_{j-1})\,\delta_{\theta_{j-1}}+
\left[1-a_j(\omega,\theta_1,\ldots,\theta_{j-1})\right] G
\]
where 
\begin{equation}
\label{eq:a}
a_j(\omega,\theta_1,\ldots,\theta_{j-1}) = \frac{\sum_{(*_j)}\Pi^{(J)}_{k}(n_1,\ldots,n_k)}{\sum_{(\Delta_j)}\Pi^{(J)}_{k}(n_1,\ldots,n_k)}
\end{equation}
where the sum at the denominator runs over the set of partitions consistent with the one generated by
 	$(\theta_1,\ldots,\theta_{j-1})$ and the one at the numerator runs over a subset of those 
 	partitions where one further has $\theta_j=\theta_{j-1}$. For $j=4$, the predictive equals
\[
\theta_{4}\mid \omega,\theta_{1}, \theta_{2},\theta_{3} \begin{cases} 
	\frac{3}{\omega+3}\delta_{\theta_{3}} + 
	\frac{\omega}{\omega+3}G &\mbox{if }\theta_{1}=\theta_{2}=\theta_{3}\\
	\frac{2}{\omega+2}\delta_{\theta_{3}} + 
	\frac{\omega}{\omega+2}G &\mbox{if }\theta_{1}\neq\theta_{2}=\theta_{3}\\
	\frac{1}{\omega+1}\delta_{\theta_{3}} + 
	\frac{\omega}{\omega+1}G&\mbox{otherwise}
	\end{cases}
\]
Explicit expressions for the function $a$, for $j=1,2,3$, can be easily computed using \eqref{eq:a} and \eqref{eq:prior2} and are provided in Section B of the Supplement.

\begin{figure}
	\centering
	\resizebox{14cm}{!}{
		\begin{tikzpicture}[-latex, auto, node distance = 2.4 cm and 3.4cm, 
		on grid,
		semithick,
		state/.style ={circle, top color = white, bottom color = white,
			draw, black, text=black, minimum width = 0.5 cm},
		fill fraction/.style n args={2}{path picture={
				\fill[#1] (path picture bounding box.south west) rectangle
				($(path picture bounding box.north west)!#2!(path picture bounding box.north east)$);}},
		transition/.style = {rectangle, draw=black!50, thick, minimum width=17cm, minimum height = 6.7cm} ]
		\node[circle, draw, scale=0.7, fill fraction={darkgray}{0.5}] (A)
		{\Large$\boldsymbol{\phi^{**}_1}$\quad $-\boldsymbol{\phi^{**}_1}$};
		\node[circle, draw, scale=0.7, fill fraction={darkgray}{0.5}] (B) [right =of A]
		{\Large$\boldsymbol{\phi^{**}_2}$\quad $-\boldsymbol{\phi^{**}_2}$};
		\node[circle, draw, scale=0.7, fill fraction={darkgray}{0.5}] (C) [right =of B] 	{\Large$\boldsymbol{\phi^{**}_3}$\quad $-\boldsymbol{\phi^{**}_3}$};
		\node[circle, draw, scale=0.7, fill fraction={darkgray}{0.5}] (D) [right =of C] 	{\Large$\boldsymbol{\phi^{**}_1}$\quad $-\boldsymbol{\phi^{**}_1}$};
		\node[circle, fill, draw, scale=0.3] (M) [right =1.5 cm of D] { };
		\node[circle, fill, draw, scale=0.3] (N) [right =0.5 cm of M] { };
		\node[circle, fill, draw, scale=0.3] (O) [right =0.5 cm of N] { };
		\node[draw=none,fill=none] (E) [above left =1.5 cm of A] {\large$\phi_{1,1}$};
		\node[draw=none,fill=none] (F) [above left =1.55 cm of B] {\large$\phi_{2,1}$};
		\node[draw=none,fill=none] (G) [below left =0.60 cm of E] {\large$\phi_{3,1}$};
		\node[draw=none,fill=none] (H) [left =1.56 cm of A] {\large$\phi_{4,1}$};
		\node[draw=none,fill=none] (I) [above left =1.55 cm of C] {\large$\phi_{5,1}$};
		\node[draw=none,fill=none] (L) [above right=1.55 cm of B] {\large$\phi_{6,1}$};
		\node[draw=none,fill=none] (P) [below left =1.35 cm of A] {  };
		\node[draw=none,fill=none] (Q) [above right=1.50 cm of A] {\large$\phi_{7,1}$};
		\node[circle, draw, scale=0.7, fill fraction={darkgray}{0.5}] (AA) 
		[below =3.2cm of A]
		{\Large$\boldsymbol{\phi^{**}_3}$\quad $-\boldsymbol{\phi^{**}_3}$};
		\node[circle, draw, scale=0.7, fill fraction={darkgray}{0.5}] (BB) 
		[below =3.2cm of B]
		{\Large$\boldsymbol{\phi^{**}_1}$\quad $-\boldsymbol{\phi^{**}_1}$};
		\node[circle, draw, scale=0.7, fill fraction={darkgray}{0.5}] (CC) 
		[below =3.2cm of C] 	{\Large$\boldsymbol{\phi^{**}_4}$\quad $-\boldsymbol{\phi^{**}_4}$};
		\node[circle, draw, scale=0.7, fill fraction={darkgray}{0.5}] (DD) 
		[below =3.2cm of D]
		{\Large$\boldsymbol{\phi^{**}_5}$\quad $-\boldsymbol{\phi^{**}_5}$};
		\node[circle, fill, draw, scale=0.3] (MM) [right =1.5 cm of DD] { };
		\node[circle, fill, draw, scale=0.3] (NN) [right =0.5 cm of MM] { };
		\node[circle, fill, draw, scale=0.3] (OO) [right =0.5 cm of NN] { };
		\node[draw=none,fill=none] (EE) [above left =1.5 cm of AA] {\large$\phi_{4,2}$};
		\node[draw=none,fill=none] (FF) [above left =1.55 cm of BB] {\large$\phi_{2,2}$};
		\node[draw=none,fill=none] (GG) [below left =0.60 cm of EE] {\large$\phi_{3,2}$};
		\node[draw=none,fill=none] (II) [above left =1.55 cm of CC] {\large$\phi_{5,2}$};
		\node[draw=none,fill=none] (LL) [above right=1.55 cm of CC] {\large$\phi_{6,2}$};
		\node[draw=none,fill=none] (PP) [below left =1.35 cm of AA] {  };
		\node[draw=none,fill=none] (QQ) [above right=1.5 cm of AA] {\large$\phi_{7,2}$};
		\node[transition] (R)  [above  left=2.2 cm of CC]{ };
	\end{tikzpicture}
}
\caption{{Chinese restaurant franchise representation of the symmetric hierarchical DP for $J=2$ populations. Each circle represents a table.}}
\label{fig:figure2}
\end{figure}
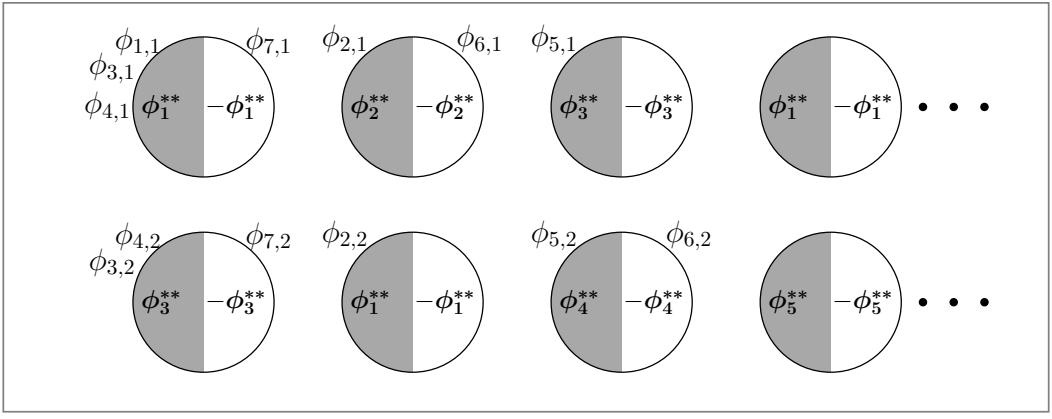

Moving to second-level partitions induced by the s-HDP, we recall that the key concept for studying random partitions on multi-sample data is the \textit{partially exchangeable partition probability function} (pEPPF). See, e.g., \cite*{lnp2014} and \citet{camerlenghi2019b}. The pEPPF returns the probability of 
a specific multi-sample partition and represents the appropriate generalization of the well-known single-sample EPPF, which in the DP case corresponds to \eqref{eq:eppf_dir}.
Discreteness of the s-HDP $(\tilde q_1,\ldots,\tilde q_m)$ in \eqref{eq:s-hdp_def} induces a partition of the elements of 
$\bmath{\phi}$ into equivalence classes identified by the distinct values. Taking into account the underlying partially exchangeable structure, such a random partition is characterized by the pEPPF
\begin{equation}
	\label{eq:peppf}
	\tilde \Pi_k^{(N)}(\bm{n}_1,\ldots,\bm{n}_J)=\mathbb{E}\left(\int_{\Phi^k}
	\prod_{j=1}^J \prod_{h=1}^k\tilde q_{j,m}^{n_{j,h}}(\ddr\phi_i)\right)
\end{equation}
where $\bm{n}_{j}=(n_{j,1},\ldots,n_{j,k})$ are non-negative integers, for any $j=1,\ldots,J$, such that $n_{j,h}$ is the number of elements in $\bm{\phi}$ corresponding to population $j$ and belonging to cluster $h$. Thus $\sum_{j=1}^J n_{j,h}\ge 1$ for any $h=1,\ldots,k$, $\sum_{h=1}^kn_{j,h}=n_j$ and $\sum_{h=1}^k \sum_{j=1}^J n_{j,h}=N$.
The determination of probability distributions of this type is challenging and only recently the first explicit instances have appeared in the literature. See e.g., \citet{lnp2014}, \citet{camerlenghi2019} and \citet{camerlenghi2019b}. With respect to the hierarchical case considered in \citet{camerlenghi2019b}, the main difference is that here we have to take into account the specific structure \eqref{eq:inv_DP} of the $\tilde q_{j,m}$. The almost sure symmetry of the process generates a natural random matching between sets in the induced partition. Therefore, instead of studying the marginal law in \eqref{eq:peppf}, we derive the joint law of the partition and of the random matching. Formally, consider a specific partition $\{A_1^{+},A_1^{-},\ldots,A_k^{+},A_k^{-}\}$ of $\bmath{\phi}$, such that, for $h=1,\ldots,k$, all the elements in $A_h^+$ belong to $\mathbb{R}^+\times\mathbb{R}^+$, all the elements in $A_h^-$ belong to $\mathbb{R}^-\times\mathbb{R}^+$ and, if $\phi_{i,j}\in A_h^+$ and $\phi_{i',j'}\in A_h^-$, then the element-wise absolute values of $\phi_{i,j}$ and $\phi_{i',j'}$ are equal. Denote with $n^+_{j,h}$ the number of elements in $A_h^{+}\cap\{\phi_{i,j},i=1,\ldots,n_j\}$ and with $n^-_{j,h}$ the number of elements in $A_h^{-}\cap\{\phi_{i,j},i=1,\ldots,n_j\}$. The probability of observing  $\{A_1^{+},A_1^{-},\ldots,A_k^{+},A_k^{-}\}$ is
\begin{equation}
	\label{eq:peppfsym}
	\ddtilde{\Pi}_{k}^{(N)}(\bmath{n_1}^+,\bmath{n_1}^-,\ldots,\bmath{n_J}^+,\bmath{n_J}^-) = \mathbb{E}\left(\int_{\Phi^k}\prod_{j=1}^J\prod_{h=1}^k{\tilde q_{j,m}}^ {n^{+}_{j,h}+n^{-}_{j,h}}(d\phi)\right)
\end{equation} 
with $\bmath{n_j^+}=(n_{j,1}^+,\ldots,n_{j,k}^+)$. As for the determination of \eqref{eq:peppfsym}, a more intuitive understanding may be gained if one considers its corresponding Chinese restaurant franchise (CRF) metaphor, which displays a variation of both the standard Chinese restaurant franchise of \cite{teh2006} and the skewed Chinese restaurant process of \cite*{iglesias2009nonparametric}. Figure~\ref{fig:figure2} provides a graphical representation. 
The scheme is as follows: there are $J$ restaurants
sharing the same menu and the customers are identified by their choice of 
$\phi_{i,j}$ but, unlike in the usual CRF, at each table two \textit{symmetric dishes} are served. Denote with $\phi^*_{t,j}=(\xi^*_{t,j},\, \sigma_{t,j}^{2*})$ and $-\phi^*_{t,j}=(-\xi^*_{t,j},\, \sigma_{t,j}^{2*})$ the two dishes served at table $t$ in restaurant $j$, with 
$\phi^{**}_{h}=(\xi^{**}_{h},\, \sigma^{**2}_{h})$ and $-\phi^{**}_{h}=(-\xi_{h},\, \sigma_{h}^{**2})$ the $h$-th pair of dishes in the menu and with $n_{j,h}^{+}$ and $n_{j,h}^{-}$ the number of customers in restaurant $j$ eating dish $\phi^{**}_{h}$ and $-\phi^{**}_{h}$, respectively. This means that two options are available to a customer entering restaurant 
$j$: she/he will either sit at an already occupied table, with probability proportional to the number of customers at that table or will sit at a new table with probability proportional to the concentration parameter $\gamma_j$. In the former case, the customer will choose the dish $\phi^*_{t,j}$ with probability $1/2$ and $-\phi^*_{t,j}$ otherwise. In the latter 
case, the customer will eat a dish served at another table of the franchise with probability proportional to half the number of tables that serve that dish, or will make a new order with probability proportional to the concentration parameter $\alpha$. In view of this scheme, the probability in \eqref{eq:peppfsym} turns out to be
\[
\ddtilde{\Pi}_{k}^{(N)}(\bmath{n_1^+},\ldots,\bmath{n_J^-}) = 2 ^ {-N}\bar\Pi_{k}^{(N)}(\bmath{n_1^+}+\bmath{n_1^-},\ldots,\bmath{n_J^+}+\bmath{n_J^-}) 
\]
and $\bar{\Pi}_k^{(N)}$ on the right-hand-side is the pEPPF of the hierarchical DP derived in \citet{camerlenghi2019b}, namely
\[
\bar{\Pi}_k^{(N)}(\bm{n}_1,\ldots,\bm{n}_k)=\Biggl(\prod_{j=1}^J \frac{\prod_{i=1}^k (\gamma_j)_{n_{j,h}}}{(\gamma_j)_{n_j}}\Biggr)\:
\sum\limits_{\bmath{\ell}}\frac{\alpha^k}{(\alpha)_{|\bmath{\ell}|}}\:
\prod_{h=1}^k (\ell_{\bullet,h} -1)!\prod_{j=1}^J P(K_{n_{j,h}}=\ell_{j,h})
\] 
where each sums runs over all $\ell_{j,h}$ in $\{1,\ldots,n_{j,h}\}$, if $n_{j,h}\ge 1$, and equals $1$ if $n_{j,h}=0$, whereas $\ell_{\bullet,h}=\sum_{j=1}^J \ell_{j,h}$ and $|\bmath{\ell}| = \sum_{j=1}^J \sum_{h=1}^k \ell_{j,h}$. Note that the latent variable $\ell_{j,h}$ is the number of tables in restaurant $j$ serving the $h$-th pair of dishes. Moreover, $K_{n_{j,h}}$ is a random variable denoting the number of distinct clusters, out of $n_{j,h}$ observations generated by a DP with parameter $\gamma_j$ and diffuse baseline $P_0$ and it is well-known that  
\[
\mathbb{P}(K_{n_{j,h}}=\ell_{j,h})=\frac{\gamma_j^{\ell_{j,h}}}
{(\gamma_j)_{n_{j,h}}}\:
|\mathfrak{s}(n_{j,h},\ell_{j,h})|
\] 
where $|\mathfrak{s}(n_{j,h},\ell_{j,h})|$ is the signless Stilring number of the first kind. 
In view of this, one can deduce the predictive distribution
\begin{equation*}
\begin{split}
	\mathbb{P}(\phi_{n_j+1,j}\in \cdot&\,|\, \bmath{\phi}) =
	\frac{\gamma_j}{i-1+\gamma_j}
	\sum\limits_{\bmath{\ell}}
	\frac{\alpha}{|\bmath{\ell}| + \alpha}\pi(\bmath{\ell}\,|\,\bmath{\phi})P_0(\cdot)\\
	&+\sum\limits_{h=1}^{k}
	\left[\frac{n_{j,h}^+ + n_{j,h}^-}{n_j+\gamma_j} + 
	\frac{\gamma_j}{n_j+\gamma_j}
	\sum\limits_{\bmath{\ell}}
	\frac{\ell_{\bullet,h}}{|\bmath{\ell}| + \alpha}\pi(\bmath{\ell}\,|\,\bmath{\phi})\right]
	\Bigg(\frac{\delta_{\phi^{**}_{h}}(\cdot)+\delta_{-\phi^{**}_{h}}(\cdot)}{2}	\Bigg)
\end{split}
\end{equation*}
where 
\begin{equation*}
\pi(\bmath{\ell}\,|\,\bmath{\phi})\propto
\frac{\alpha^k}{(\alpha)_{|\bmath{\ell}|}}
\prod_{h=1}^k (\ell_{\bullet,h} -1)!\prod_{j=1}^J
\frac{\gamma_j^{\ell_{j,h}}}
{(\gamma_j)_{n_{j,h}^+ + n_{j,h}^-}}\:
|\mathfrak{s}(n_{j,h}^+ + n_{j,h}^-,\ell_{j,h})|\mathds{1}_{\{1,\ldots,n_{j,h}^+ + n_{j,h}^-\}}(\ell_{j,h})
\end{equation*}
is the posterior distribution of the latent variables $\ell_{j,h}$'s, and $\mathds{1}_A$ is the indicator function of set $A$.

\section{Posterior inference}
\label{s:inference}
The findings of the previous section are the key ingredients to perform posterior inference with a marginal Gibbs sampler.
The output of the sampler is structured into three levels: the first produces posterior probabilities on partitions of disease-specific locations; the second generates density estimates; the third provides clusters of patients. 
For notational simplicity, we omit
the dependence on $m$, except for the description of the sampling step that generates $\omega$.
Recall that ${\bmath{\theta}}=(\theta_{1},\ldots,\theta_{J}) $ and 
$\bmath{\phi}=\{(\phi_{1,j},\ldots,\phi_{n_j,j}):\,j=1,\ldots,J\}$, with $\phi_{i,j}=(\xi_{i,j},\sigma_{i,j}^2)$. The target distribution of 
the sampler is the joint distribution of $\bm{\theta}$, $\bm{\phi}$ and $\omega$ conditionally on the observed data $\bm{X}$.

\textbf{Sampling $\boldsymbol{\phi}$.} In view of the CRF representation of the 
s-HDP, $t_{i,j}$ stands for the label of the table where the $i$-th customer in restaurant $j$ sits 
and $h_{t,j}$ for the dish label served at table $t$ in restaurant $j$ and with $\bmath{t}$ and $\bmath{h}$ we denote the corresponding arrays. 
Moreover, define the assignment variable
$s_{i,j}=\mathds{1}(\phi_{i,j}=\phi^*_{t_{i,j},j})-
\mathds{1}(\phi_{i,j}=-\phi^*_{t_{i,j},j})$ and $\bmath{s}$ is the corresponding arrays. In order to generate $\bmath{\phi}$, we need to sample 
\begin{itemize}
\item[(i)]  $(t_{i,j}, s_{i,j})$ for $i=1,\ldots,n_j$ and $j=1,\ldots,J$;
\item[(ii)] $h_{t,j}$ for $t \in \bmath{t}$ and $j=1,\ldots,J$;
\item[(ii)] $\phi^{**}_{h}$ for $h \in \bmath{h}$. 
\end{itemize}

Note that, using the latent allocation indicators in $\bmath{t}$ and $\bmath{h}$, the sampling scheme is more efficient than sampling directly from the full conditional of each $\phi_{i,j}$, since the algorithm can update more than one parameter simultaneously \citep{neal2000}. Define $\ep_{i,j} = X_{i,j} - \theta_{j}$ and denote with $h(\ep_{i,j,m}|\phi^*)$ the conditional normal density of $\ep_{i,j}$
given $\phi^*=(\xi^*,\sigma^{2*})$, while the marginal density
is
\[
\bar h (\ep_{i,j}) = \int h(\ep_{i,j}|\phi) P_{0}(d\phi)
\]

To sample $(t_{i,j}, s_{i,j})$ from their joint full conditional, we first sample $t_{i,j}$ from
\[
P(t_{i,j}=t\mid \bmath{t}^{-(i,j)}, \bmath{h}^{-(i,j)}, \bmath{\phi}^{*-({i,j})},\bmath{\phi}^{**-({i,j})},\ep_{i,j}) \propto
\begin{cases} n_{t,j}^{-(i,j)}\,p_{\mbox{\footnotesize old}}(\ep_{i,j}|\phi^*_{t,j})
& \mbox{if } t\in \bmath{t}^{-(i,j)}
\\ 
\gamma_{j}\,p_{\mbox{\footnotesize new}}(\ep_{i,j}|{\bmath{\phi}^{**-({i,j})}})
& \mbox{if } t=t^{\mbox{new}}\end{cases}
\]
where $\bmath{t}^{-(i,j)}$, $\bmath{h}^{-(i,j)}$ $\bmath{\phi}^{*-({i,j})}$,$\bmath{\phi}^{**-({i,j})}$ coincide with the arrays $\bmath{t}$, $\bmath{h}$ $\bmath{\phi}^*$,$\bmath{\phi}^{**}$ after having removed the entries corresponding to the $i$-th customer in restaurant $j$. Moreover
\[
p_{\mbox{\footnotesize old}}(\ep_{i,j} | \phi^*_{t,j}) 
=
\frac{1}{2}h(\ep_{i,j}|\phi^*_{t,j})  + \frac{1}{2}h(\ep_{i,j}|-\phi^*_{t,j})
\]
and
\[
p_{\mbox{\footnotesize new}}(\ep_{i,j}|{\bmath{\phi}^{**-({i,j})}}) 
= \sum\limits_{h=1}^{k^{-(i,j)}}\frac{\ell_{\bullet,h}}{|\bmath{\ell}|+ \alpha}\left\{\frac{1}{2}h(\ep_{i,j}|\phi^{**}_h)+\frac{1}{2}h(\ep_{i,j}|-\phi^{**}_h)	\right\}+\frac{\alpha}{|\bmath{\ell}|+ \alpha} \bar h (\ep_{i,j})
\]	
Then we sample $s_{i,j}$ from its full conditional
\[
p(s_{i,j}=s\mid\bmath{\phi}^*, t_{i,j},\epsilon_{i,j}) \propto \begin{cases} h(\ep_{i,j}|\phi^*_{t_{i,j}}) & \mbox{if } s=1 \\ h(\ep_{i,j}|-\phi^*_{t_{i,j}}) & \mbox{if } s=-1\end{cases}
\]
The conditional distribution of $h_{t,j}$ is
\[
p(h_{t,j}=h\mid \bmath{t},\bmath{h}^{-(t,j)},\bmath{\phi}^{**-(t,j)},\bmath{s},\bmath{\ep}) 
\propto
\begin{cases} \ell_{\bullet,h}^{-(t,j)}\prod\limits_{\{(i,j):\:t_{i,j}=t\}}
h(s_{i,j}\,\ep_{i,j}|\phi_{h})
& \mbox{if } h \in \bmath{h}^{-(t,j)} \\ 
\alpha\,\displaystyle\int \prod\limits_{\{(i,j):\:t_{i,j}=t\}} h(s_{i,j}\,\ep_{i,j}|\phi) P_{0}(d\phi) & \mbox{if } h=h^{new}\end{cases}
\]

Finally, when $P_{0}$ is conjugate with respect to the Gaussian kernel, the full conditional distribution of $\phi_h^{**}$ is obtained in closed form as posterior distribution of a Gaussian model, using as observations the collection $\{\,(s_{i,j}\,\epsilon_{i,j}):\: h	_{t_{i,j},j} = h\}$. 

\textbf{Sampling $\boldsymbol{\theta}$.} When sampling the disease-specific location parameters, one can rely on 
a Chinese restaurant process restricted to those partitions that are consistent with the ordering of the diseases.
Thus, in order to generate $\bmath{\theta}$, we first sample the labels $\bmath{t}_{\theta}=\{t_{1},\ldots,t_{J}\}$, where $t_{j}$ is the label of the table where the $j$-th customer sits. Then, we sample the dish $\theta^*_{t}$ associated to table $t$ for all $t\in\bmath{t}_{\theta}$. If $z_{i,j} = X_{i,j} - \xi_{i,j}$, the conditional density of $\bm{z}_{j}=(z_{1,j},\ldots,z_{n_j,j})$ associated to the location parameter $\theta^*$,
given $\bm{\sigma}_{j}=(\sigma_{1,j},\ldots,\sigma_{n_j,j})$, is
\[
f_{\theta^*}(\bm{z}_{j}|\bm{\sigma}_{j})=\frac{1}{\sqrt{2\pi}\prod\limits_{i=1}^{n_j} \sigma_{i,j}}\:
\exp\left\{-\frac{1}{2}\sum\limits_{i=1}^{n_j}\frac{(z_{i,j}-\theta^*)^2}{\sigma_{i,j}^2}
\right\}
\]
Under the prior in \eqref{eqnew}, the full conditional distribution of $\bmath{t}_{\theta}$ is provided by
\[
\begin{split}
	p(t_{j}=t\mid t_1,\ldots,&t_{j-1},\theta_{j-1},\bm{z}_{j}, \bm{\sigma}_{j})\\
	&\propto
	\begin{cases} 
		a(\omega,\theta_1,\ldots,\theta_{j-1})\,f_{\theta_{j-1}}(\bm{z}_{j}|\bm{\sigma}_{j}) 
		& \mbox{if } t=t_j \\[4pt] 
		[1 - a(\omega,\theta_1,\ldots,\theta_{j-1})] \, 
		\displaystyle\int f_{\theta}(\bm{z}_{j}|\bm{\sigma}_{j})\, G (d\theta)& \mbox{if } t=t^{\mbox{\footnotesize new}}\\[4pt]
		0&\mbox{otherwise}
	\end{cases}
\end{split}
\]
Finally, when $G$ is conjugate with respect to the Gaussian kernel, the full conditional distribution of $\theta_{t}^{*}$, given $\{\bm{z}_{j}:\: t_{j}=t\}$, is obtained in closed form using conjugacy of the Normal-Normal model.

\textbf{Sampling the concentration parameter.} Finally, 
the concentration parameter $\omega$ can be sampled through an importance sampling step using as importance distribution the prior $p_{\omega}$ over $\omega$. Denoting with $M_m$ the selected partition for $\bm{\theta}_{m}$ and with $T_m$ the number of clusters in $M_m$, we have 
\[
p(\omega\,|\,M_m:m=1,\ldots,M)\propto p_{\omega} (\omega)  \,\frac{\omega^ {\sum_{m=1}^M T_m - M}}{(\omega + 2)^{M} (\omega^2 + \omega + 3)^{M}}.
\]

\section{Results}
\label{s:results}
\subsection{Simulation studies}
\label{ss:sim}
We perform a series of simulation studies with two main goals. First, we aim to highlight the drawbacks of clustering based on the entire distribution with respect to our proposal in the context of small sample sizes. Second, we check the model's ability of detecting the presence of underlying relevant factors in the sense described in Section~\ref{ss:s-HDP}. 

To accomplish the first goal, we compare the results obtained using our model with the nested Dirichlet process (NDP) \citep{rodriguez2008}, arguably the most popular Bayesian model to cluster populations. Mimicking the real hypertensive dataset, we simulate data for 4 samples, ideally corresponding to four diseases, with respective sample sizes 
of 50, 19, 9 and 22, which correspond to the sample sizes of the real data investigated in Section \ref{ss:real}. Since the NDP does not allow to treat jointly multiple response variables, we consider only one response variable to ensure a fair comparison. The observations are sampled from the following distributions and 100 simulation studies are performed.
\begin{equation*}
\begin{split}
	X_{i,1}\,\overset{iid}{\sim}0.5\,\mathcal{N}(\,0,\,0.5\,)+0.5\,\mathcal{N}(\,2,\,0.5\,) \qquad&\text{for}\enskip i=1,\ldots,n_1\\
	X_{i,2}\,\overset{iid}{\sim}0.5\,\mathcal{N}(\,2,\,0.5\,)+0.5\,\mathcal{N}(\,4,\,0.5\,)\qquad&\text{for}\enskip i=1,\ldots,n_2\\
	X_{i,3}\,\overset{iid}{\sim}0.5\,\mathcal{N}(\,4,\,0.5\,)+0.5\,\mathcal{N}(\,6,\,0.5\,) \qquad&\text{for}\enskip i=1,\ldots,n_3\\
	X_{i,4}\,\overset{iid}{\sim}0.5\,\mathcal{N}(\,6,\,0.5\,)+0.5\,\mathcal{N}(\,8,\,0.5\,) \qquad&\text{for}\enskip i=1,\ldots,n_4\\
\end{split}
\end{equation*}
Note that the true data generating process corresponds to samples from distinct distributions with pairwise sharing of a mixture component. Alternative scenarios are considered in the additional simulation studies that can be found in Section D of the Supplement.

The implementation of the NDP was carried out through the marginal sampling scheme proposed in \cite{zuanetti2018clustering}, which is suitably extended 
to accommodate hyperpriors on the concentration parameters of the NDP. To simplify the choice of the hyperparameters, as suggested by \citet[p.~535 and p.~551--554]{gelman2013bayesian} we estimate both models over standardized data. For our model, we set $G_m=\mathcal{N}(0,\,1)$ and $P_{0,m}=\text{NIG}(\mu=0, \,\tau=1,\, \alpha=2,\,\beta=4)$. Here, $\text{NIG}(\mu, \,\tau,\, \alpha,\,\beta)$ indicates a normal inverse gamma distribution. The base distribution for the NDP is $\text{NIG}(\mu=0, \,\tau=0.01,\, \alpha=3,\,\beta=3)$, as in \cite{rodriguez2008}. Finally, we use gamma priors with shape 3 and rate 3 for all concentration parameters, which is a common choice.
For each simulation study, we perform 10,000 iterations of the MCMC algorithms with the first 5,000 used as burn-in.

\begin{table}[t]
		\caption{Simulation studies summaries.}
	\label{tab:table1}
\begin{center}
		\begin{tabular}{lcccccc}
			&\multicolumn{3}{c}{\textbf{sHDP}}& \multicolumn{3}{c}{\textbf{NDP}}\\\hline\hline
			&MAP&Average&Median&MAP&Average&Median\\
			Partitions& count& post. prob.&post. prob.& count& post. prob.& post. prob.\\\hline
			\{\textcolor{aoe}{1},\textcolor{burntorange}{2},\textcolor{bostonuniversityred}{3},\textcolor{burgundy}{4}\}&0&0.000&0.000&0&0.000&0.000\\\hline
			\{\textcolor{aoe}{1}\}\{\textcolor{burntorange}{2},\textcolor{bostonuniversityred}{3},\textcolor{burgundy}{4}\}&0&0.000&0.000&2&0.020&0.000\\\hline
			\{\textcolor{aoe}{1},\textcolor{burntorange}{2}\}\{\textcolor{bostonuniversityred}{3},\textcolor{burgundy}{4}\}&0&0.000&0.000&\textbf{72}&\textbf{0.695}&\textbf{0.860}\\\hline
				\rowcolor{shadecolor}\{\textcolor{aoe}{1},\textcolor{bostonuniversityred}{3},\textcolor{burgundy}{4}\}\{\textcolor{burntorange}{2}\}&0&0.000&0.000&0&0.000&0.000\\\hline
			\{\textcolor{aoe}{1}\}\{\textcolor{burntorange}{2}\}\{\textcolor{bostonuniversityred}{3},\textcolor{burgundy}{4}\}&0&0.027&0.007&3&0.035&0.000\\\hline
			\{\textcolor{aoe}{1},\textcolor{burntorange}{2},\textcolor{bostonuniversityred}{3}\}\{\textcolor{burgundy}{4}\}&0&0.000&0.000&5&0.061&0.000\\\hline
				\rowcolor{shadecolor}\{\textcolor{aoe}{1},\textcolor{burgundy}{4}\}\{\textcolor{burntorange}{2},\textcolor{bostonuniversityred}{3}\}&0&0.000&0.000&0&0.000&0.000\\\hline
			\{\textcolor{aoe}{1}\}\{\textcolor{burntorange}{2},\textcolor{bostonuniversityred}{3}\}\{\textcolor{burgundy}{4}\}&1&0.054&0.015&0&0.014&0.000\\\hline
				\rowcolor{shadecolor}\{\textcolor{aoe}{1},\textcolor{bostonuniversityred}{3}\}\{\textcolor{burntorange}{2},\textcolor{burgundy}{4}\}&0&0.000&0.000&0&0.000&0.000\\\hline
				\rowcolor{shadecolor}\{\textcolor{aoe}{1},\textcolor{burntorange}{2},\textcolor{burgundy}{4}\}\{\textcolor{bostonuniversityred}{3}\}&0&0.000&0.000&0&0.000&0.000\\\hline
				\rowcolor{shadecolor}\{\textcolor{aoe}{1}\}\{\textcolor{burntorange}{2},\textcolor{burgundy}{4}\}\{\textcolor{bostonuniversityred}{3}\}&0&0.000&0.000&0&0.000&0.000\\\hline
			\{\textcolor{aoe}{1},\textcolor{burntorange}{2}\}\{\textcolor{bostonuniversityred}{3}\}\{\textcolor{burgundy}{4}\}&0&0.004&0.000&18&0.175&0.032\\\hline
				\rowcolor{shadecolor}\{\textcolor{aoe}{1},\textcolor{bostonuniversityred}{3}\}\{\textcolor{burntorange}{2}\}\{\textcolor{burgundy}{4}\}&0&0.000&0.000&0&0.000&0.000\\\hline
				\rowcolor{shadecolor}\{\textcolor{aoe}{1},\textcolor{burgundy}{4}\}\{\textcolor{burntorange}{2}\}\{\textcolor{bostonuniversityred}{3}\}&0&0.000&0.000&0&0.000&0.000\\\hline
			\{\textcolor{aoe}{1}\}\{\textcolor{burntorange}{2}\}\{\textcolor{bostonuniversityred}{3}\}\{\textcolor{burgundy}{4}\}&\textbf{99}&\textbf{0.915}&\textbf{0.954}&0&0.000&0.000
	\end{tabular}
\end{center}
\vspace{-\baselineskip}
\end{table}

Table~\ref{tab:table1} displays summaries of the results on population clustering, darker rows correspond to partitions that are not consistent with the natural ordering of the diseases. The true clustering structure is given by the finest partition. 
As already observed in \cite{rodriguez2008}, the NDP tends to identify fewer, rather than more clusters, due to the presence of small sample sizes. Using the \textit{maximum a posteriori} estimate, our model correctly 
identifies the partition in 99 out of 100 simulation studies and a partition with three elements or more in 100 out of 100 simulation studies. The 
same counts for the NDP are, respectively, 0 out of 100 and 21 out of 100. Analogous conclusions can be drawn looking at posterior probability averages and medians across the 100 simulation studies (see Table~\ref{tab:table1}) leaving 
no doubt about the model to be preferred under this scenario. 

\begin{figure}
	\centering
	\begin{subfigure}{7cm}
		\centering
		\includegraphics[ width=7cm]{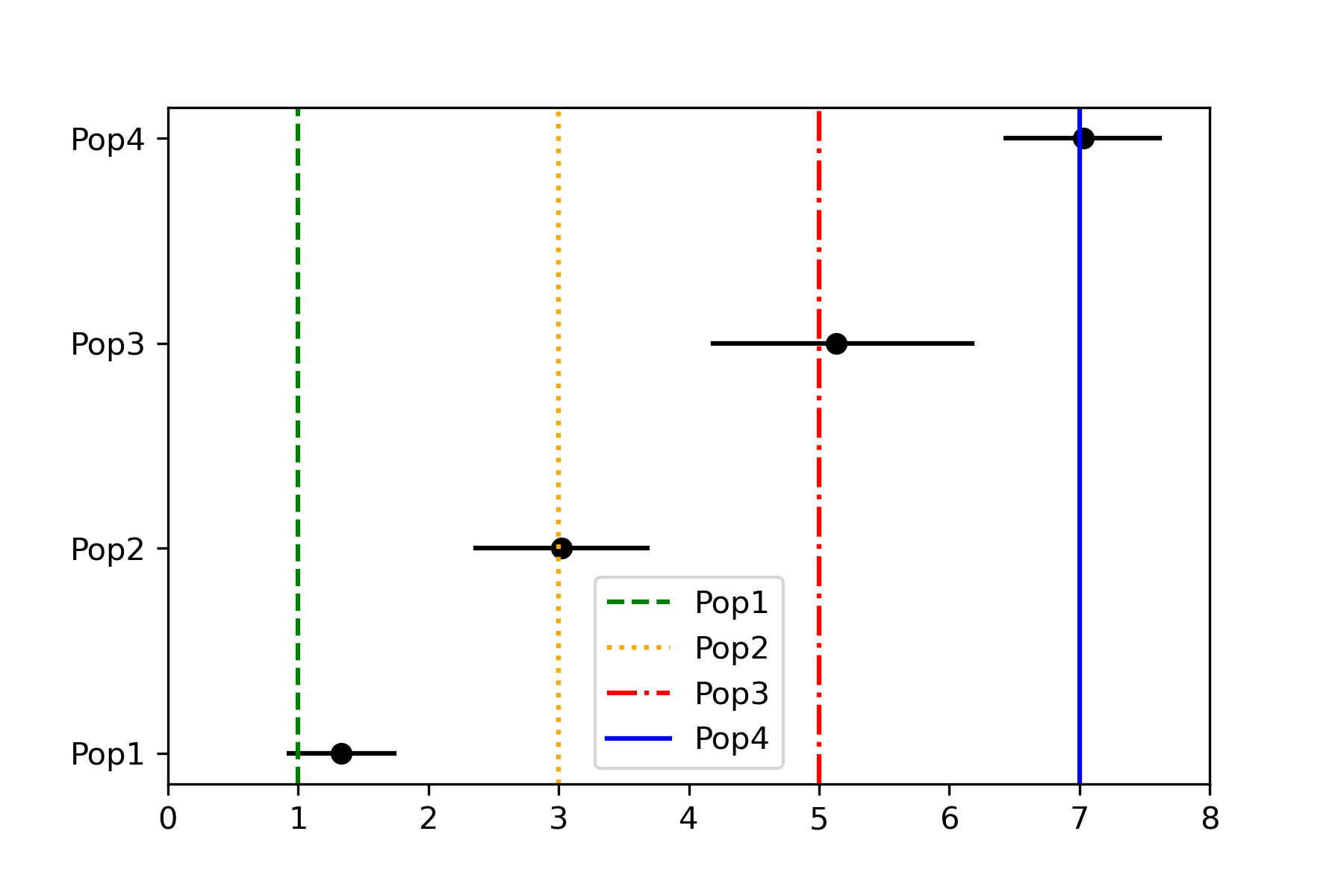}
		\caption{95\% credible intervals for population-specific locations}
		\label{fig:fig3a}
	\end{subfigure}\hspace{0.2cm}%
	\begin{subfigure}{7cm}
		\centering
		\includegraphics[width=7cm]{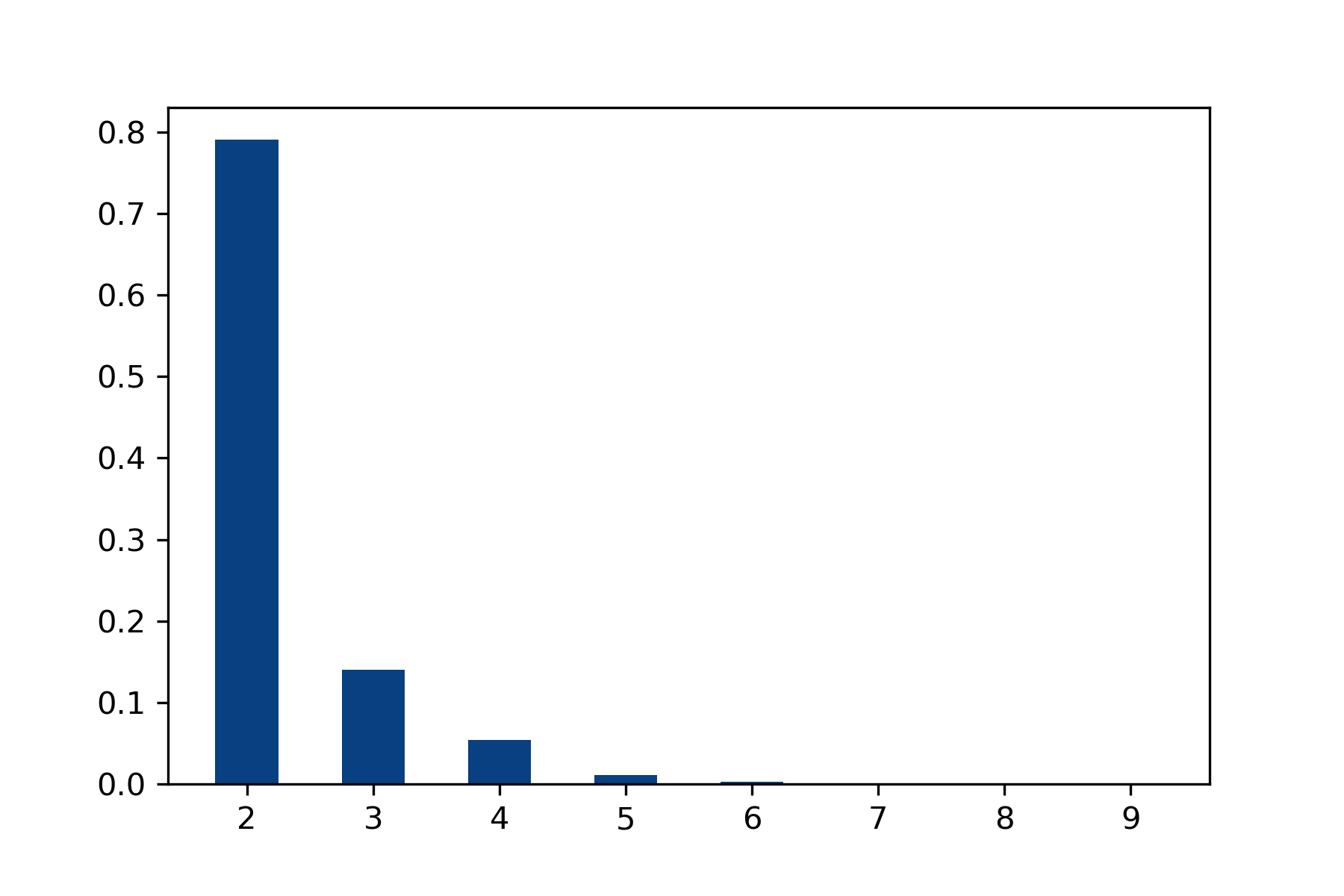}
		\caption{Number of second-level clusters.}
		\label{fig:fig3b}
	\end{subfigure}
	\begin{subfigure}{7cm}
		\centering
		\includegraphics[ width=7cm, trim={2cm 0 2cm 0}]{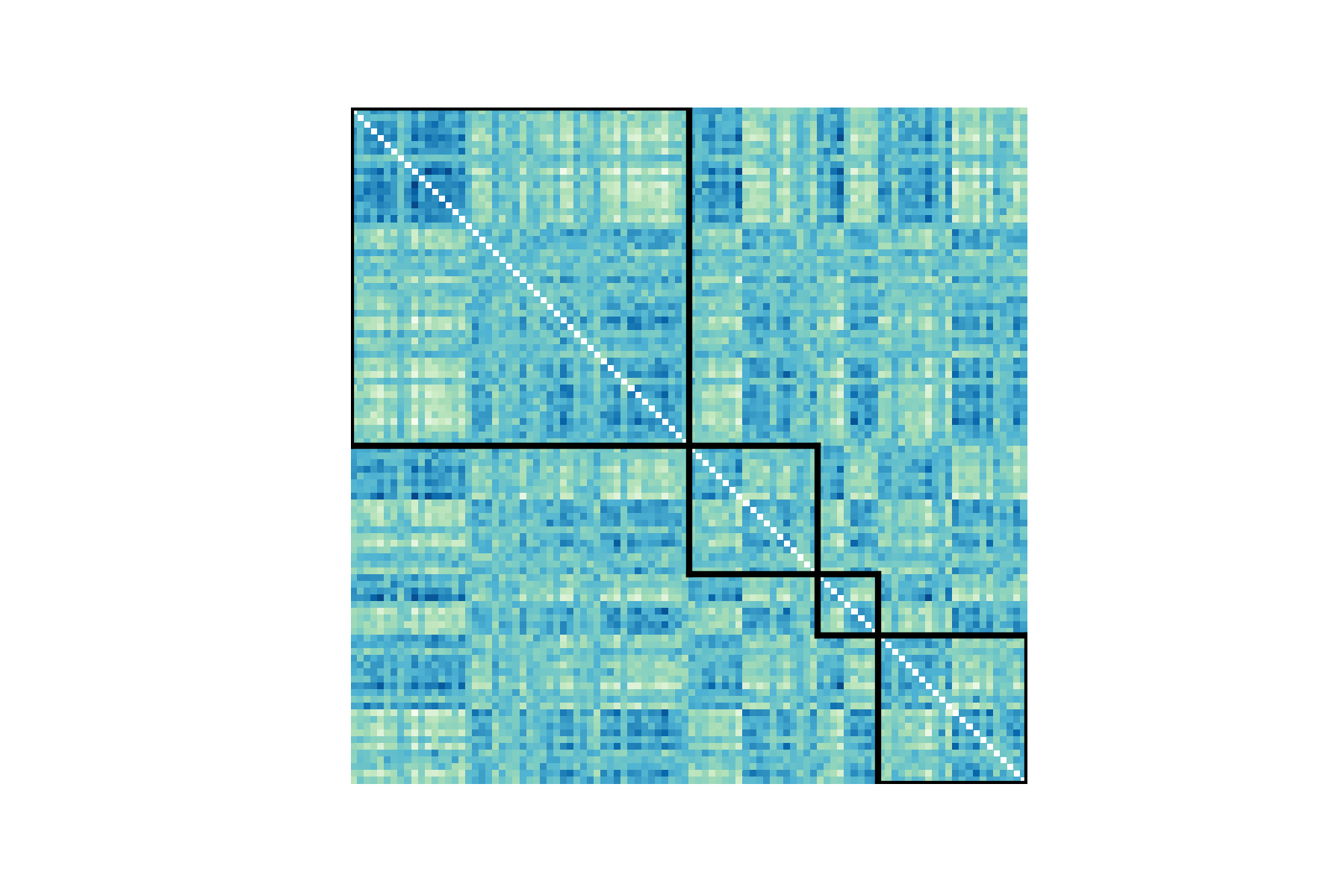}
		\caption{Co-clustering.}
		\label{fig:fig3c}
	\end{subfigure}\hspace{0.2cm}%
	\begin{subfigure}{7cm}
		\centering
		\includegraphics[width=7cm,, trim={2cm 0 2cm 0}]{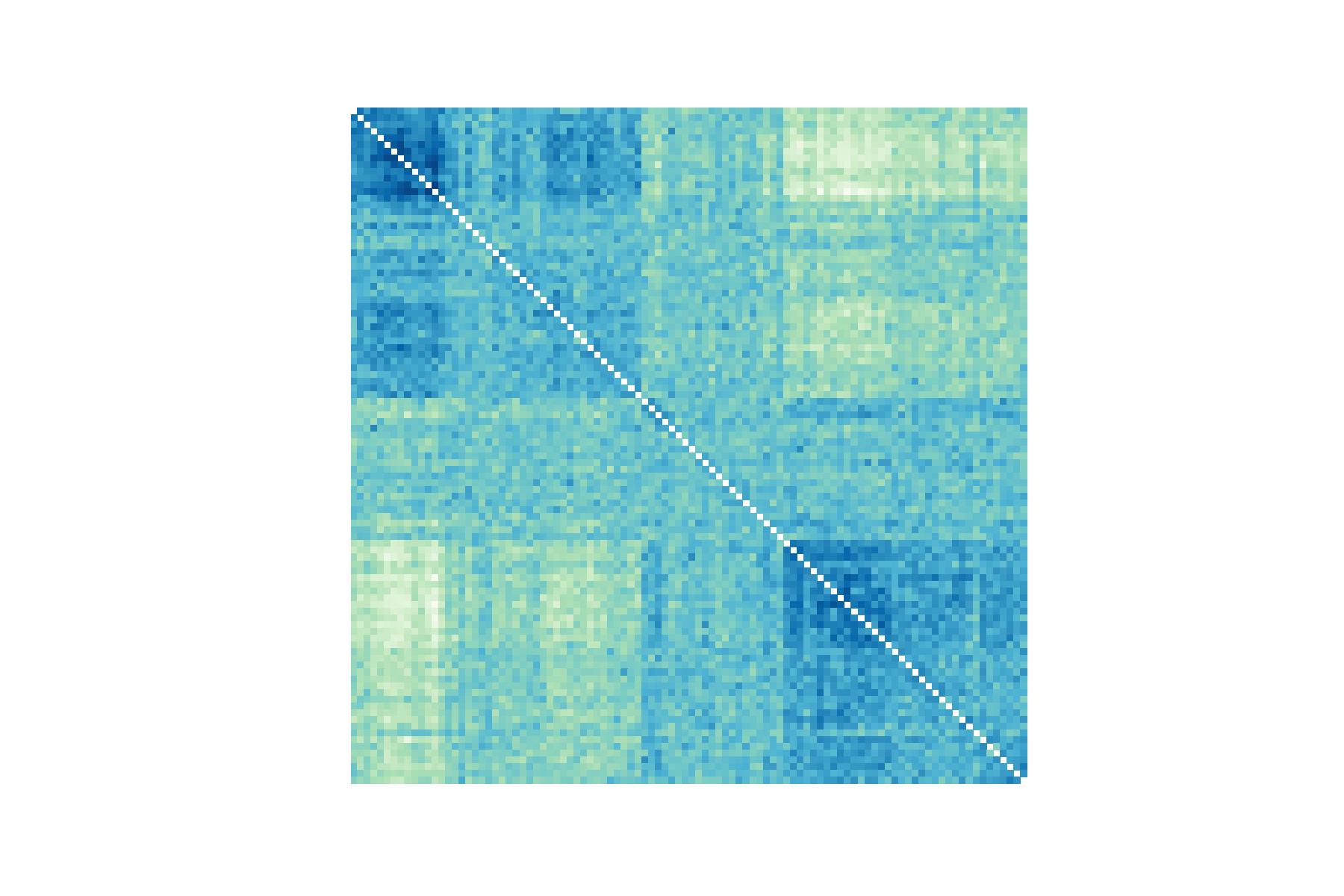}
		\caption{Co-clustering.}
		\label{fig:fig3d}
	\end{subfigure}
\caption{ Panel (a): Mean point estimates and 95\% credible intervals for the four populations, vertical lines correspond to true values. Panel (b): Posterior distribution on the number of second-level clusters. Panels (c) and (d): heatmaps of second-level clustering, darker colors correspond to higher probability of co-clustering; in (c) patients are ordered based on the diagnosis and the four black squares highlight the within-sample probabilities and in (d) patients are reordered based on co-clustering probabilities.}
\end{figure}

Finally, we randomly select three simulation studies among the 100 to better understand the performance in estimating the other model parameters. Here we comment on one of the studies, the other two leading to similar results are reported in Section D.1.1 of the Supplement. 
Figure~\ref{fig:fig3a} shows point estimates and credible intervals for the population-specific location parameters $\theta_1,\theta_2,\theta_3,\theta_4$. The true means belong to the 95\% credible intervals.

Moreover, it turns out that the model is able to detect the presence of two clusters of subjects leading to a posterior distribution for the number of clusters that is 
rather concentrated on the true value, see Figure~\ref{fig:fig3b}--\ref{fig:fig3d}. Moreover, the point estimate for the subject partition, obtained minimizing the Binder loss function, also contains two clusters, proving the ability of the model to detect the underlying relevant factor. In Section D of the Supplement, a number of additional simulation studies are conducted, both using alternative specifications over the disorder-specific parameters and different data generating mechanisms: the results highlight a good performance of the model, which appears also able to detect outliers, to highlight non-location effects of the disorders and to produce reliable outputs even under deviation from symmetry.

\subsection{Impact of hypertensive disorders on maternal cardiac dysfunction}
\label{ss:real}
Our analysis is based on the dataset of \cite{data}, which can be obtained from https://data.mendeley.com/datasets/d72zr4xggx/1. The dataset contains observations for $10$ cardiac function measurements collected through a prospective case-control study on women in the third semester of pregnancy divided in $n_1=50$ control cases (C), $n_2=19$ patients with gestational hypertension (G), $n_3=9$ patients with mild preeclampsia (M) and $n_4=22$ patients with severe preeclampsia (S). The cases are 
women admitted from 2012 to 2014 to King George Hospital in Visakhapatnam, India. The healthy sample is composed by normotensive pregnant women. All women with hypertension were on antihypertensive treatment with oral Labetalol or Nifedipine. Women with severe hypertension were treated with either oral nifedipine and parenteral labetalol or a combination. For more details on the dataset, we refer to \cite{tatapudi2017}. The prior specification is the same as in the previous section. Sections E.2 and E.3 of the Supplement contain a prior-sensitivity analysis and show rather robust results w.r.t. different prior specifications. Inference is based on 10,000 MCMC iterations with the 
first half used as burn-in. 

\begin{table}
\caption{Posterior probabilities over partitions of means. Maximum a posteriori probabilities are in \textbf{bold}.}
\label{tab:table2}
\begin{center}
\resizebox{\textwidth}{!}{
		\begin{tabular}{lcccccccccc}
			partitions&CI&CWI&LVMI&IVST&LVPW&EF&FS&EW&AW&E/A\\\hline\hline
			\{\textcolor{aoe}{C},\textcolor{burntorange}{G},\textcolor{bostonuniversityred}{M},\textcolor{burgundy}{S}\}
			&0.021&0.000&0.000&0.000&0.000&\textbf{0.365}&\textbf{0.303}&0.096&0.000&0.000\\
			\{\textcolor{aoe}{C}\}\{\textcolor{burntorange}{G},\textcolor{bostonuniversityred}{M},\textcolor{burgundy}{S}\}
			&0.002&\textbf{0.546}&0.001&0.083&0.016&0.078&0.190&0.021&0.036&0.000\\
			\{\textcolor{aoe}{C},\textcolor{burntorange}{G}\}\{\textcolor{bostonuniversityred}{M},\textcolor{burgundy}{S}\}
			&0.002&0.000&0.001&0.000&0.000&0.037&0.038&0.072&0.076&0.049\\
				\rowcolor{shadecolor}\{\textcolor{aoe}{C},\textcolor{bostonuniversityred}{M},\textcolor{burgundy}{S}\}\{\textcolor{burntorange}{G}\}
			&0.000&0.000&0.000&0.000&0.000&0.000&0.000&0.000&0.000&0.000\\
			\{\textcolor{aoe}{C}\}\{\textcolor{burntorange}{G}\}\{\textcolor{bostonuniversityred}{M},\textcolor{burgundy}{S}\}
			&0.001&0.139&0.001&0.019&0.024&0.028&0.078&0.042&0.232&0.055\\
			\{\textcolor{aoe}{C},\textcolor{burntorange}{G},\textcolor{bostonuniversityred}{M}\}\{\textcolor{burgundy}{S}\}
			&\textbf{0.463}&0.000&\textbf{0.595}&0.000&0.000&0.276&0.045&\textbf{0.498}&0.020&0.002\\
				\rowcolor{shadecolor}\{\textcolor{aoe}{C},\textcolor{burgundy}{S}\}\{\textcolor{burntorange}{G},\textcolor{bostonuniversityred}{M}\}
			&0.000&0.000&0.000&0.000&0.000&0.000&0.000&0.000&0.000&0.000\\
			\{\textcolor{aoe}{C}\}\{\textcolor{burntorange}{G},\textcolor{bostonuniversityred}{M}\}\{\textcolor{burgundy}{S}\}
			&0.146&0.099&0.188&\textbf{0.551}&\textbf{0.672}&0.074&0.164&0.092&0.260&0.033\\
				\rowcolor{shadecolor}\{\textcolor{aoe}{C},\textcolor{bostonuniversityred}{M}\}\{\textcolor{burntorange}{G},\textcolor{burgundy}{S}\}
			&0.000&0.000&0.000&0.000&0.000&0.000&0.000&0.000&0.000&0.000\\
				\rowcolor{shadecolor}\{\textcolor{aoe}{C},\textcolor{burntorange}{G},\textcolor{burgundy}{S}\}\{\textcolor{bostonuniversityred}{M}\}
			&0.000&0.000&0.000&0.000&0.000&0.000&0.000&0.000&0.000&0.000\\
				\rowcolor{shadecolor}\{\textcolor{aoe}{C}\}\{\textcolor{burntorange}{G},\textcolor{burgundy}{S}\}\{\textcolor{bostonuniversityred}{M}\}
			&0.000&0.000&0.000&0.000&0.000&0.000&0.000&0.000&0.000&0.000\\
			\{\textcolor{aoe}{C},\textcolor{burntorange}{G}\}\{\textcolor{bostonuniversityred}{M}\}\{\textcolor{burgundy}{S}\}
			&0.233&0.000&0.107&0.000&0.000&0.083&0.062&0.114&0.091&0.371\\
				\rowcolor{shadecolor}\{\textcolor{aoe}{C},\textcolor{bostonuniversityred}{M}\}\{\textcolor{burntorange}{G}\}\{\textcolor{burgundy}{S}\}
			&0.000&0.000&0.000&0.000&0.000&0.000&0.000&0.000&0.000&0.000\\
				\rowcolor{shadecolor}\{\textcolor{aoe}{C},\textcolor{burgundy}{S}\}\{\textcolor{burntorange}{G}\}\{\textcolor{bostonuniversityred}{M}\}
			&0.000&0.000&0.000&0.000&0.000&0.000&0.000&0.000&0.000&0.000\\
			\{\textcolor{aoe}{C}\}\{\textcolor{burntorange}{G}\}\{\textcolor{bostonuniversityred}{M}\}\{\textcolor{burgundy}{S}\}
			&0.133&0.216&0.108&0.347&0.288&0.060&0.121&0.065&\textbf{0.287}&\textbf{0.491}\\
			\hline
			$\sum\log_{15} \left(p_i ^{-p_i}\right)$ &0.501&0.430&0.415&0.361&0.289&0.632&0.688&0.598&0.613&0.424
	\end{tabular}}
\end{center}
\end{table} 
Table \ref{tab:table2} displays the posterior distributions for the partitions of unknown disease-specific means along with the corresponding entropy measurements, that can be used as measures of uncertainty. 
First note that if one takes also the ordering among distinct disease-specific locations into account, the posterior partition probabilities are, as desired, concentrated on specific orders of the associated unique values for all ten cardiac indexes. For instance, we have 
$\mathbb{P}(\{\theta_{C,CI}=\theta_{G,CI}=\theta_{M,CI}\}\{\theta_{S,CI}\}\mid X) =\mathbb{P}(\theta_{C,CI}=\theta_{G,CI}=\theta_{M,CI}>\theta_{S,CI}\mid X) = 0.463$. The ordered partitions with the highest posterior probability are displayed in Table~\ref{tab:tableord}. 
\begin{table}
\caption{Posterior probabilities over ordered partitions of means.}
\label{tab:tableord}
\begin{center}
	\begin{tabular}{lcc} 
		& ordered partition with& \\
		cardiac index& highest posterior probability&posterior prob\\
		\hline\hline
		CI&\{\textcolor{aoe}{C},\textcolor{burntorange}{G},\textcolor{bostonuniversityred}{M}\}$>$\{\textcolor{burgundy}{S}\}&0.463\\
		CWI&\{\textcolor{aoe}{C}\}$<$\{\textcolor{burntorange}{G},\textcolor{bostonuniversityred}{M},\textcolor{burgundy}{S}\}&0.546\\
		LVMI&\{\textcolor{aoe}{C},\textcolor{burntorange}{G},\textcolor{bostonuniversityred}{M}\}$<$\{\textcolor{burgundy}{S}\}&0.595\\
		IVST&\{\textcolor{aoe}{C}\}$<$\{\textcolor{burntorange}{G},\textcolor{bostonuniversityred}{M}\}$<$\{\textcolor{burgundy}{S}\}&0.548\\
		LVPW&\{\textcolor{aoe}{C}\}$<$\{\textcolor{burntorange}{G},\textcolor{bostonuniversityred}{M}\}$<$\{\textcolor{burgundy}{S}\}&0.671\\
		EF&\{\textcolor{aoe}{C},\textcolor{burntorange}{G},\textcolor{bostonuniversityred}{M},\textcolor{burgundy}{S}\}&0.365\\
		FS&\{\textcolor{aoe}{C},\textcolor{burntorange}{G},\textcolor{bostonuniversityred}{M},\textcolor{burgundy}{S}\}&0.303\\
		EW&\{\textcolor{aoe}{C},\textcolor{burntorange}{G},\textcolor{bostonuniversityred}{M}\}$>$\{\textcolor{burgundy}{S}\}&0.497\\
		AW&\{\textcolor{aoe}{C}\}$<$\{\textcolor{burntorange}{G},\textcolor{bostonuniversityred}{M}\}$<$\{\textcolor{burgundy}{S}\}&0.256\\
		E/A&\{\textcolor{aoe}{C}\}$>$\{\textcolor{burntorange}{G}\}$>$\{\textcolor{bostonuniversityred}{M}\}$>$\{\textcolor{burgundy}{S}\}&0.466
		\end{tabular}
	\end{center}
\end{table}

\begin{figure}
	\centering
	\begin{subfigure}{.45\textwidth}
		\centering
		\includegraphics[width=\linewidth]{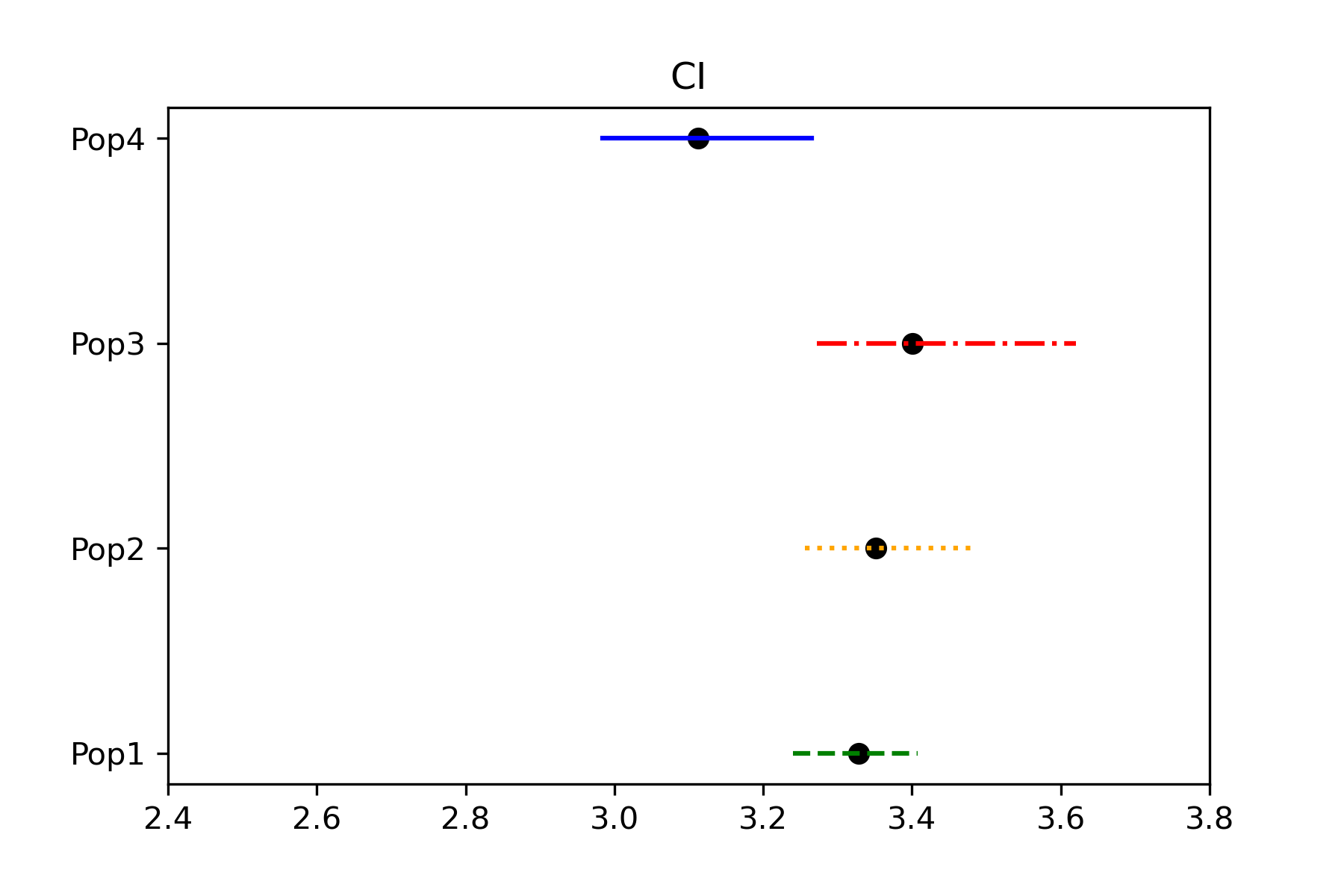}
	\end{subfigure}\hspace{0.05\textwidth}%
	\begin{subfigure}{.45\textwidth}
	\centering
	\includegraphics[width=\linewidth]{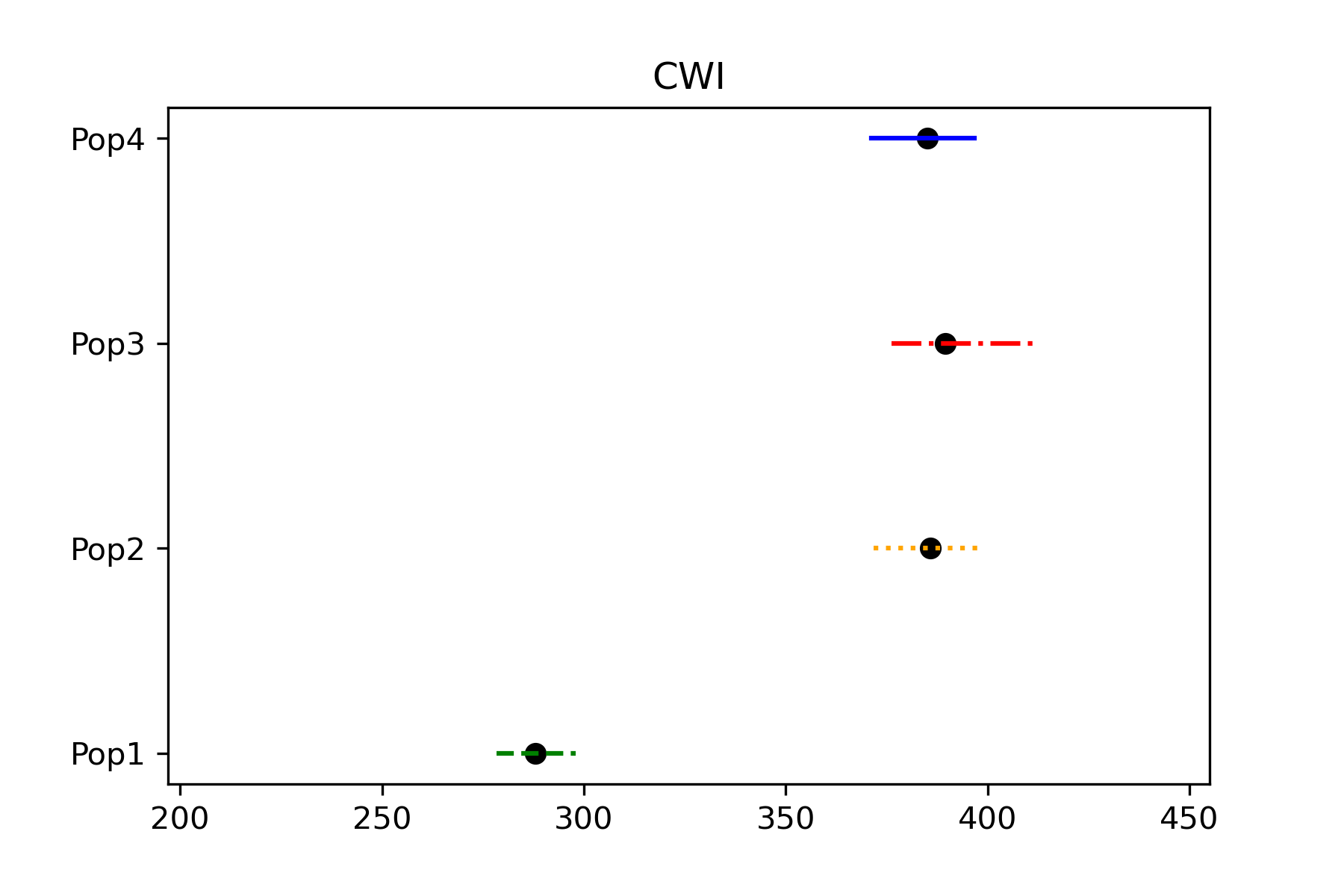}
	\end{subfigure}
\caption{95\% credible intervals for population-specific locations for CI and CWI}
\label{fig:fig5}
\end{figure}

Considering the posterior probabilities summarized in Table~\ref{tab:table2} and in Table~\ref{tab:tableord}, we find that the cardiac index (CI) is reduced in severe preeclampsia compared to all other patients, indicating reduced myocardial contractility in the presence of the most severe disorder. 
The cardiac work index (CWI) is a good indicator to distinguish between cases and control, but not among cases. The left ventricular mass index (LVMI) is increased in severe preeclampsia patients compared to other pregnant women, indicating ventricular remodelling. While inter ventricular septal thickness (IVST) and left ventricular posterior wall thickness (LVPW) differ both between cases and controls and between severe preeclampsia and other disorders, indicating a progressive increase in the indexes with the severity of the disorder. The posterior probabilities associated to 
indexes of systolic function such as ejection fraction (EF) and fraction shortening (FS) are relatively concentrated on the partition of complete homogeneity, letting us to conclude that no differences are present among 
patients.
As for the parameters of the diastolic function, the posterior distribution for the E-wave indicator identifies a modified index in severe preeclampsia patients, while the mean E/A ratio indicates a 
decreasing diastolic function with the severity of the disorder. The posterior for the A-wave index is actually concentrated on three distinct partitions, leaving a relatively high uncertainty regarding the modifications of the index. However, considering jointly the three partitions with the highest posterior probability, differences are detected between control and cases with a total posterior probability equal to 0.779. Figure~\ref{fig:fig5} shows point estimates and credible intervals for disorder-specific location parameters for the first two cardiac indexes. Analogous plots for all cardiac indexes can be found in Section E.1 of the Supplement.

Table~\ref{tab:table3} shows the results obtained using the prior in \eqref{eq:eppf_dir}, instead of \eqref{eq:prior2}. We remark that for all ten cardiac indexes, the posterior associates negligible probabilities to partitions that are in contrast with the natural order of the diagnoses. This is particularly reassuring in that the model, even without imposing such an order a priori, is able to single it out systematically across cardiac indexes.
Moreover, we observe how the partitions identified by MAP are the same of Table~\ref{tab:table2} for all cardiac index except AW. However, even under this alternative prior, the A-wave index is concentrated on the same three distinct partitions leading to the conclusion that there exists a difference between cases and control.
\begin{table}
	\caption{Posterior probabilities over partitions of means. Maximum a posteriori probabilities are in \textbf{bold}.}
	\label{tab:table3}
	\vspace{-0.3cm}
	\begin{center}
\resizebox{\textwidth}{!}{
		\begin{tabular}{lcccccccccc}
			partitions&CI&CWI&LVMI&IVST&LVPW&EF&FS&EW&AW&E/A\\\hline\hline
			\{\textcolor{aoe}{C},\textcolor{burntorange}{G},\textcolor{bostonuniversityred}{M},\textcolor{burgundy}{S}\}
			&0.019&0.000&0.000&0.000&0.000&\textbf{0.332}&\textbf{0.247}&0.078&0.000&0.000\\
			\{\textcolor{aoe}{C}\}\{\textcolor{burntorange}{G},\textcolor{bostonuniversityred}{M},\textcolor{burgundy}{S}\}
			&0.002&\textbf{0.643}&0.001&0.114&0.031&0.065&0.130&0.048&0.080&0.000\\
			\{\textcolor{aoe}{C},\textcolor{burntorange}{G}\}\{\textcolor{bostonuniversityred}{M},\textcolor{burgundy}{S}\}
			&0.004&0.000&0.003&0.000&0.000&0.044&0.019&0.152&0.073&0.103\\
				\rowcolor{shadecolor}\{\textcolor{aoe}{C},\textcolor{bostonuniversityred}{M},\textcolor{burgundy}{S}\}\{\textcolor{burntorange}{G}\}
			&0.004&0.000&0.000&0.000&0.000&0.037&0.105&0.013&0.000&0.000\\
			\{\textcolor{aoe}{C}\}\{\textcolor{burntorange}{G}\}\{\textcolor{bostonuniversityred}{M},\textcolor{burgundy}{S}\}
			&0.002&0.065&0.002&0.047&0.078&0.027&0.036&0.063&\textbf{0.424}&0.167\\
			\{\textcolor{aoe}{C},\textcolor{burntorange}{G},\textcolor{bostonuniversityred}{M}\}\{\textcolor{burgundy}{S}\}
			&\textbf{0.316}&0.000&\textbf{0.527}&0.000&0.000&0.178&0.032&\textbf{0.288}&0.002&0.000\\
				\rowcolor{shadecolor}\{\textcolor{aoe}{C},\textcolor{burgundy}{S}\}\{\textcolor{burntorange}{G},\textcolor{bostonuniversityred}{M}\}
			&0.023&0.000&0.000&0.000&0.000&0.019&0.103&0.006&0.000&0.000\\
			\{\textcolor{aoe}{C}\}\{\textcolor{burntorange}{G},\textcolor{bostonuniversityred}{M}\}\{\textcolor{burgundy}{S}\}
			&0.173&0.089&0.124&\textbf{0.472}&\textbf{0.594}&0.033&0.054&0.064&0.140&0.042\\
				\rowcolor{shadecolor}\{\textcolor{aoe}{C},\textcolor{bostonuniversityred}{M}\}\{\textcolor{burntorange}{G},\textcolor{burgundy}{S}\}
			&0.002&0.000&0.001&0.003&0.000&0.044&0.031&0.017&0.000&0.000\\
				\rowcolor{shadecolor}\{\textcolor{aoe}{C},\textcolor{burntorange}{G},\textcolor{burgundy}{S}\}\{\textcolor{bostonuniversityred}{M}\}
			&0.018&0.000&0.000&0.000&0.000&0.061&0.067&0.016&0.000&0.000\\
				\rowcolor{shadecolor}\{\textcolor{aoe}{C}\}\{\textcolor{burntorange}{G},\textcolor{burgundy}{S}\}\{\textcolor{bostonuniversityred}{M}\}
			&0.005&0.163&0.001&0.095&0.006&0.028&0.040&0.015&0.016&0.000\\
			\{\textcolor{aoe}{C},\textcolor{burntorange}{G}\}\{\textcolor{bostonuniversityred}{M}\}\{\textcolor{burgundy}{S}\}
			&0.213&0.000&0.124&0.000&0.000&0.052&0.014&0.121&0.036&0.241\\
				\rowcolor{shadecolor}\{\textcolor{aoe}{C},\textcolor{bostonuniversityred}{M}\}\{\textcolor{burntorange}{G}\}\{\textcolor{burgundy}{S}\}
			&0.074&0.000&0.137&0.003&0.000&0.041&0.022&0.055&0.001&0.000\\
				\rowcolor{shadecolor}\{\textcolor{aoe}{C},\textcolor{burgundy}{S}\}\{\textcolor{burntorange}{G}\}\{\textcolor{bostonuniversityred}{M}\}
			&0.014&0.000&0.000&0.000&0.000&0.011&0.067&0.004&0.000&0.000\\
			\{\textcolor{aoe}{C}\}\{\textcolor{burntorange}{G}\}\{\textcolor{bostonuniversityred}{M}\}\{\textcolor{burgundy}{S}\}
			&0.133&0.040&0.079&0.265&0.291&0.029&0.033&0.059&0.229&\textbf{0.448}\\
			\hline
			$\sum\log_{15} \left(p_i ^{-p_i}\right)$&0.687&0.407&0.509&0.501&0.371&0.828&0.886&0.823&0.582&0.505
		\end{tabular}}
	\end{center}
\vspace{-0.1cm}
\end{table}

\begin{figure}[t]
	\vspace{-0.1 cm}
	\centering 
	\begin{subfigure}{.4\textwidth}
		\centering
		\includegraphics[ width=\linewidth]{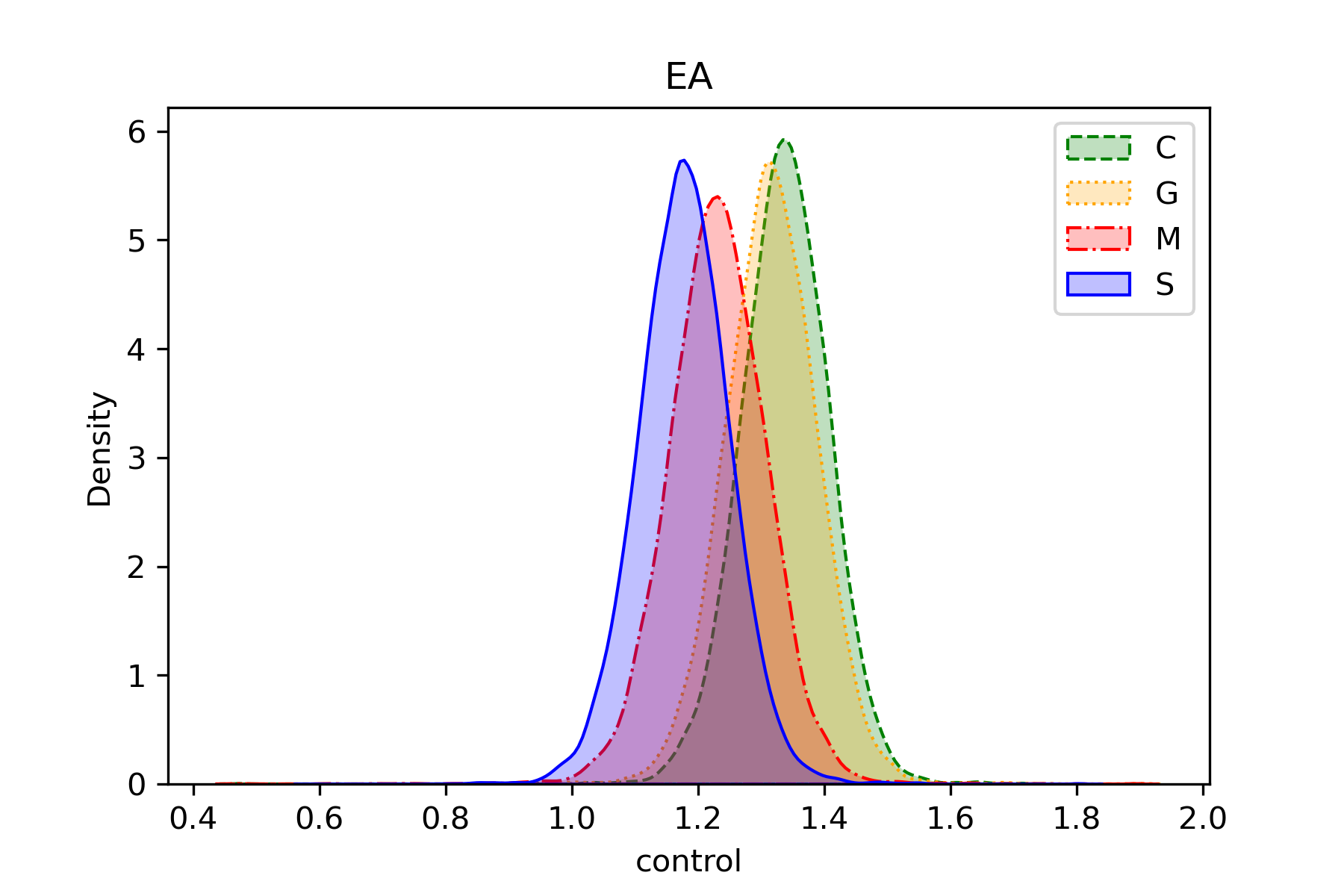}
		\caption{density estimation}
	\end{subfigure}\hspace{0.05\textwidth}%
	\begin{subfigure}{.25\textwidth}
		\centering
		\includegraphics[trim={3cm 1cm 3cm 1cm}, clip,  width=\linewidth]{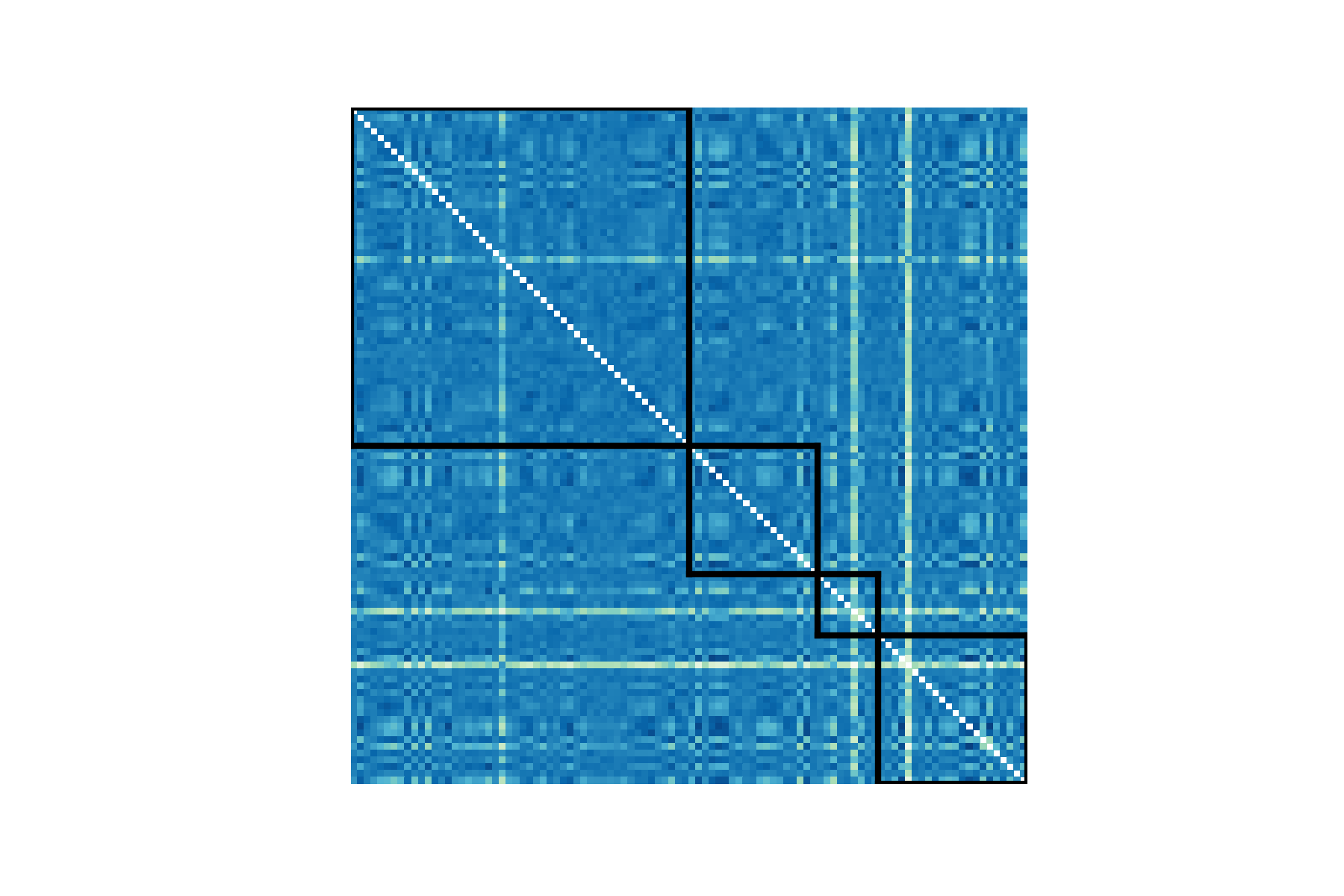}
		\caption{co-clustering}\label{fig:fig4b}
	\end{subfigure}\hspace{0.05\textwidth}%
	\begin{subfigure}{.25\textwidth}
		\centering
		\includegraphics[trim={3cm 1cm 3cm 1cm}, clip,  width=\linewidth]{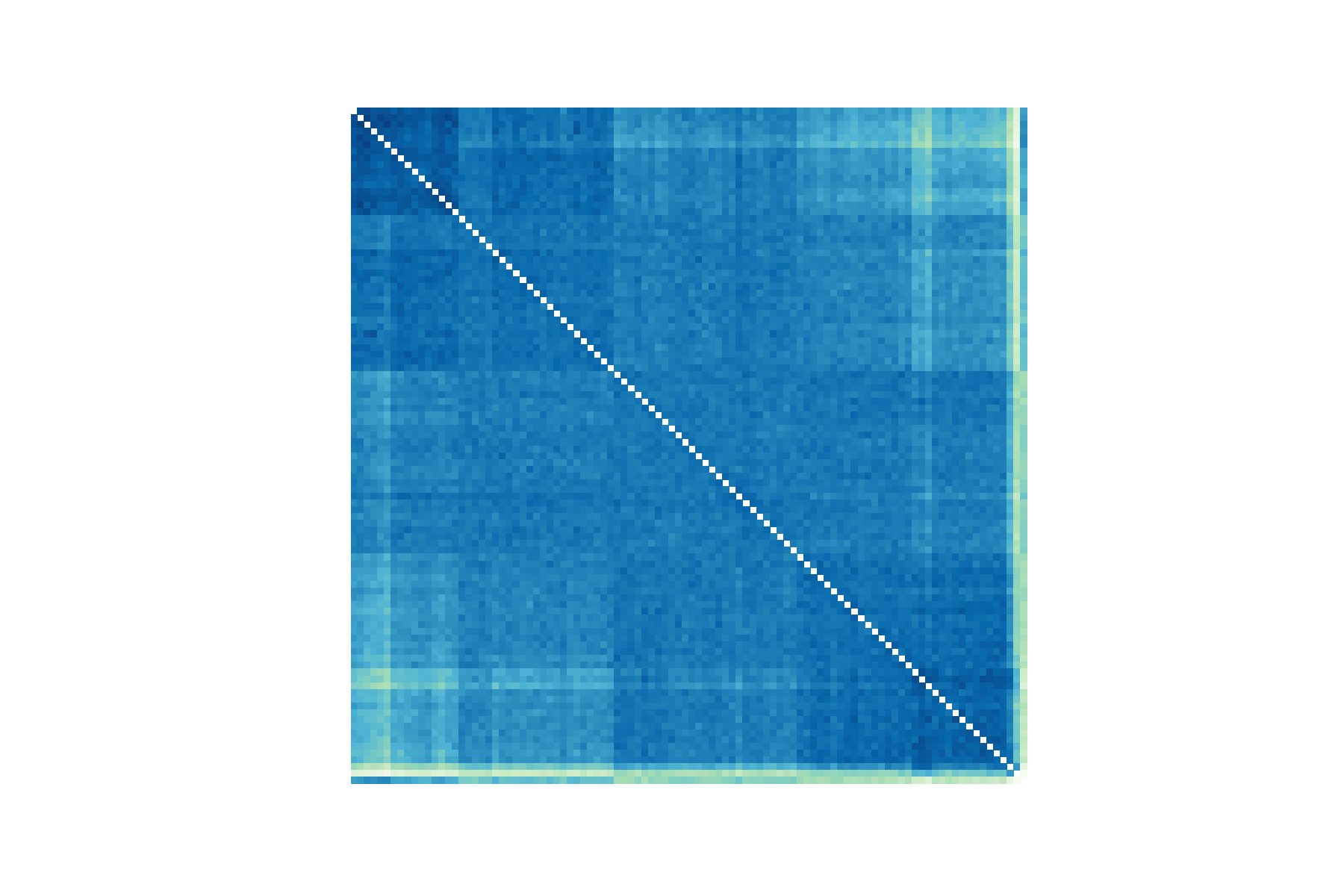}
		\caption{co-clustering}
	\end{subfigure}
	\begin{subfigure}{.4\textwidth}
		\centering
		\includegraphics[ width=\linewidth]{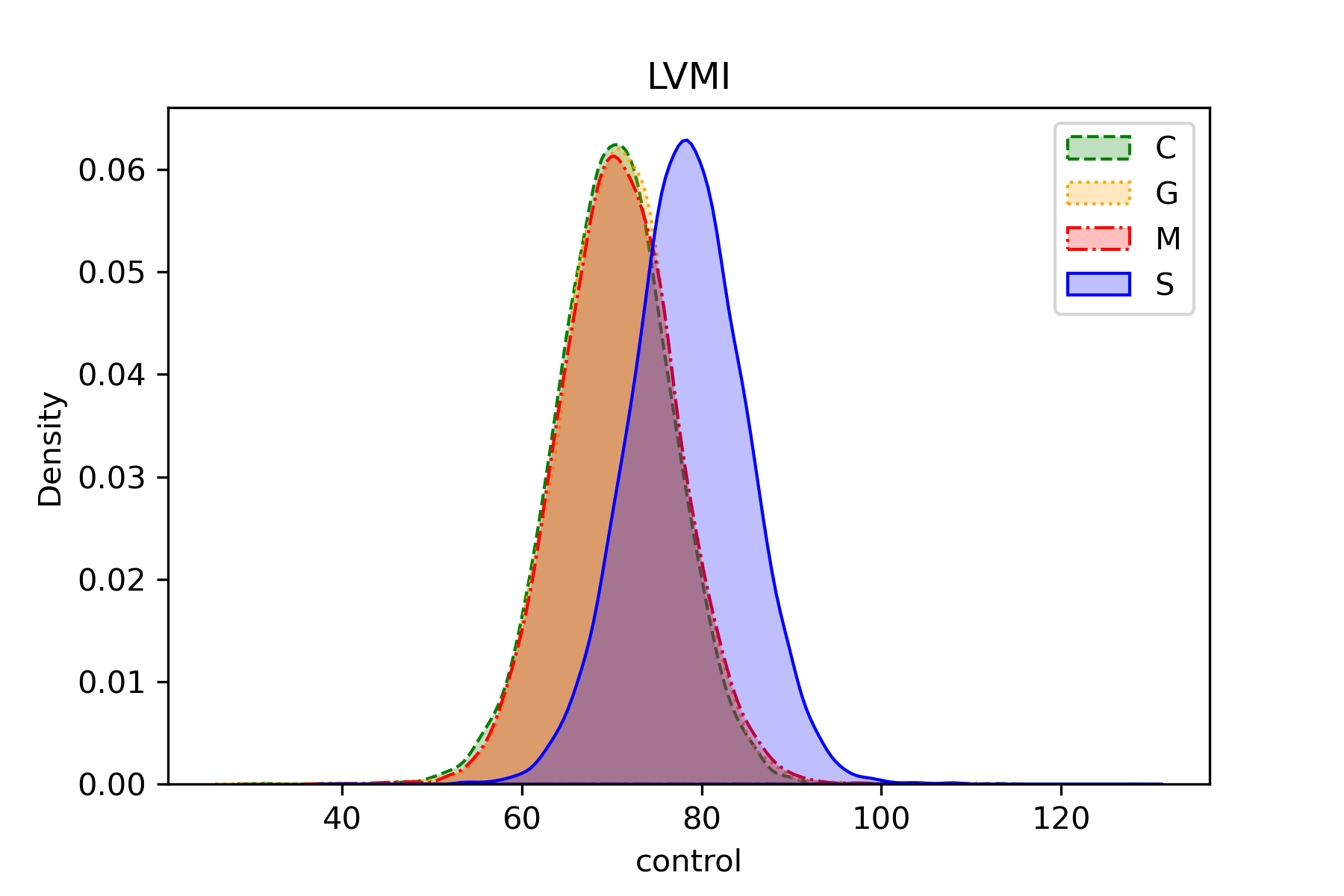}
		\caption{density estimation}
	\end{subfigure}\hspace{0.05\textwidth}%
	\begin{subfigure}{.25\textwidth}
		\centering
		\includegraphics[trim={3cm 1cm 3cm 1cm}, clip,  width=\linewidth]{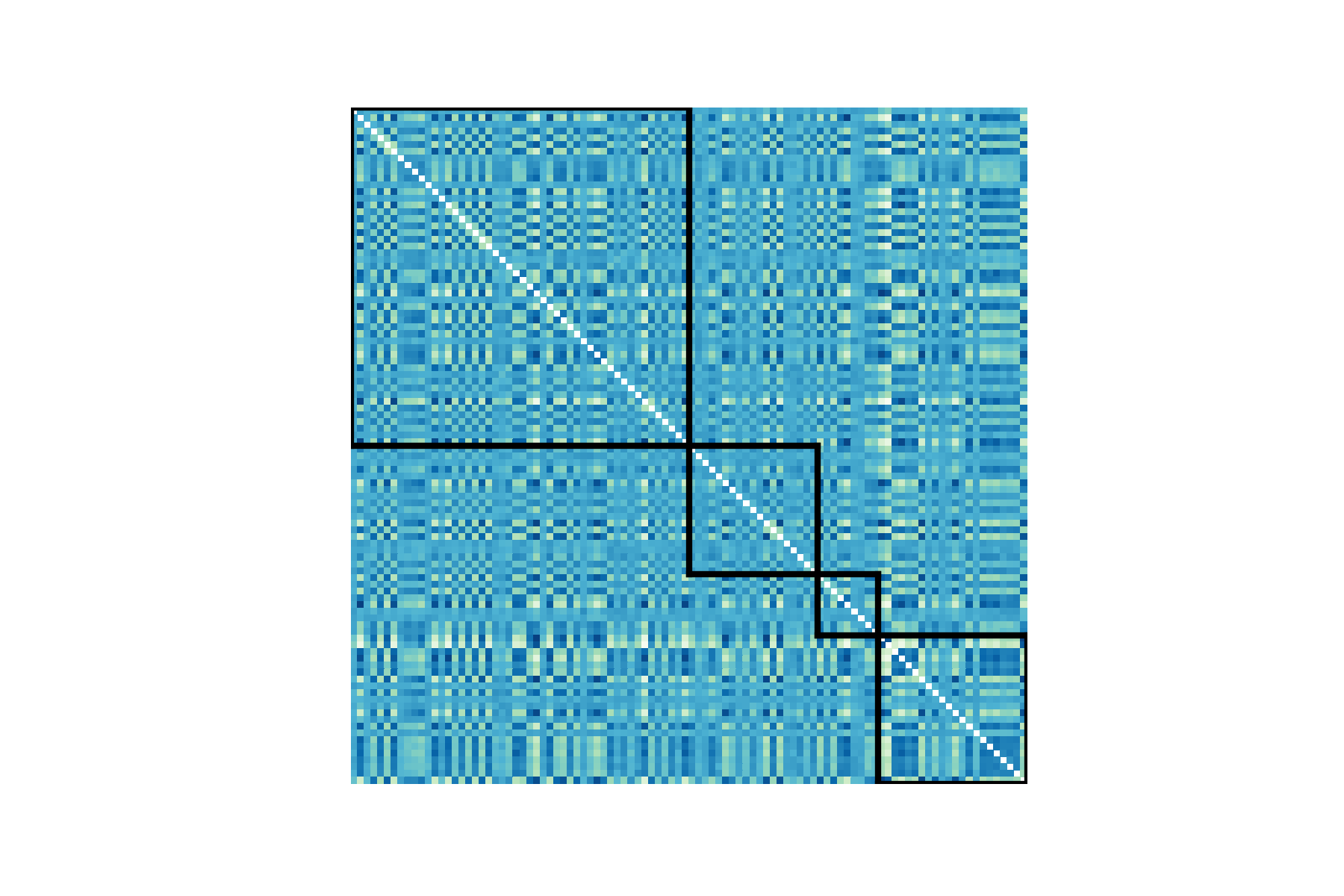}
		\caption{co-clustering}\label{fig:fig4e}
	\end{subfigure}\hspace{0.05\textwidth}%
	\begin{subfigure}{.25\textwidth}
		\centering
		\includegraphics[trim={3cm 1cm 3cm 1cm}, clip, width=\linewidth]{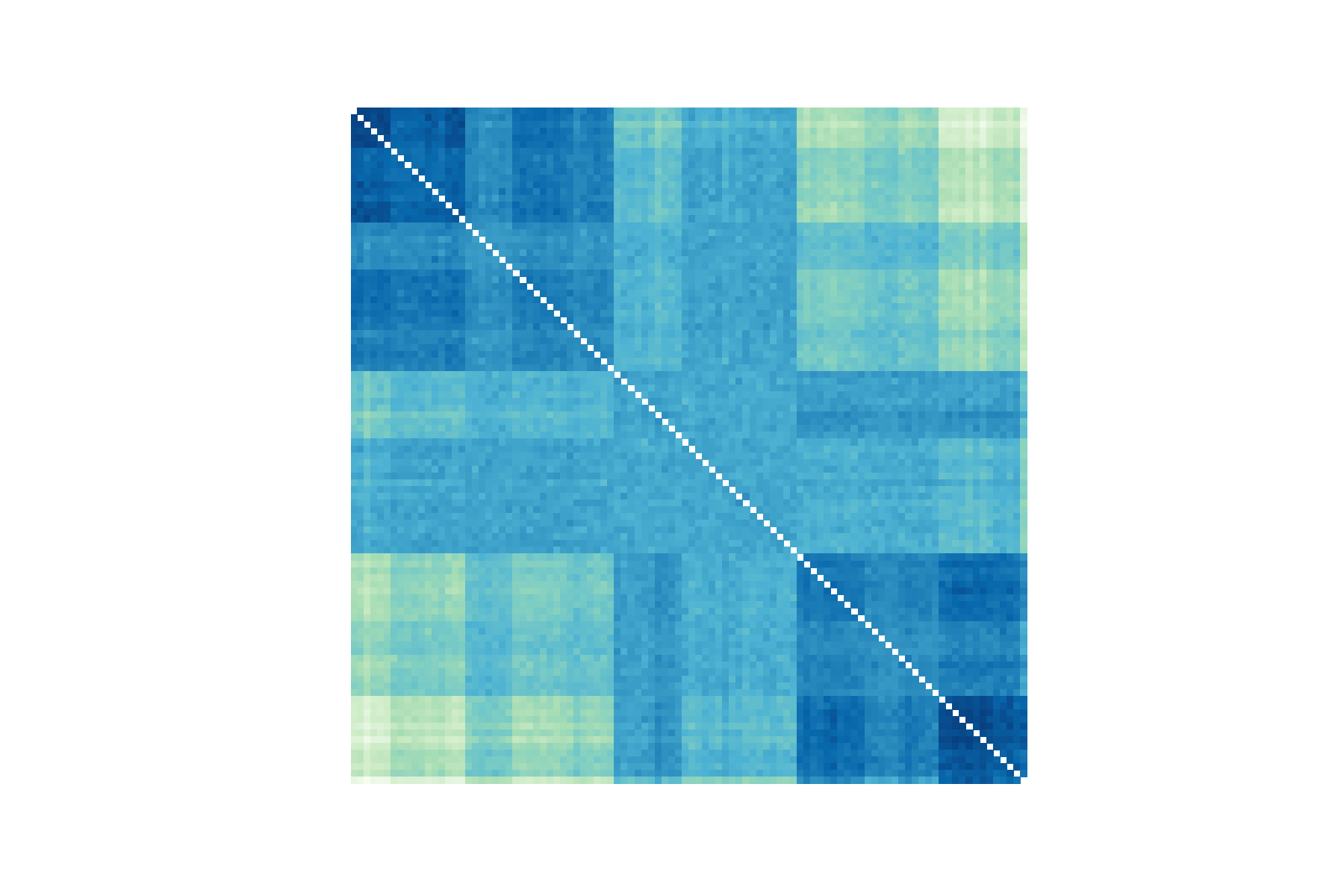}
		\caption{co-clustering}\label{fig:fig4f}
	\end{subfigure}
\vspace{-0.1 cm}
\caption{ Panels (a) and (d): density estimates. Panels (b)--(c) and (e)--(f): heatmaps of the posterior probabilities of co-clustering; in (b) and (e) patients are ordered based on the diagnosis and the four black squares highlight the within-sample probabilities; in (c) and (f) patients are reordered based on co-clustering probabilities.}
\label{fig:fig4}
\vspace{-0.1 cm}
\end{figure} 

As far as prediction and second-level clustering are concerned, Figure~\ref{fig:fig4} displays the density estimates and the heatmap of co-clustering probabilities between pairs of patients for the E/A ratio and LVMI. Figure~\ref{fig:fig4b} shows that co-clustering probabilities are similar within and across diagnoses, indicating that the effect of the diseases on the distribution of the cardiac index is mostly explained through shifts between disease-specific locations. Moreover Figure~\ref{fig:fig4b} suggests the presence of three outliers that have low probability of co-clustering with all the other subjects and that would be ignored by the model using a more traditional ANOVA structure. On the other hand, Figure~\ref{fig:fig4e} shows a slightly different pattern for co-clustering probabilities in the fourth square, which suggests that the heterogeneity between severe preeclampsia patients and the other patients is not entirely explained by shifts in disease-specific locations. Finally, Figure~\ref{fig:fig4f} suggests the presence of an underlying relevant factor. The corresponding figures for all ten response variables are reported in Section E.1 of the Supplement and can be used for prediction and for a graphical analysis aimed at controlling the presence of underlying relevant factors, outliers and differences across diseases distinct from shifts between disease-specific locations.

Our results are coherent with almost all of the findings in \cite{tatapudi2017}, where results were obtained through a series of independent frequentist tests. However, importantly, we are able to provide more insights thanks to the simultaneous comparison approach and the latent clustering of subjects. For instance, considering the response LVMI, \cite{tatapudi2017} detected a significant increase in cases compared to controls and an increase in severe preeclampsia compared to gestational hypertensive and mild preeclampsia patients. Such results do not clarify whether a modification exists between the control group and gestational hypertensive patients or between the latter and mild preeclampsia patients. Moreover, in contrast to our analysis, their results do not provide any information concerning the presence of underlying common factors, outliers or distributional effects (different from shifts in locations).

\section{Concluding remarks}
\label{s:conclusion}
We designed a Bayesian nonparametric model to detect clusters of hypertensive disorders over different cardiac function indexes and found modified cardiac functions in hypertensive patients compared to healthy subjects as well as progressively increased alterations with the severity of the disorder. The proposed model has application potential also beyond the 
considered setup when the goal is to cluster populations according to multivariate information: it borrows strength across response variables, preserves the flexibility intrinsic to nonparametric models, and correctly detects partitions of populations even in presence of small sample sizes, 
when alternative distribution-based clustering models tend to underestimate the number of clusters. The key component of the model is the s-HDP, a hierarchical nonparametric structure for the error terms that offers flexibility and serves as a tool to investigate the presence of unobserved factors, outliers and effects other than changes in locations.  Interesting extensions of the model include generalizations to other types of invariances in order to accommodate identifiability in generalized linear models, for instance in presence of count data and a log link function, as well as generalizations to other types of processes, beyond the Dirichlet process.

\section*{Acknowledgements}
Most of the paper was completed while B. Franzolini was a Ph.D. student at the Bocconi University, Milan.
 A. Lijoi and I. Pr\"unster are partially supported by MIUR, PRIN Project 2015SNS29B.

\newpage
\newgeometry{top=20mm, bottom=20mm, left=20mm, right=20mm,
foot=10mm, marginparsep=0mm}

\fancypagestyle{mypagestyle}{
	\fancyhf{}
	\fancyfoot[C]{Supplement -- \thepage}  
	\renewcommand{\headrulewidth}{0pt}
	\renewcommand{\footrulewidth}{0pt}
}
\makeatletter
\let\ps@plain\ps@mypagestyle
\makeatother

\pagestyle{mypagestyle}
\setcounter{page}{1}

\part{\large Supplement to \emph{Model Selection for  Maternal Hypertensive Disorders with Symmetric Hierarchical Dirichlet Processes}}

\normalsize

\fontsize{10.5}{11}\selectfont
\renewcommand{\thesection}{\Alph{section}}

\section{A review of the invariant Dirichlet process} 
We provide a brief review of the invariant Dirichlet process (IDP), introduced by \cite{dalal1979}, which serves as building block for the proposed model. After recalling the definition, we present two representations of the process: the first is the analog of the stick-breaking construction of the Dirichlet Process (DP), whereas the second is an extension of the generalized P\'olya urn scheme of \cite{blackwell1973}.

\vspace{3pt}

\noindent Let $(E,\mathcal{E})$ be any measurable Euclidean space and $\mathcal{G}=\{g_1,\ldots,g_L\}$ be a finite group of measurable transformations on $(E, \mathcal{E})$.

\begin{definition}[Invariant Probability]A probability measure $P_0$ on $(E, \mathcal{E})$ is a $\mathcal{G}$-invariant probability distribution, if
	$P_0(A) =  P_0(g_l(A))$, for any $A \in \mathcal{E}$ 
	and $l=1,\ldots,L$.
\end{definition}
\vspace{2pt}

\begin{definition}[Invariant Random Probability]A random probability $\tilde p$ on $(E, \mathcal{E})$ is said $\mathcal{G}$-invariant, if it is almost surely $\mathcal{G}$-invariant. 
\end{definition}

\vspace{2pt}

\begin{definition}[Invariant Partition]
	A measurable partition $A_1,A_2,\ldots,A_K$ of $E$ is a $\mathcal{G}$\textit{-invariant partition}, if
	$A_k = g_l(A_k)$, $\forall k=1,\ldots,K$ and $\forall l=1,\ldots,L$.
\end{definition}

\begin{definition}[Invariant Dirichlet Process] A random probability $\tilde p$ is an IDP with group of transformations $\mathcal{G}$, if 
	\begin{enumerate}
		\item $\tilde p$ is almost surely $\mathcal{G}$-invariant
		\[
		\tilde p(A) = \tilde p(g_l(A)) \quad \text{for } l=1,\ldots,L \quad \text{a.s.}
		\]
		\item there exists a $\mathcal{G}$-invariant probability distribution $P_0$ on $(E, \mathcal{E})$ and $\alpha\in\mathbb{R}^+$, such that for any $k\in\mathbb{N}$ and any $\mathcal{G}$-invariant measurable partition $A_1,\ldots,A_k$
		\[
		\big(\tilde p(A_1),\ldots,\tilde p(A_k)\big) \sim \text{D}_{k-1}(\alpha\, P_0(A_1),\ldots,\alpha\, P_0(A_k))
		\]
	\end{enumerate}
	where $\alpha$ is the concentration parameter and $P_0$ the baseline probability measure.
\end{definition}
\noindent The notation $\tilde p\sim \text{IDP}(\alpha,\,P_0,\,\mathcal{G})$ indicates that the random measure $\tilde p$ is distributed according to an IDP. Note that, if $\tilde p\sim \text{DP}(\alpha,\,P_0)$, then $\tilde p$ is not an IDP, since it is not an invariant random probability. Vice versa, if $\tilde p\sim \text{IDP}(\alpha,\,P_0,\,\mathcal{G})$, then $\tilde p$ is not a DP, since its finite dimensional distributions over non $\mathcal{G}$-invariant partitions are not Dirichlet. However there is a strong relationship between the two processes as shown in \cite{dalal1979}.
\begin{theorem}[\citealp{dalal1979}]
	Let $\tilde q\sim \mbox{\rm DP}(\alpha,\,P_0)$ and $\tilde p\sim \mbox{\rm IDP}(\alpha,\,P_0,\,\mathcal{G})$. Define
	\[
	q^*(\cdot)=\frac{1}{L}\sum\limits_{l=1}^L \tilde q(g_l(\cdots))
	\]
	then 
	\[
	\tilde p \overset{d}{=}q^*
	\]
\end{theorem}
\cite{tiwari1988convergence} provided also a constructive definition for the IDP, which is the analogue  of the stick-breaking representation of \cite{sethuraman1994} for the DP.
\begin{proposition}[\citealp{tiwari1988convergence}]
	If $\tilde p\sim \text{IDP}(\alpha,\,P_0,\,\mathcal{G})$, then 
	\[
	\tilde p=\sum\limits_{h=1}^{\infty}\pi_h \sum_{l=1}^{L}\delta_{g_l(\phi^*_h)}
	\]
	\[
	\text{with}\quad \quad  \pi_h=\frac{\pi'_h}{L}\prod_{r=1}^{h-1}(1-\pi'_r)\quad \quad \pi'_r\sim\text{Beta}(1,\,\alpha) \quad \quad \phi^*_h\overset{iid}{\sim}P_0.
	\]
\end{proposition}
Moreover, if 
$\phi_{i}
\mid \tilde Q\overset{iid}{\sim}\tilde Q$ with $\tilde Q\sim \mbox{IDP}(\alpha,\, P_0,\,\mathcal{G})$,
by integrating out $\tilde Q$, we get the corresponding generalized P\'olya urn representation for the process

\[
\phi_{1}\sim P_0
\] 
\[
\phi_{i}\mid \phi_{1},\ldots \phi_{i-1} 
\sim \sum\limits_{j=1}^{i-1} 
\frac{1}{i-1+\alpha}\Bigg(\frac{1}{L}\sum\limits_{l=1}^{L}
\delta_{g_l(\phi_j)}\Bigg)+\frac{\alpha}{i-1+\alpha}P_0
\]
For more details about IDPs we refer to \cite{dalal1979}, \cite{dalal1979nonparametric}, \cite{hannum1983robustness}, \cite{doss1984}, \cite{diaconis1986}, \cite{tiwari1988convergence}, \cite*{ferguson1992bayesian} and \cite{ghosal1999}.

\section{Predictive distribution for disease-specific locations}
We recall that the prior over partitions is given by 

\[
\mathbb{P}(M_b^m\mid \omega) \propto \begin{cases}
	\omega^{k-1}\,\prod_{i=1}^k (n_i-1)!&\text{if $M_b^m$ is compatible with the natural order}\\
	0 & \text{otherwise}\end{cases}
\]
where $k$ is the number of distinct clusters according to the partition $M_b^m$ and $n_1,\ldots,n_k$ are the clusters' frequencies. 
Thus, being $J=4$, one obtains the probabilities in Figure \ref{fig:probpart}, starting from which it is possible to compute the joint distribution of $(\theta_{1,m},\ldots,\theta_{4,m})$ conditional on $\omega$
\[
\begin{split}
	\theta_{1,m}\mid \omega &\sim G_m \\
	&\\
	\theta_{2,m}\mid \theta_{1,m},\omega & \sim \frac{\omega^2+3\omega+6}{(\omega+2)(\omega^2+\omega+3)}\delta_{\theta_{1,m}} + 
	\frac{\omega^3+2\omega^2+2\omega}{(\omega+2)(\omega^2+\omega+3)}G_m\\
	&\\
	\theta_{3,m}\mid \theta_{1,m}, \theta_{2,m},\omega & \sim\begin{cases} 
		\frac{2\omega+6}{\omega^2+3\omega+6}\delta_{\theta_{2,m}} + 
		\frac{\omega^2+\omega}{\omega^2+3\omega+6}G_m& \mbox{if }\theta_{1,m}=\theta_{2,m}\\
		\frac{\omega+2}{\omega^2+2\omega+2}\delta_{\theta_{2,m}} + 
		\frac{\omega^2+\omega}{\omega^2+2\omega+2}G_m& \mbox{if }\theta_{1,m}\neq\theta_{2,m}
	\end{cases}\\
	&\\
	\theta_{4,m}\mid \theta_{1,m}, \theta_{2,m},\theta_{3,m},\omega & \sim \begin{cases} 
		\frac{3}{\omega+3}\delta_{\theta_{3,m}} + 
		\frac{\omega}{\omega+3}G_m &\mbox{if }\theta_{1,m}=\theta_{2,m}=\theta_{3,m}\\
		\frac{2}{\omega+2}\delta_{\theta_{3,m}} + 
		\frac{\omega}{\omega+2}G_m &\mbox{if }\theta_{1,m}\neq\theta_{2,m}=\theta_{3,m}\\
		\frac{1}{\omega+1}\delta_{\theta_{3,m}} + 
		\frac{\omega}{\omega+1}G_m&\mbox{otherwise}
	\end{cases}
\end{split}
\]
\begin{figure}
	\begin{center}
		\begin{tikzpicture}[-latex ,auto ,node distance =0.3 cm and 4.2cm ,on grid ,
			semithick ,
			state/.style ={ circle ,top color =white , bottom color = white ,
				draw,black , text=black , minimum width =1 cm},
			transition/.style = {rectangle, draw=black!50, thick, minimum width=6.1cm, minimum height = 5.8cm},
			transition3/.style = {rectangle, draw=black!50, thick, minimum width=4cm, minimum height = 3.8cm}, 
			transition2/.style = {rectangle, draw=black!50, thick, minimum width=6.5cm, minimum height = 8cm}, scale=0.8]
			\node[draw=none] (A){$\mathbb{P}[\{\textcolor{aoe}{\theta_1},\textcolor{burntorange}{\theta_2},\textcolor{bostonuniversityred}{\theta_3},\textcolor{burgundy}{\theta_4}\}]\propto 3!$};
			\node[draw=none] (B) [below =of A] { };
			\node[draw=none] (C) [below =of B]{$\mathbb{P}[\{\textcolor{aoe}{\theta_1},\textcolor{burntorange}{\theta_2},\textcolor{bostonuniversityred}{\theta_3}\},\{\textcolor{burgundy}{\theta_4}\}]\propto 2!\,\omega$};
			\node[draw=none] (D) [below =of C] { };
			\node[draw=none] (E) [below =of D]{$\mathbb{P}[\{\textcolor{aoe}{\theta_1},\textcolor{burntorange}{\theta_2}\},\{\textcolor{bostonuniversityred}{\theta_3},\textcolor{burgundy}{\theta_4}\}]\propto \omega$};
			\node[draw=none] (F) [below =of E] { };
			\node[draw=none] (G) [below =of F]{$\mathbb{P}[\{\textcolor{aoe}{\theta_1},\textcolor{burntorange}{\theta_2}\},\{\textcolor{bostonuniversityred}{\theta_3}\},\{\textcolor{burgundy}{\theta_4}\}]\propto \omega^2$};
			\node[draw=none] (H) [below =of G] { };
			\node[draw=none] (I) [below =of H]{{$\mathbb{P}[\{\textcolor{aoe}{\theta_1}\},\{\textcolor{burntorange}{\theta_2},\textcolor{bostonuniversityred}{\theta_3},\textcolor{burgundy}{\theta_4}\}]\propto 2!\,\omega$}};
			\node[draw=none] (L) [below =of I] { };
			\node[draw=none] (M) [below =of L]{$\mathbb{P}[\{\textcolor{aoe}{\theta_1}\},\{\textcolor{burntorange}{\theta_2},\textcolor{bostonuniversityred}{\theta_3}\},\{\textcolor{burgundy}{\theta_4}\}]\propto \omega^2$};
			\node[draw=none] (N) [below =of M] { };
			\node[draw=none] (O) [below =of N]{$\mathbb{P}[\{\textcolor{aoe}{\theta_1}\},\{\textcolor{burntorange}{\theta_2}\},\{\textcolor{bostonuniversityred}{\theta_3},\textcolor{burgundy}{\theta_4}\}]\propto \omega^2$};
			\node[draw=none] (P) [below =of O] { };
			\node[draw=none] (Q) [below =of P]{$\mathbb{P}[\{\textcolor{aoe}{\theta_1}\},\{\textcolor{burntorange}{\theta_2}\},\{\textcolor{bostonuniversityred}{\theta_3}\},\{\textcolor{burgundy}{\theta_4}\}]\propto \omega^3$};

			\node[draw=none] (b) [right =of B] {$\mathbb{P}[\textcolor{aoe}{\theta_1}=\textcolor{burntorange}{\theta_2}=\textcolor{bostonuniversityred}{\theta_3}]$};
			\node[draw=none] (c) [right =of C]{$\propto 2!\,\omega+3!$};
			\node[draw=none] (d) [right =of D] { };
			\node[draw=none] (f) [right =of F] {$\mathbb{P}[\textcolor{aoe}{\theta_1}=\textcolor{burntorange}{\theta_2}\neq\textcolor{bostonuniversityred}{\theta_3}]$};
			\node[draw=none] (g) [right =of G]{$\propto \omega^2+\omega$};
			\node[draw=none] (h) [right =of H] { };
			\node[draw=none] (l) [right =of L] {$\mathbb{P}[\textcolor{aoe}{\theta_1}\neq\textcolor{burntorange}{\theta_2}=\textcolor{bostonuniversityred}{\theta_3}]$ };
			\node[draw=none] (m) [right =of M]{$\propto \omega^2+2!\omega$};
			\node[draw=none] (n) [right =of N] { };
			\node[draw=none] (p) [right =of P] {$\mathbb{P}[\textcolor{aoe}{\theta_1}\neq\textcolor{burntorange}{\theta_2}\neq\textcolor{bostonuniversityred}{\theta_3}$]};
			\node[draw=none] (pp) [right =of Q]{$\propto \omega^3+\omega^2$};
			
			\node[draw=none] (dd) [right =of d] {$\mathbb{P}[\textcolor{aoe}{\theta_1}=\textcolor{burntorange}{\theta_2}]$};
			\node[draw=none] (xx) [right =8.5cm of E]{$\propto\omega^2+3\,\omega+3!$};
			\node[draw=none] (nn) [right =of n] {$\mathbb{P}[\textcolor{aoe}{\theta_1}\neq\textcolor{burntorange}{\theta_2}]$};
			\node[draw=none] (yy) [right =8.5cm of O]{$\propto\omega^3+2\,\omega^2+2\,\omega$};
			
			\path (A) edge node {} (b);
			\path (C) edge node[below] {} (b);
			\path (E) edge node {} (f);
			\path (G) edge node[below] {} (f);
			\path (I) edge node {} (l);
			\path (M) edge node[below] {} (l);
			\path (O) edge node {} (p);
			\path (Q) edge node[below] {} (p);
			\path (b) edge node {} (dd);
			\path (f) edge node[below] {} (dd);
			\path (l) edge node {} (nn);
			\path (p) edge node[below] {} (nn);
		\end{tikzpicture}
		\caption{\label{fig:probpart}\small A priori partitions' probabilities joint (on the left) and marginals}
	\end{center}
\end{figure}

\section{Alternative priors over disorder-specific locations}
For comparison purposes and prior sensitivity analysis we consider also two alternative priors over the disorder-specific locations: a uniform prior, which does not penalize multiplicity but incorporates the prior information on the severity of disorders, and a mixture of Dirichlet processes (DPs), which penalizes for multiplicity but does not reflect prior information. 

\subsection{Uniform prior}
The uniform prior is obtained associating zero-probability to nonsensical partitions and a uniform prior over the remaining, i.e.
\begin{equation*}
	\mathbb{P}(M_b^m) \propto \begin{cases}
		\frac{1}{8}&\text{if $M_b^m$ is compatible with the natural order}\\
		0 & \text{otherwise}\end{cases}
\end{equation*}   
The predictive distributions are
\[
\theta_j\,|\theta_1,\ldots,\theta_{j-1}\:\sim\:
\frac{1}{2}\,\delta_{\theta_{j-1}}+
\frac{1}{2}G
\]
and the full conditional distribution of $t_{j}$ is
\[
p(t_{j}=t\mid \bmath{t}_{\theta}^{(-j)},\bmath{\theta}^{{(-j)}},\bm{z}_{j}, \bm{\sigma}_{j}) \propto
\begin{cases} 
	f_{\theta_{j-1}}(\bm{z}_{j}|\bm{\sigma}_{j}) 
	& \mbox{if } t=t_{j-1} \\[4pt] 
	\displaystyle\int f_{\theta}(\bm{z}_{j}|\bm{\sigma}_{j})\, G (d\theta)& \mbox{if } t=t^{\mbox{\footnotesize new}}\end{cases}
\]
Note that with this prior there is not a common concentration parameter and therefore there is no borrowing of information across cardiac indexes as well as no Occam’s razor effect. 

\subsection{Mixture of DPs prior}
Using mixtures of DPs as prior, the locations $(\theta_1,\ldots,\theta_J)$, conditionally on $\omega$, are from a DP and the law of the partition (described in Section 2.1 of the main paper) yields the well-known predictive distributions
\[
\theta_j\,|\,\omega,\theta_1,\ldots,\theta_{j-1}\:\sim\:
\sum_{t=1}^{T_{j-1}}\frac{n_t}{j-1+\omega}\,\delta_{\theta^*_{t}}+
\frac{\omega}{j-1+\omega} G
\]
with $T_{j-1}$ the number of distinct values $\theta^*_t$ in $(\theta_1,\ldots,\theta_{j-1})$ and $n_t=\mbox{card}\{i\in\{1,\ldots,j-1\}:\: \theta_i=\theta_t^*\}$.
From this, one easily deduces that the conditional prior odds against two populations sharing the same location is
\begin{equation*}
	\label{eq:odds}
	\frac{\mathbb{P}(\theta_{j,m}\neq\theta_{j'm}\mid\omega)}{\mathbb{P}(\theta_{j,m}=\theta_{j'm}\mid\omega)}=\frac{\Pi_2^{(2)}(1,1)}{\Pi_1^{(2)}(2)}
	=\omega 
\end{equation*}
Under the mixture of DP prior, the full conditional distribution of $t_{j}$ is
\[
p(t_{j}=t\mid \bmath{t}_{\theta}^{(-j)},\bmath{\theta}^{{*(-j)}},\bm{z}_{j}, \bm{\sigma}_{j}) \propto
\begin{cases} 
	n_{t}^{-j}\,f_{\theta^*_t}(\bm{z}_{j}|\bm{\sigma}_{j}) 
	& \mbox{if } t\in \bm{t}^{(-j)} \\[4pt] 
	\omega \, 
	\displaystyle\int f_{\theta}(\bm{z}_{j}|\bm{\sigma}_{j})\, G (d\theta)& \mbox{if } t=t^{\mbox{\footnotesize new}}\end{cases}
\]
where $\bm{t}^{(-j)}=\{t_{j'}:\:j'\neq j\}$,  $\bm{\theta}^{{*(-j)}}=\{\theta^{*}_t:\:t \in \bm{t}^{(-j)}\}$ and $n_{t}^{-j}$ denotes the number of customers already allocated to table $t$, after removing the $j$-th 
customer.

Moreover, if the prior $p_\omega$ for the concentration parameter is chosen to be gamma with shape $a$ and rate $b$, the full conditional for the parameter $\omega$ can be obtained by generalizing the result for a single mixture of DPs in \cite{escobar1995}, as follows. Denote with $T_m$ the number of distinct values of $\boldsymbol{\theta}_m=\{\theta_{1,m},\ldots,\theta_{d,m}\}$, for $m=1,\ldots,M$ and note that $\omega$ depends on the data only through $T_1,\ldots,T_M$. The full conditional distribution of $\omega$ is: 
\begin{equation*}
	\begin{split}
		p(\omega \mid T_1,\ldots,T_M)&\propto p_{\omega}(\omega) \cdot \prod\limits_{m=1}^M p(T_m \mid \omega)\\
		&\propto p_{\omega}(\omega) \cdot \prod\limits_{m=1}^M \left[c_d(T_m)\, d!\, \omega^{T_m}\, \frac{\Gamma(\omega)}{\Gamma(\omega+d)}\right]
	\end{split}
\end{equation*}
where $p_{\omega}(\omega)$ is the prior density of $\omega$ and $c_d(T_m) = p(T_m\mid\omega=1)$. Therefore
\begin{equation*}
	p(\omega \mid T_1,\ldots,T_m)\propto p_{\omega}(\omega) \cdot \omega^{\sum_m T_m-M}\, (\omega+d)^M\, \prod\limits_{m=1}^M\left[\int\limits_{0}^1u^{\omega}(1-u)^{d-1}du\right]
\end{equation*}
Defining $M$ auxiliary random variables $u_m$ for $m=1,\ldots,M$ such that  $u_m\mid\omega\,\overset{iid}{\sim}\,Beta(\,\omega+1,\,d)$.
It follows that
\begin{center}
	\begin{tabular}{rcl}
		$p(\omega,\, u_1,\ldots,u_M \mid T_1,\ldots,T_m)$&$\propto$&$ p_{\omega}(\omega) \cdot \omega^{\sum_m T_m-M}\, (\omega+d)^M\, \prod\limits_{m=1}^Mu_m^{\omega}\prod\limits_{m=1}^M(1-u_m)^{d-1}$\\
	\end{tabular}
\end{center}
Finally if $p_{\omega} \equiv \text{Gamma}(a,\,b)$, then
\begin{equation*}
	\begin{split}
		p(\omega\mid u_1,\ldots,u_M, T_1,\ldots,T_m)\propto\omega^{a+\sum_m T_m-M-1}\, (\omega+d)^M\, \exp\left\{-\omega(b-\sum\limits_{m=1}^M \log(u_m))\right\}\\
		\propto\sum\limits_{v=0}^{M}{M\choose v}  \frac{d^{v}\,\Gamma\left(a+\sum\limits_{m=1}^M T_m -v\right)}
		{\left(b-\sum\limits_{m=1}^M \text{log}(u_m)\right)^{a+\sum\limits_{m=1}^M T_m -v}}\,\times
		\text{Gamma}\left(a+\sum\limits_{m=1}^M T_m-v,\,b-\sum\limits_{m=1}^M \text{log}(u_m)\right)
	\end{split}
\end{equation*}
so that the conditional distribution of $\omega$ is a mixture of $M+1$ Gamma distributions. The sampling of $\omega$ becomes

\begin{itemize}
	\item[(i)] Sample $u_m$, for $m=1,\ldots,M$, independently from
	a $\text{Beta}(\omega+1,\,J) $,
	where $J$ is the number of populations.
	\item[(ii)]  Sample $v_{\omega}$ from
	\[
	p(v_{\omega} = v \,|\, u_1,\ldots,u_m) = 
	{M\choose v}  d^{v}\,\Gamma\Big(a+\sum_{m=1}^M T_m -v\Big)\:\Big(b-\sum_{m=1}^M \text{log}(u_m)\Big)^{v} 
	\]
	for $v\in\{0,\ldots,M\}$, where $T_m$ is the number of distinct  values in $\bm{\theta}_{m}$, for $m=1,\ldots,M$.
	\item[(iii)]  Sample $\omega$ from
	$\text{Gamma}\Big(a+\sum_{m=1}^M T_m-v,\,b-\sum_{m=1}^M \text{log}(u_m)\Big)$.
\end{itemize}

\section{Additional simulations studies}

This section provides additional results obtained from simulation studies. It is divided in four sub-sections based on the data generating process (DGP) used to simulate observations.

\subsection{Generating mechanism with underlying relevant factor}
The DGP used in this sub-section is the same considered in Section 5.1 of the main paper, i.e.
\begin{equation*}
	\begin{split}
		X_{i,1}\,\overset{iid}{\sim}0.5\,\mathcal{N}(\,0,\,0.5\,)+0.5\,\mathcal{N}(\,2,\,0.5\,) \qquad&\text{for}\enskip i=1,\ldots,n_1\\
		X_{i,2}\,\overset{iid}{\sim}0.5\,\mathcal{N}(\,2,\,0.5\,)+0.5\,\mathcal{N}(\,4,\,0.5\,)\qquad&\text{for}\enskip i=1,\ldots,n_2\\
		X_{i,3}\,\overset{iid}{\sim}0.5\,\mathcal{N}(\,4,\,0.5\,)+0.5\,\mathcal{N}(\,6,\,0.5\,) \qquad&\text{for}\enskip i=1,\ldots,n_3\\
		X_{i,4}\,\overset{iid}{\sim}0.5\,\mathcal{N}(\,6,\,0.5\,)+0.5\,\mathcal{N}(\,8,\,0.5\,) \qquad&\text{for}\enskip i=1,\ldots,n_4\\
	\end{split}
\end{equation*}

\subsubsection{Inferential results from two additional randomly selected studies}
In Figures~\ref{fig2} and \ref{fig3} below, we display the plots regarding the inference for two additional randomly selected simulation studies among the 100  of Section 5.1.
Like for the simulation study already discussed in Section 5.1, the true means belong to the 95\% credible intervals and the model correctly identifies the two clusters.
\begin{figure}[H]
	\centering
	\begin{subfigure}{7cm}
		\centering
		\includegraphics[ width=7cm]{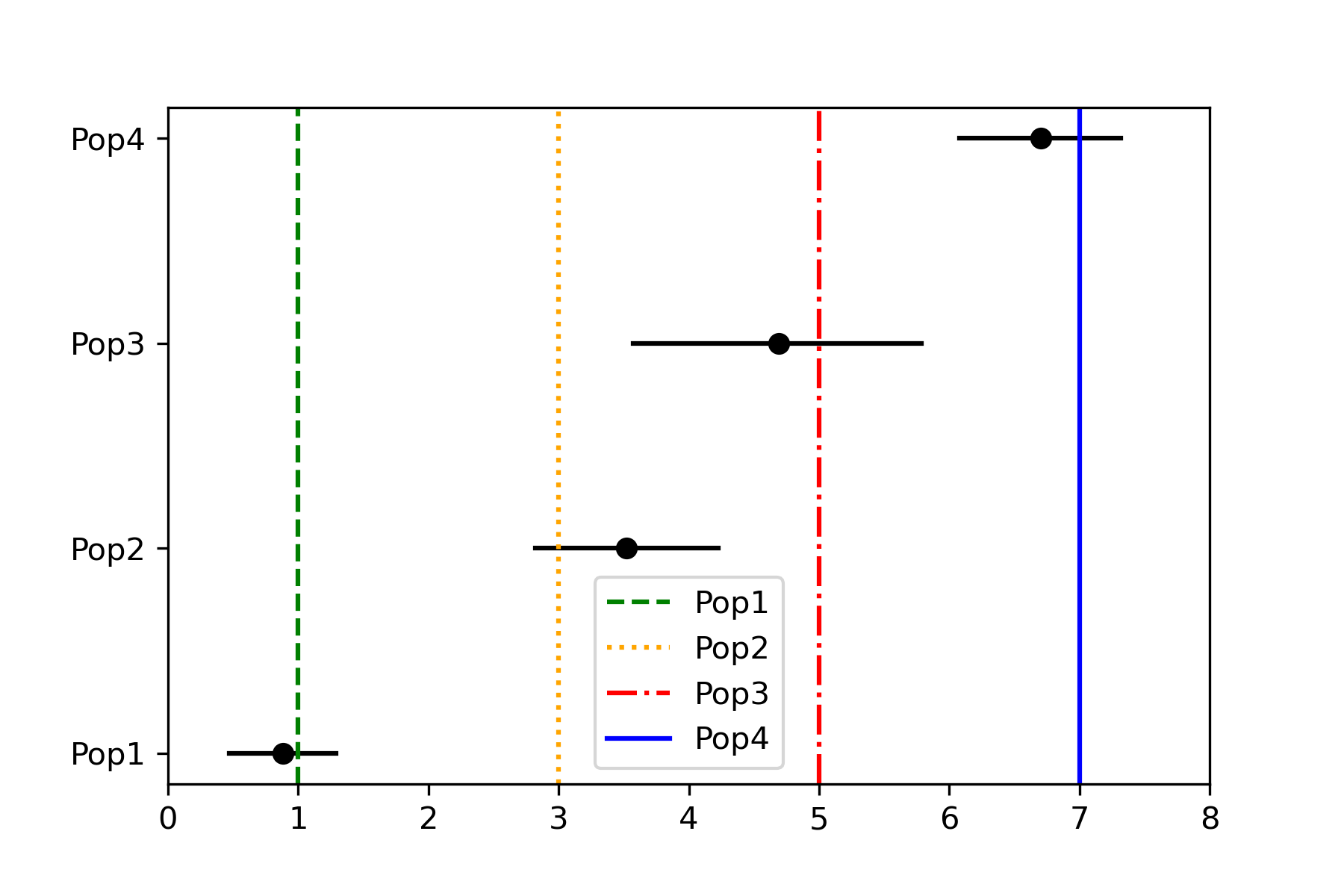}
		\caption{Inference on location parameters}
	\end{subfigure}\hspace{0.2cm}%
	\begin{subfigure}{7cm}
		\centering
		\includegraphics[width=7cm]{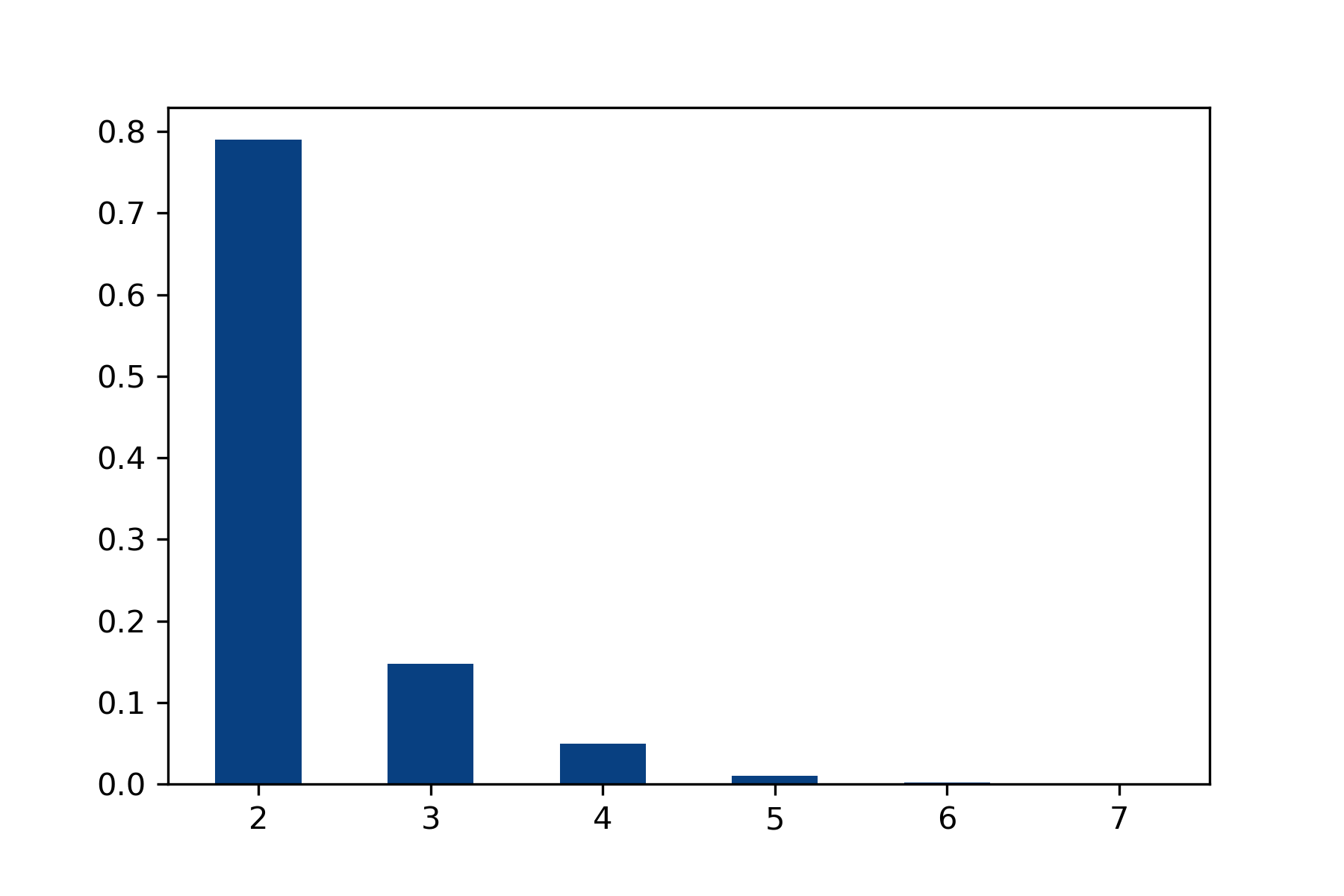}
		\caption{Number of second-level clusters.}
	\end{subfigure}
	\begin{subfigure}{7cm}
		\centering
		\includegraphics[ width=7cm, trim={2cm 0 2cm 0}]{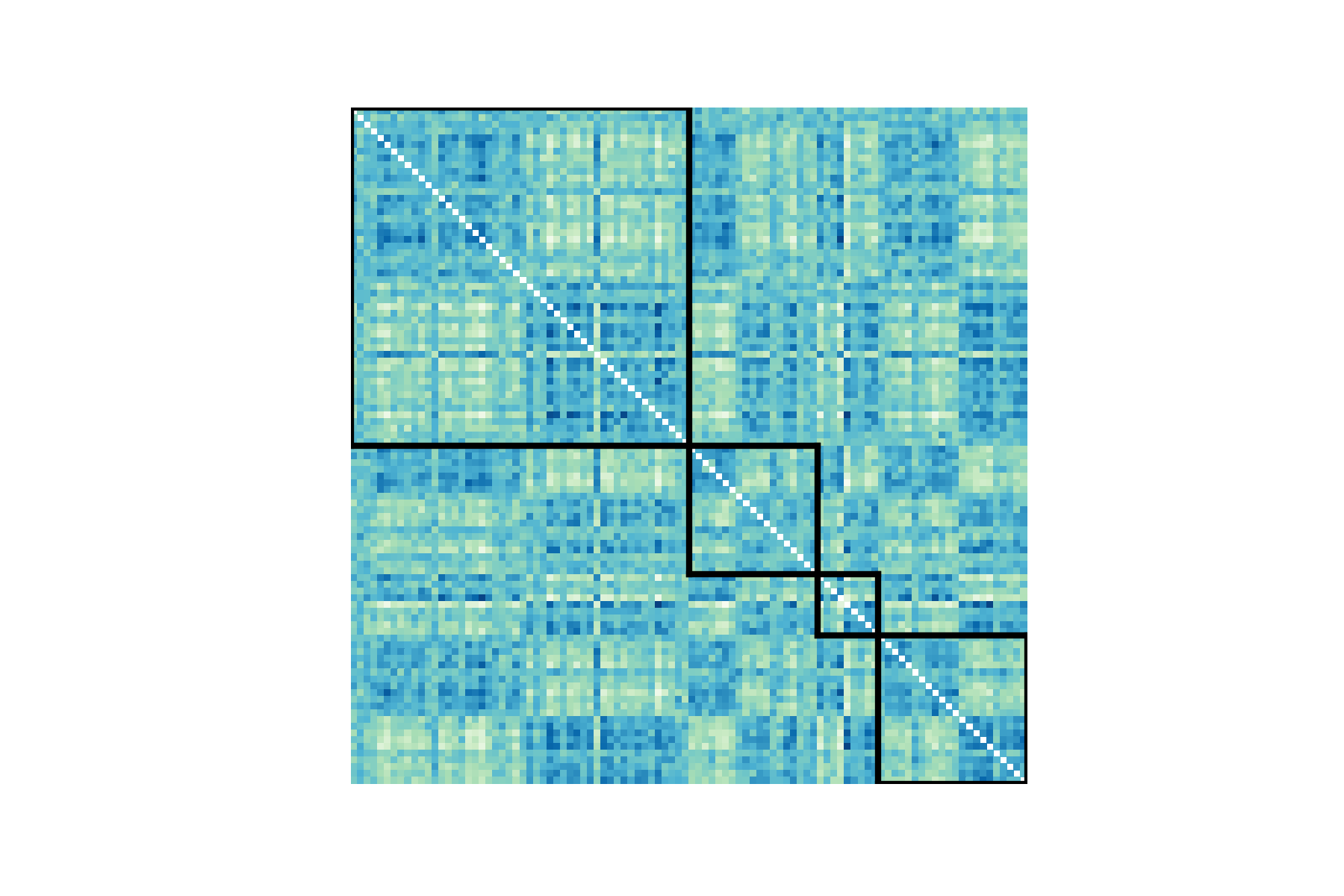}
		\caption{Co-clustering.}
	\end{subfigure}\hspace{0.2cm}%
	\begin{subfigure}{7cm}
		\centering
		\includegraphics[width=7cm,, trim={2cm 0 2cm 0}]{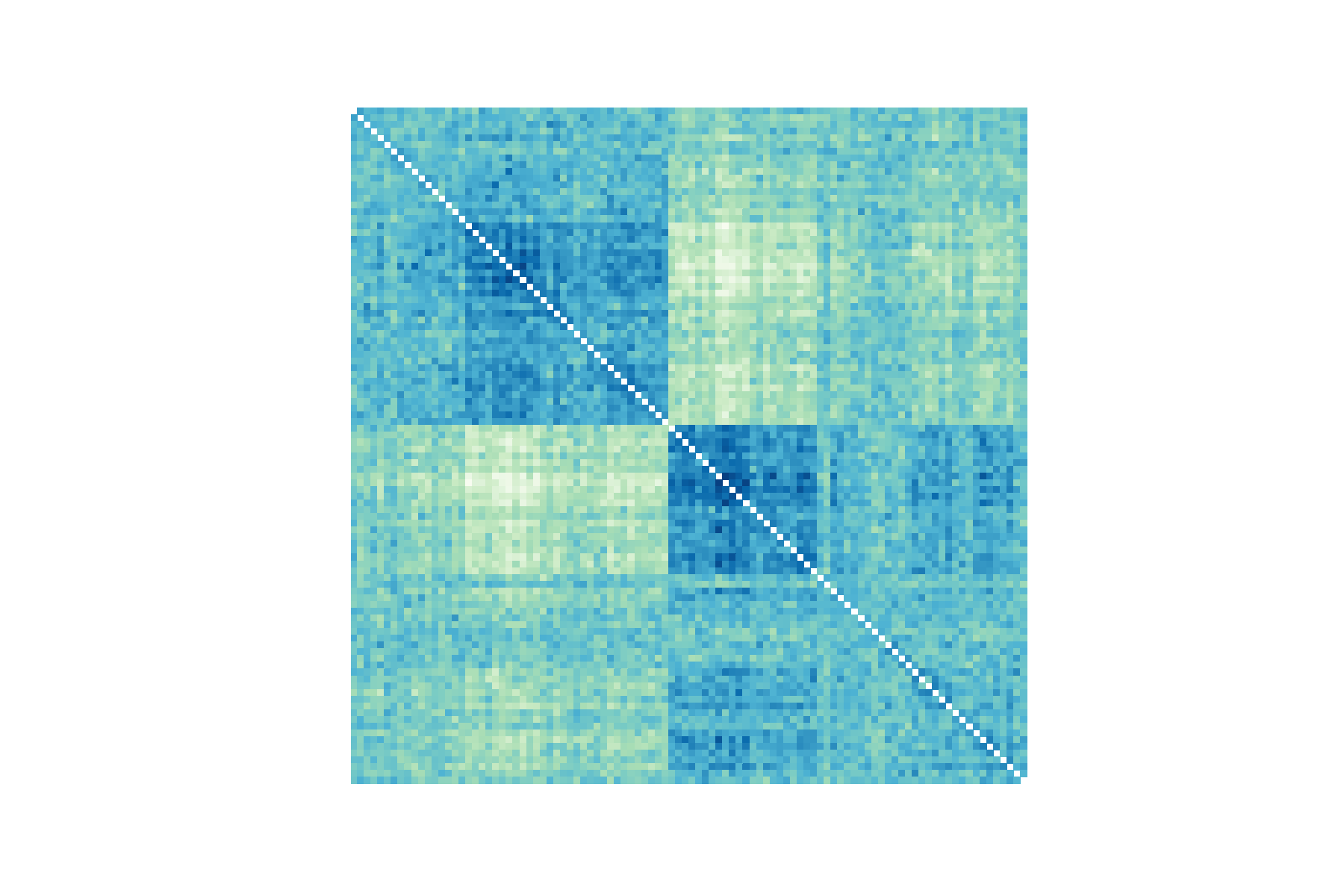}
		\caption{Co-clustering.}
	\end{subfigure}
	\caption{ Panel (a): Results of the 37th simulation study. Mean point estimates and 95\% credible intervals for the four populations, vertical lines correspond to true values. Panel (b): Posterior distribution on the number of second-level clusters. Panels (c) and (d): heatmaps of second level clustering, darker colors correspond to higher probability of co-clustering; in (c) patients are ordered based on the diagnosis and the four black squares highlight the within-sample probabilities and in (d) patients are reordered based on co-clustering probabilities.}
	\label{fig2}
\end{figure}
\begin{figure}
	\centering
	\begin{subfigure}{7cm}
		\centering
		\includegraphics[ width=7cm]{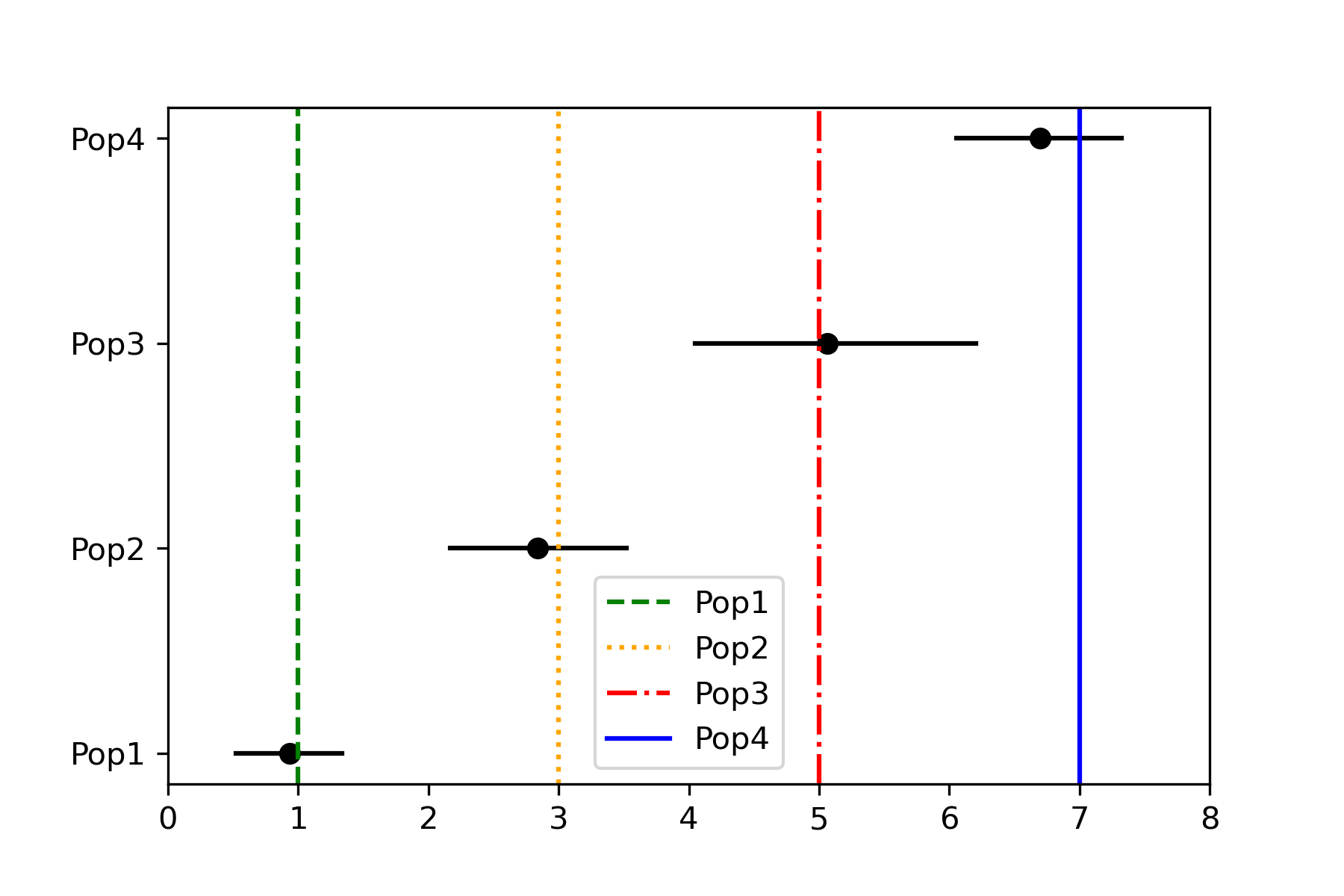}
		\caption{Inference on location parameters}
	\end{subfigure}\hspace{0.2cm}%
	\begin{subfigure}{7cm}
		\centering
		\includegraphics[width=7cm]{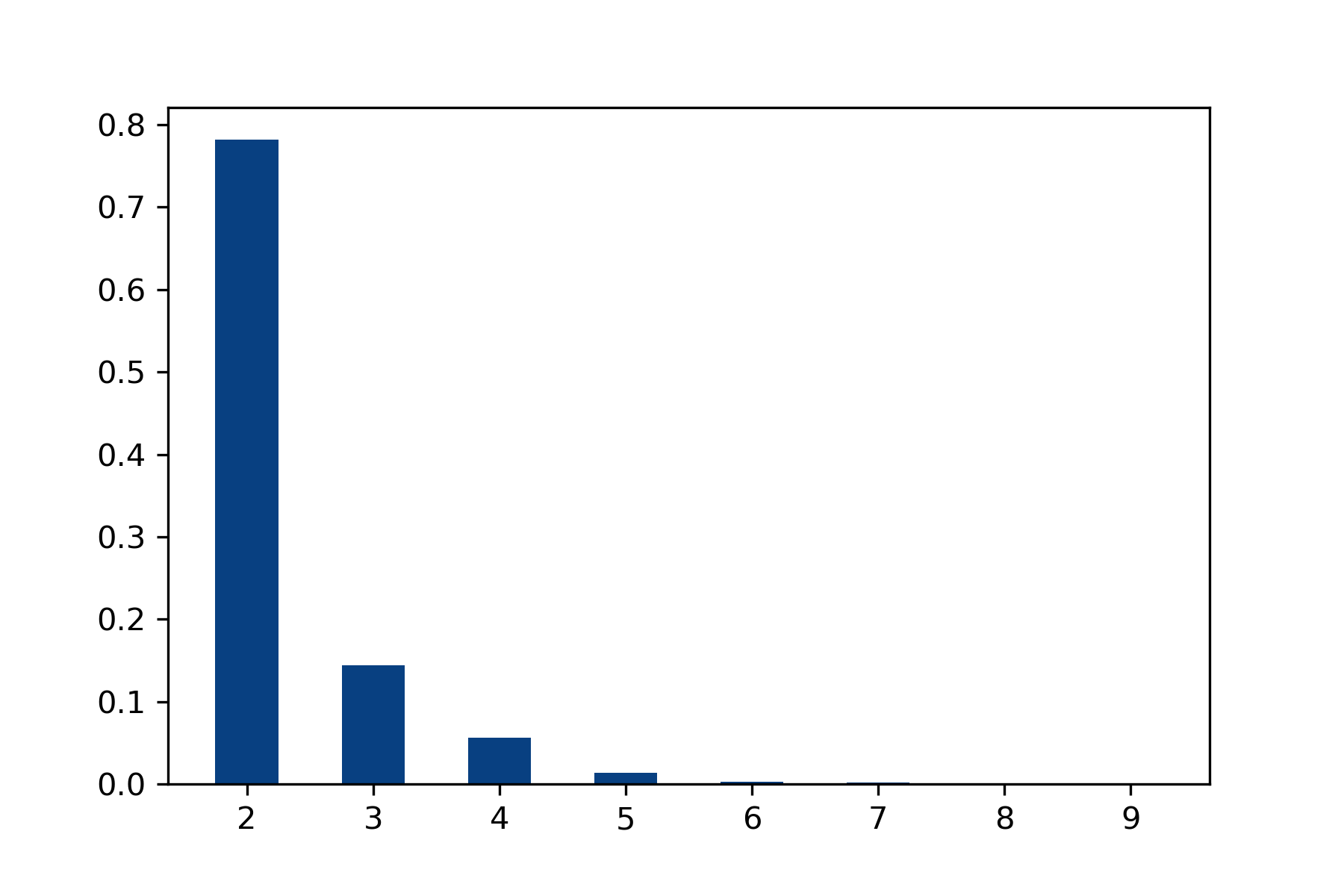}
		\caption{Number of second-level clusters.}
	\end{subfigure}
	\begin{subfigure}{7cm}
		\centering
		\includegraphics[ width=7cm, trim={2cm 0 2cm 0}]{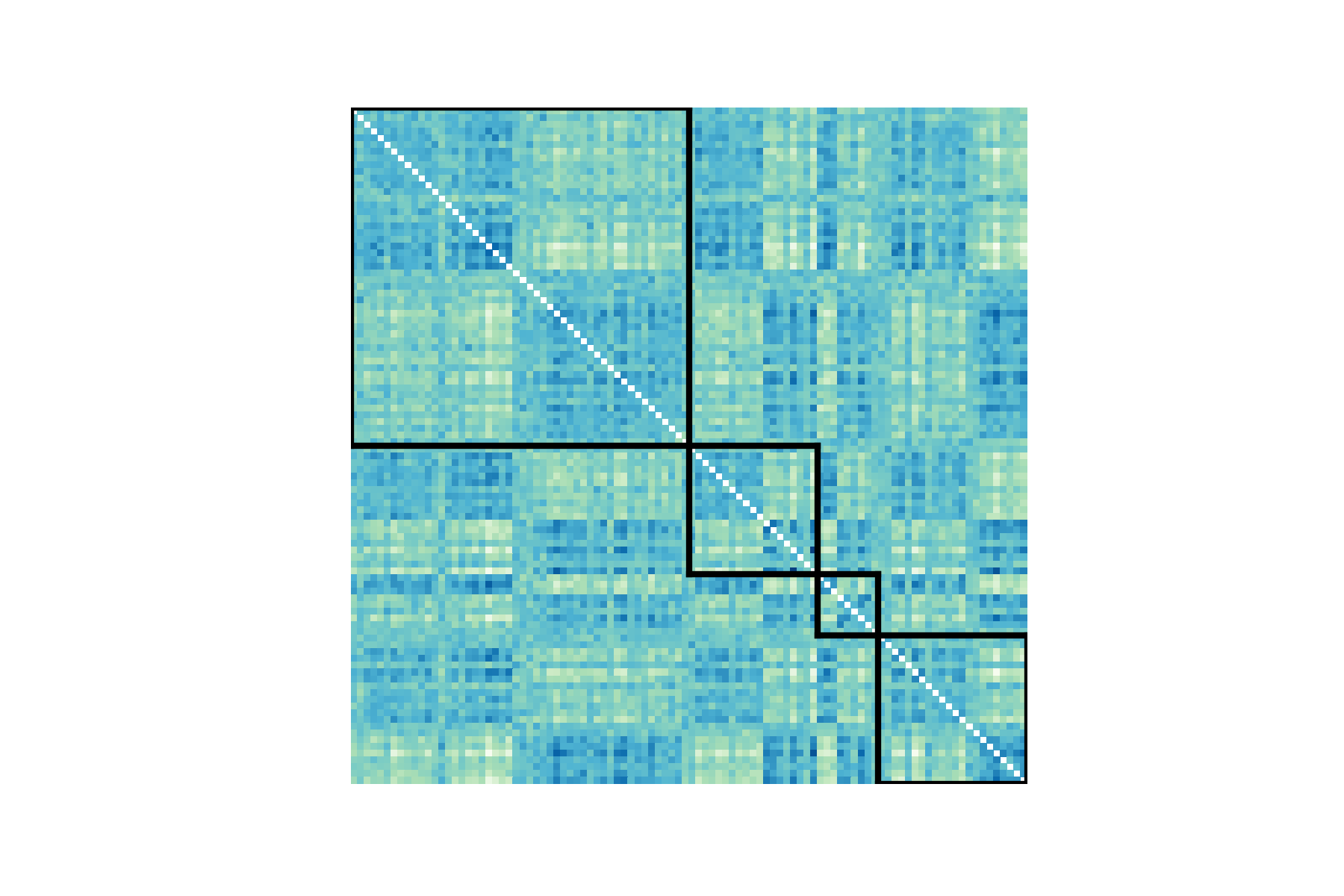}
		\caption{Co-clustering.}
	\end{subfigure}\hspace{0.2cm}%
	\begin{subfigure}{7cm}
		\centering
		\includegraphics[width=7cm,, trim={2cm 0 2cm 0}]{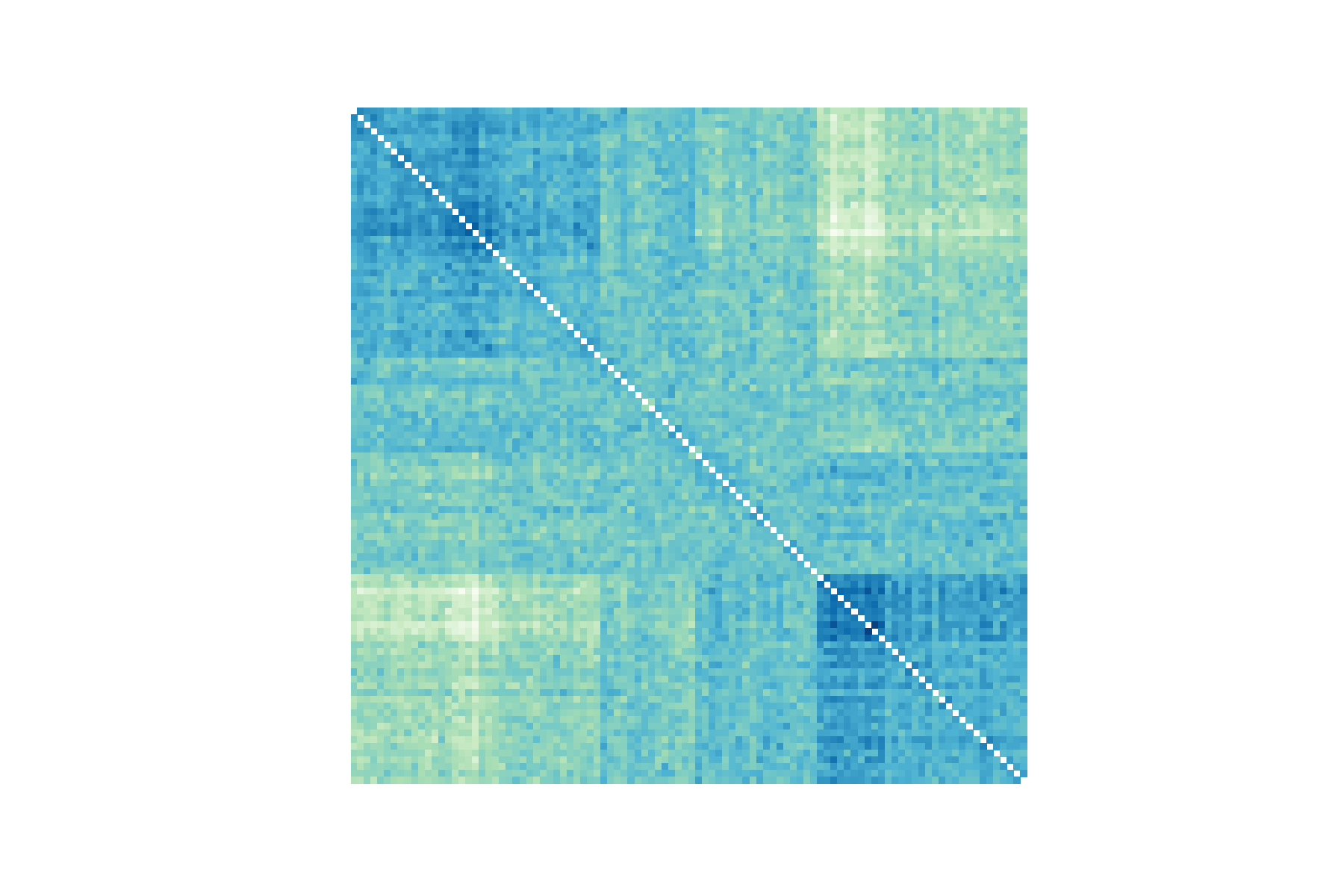}
		\caption{Co-clustering.}
	\end{subfigure}
	\caption{ Results of the 9th simulation study. Panel (a): Mean point estimates and 95\% credible intervals for the four populations, vertical lines correspond to true values. Panel (b): Posterior distribution on the number of second-level clusters. Panels (c) and (d): heatmaps of second level clustering, darker colors correspond to higher probability of co-clustering; in (c) patients are ordered based on the diagnosis and the four black squares highlight the within-sample probabilities and in (d) patients are reordered based on co-clustering probabilities.}
	\label{fig3}
\end{figure}

\subsubsection{Inference results with alternative priors}
Below we display the results obtained on the simulated data of Section 5.1 of the main paper using the alternative priors described in Section C. 
\vspace{-\baselineskip}
\begin{table}[H]
	\caption{Simulation studies summaries.}
	\vspace{-\baselineskip}
	\begin{center}
		\begin{tabular}{lcccccc}
			&\multicolumn{3}{c}{\textbf{sHDP-with mixture of DPs}}& \multicolumn{3}{c}{\textbf{sHDP-with unifor prior}}\\\hline\hline
			&MAP&Average&Median&MAP&Average&Median\\
			Partitions& count& post. prob.&post. prob.& count& post. prob.& post. prob.\\\hline
			\{\textcolor{aoe}{1},\textcolor{burntorange}{2},\textcolor{bostonuniversityred}{3},\textcolor{burgundy}{4}\}&0&0.000&0.000&0&0.000&0.000\\\hline
			\{\textcolor{aoe}{1}\}\{\textcolor{burntorange}{2},\textcolor{bostonuniversityred}{3},\textcolor{burgundy}{4}\}&0&0.000&0.000&0&0.000&0.000\\\hline
			\{\textcolor{aoe}{1},\textcolor{burntorange}{2}\}\{\textcolor{bostonuniversityred}{3},\textcolor{burgundy}{4}\}&0&0.000&0.000&0&0.000&0.000\\\hline
			\{\textcolor{aoe}{1},\textcolor{bostonuniversityred}{3},\textcolor{burgundy}{4}\}\{\textcolor{burntorange}{2}\}&0&0.000&0.000&0&0.000&0.000\\\hline
			\{\textcolor{aoe}{1}\}\{\textcolor{burntorange}{2}\}\{\textcolor{bostonuniversityred}{3},\textcolor{burgundy}{4}\}&5&0.083&0.022&0&0.030&0.009\\\hline
			\{\textcolor{aoe}{1},\textcolor{burntorange}{2},\textcolor{bostonuniversityred}{3}\}\{\textcolor{burgundy}{4}\}&0&0.000&0.000&0&0.001&0.000\\\hline
			\{\textcolor{aoe}{1},\textcolor{burgundy}{4}\}\{\textcolor{burntorange}{2},\textcolor{bostonuniversityred}{3}\}&0&0.000&0.000&0&0.000&0.000\\\hline
			\{\textcolor{aoe}{1}\}\{\textcolor{burntorange}{2},\textcolor{bostonuniversityred}{3}\}\{\textcolor{burgundy}{4}\}&2&0.056&0.012&1&0.051&0.014\\\hline
			\{\textcolor{aoe}{1},\textcolor{bostonuniversityred}{3}\}\{\textcolor{burntorange}{2},\textcolor{burgundy}{4}\}&0&0.000&0.000&0&0.000&0.000\\\hline
			\{\textcolor{aoe}{1},\textcolor{burntorange}{2},\textcolor{burgundy}{4}\}\{\textcolor{bostonuniversityred}{3}\}&0&0.000&0.000&0&0.000&0.000\\\hline
			\{\textcolor{aoe}{1}\}\{\textcolor{burntorange}{2},\textcolor{burgundy}{4}\}\{\textcolor{bostonuniversityred}{3}\}&0&0.000&0.000&0&0.000&0.000\\\hline
			\{\textcolor{aoe}{1},\textcolor{burntorange}{2}\}\{\textcolor{bostonuniversityred}{3}\}\{\textcolor{burgundy}{4}\}&0&0.002&0.000&0&0.003&0.000\\\hline
			\{\textcolor{aoe}{1},\textcolor{bostonuniversityred}{3}\}\{\textcolor{burntorange}{2}\}\{\textcolor{burgundy}{4}\}&0&0.000&0.000&0&0.000&0.000\\\hline
			\{\textcolor{aoe}{1},\textcolor{burgundy}{4}\}\{\textcolor{burntorange}{2}\}\{\textcolor{bostonuniversityred}{3}\}&0&0.000&0.000&0&0.000&0.000\\\hline
			\{\textcolor{aoe}{1}\}\{\textcolor{burntorange}{2}\}\{\textcolor{bostonuniversityred}{3}\}\{\textcolor{burgundy}{4}\}&\textbf{93}&\textbf{0.859}&\textbf{0.918}&\textbf{99}&\textbf{0.916}&\textbf{0.956}
		\end{tabular}
	\end{center}
	\vspace{-\baselineskip}
\end{table}

Both models perform better than the NDP, whose results are in Table 1 of the main paper, confirming the advantages of location-based clustering in presence of small sample sizes, when compared to distribution-based clustering. Moreover, sHDP-with mixture of DPs has a slightly worse performance with respect to our main proposal; this was expected, since the corresponding prior incorporates less information and ignores the natural order of the four populations. 

\subsection{Generating mechanism with outliers}
We present here a simulation study with a twofold goal: (1) compare again the location--based clustering approach of our proposal with the distribution--based clustering approach of the nested Dirichlet process (NDP) under a different DGP; (2) study the performance of our model in presence of outliers. The simulated data have been sampled according to the following DGP
\begin{equation*}
	\begin{tabular}{l l l}
		\textbf{DGP 1: } & $X_{i,1}\,\overset{iid}{\sim}\mathcal{N}(\,0,\,0.5\,) \quad$&$\text{for}\enskip i=1,\ldots,n_1-1$\\
		&$X_{n_1,1}\,\sim\mathcal{N}(\,4,\,0.5\,)\quad$&\\
		&$X_{i,2}\,\overset{iid}{\sim}\mathcal{N}(\,1,\,0.5\,) \quad$&$\text{for}\enskip i=1,\ldots,n_2$\\
		&$X_{i,3}\,\overset{iid}{\sim}\mathcal{N}(\,1,\,0.5\,) \quad$&$\text{for}\enskip i=1,\ldots,n_3$\\
		&$X_{i,4}\,\overset{iid}{\sim}\mathcal{N}(\,2,\,0.5\,) \quad$&$\text{for}\enskip i=1,\ldots,n_4$
	\end{tabular}
\end{equation*}
Thus, the true partition is $\{1\},\{2,3\},\{4\}$. Moreover, there is one outlier in the first sample.

\begin{table}[H]
	\caption{Posterior probabilities over the space of partitions.}
	\label{tab:A}
	\begin{center}
		\begin{tabular}{l|cc}
			&s-HDP&NDP\\\hline
			\{\textcolor{aoe}{1},\textcolor{burntorange}{2},\textcolor{bostonuniversityred}{3},\textcolor{burgundy}{4}\}&0&0\\\hline
			\{\textcolor{aoe}{1}\}\{\textcolor{burntorange}{2},\textcolor{bostonuniversityred}{3},\textcolor{burgundy}{4}\}&0&0\\\hline
			\{\textcolor{aoe}{1},\textcolor{burntorange}{2}\}\{\textcolor{bostonuniversityred}{3},\textcolor{burgundy}{4}\}&0&0\\\hline
			\{\textcolor{aoe}{1},\textcolor{bostonuniversityred}{3},\textcolor{burgundy}{4}\}\{\textcolor{burntorange}{2}\}&0&0\\\hline
			\{\textcolor{aoe}{1}\}\{\textcolor{burntorange}{2}\}\{\textcolor{bostonuniversityred}{3},\textcolor{burgundy}{4}\}&0&0\\\hline
			\{\textcolor{aoe}{1},\textcolor{burntorange}{2},\textcolor{bostonuniversityred}{3}\}\{\textcolor{burgundy}{4}\}&0.013&\textbf{0.980}\\\hline
			\{\textcolor{aoe}{1},\textcolor{burgundy}{4}\}\{\textcolor{burntorange}{2},\textcolor{bostonuniversityred}{3}\}&0&0\\\hline
			\{\textcolor{aoe}{1}\}\{\textcolor{burntorange}{2},\textcolor{bostonuniversityred}{3}\}\{\textcolor{burgundy}{4}\}&\textbf{0.771}&0.010\\\hline
			\{\textcolor{aoe}{1},\textcolor{bostonuniversityred}{3}\}\{\textcolor{burntorange}{2},\textcolor{burgundy}{4}\}&0&0\\\hline
			\{\textcolor{aoe}{1},\textcolor{burntorange}{2},\textcolor{burgundy}{4}\}\{\textcolor{bostonuniversityred}{3}\}&0&0\\\hline
			\{\textcolor{aoe}{1}\}\{\textcolor{burntorange}{2},\textcolor{burgundy}{4}\}\{\textcolor{bostonuniversityred}{3}\}&0&0\\\hline
			\{\textcolor{aoe}{1},\textcolor{burntorange}{2}\}\{\textcolor{bostonuniversityred}{3}\}\{\textcolor{burgundy}{4}\}&0.006&0.020\\\hline
			\{\textcolor{aoe}{1},\textcolor{bostonuniversityred}{3}\}\{\textcolor{burntorange}{2}\}\{\textcolor{burgundy}{4}\}&0&0\\\hline
			\{\textcolor{aoe}{1},\textcolor{burgundy}{4}\}\{\textcolor{burntorange}{2}\}\{\textcolor{bostonuniversityred}{3}\}&0&0\\\hline
			\{\textcolor{aoe}{1}\}\{\textcolor{burntorange}{2}\}\{\textcolor{bostonuniversityred}{3}\}\{\textcolor{burgundy}{4}\}&0.210&0\\\hline
		\end{tabular}
	\end{center}
	\vspace{-\baselineskip}
\end{table}
Table~\ref{tab:A} displays the posterior probabilities obtained using our model (s-HDP) and the NDP. Our model largely outperforms the competitor.

\begin{figure}[H]
	\vspace{-\baselineskip}
	\centering
	\begin{subfigure}{7cm}
		\centering
		\includegraphics[ width=7cm, trim={2cm 0 2cm 0}]{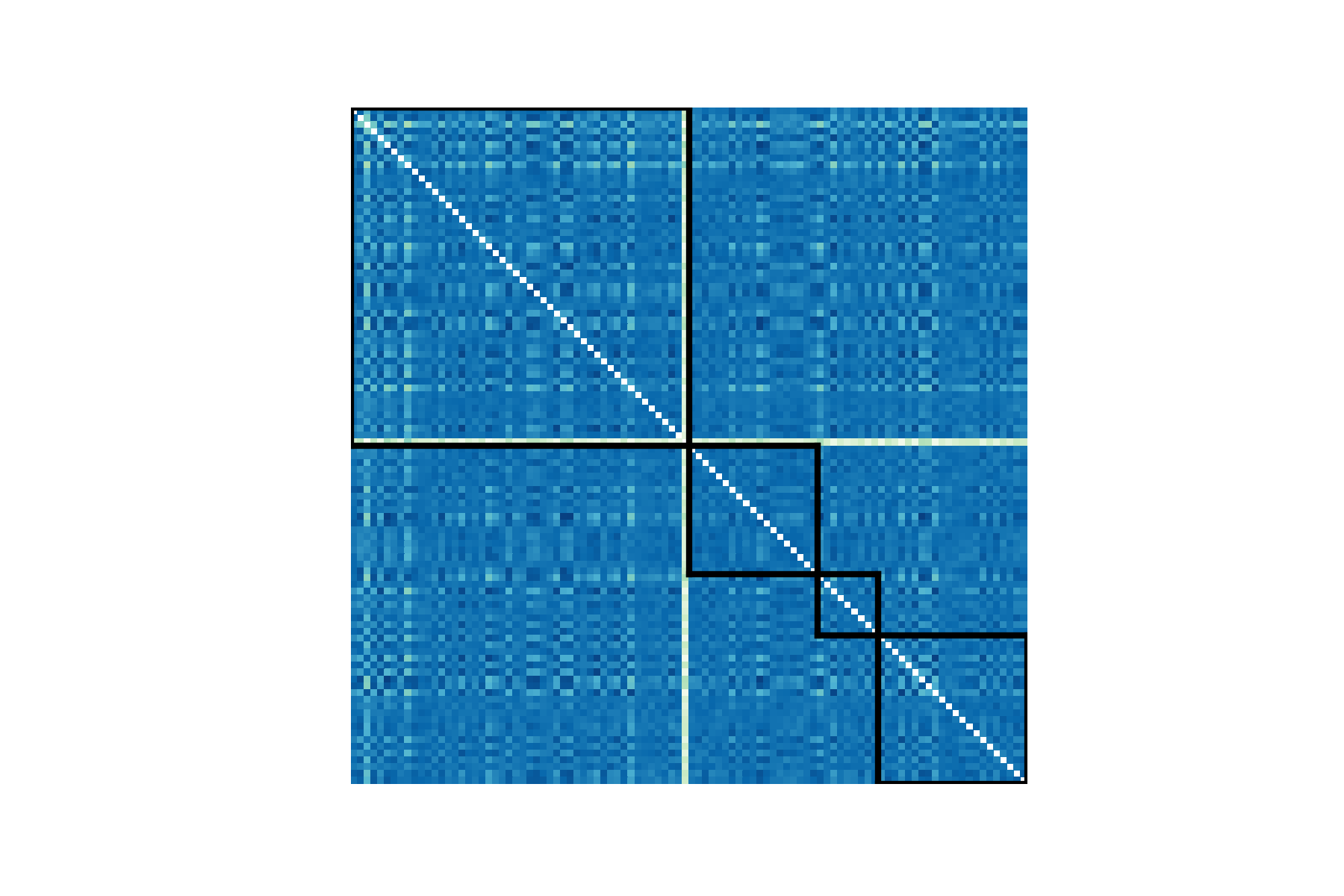}
		\caption{Co-clustering.}
		\label{fig:figAa}
	\end{subfigure}\hspace{0.2cm}%
	\begin{subfigure}{7cm}
		\centering
		\includegraphics[width=7cm,, trim={2cm 0 2cm 0}]{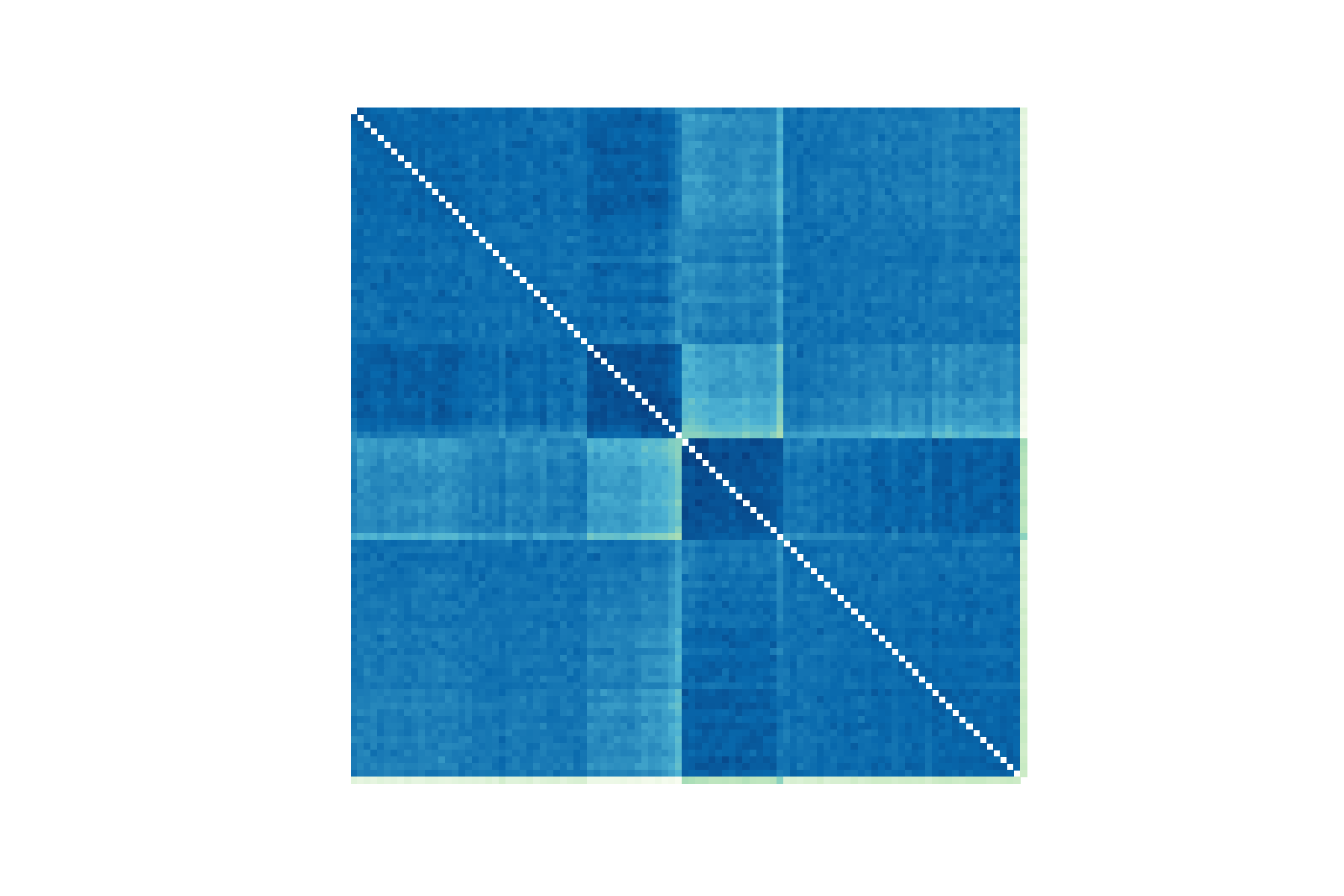}
		\caption{Co-clustering.}
		\label{fig:figAb}
	\end{subfigure}
	\caption{Posterior similarity matrices for the simulation study under DGP 1. In (a) patients are ordered based on the diagnosis and the four black squares highlight the within-sample probabilities; in (b) patients are reordered based on co-clustering probabilities.}
	\label{fig:A}
	\vspace{-\baselineskip}
\end{figure}

Figure~\ref{fig:A} shows the posterior co-clustering probabilities obtained in the simulations study. Our proposal is able to correctly identify the outlier.

\subsection{Generating mechanism with non-location effects}
\begin{figure}[H]
	\vspace{-0.5\baselineskip}
	\centering
	\begin{subfigure}{7cm}
		\centering
		\includegraphics[ width=7cm, trim={2cm 0 2cm 0}]{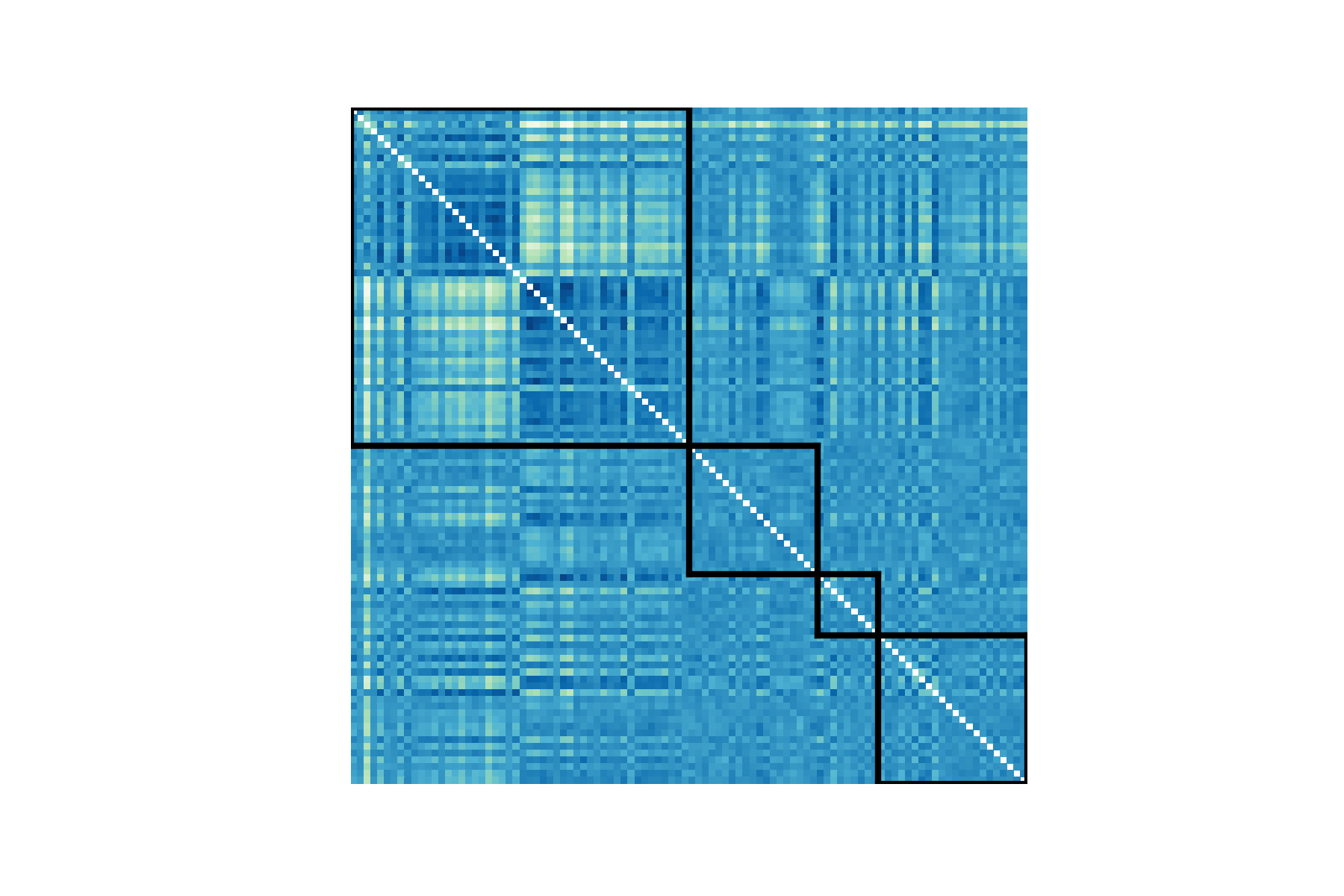}
		\caption{Co-clustering.}
		\label{fig:figBa}
	\end{subfigure}\hspace{0.2cm}%
	\begin{subfigure}{7cm}
		\centering
		\includegraphics[width=7cm,, trim={2cm 0 2cm 0}]{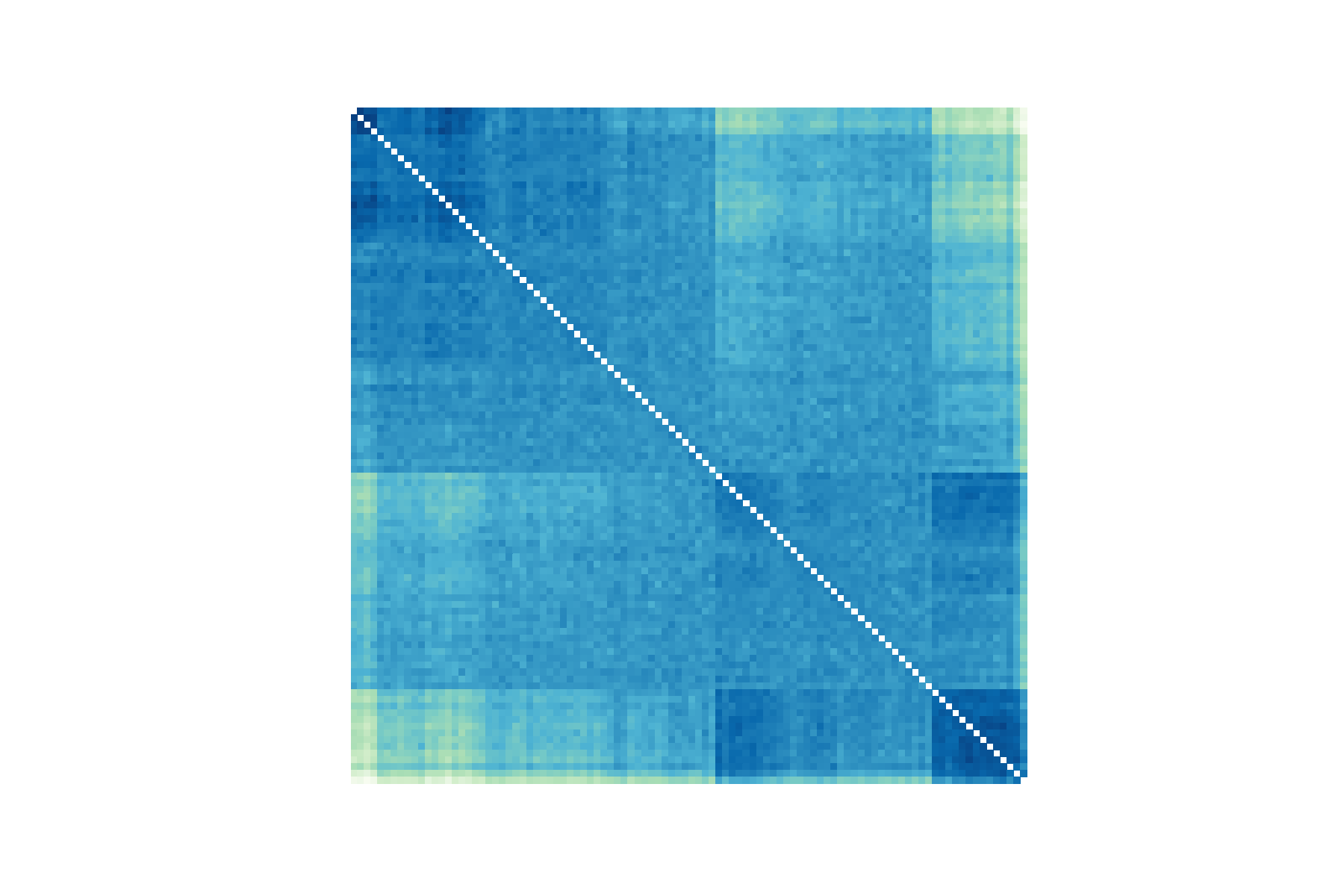}
		\caption{Co-clustering.}
		\label{fig:figBb}
	\end{subfigure}
	\caption{Posterior similarity matrices for the simulation study under DGP 2. In (a) patients are ordered based on the diagnosis and the four black squares highlight the within-sample probabilities; in (b) patients are reordered based on co-clustering probabilities.}
	\label{fig:B}
\end{figure}

We present a simulation study with a twofold goal: (1) compare again the location--based clustering approach of our proposal with the distribution--based clustering approach of the nested Dirichlet process (NDP) under a different DGP; (2) study the performance of our model for the case of heterogeneity between populations not being fully explained by shift in locations. The simulated data have been sampled according to the following DGP.

\begin{equation*}
	\begin{tabular}{l l l}
		\textbf{DGP 2:} & $X_{i,1}\,\overset{iid}{\sim}\,0.5\,\mathcal{N}(\,-1,\,0.5\,) + 0.5\,\mathcal{N}(\,1,\,0.5\,) \qquad$&$\text{for}\enskip i=1,\ldots,n_1$\\
		&$X_{i,2}\,\overset{iid}{\sim}\mathcal{N}(\,1,\,0.5\,) \qquad$&$\text{for}\enskip i=1,\ldots,n_2$\\
		&$X_{i,3}\,\overset{iid}{\sim}\mathcal{N}(\,1,\,0.5\,) \qquad$&$\text{for}\enskip i=1,\ldots,n_3$\\
		&$X_{i,4}\,\overset{iid}{\sim}\mathcal{N}(\,2,\,0.5\,) \qquad$&$\text{for}\enskip i=1,\ldots,n_4$\\
	\end{tabular}
\end{equation*}
Thus, the true partition is $\{1\},\{2,3\},\{4\}$. Moreover, the relative effect of the first population w.r.t.~the others is not fully explained by the shift of the location, since the whole distribution is different and not only the mean. Table~\ref{tab:B} displays the posterior probabilities obtained using our model (s-HDP) and the NDP. Our model largely outperforms the competitor. Figure~\ref{fig:B} shows the posterior co-clustering probabilities. Our proposal is able to correctly identify the non--location effect (see Figure~\ref{fig:B}(a)).

\begin{table}[H]
	\caption{Posterior probabilities over the space of partitions.}
	\label{tab:B}
	\begin{center}
		\vspace{-\baselineskip}
		\begin{tabular}{l|cc}
			&s-HDP&NDP\\\hline
			\{\textcolor{aoe}{1},\textcolor{burntorange}{2},\textcolor{bostonuniversityred}{3},\textcolor{burgundy}{4}\}&0&0\\\hline
			\{\textcolor{aoe}{1}\}\{\textcolor{burntorange}{2},\textcolor{bostonuniversityred}{3},\textcolor{burgundy}{4}\}&0.001&0\\\hline
			\{\textcolor{aoe}{1},\textcolor{burntorange}{2}\}\{\textcolor{bostonuniversityred}{3},\textcolor{burgundy}{4}\}&0&0\\\hline
			\{\textcolor{aoe}{1},\textcolor{bostonuniversityred}{3},\textcolor{burgundy}{4}\}\{\textcolor{burntorange}{2}\}&0&0\\\hline
			\{\textcolor{aoe}{1}\}\{\textcolor{burntorange}{2}\}\{\textcolor{bostonuniversityred}{3},\textcolor{burgundy}{4}\}&0.001&0\\\hline
			\{\textcolor{aoe}{1},\textcolor{burntorange}{2},\textcolor{bostonuniversityred}{3}\}\{\textcolor{burgundy}{4}\}&0.058&\textbf{0.98}\\\hline
			\{\textcolor{aoe}{1},\textcolor{burgundy}{4}\}\{\textcolor{burntorange}{2},\textcolor{bostonuniversityred}{3}\}&0&0\\\hline
			\{\textcolor{aoe}{1}\}\{\textcolor{burntorange}{2},\textcolor{bostonuniversityred}{3}\}\{\textcolor{burgundy}{4}\}&\textbf{0.706}&0.01\\\hline
			\{\textcolor{aoe}{1},\textcolor{bostonuniversityred}{3}\}\{\textcolor{burntorange}{2},\textcolor{burgundy}{4}\}&0&0\\\hline
			\{\textcolor{aoe}{1},\textcolor{burntorange}{2},\textcolor{burgundy}{4}\}\{\textcolor{bostonuniversityred}{3}\}&0&0\\\hline
			\{\textcolor{aoe}{1}\}\{\textcolor{burntorange}{2},\textcolor{burgundy}{4}\}\{\textcolor{bostonuniversityred}{3}\}&0&0\\\hline
			\{\textcolor{aoe}{1},\textcolor{burntorange}{2}\}\{\textcolor{bostonuniversityred}{3}\}\{\textcolor{burgundy}{4}\}&0.019&0.02\\\hline
			\{\textcolor{aoe}{1},\textcolor{bostonuniversityred}{3}\}\{\textcolor{burntorange}{2}\}\{\textcolor{burgundy}{4}\}&0&0\\\hline
			\{\textcolor{aoe}{1},\textcolor{burgundy}{4}\}\{\textcolor{burntorange}{2}\}\{\textcolor{bostonuniversityred}{3}\}&0&0\\\hline
			\{\textcolor{aoe}{1}\}\{\textcolor{burntorange}{2}\}\{\textcolor{bostonuniversityred}{3}\}\{\textcolor{burgundy}{4}\}&0.214&0\\\hline
		\end{tabular}
	\end{center}
	\vspace{-\baselineskip}
\end{table}

\subsection{Simulation studies under non-symmetric data generating process}
\begin{figure}[H]
	\vspace{-\baselineskip}
	\centering
	\begin{subfigure}{7cm}
		\centering
		\includegraphics[ width=7cm, trim={2cm 0 2cm 0}]{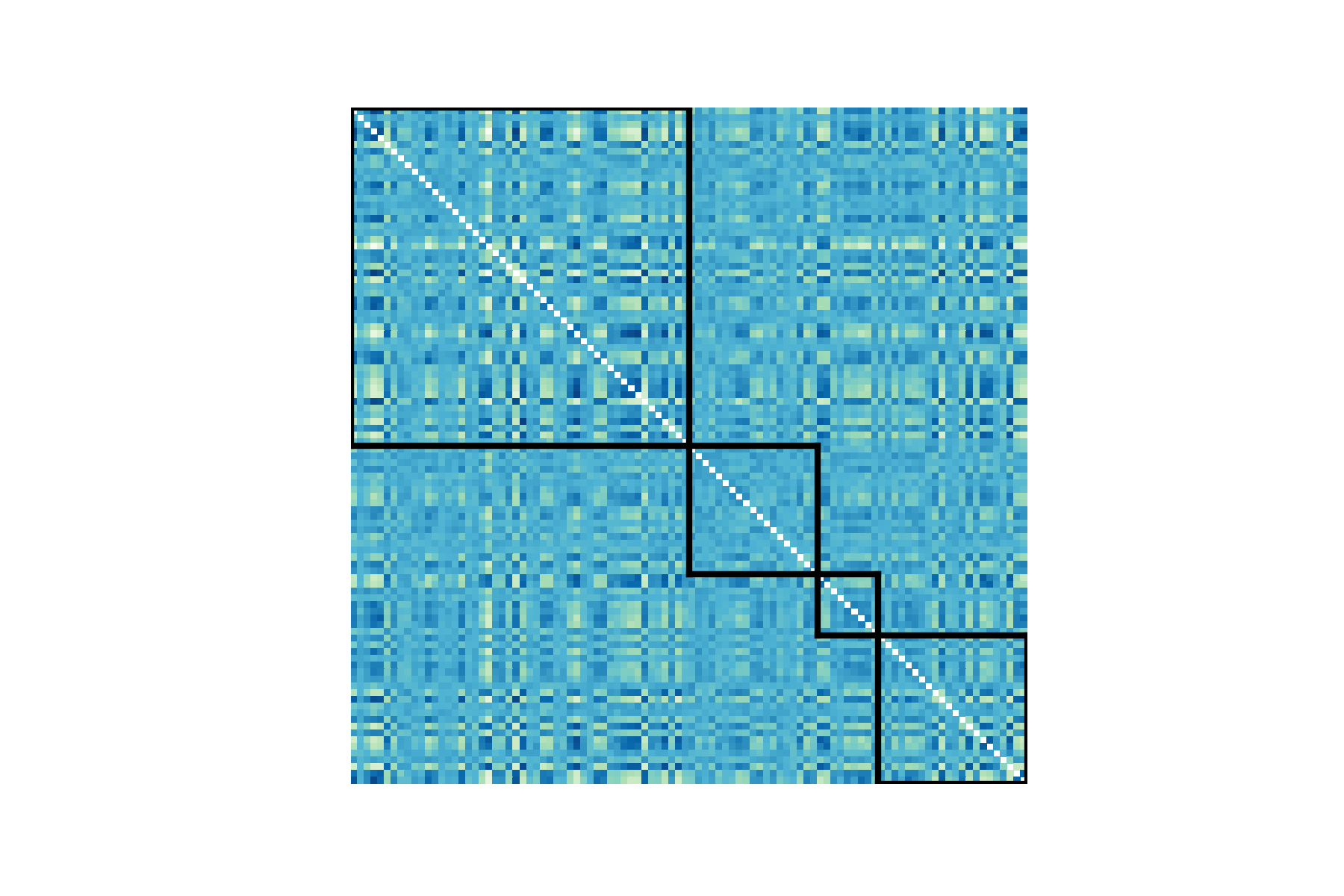}
		\caption{Co-clustering DGP 3.}
		\label{fig:figCa}
	\end{subfigure}\hspace{0.2cm}%
	\begin{subfigure}{7cm}
		\centering
		\includegraphics[width=7cm,, trim={2cm 0 2cm 0}]{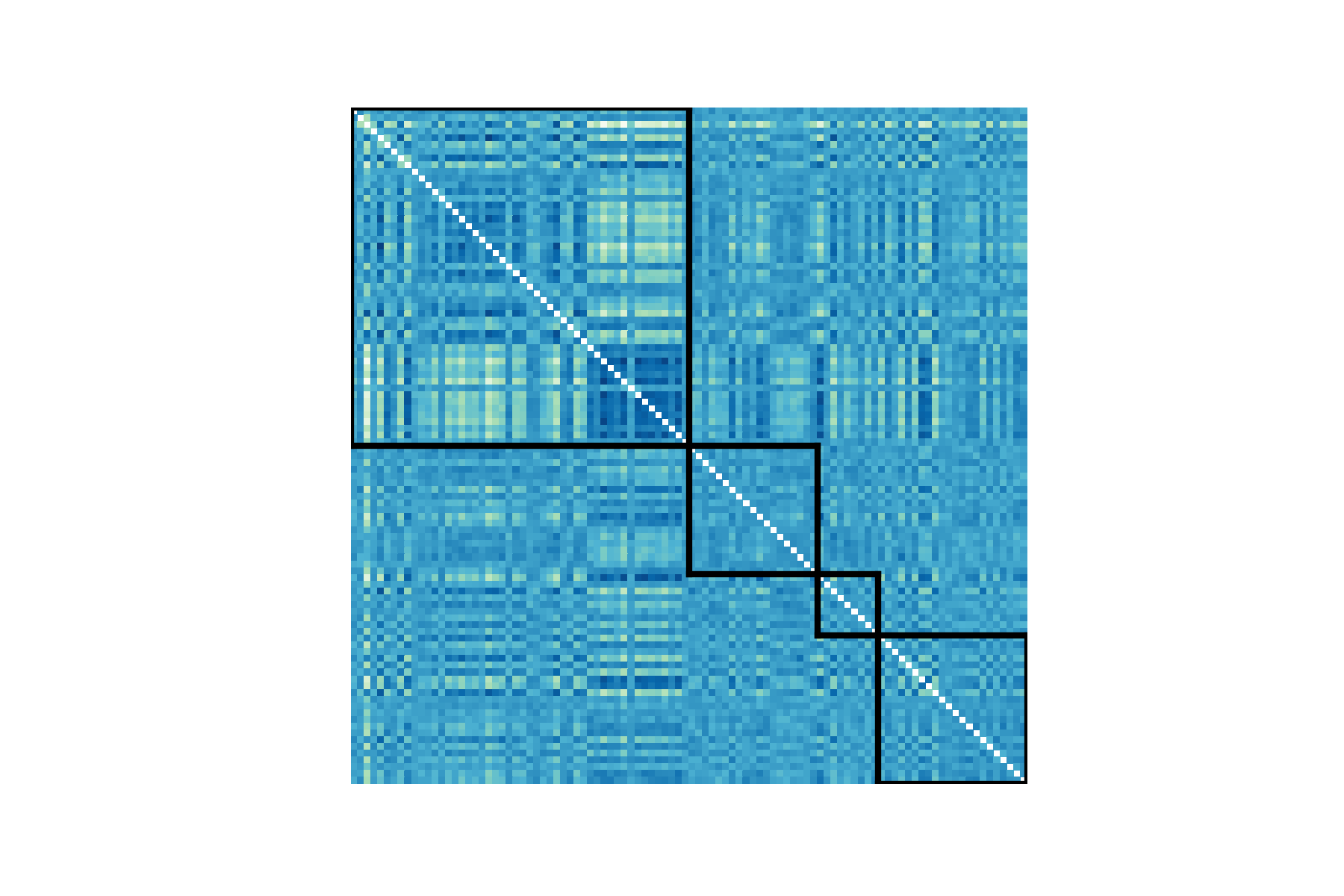}
		\caption{Co-clustering DGP 4.}
		\label{fig:figCb}
	\end{subfigure}
	\begin{subfigure}{7cm}
		\centering
		\includegraphics[width=7cm,, trim={2cm 0 2cm 0}]{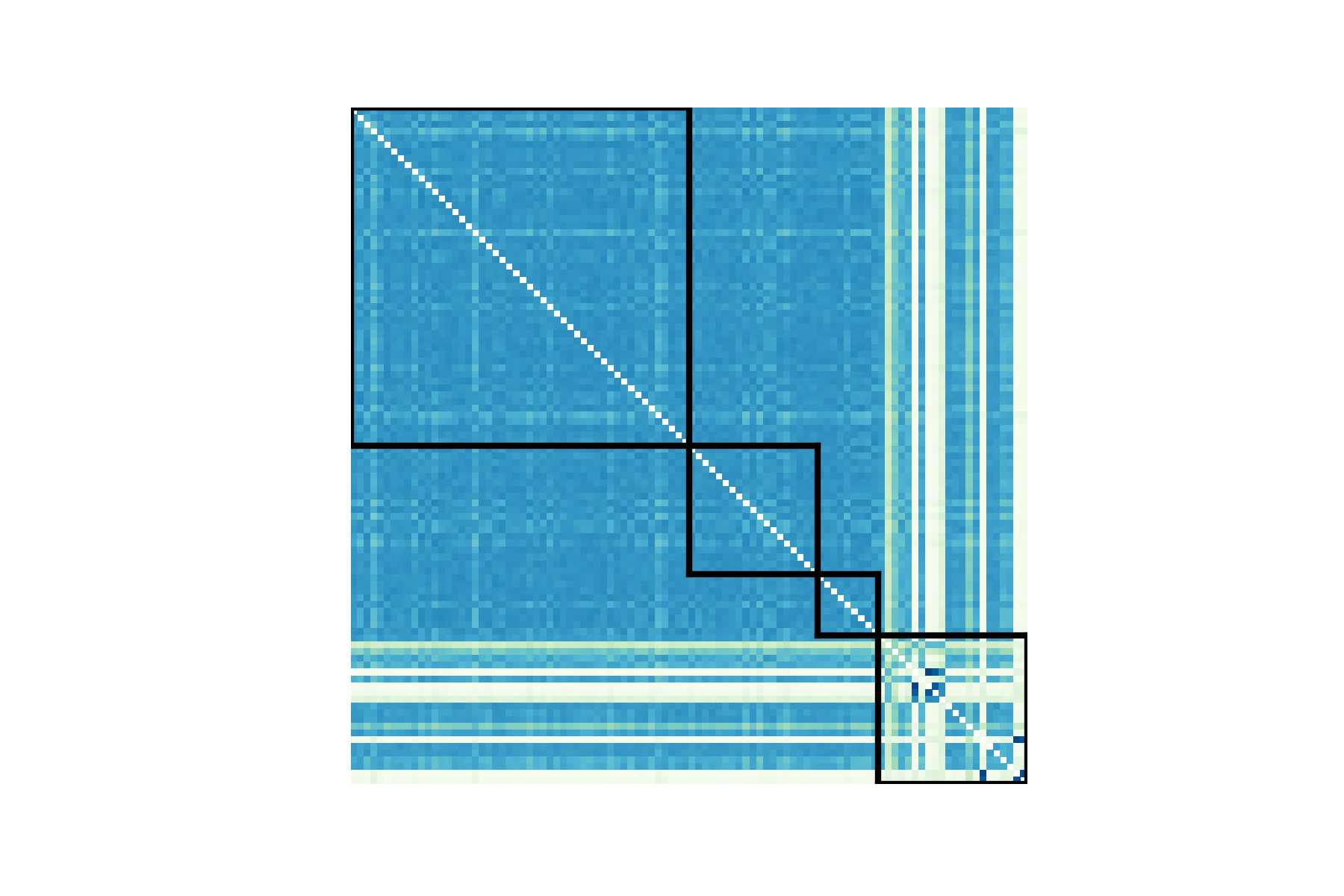}
		\caption{Co-clustering DGP 5.}
		\label{fig:figCc}
	\end{subfigure}
	\caption{Posterior similarity matrices under DGP 3-4-5. Patients are ordered based on the diagnosis and the four black squares highlight the within-sample probabilities.}
	\label{fig:C}
\end{figure}

Here we report three simulation studies to check the performance of the model in presence of deviations from symmetry. The simulated data have been sampled according to the following DGPs.

\begin{equation*}
	\begin{tabular}{l l l}
		\textbf{DGP 3:} & $X_{i,1}\,\overset{iid}{\sim}\mathcal{N}(0,\,0.5\,)$&$\text{for}\enskip i=1,\ldots,n_1$\\
		&$X_{i,2}\,\overset{iid}{\sim}\text{Gamma}(\,3,\,3) \qquad$&$\text{for}\enskip i=1,\ldots,n_2$\\
		&$X_{i,3}\,\overset{iid}{\sim}\mathcal{N}(\,1,\,0.5\,) \qquad$&$\text{for}\enskip i=1,\ldots,n_3$\\
		&$X_{i,4}\,\overset{iid}{\sim}\mathcal{N}(\,2,\,0.5\,) \qquad$&$\text{for}\enskip i=1,\ldots,n_4$\\
	\end{tabular}
\end{equation*}

\begin{equation*}
	\begin{tabular}{l l l}
		\textbf{DGP 4:} & $X_{i,1}\,\overset{iid}{\sim}\,0.7\,\mathcal{N}(\,-1,\,0.5\,) + 0.3\,\mathcal{N}(\,1,\,0.5\,) \qquad$&$\text{for}\enskip i=1,\ldots,n_1$\\
		&$X_{i,2}\,\overset{iid}{\sim}\mathcal{N}(\,1,\,0.5\,) \qquad$&$\text{for}\enskip i=1,\ldots,n_2$\\
		&$X_{i,3}\,\overset{iid}{\sim}\mathcal{N}(\,1,\,0.5\,) \qquad$&$\text{for}\enskip i=1,\ldots,n_3$\\
		&$X_{i,4}\,\overset{iid}{\sim}\mathcal{N}(\,2,\,0.5\,) \qquad$&$\text{for}\enskip i=1,\ldots,n_4$\\
	\end{tabular}
\end{equation*}

\begin{equation*}
	\begin{tabular}{l l l}
		\textbf{DGP 5:} & $X_{i,1}\,\overset{iid}{\sim}\text{Gamma}(\,10,\,10)  \qquad$&$\text{for}\enskip i=1,\ldots,n_1$\\
		&$X_{i,2}\,\overset{iid}{\sim}\text{Gamma}(\,10,\,10)  \qquad$&$\text{for}\enskip i=1,\ldots,n_2$\\
		&$X_{i,3}\,\overset{iid}{\sim}\text{Gamma}(\,10,\,10)  \qquad$&$\text{for}\enskip i=1,\ldots,n_3$\\
		&$X_{i,4}\,\overset{iid}{\sim}0.5\,\mathcal{N}(\,0,\,0.5\,)+0.5\,\mathcal{N}(\,2,\,0.5\,) \qquad$&$\text{for}\enskip i=1,\ldots,n_4$\\
	\end{tabular}
\end{equation*}
Under all DGPs the model is misspecified due to lack of symmetry in one or more populations. Under DGP 3 and DGP 4 the true partition is $\{1\},\{2,3\},\{4\}$, while under DGP 5 it is $\{1,2,3,4\}$. In DGP 3 the second population differs from the others also in distribution (what we called non-location effect), the same is true for the first and the fourth populations respectively under DGP 4 and DGP 5. Table~\ref{tab:C} shows that the model is able to detect the right clustering of the population-specific locations under all three DGPs. Moreover, Figure~\ref{fig:C} shows co-clustering probabilities that differ in correspondence of the populations affected by non-location effects, more or less evidently based on the DGP used to generate the data. These results under misspecification are reassuring: the model appears robust in estimating the  partitions of the locations and, moreover, the different within-population patterns of co-clustering probabilities still highlight heterogeneities different than shifts in population-specific locations.

\begin{table}[H]
	\caption{Posterior probabilities over the space of partitions.}
	\label{tab:C}
	\begin{center}
		\begin{tabular}{l|ccc}
			&DGP 3 & DGP 4 & DGP 5\\\hline
			\{\textcolor{aoe}{1},\textcolor{burntorange}{2},\textcolor{bostonuniversityred}{3},\textcolor{burgundy}{4}\}&0&0.001&\textbf{0.494}\\\hline
			\{\textcolor{aoe}{1}\}\{\textcolor{burntorange}{2},\textcolor{bostonuniversityred}{3},\textcolor{burgundy}{4}\}&0&0&0.023\\\hline
			\{\textcolor{aoe}{1},\textcolor{burntorange}{2}\}\{\textcolor{bostonuniversityred}{3},\textcolor{burgundy}{4}\}&0&0&0.014\\\hline
			\{\textcolor{aoe}{1},\textcolor{bostonuniversityred}{3},\textcolor{burgundy}{4}\}\{\textcolor{burntorange}{2}\}&0&0&0\\\hline
			\{\textcolor{aoe}{1}\}\{\textcolor{burntorange}{2}\}\{\textcolor{bostonuniversityred}{3},\textcolor{burgundy}{4}\}&0&0&0.004\\\hline
			\{\textcolor{aoe}{1},\textcolor{burntorange}{2},\textcolor{bostonuniversityred}{3}\}\{\textcolor{burgundy}{4}\}&0.016&0&0.375\\\hline
			\{\textcolor{aoe}{1},\textcolor{burgundy}{4}\}\{\textcolor{burntorange}{2},\textcolor{bostonuniversityred}{3}\}&0&0&0\\\hline
			\{\textcolor{aoe}{1}\}\{\textcolor{burntorange}{2},\textcolor{bostonuniversityred}{3}\}\{\textcolor{burgundy}{4}\}&\textbf{0.736}&\textbf{0.788}&0.047\\\hline
			\{\textcolor{aoe}{1},\textcolor{bostonuniversityred}{3}\}\{\textcolor{burntorange}{2},\textcolor{burgundy}{4}\}&0&0&0\\\hline
			\{\textcolor{aoe}{1},\textcolor{burntorange}{2},\textcolor{burgundy}{4}\}\{\textcolor{bostonuniversityred}{3}\}&0&0&0\\\hline
			\{\textcolor{aoe}{1}\}\{\textcolor{burntorange}{2},\textcolor{burgundy}{4}\}\{\textcolor{bostonuniversityred}{3}\}&0&0&0\\\hline
			\{\textcolor{aoe}{1},\textcolor{burntorange}{2}\}\{\textcolor{bostonuniversityred}{3}\}\{\textcolor{burgundy}{4}\}&0.015&0&0.030\\\hline
			\{\textcolor{aoe}{1},\textcolor{bostonuniversityred}{3}\}\{\textcolor{burntorange}{2}\}\{\textcolor{burgundy}{4}\}&0&0&0\\\hline
			\{\textcolor{aoe}{1},\textcolor{burgundy}{4}\}\{\textcolor{burntorange}{2}\}\{\textcolor{bostonuniversityred}{3}\}&0&0&0\\\hline
			\{\textcolor{aoe}{1}\}\{\textcolor{burntorange}{2}\}\{\textcolor{bostonuniversityred}{3}\}\{\textcolor{burgundy}{4}\}&0.232&0.211&0.012\\\hline
		\end{tabular}
	\end{center}
	\vspace{-\baselineskip}
\end{table}

\section{Hypertensive dataset}
\subsection{Additional results}
The figures below report the density estimates, the heatmaps of co-clustering probabilities between pairs of patients and population-specific credible intervals for all ten response variables.
\begin{figure}[H]
	\centering
	\begin{subfigure}{.4\textwidth}
		\centering
		\includegraphics[width=\linewidth]{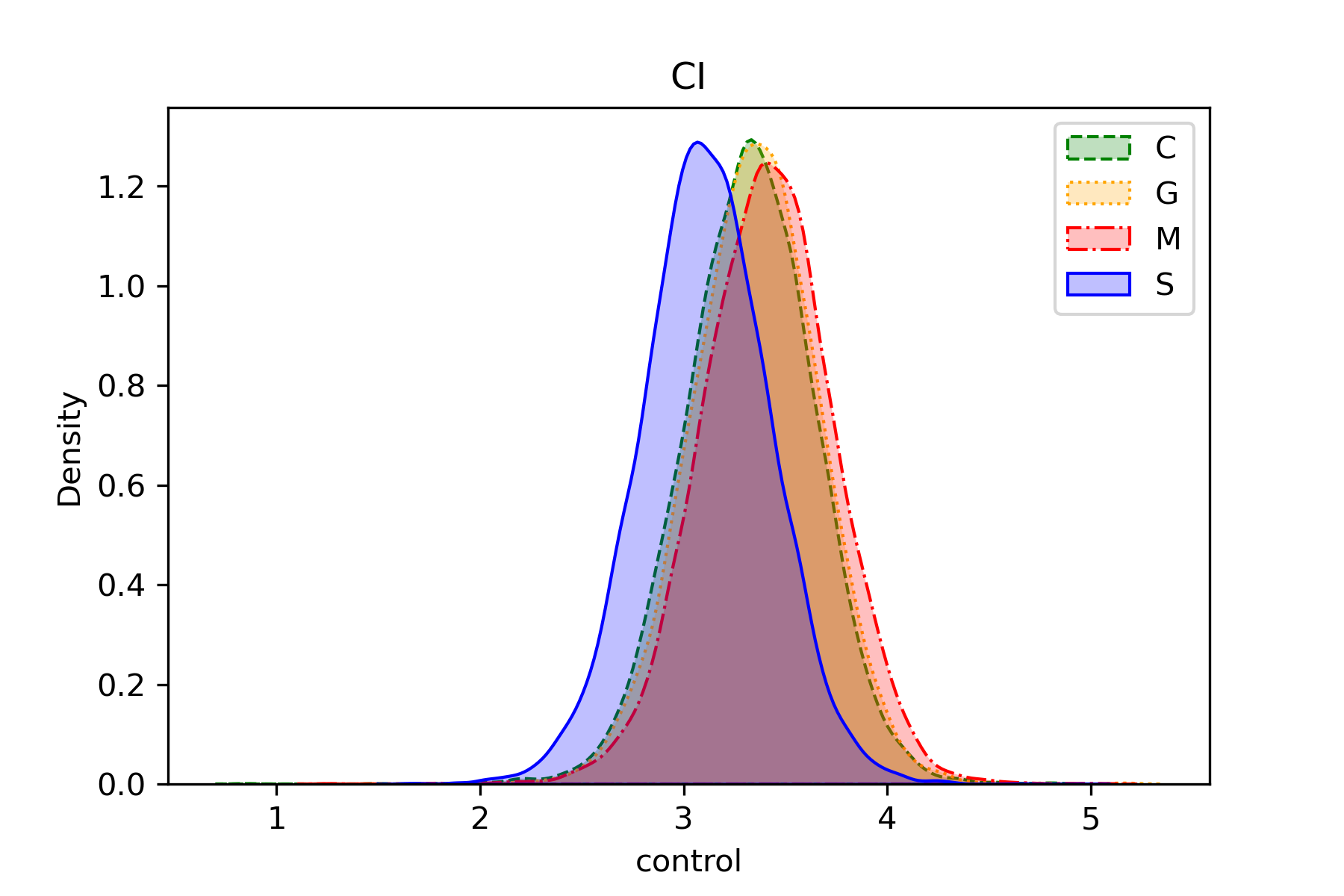}
		\caption{density estimation}
	\end{subfigure}\hspace{0.05\textwidth}%
	\begin{subfigure}{.25\textwidth}
		\centering
		\includegraphics[ trim={3cm 1cm 3cm 1cm}, clip, width=\linewidth]{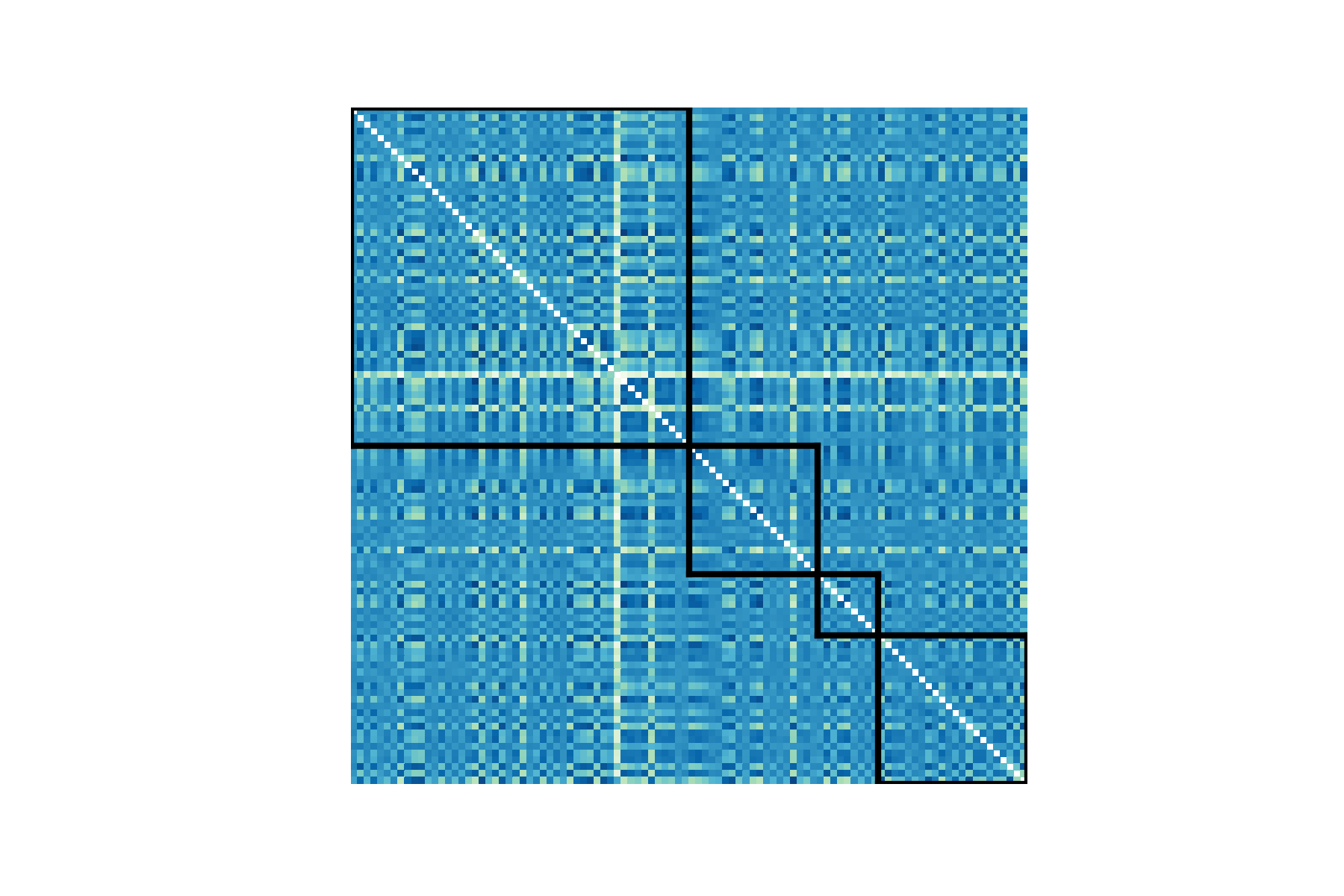}
		\caption{co-clustering}
	\end{subfigure}\hspace{0.05\textwidth}%
	\begin{subfigure}{.25\textwidth}
		\centering
		\includegraphics[ trim={3cm 1cm 3cm 1cm}, clip, width=\linewidth]{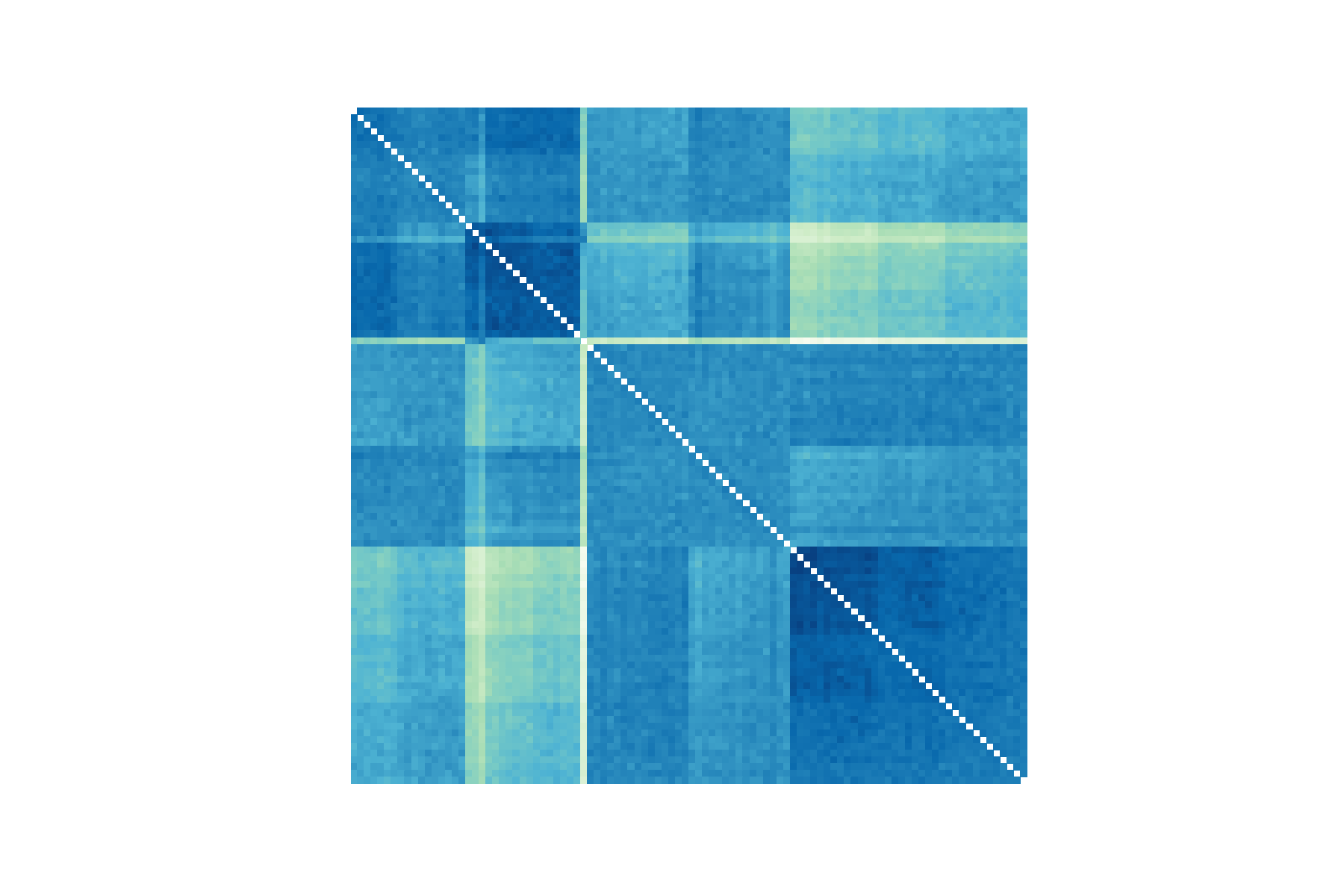}
		\caption{co-clustering}
	\end{subfigure}
\end{figure}

\begin{figure}[H]
	\centering
	\begin{subfigure}{.4\textwidth}
		\centering
		\includegraphics[ width=\linewidth]{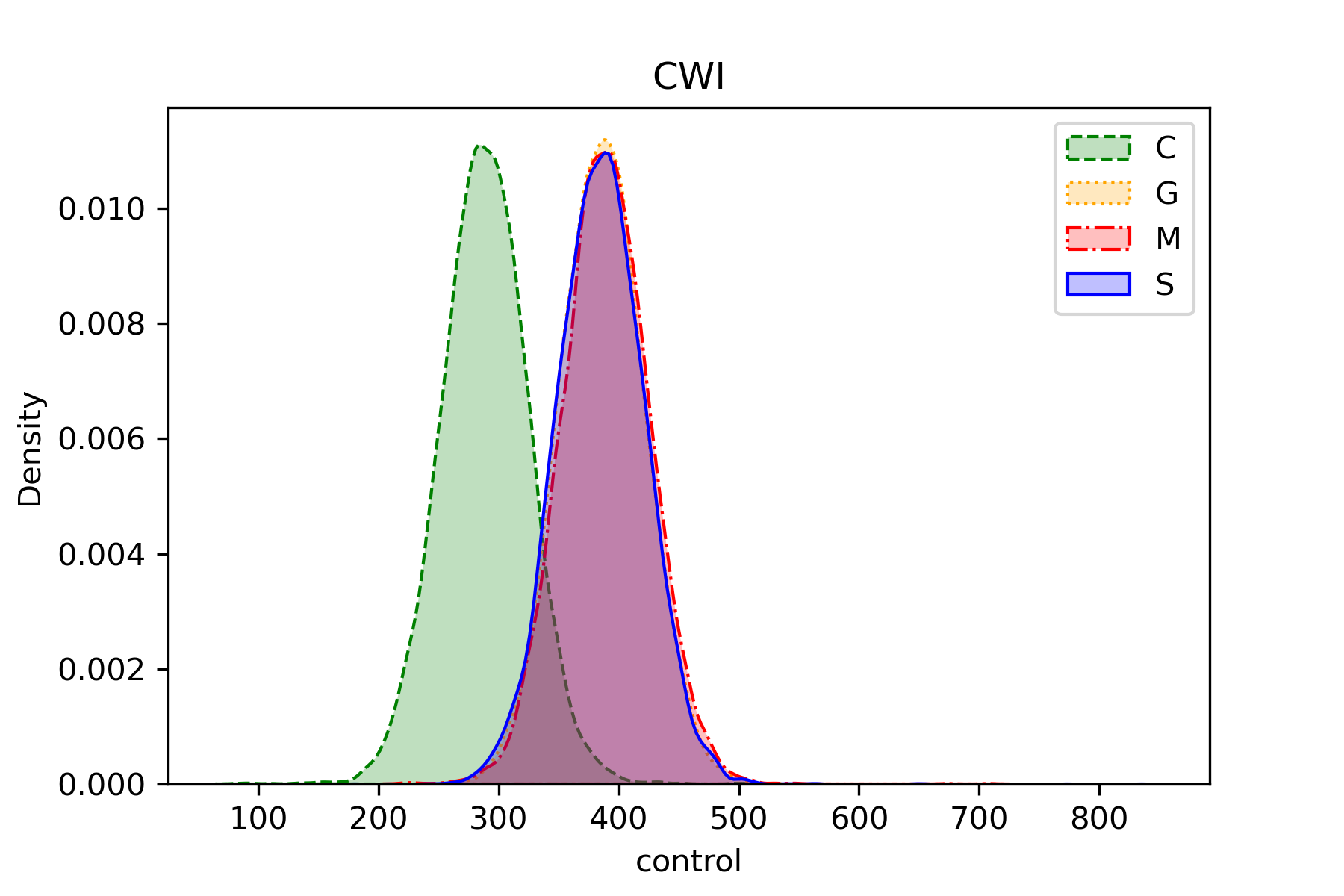}
		\caption{density estimation}
	\end{subfigure}\hspace{0.05\textwidth}%
	\begin{subfigure}{.25\textwidth}
		\centering
		\includegraphics[trim={3cm 1cm 3cm 1cm}, clip, width=\linewidth ]{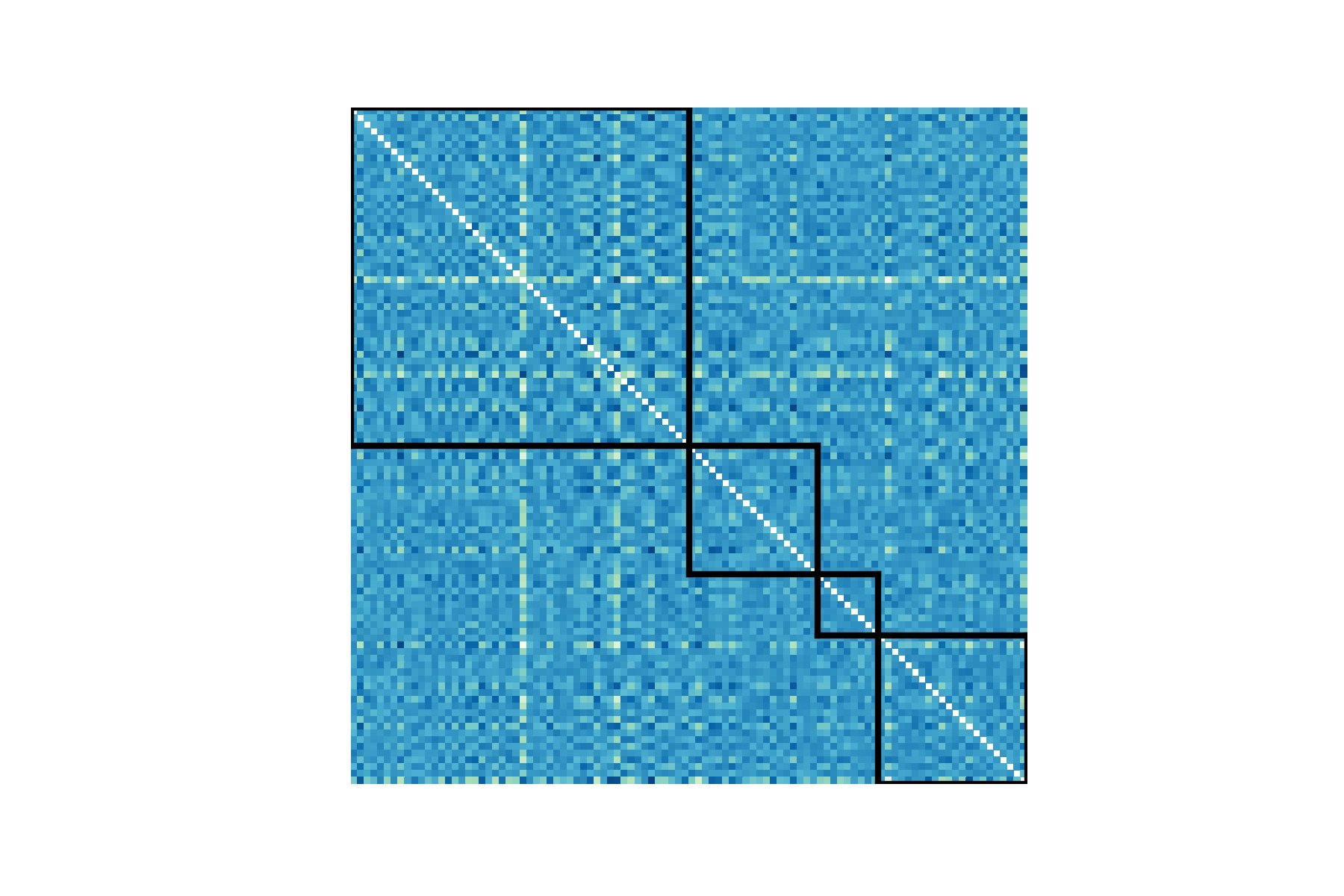}
		\caption{co-clustering}
	\end{subfigure}\hspace{0.05\textwidth}%
	\begin{subfigure}{.25\textwidth}
		\centering
		\includegraphics[trim={3cm 1cm 3cm 1cm}, clip, width=\linewidth]{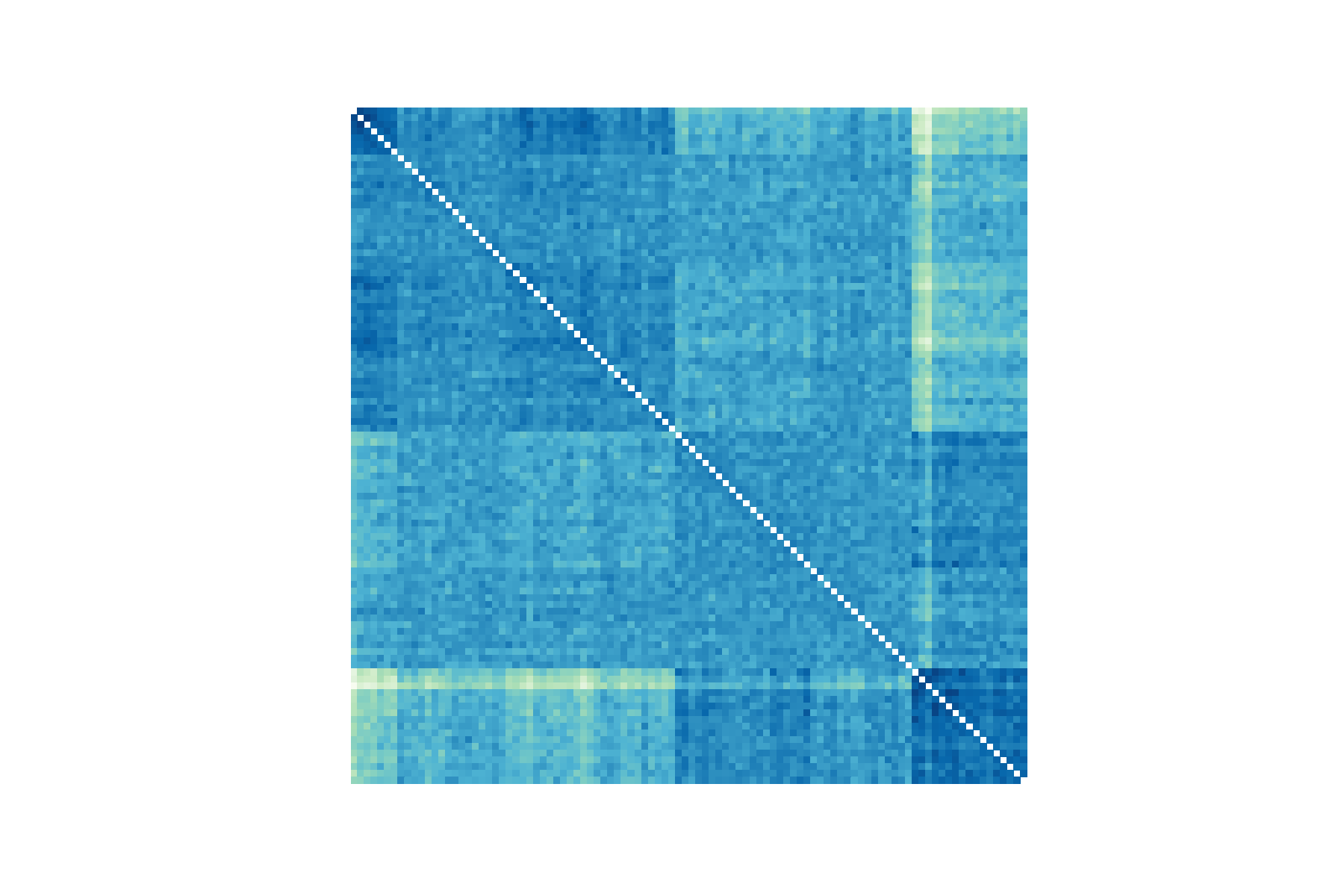}
		\caption{co-clustering}
	\end{subfigure}
\end{figure}

\begin{figure}[H]
	\centering
	\begin{subfigure}{.4\textwidth}
		\centering
		\includegraphics[ width=\linewidth]{figures/LVMI}
		\caption{density estimation}
	\end{subfigure}\hspace{0.05\textwidth}%
	\begin{subfigure}{.25\textwidth}
		\centering
		\includegraphics[trim={3cm 1cm 3cm 1cm}, clip,  width=\linewidth]{figures/cluster_LVMI}
		\caption{co-clustering}
	\end{subfigure}\hspace{0.05\textwidth}%
	\begin{subfigure}{.25\textwidth}
		\centering
		\includegraphics[trim={3cm 1cm 3cm 1cm}, clip, width=\linewidth]{figures/cluster_ord_LVMI}
		\caption{co-clustering}
	\end{subfigure}
\end{figure}

\begin{figure}[H]
	\centering
	\begin{subfigure}{.4\textwidth}
		\centering
		\includegraphics[width=\linewidth]{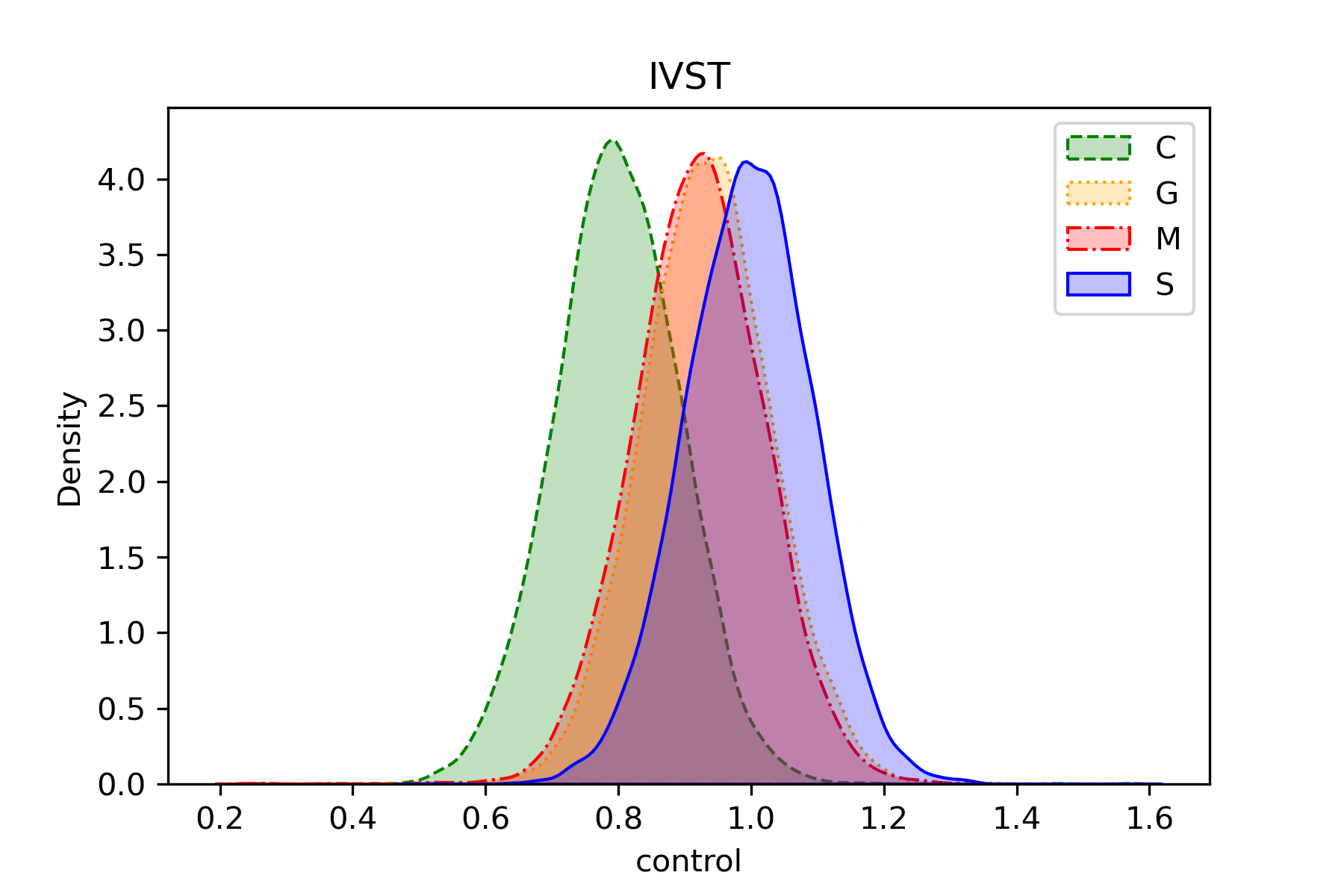}
		\caption{density estimation}
	\end{subfigure}\hspace{0.05\textwidth}%
	\begin{subfigure}{.25\textwidth}
		\centering
		\includegraphics[trim={3cm 1cm 3cm 1cm}, clip,  width=\linewidth]{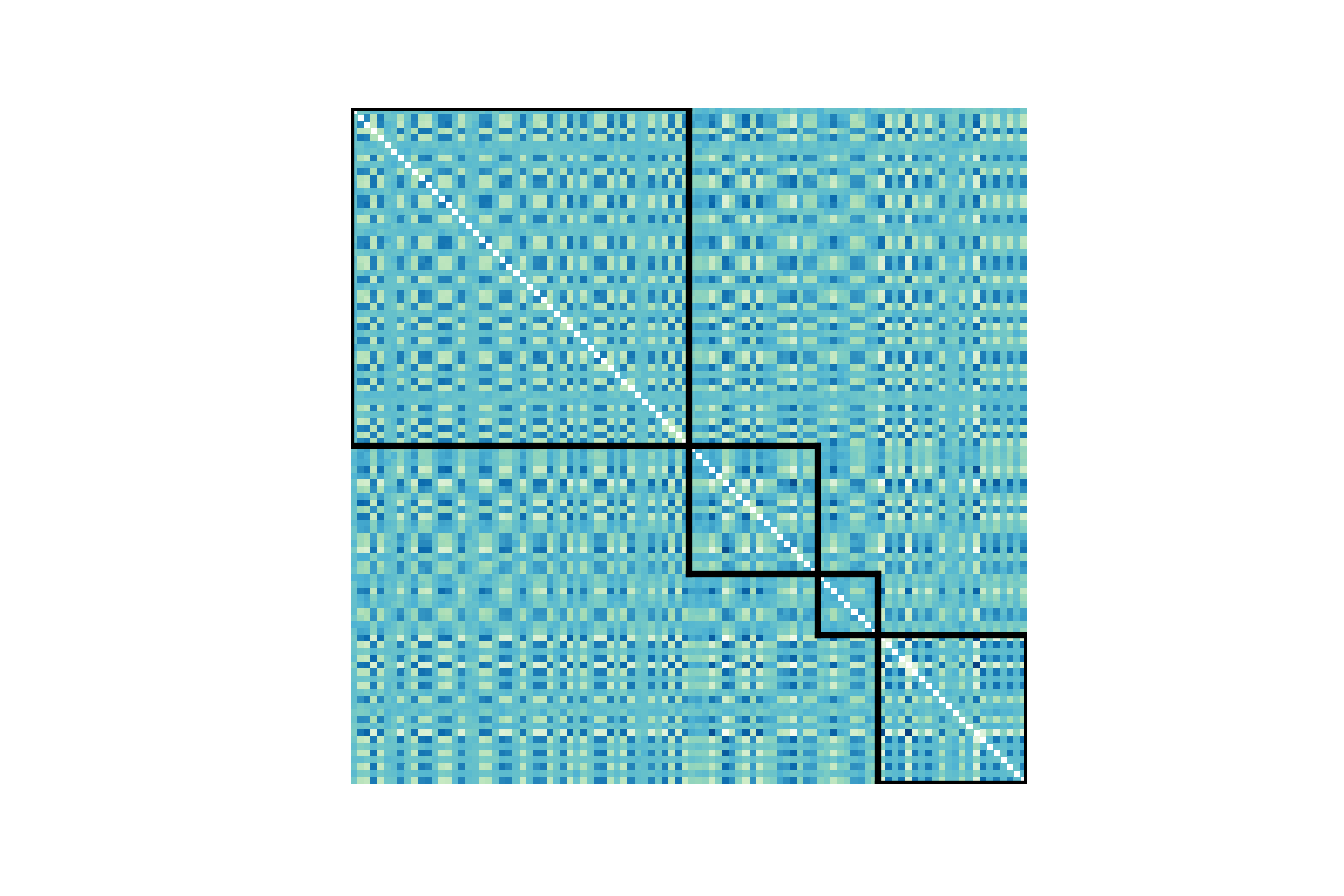}
		\caption{co-clustering}
	\end{subfigure}\hspace{0.05\textwidth}%
	\begin{subfigure}{.25\textwidth}
		\centering
		\includegraphics[trim={3cm 1cm 3cm 1cm}, clip,  width=\linewidth]{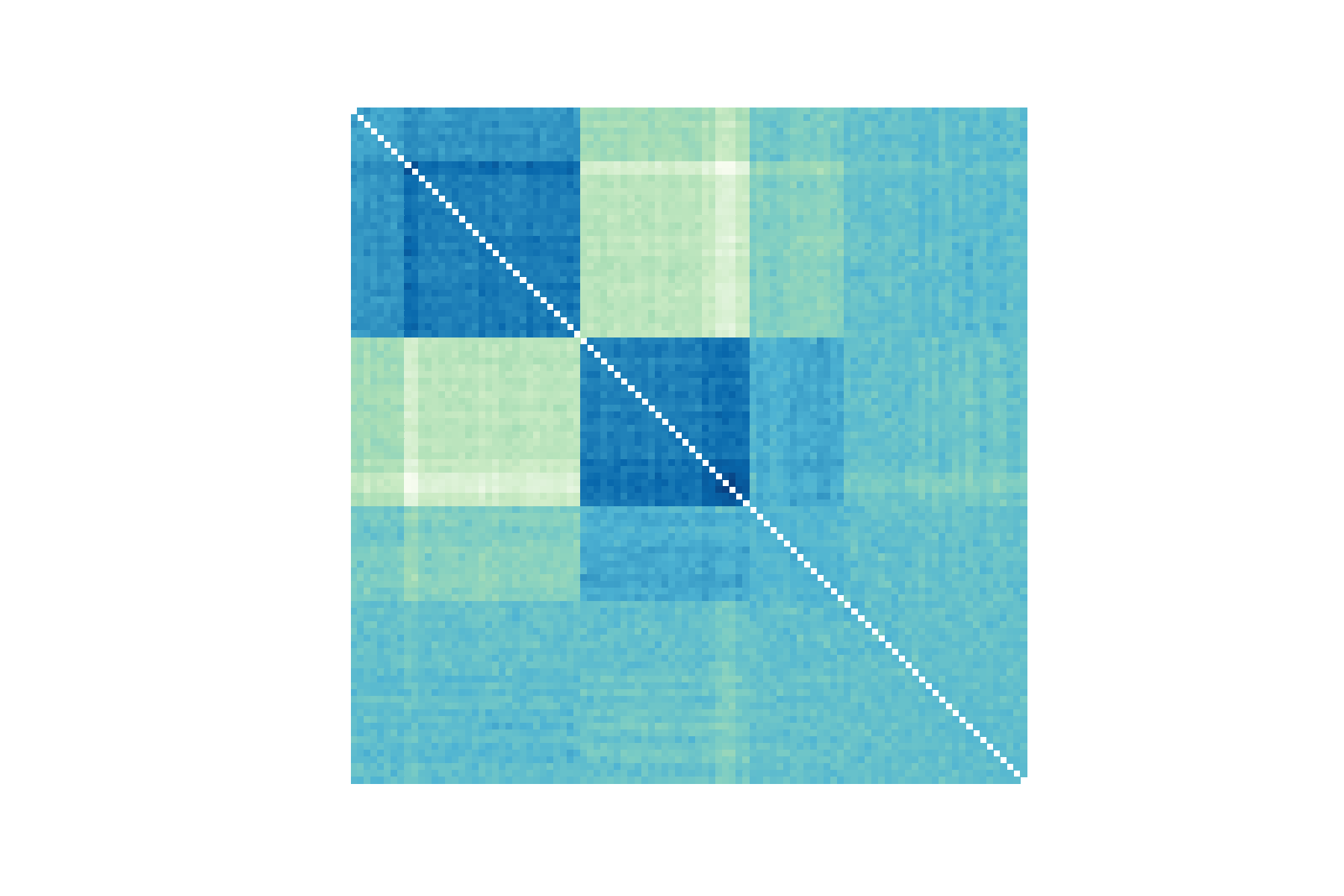}
		\caption{co-clustering}
	\end{subfigure}
\end{figure}

\begin{figure}[H]
	\centering
	\begin{subfigure}{.4\textwidth}
		\centering
		\includegraphics[ width=\linewidth]{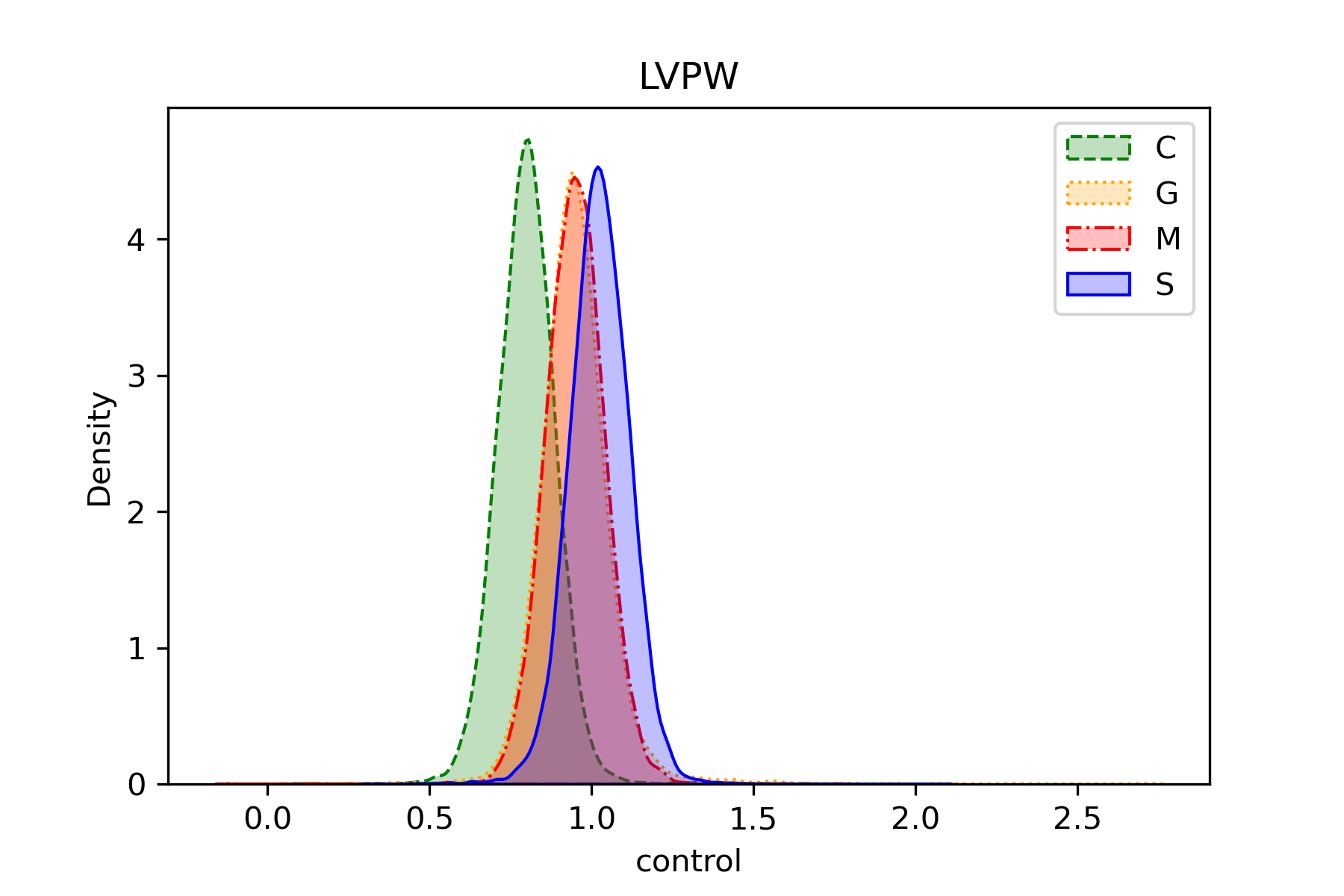}
		\caption{density estimation}
	\end{subfigure}\hspace{0.05\textwidth}%
	\begin{subfigure}{.25\textwidth}
		\centering
		\includegraphics[trim={3cm 1cm 3cm 1cm}, clip,  width=\linewidth]{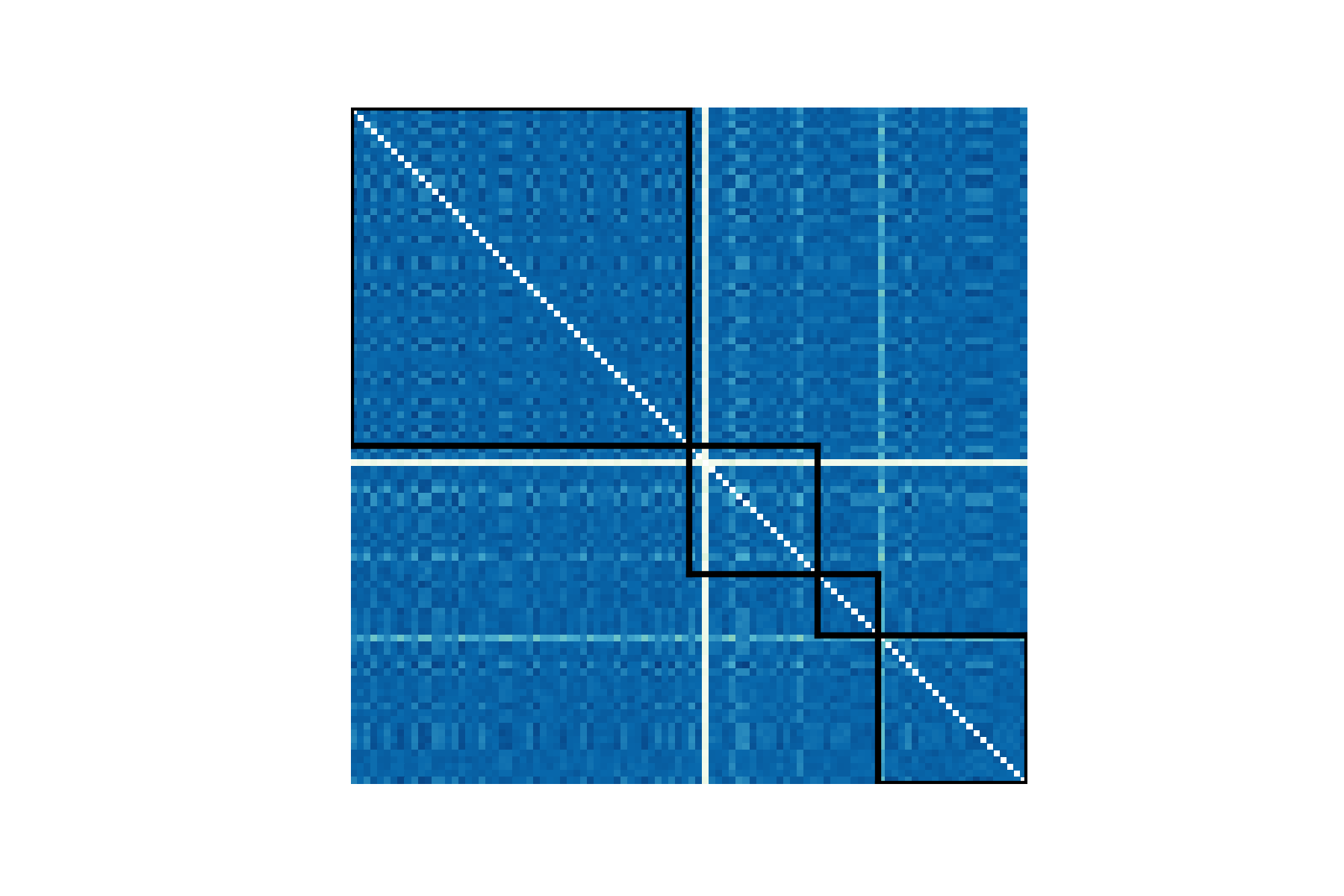}
		\caption{co-clustering}
	\end{subfigure}\hspace{0.05\textwidth}%
	\begin{subfigure}{.25\textwidth}
		\centering
		\includegraphics[trim={3cm 1cm 3cm 1cm}, clip,  width=\linewidth]{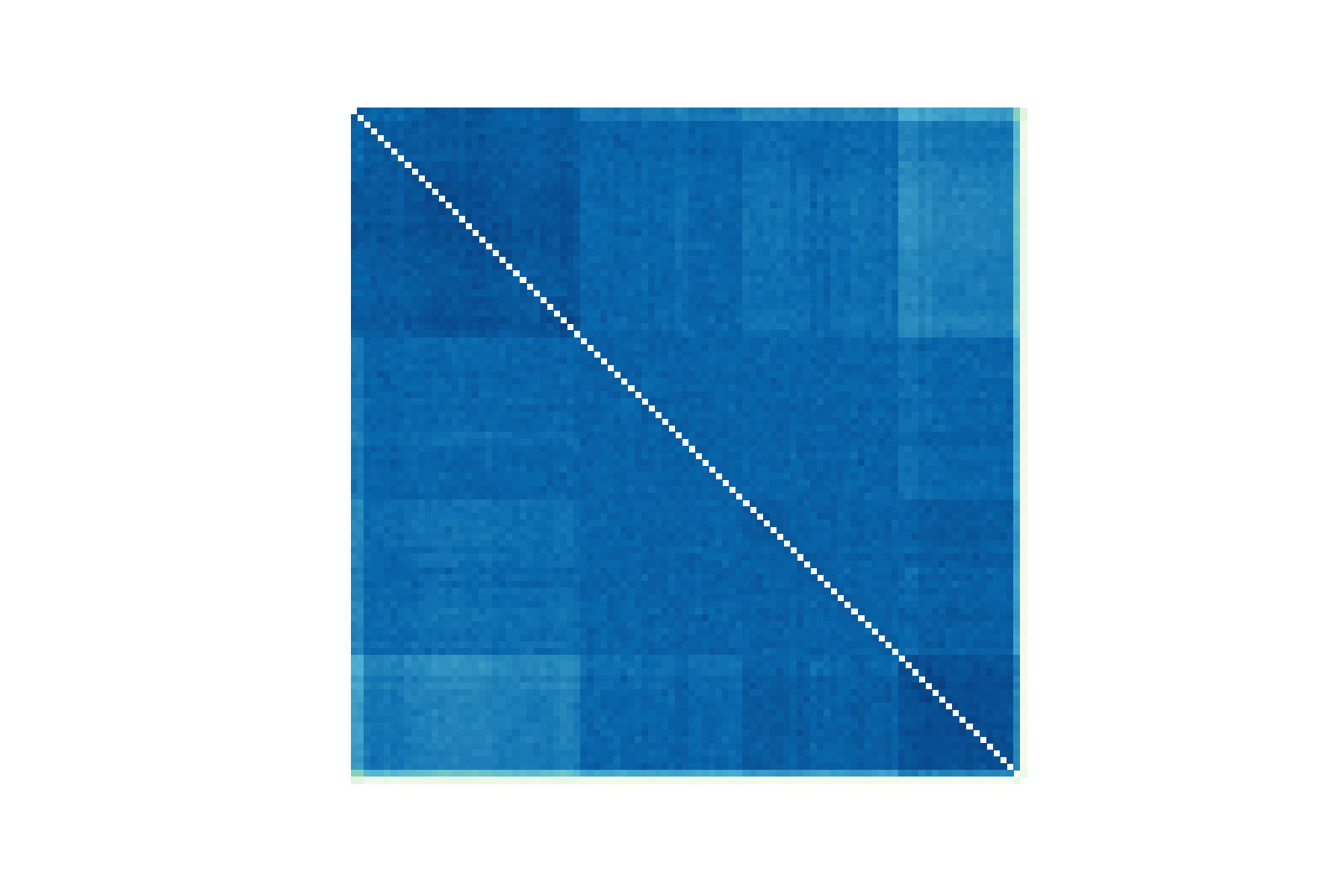}
		\caption{co-clustering}
	\end{subfigure}
\end{figure}

\begin{figure}[H]
	\centering
	\begin{subfigure}{.4\textwidth}
		\centering
		\includegraphics[ width=\linewidth]{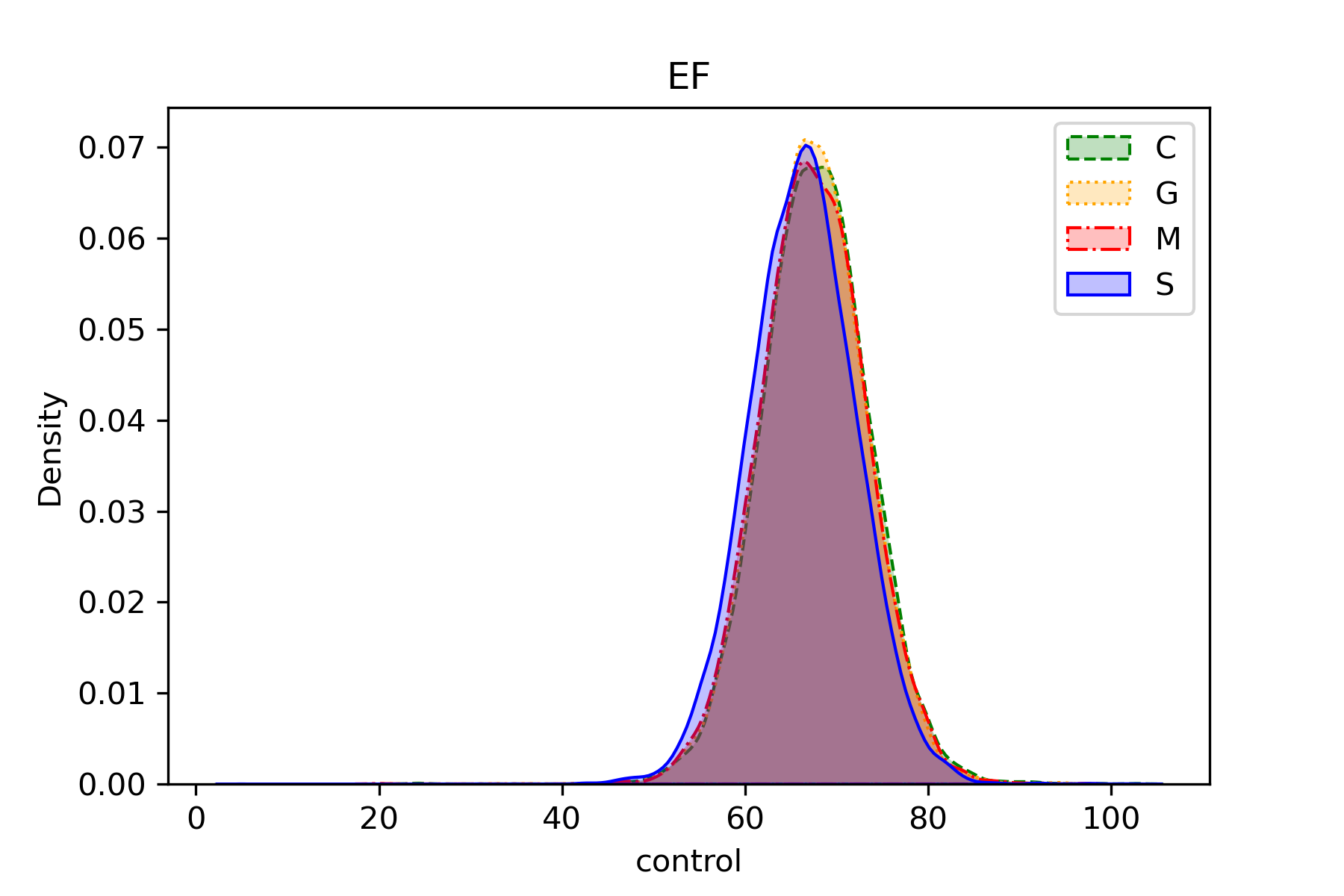}
		\caption{density estimation}
	\end{subfigure}\hspace{0.05\textwidth}%
	\begin{subfigure}{.25\textwidth}
		\centering
		\includegraphics[trim={3cm 1cm 3cm 1cm}, clip,  width=\linewidth]{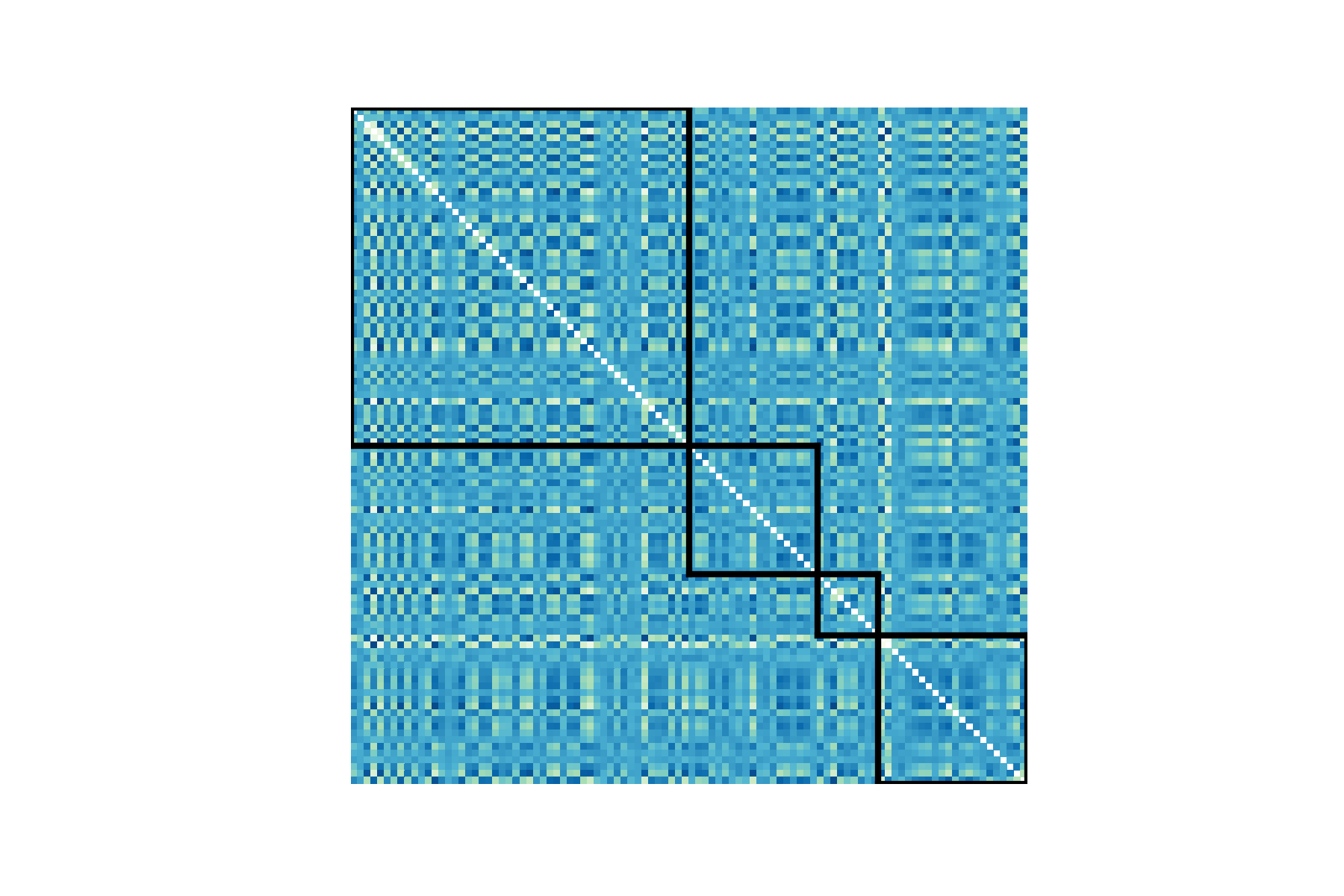}
		\caption{co-clustering}
	\end{subfigure}\hspace{0.05\textwidth}%
	\begin{subfigure}{.25\textwidth}
		\centering
		\includegraphics[trim={3cm 1cm 3cm 1cm}, clip,  width=\linewidth]{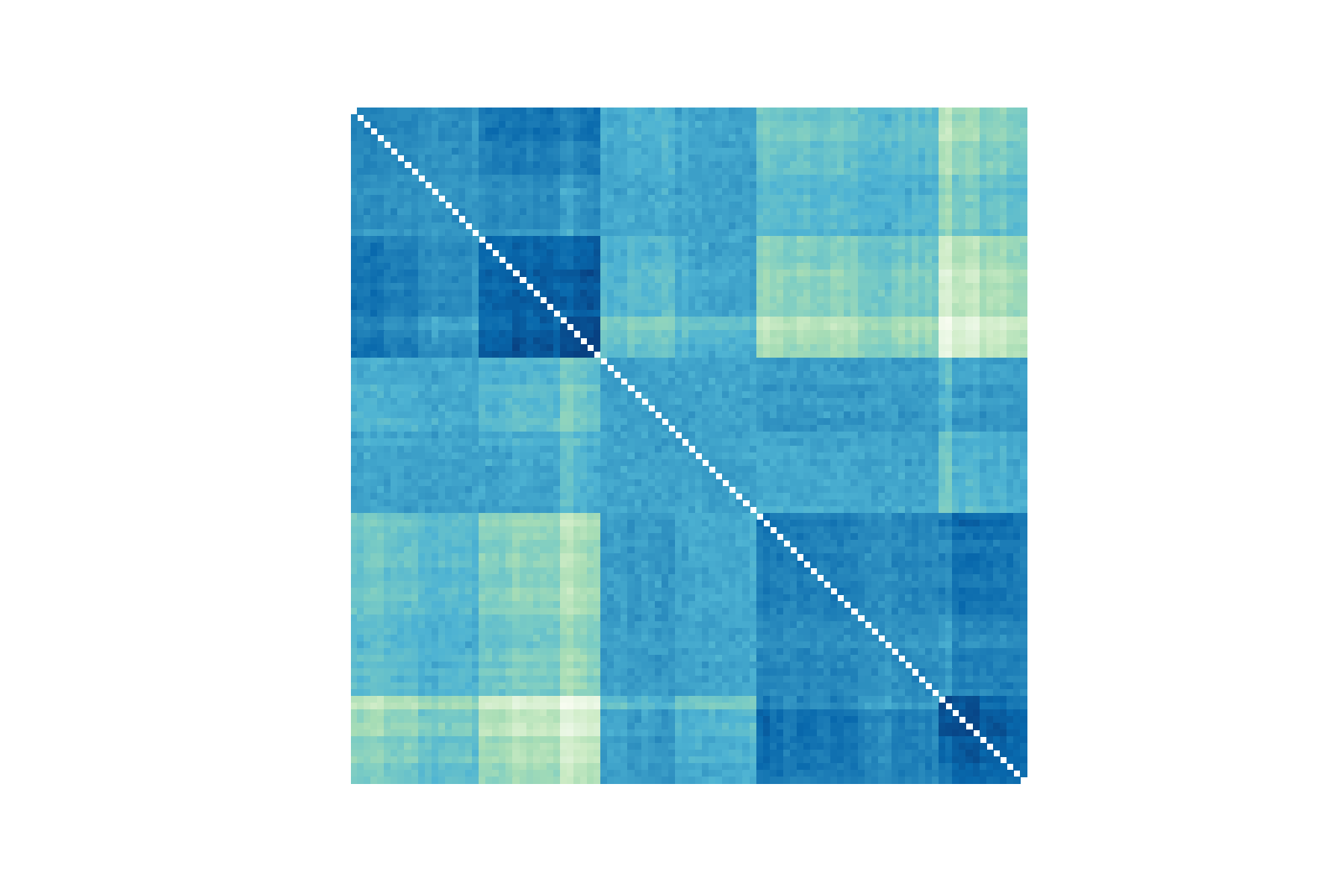}
		\caption{co-clustering}
	\end{subfigure}
\end{figure}

\begin{figure}[H]
	\centering
	\begin{subfigure}{.4\textwidth}
		\centering
		\includegraphics[ width=\linewidth]{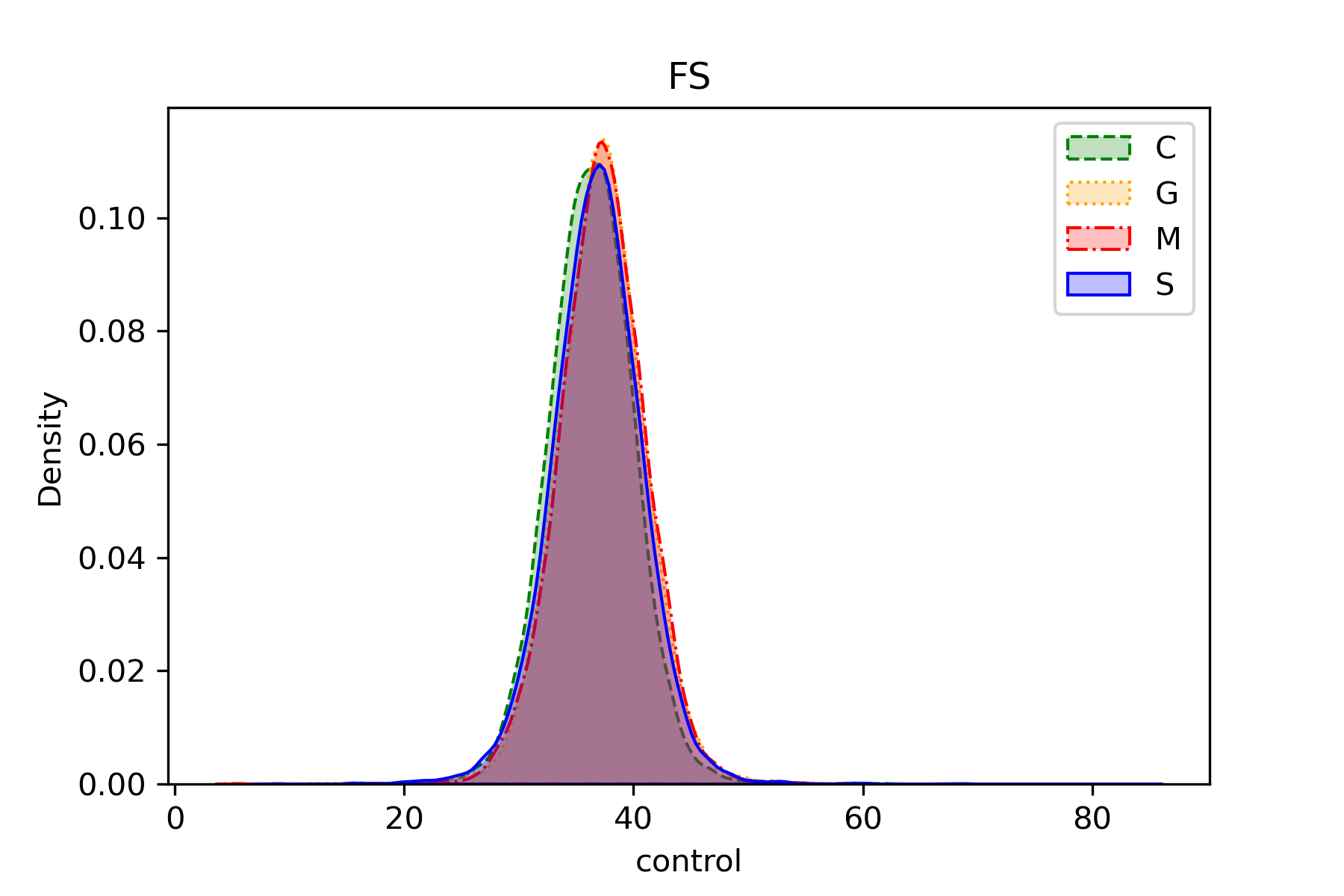}
		\caption{density estimation}
	\end{subfigure}\hspace{0.05\textwidth}%
	\begin{subfigure}{.25\textwidth}
		\centering
		\includegraphics[trim={3cm 1cm 3cm 1cm}, clip,  width=\linewidth]{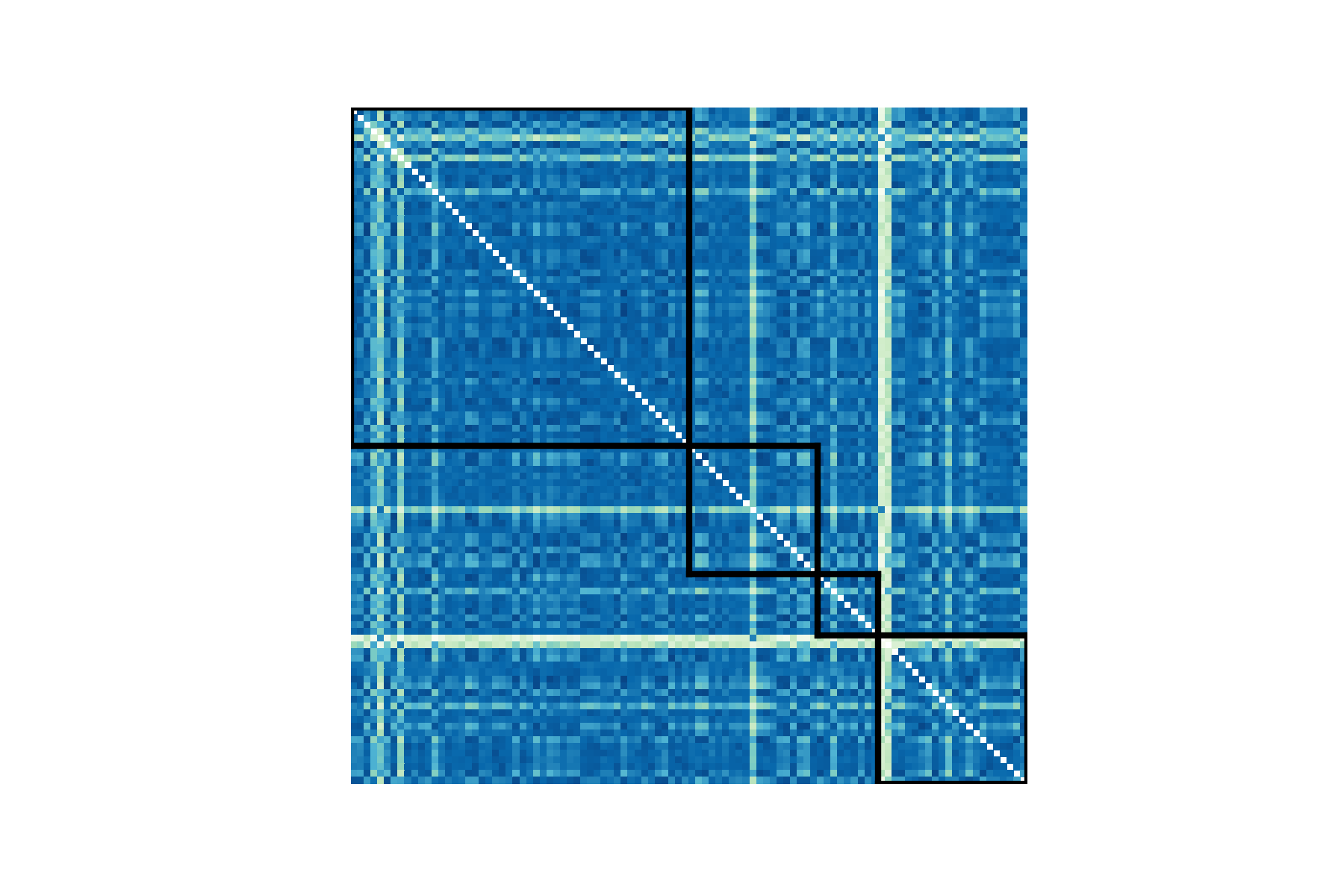}
		\caption{co-clustering}
	\end{subfigure}\hspace{0.05\textwidth}%
	\begin{subfigure}{.25\textwidth}
		\centering
		\includegraphics[trim={3cm 1cm 3cm 1cm}, clip,  width=\linewidth]{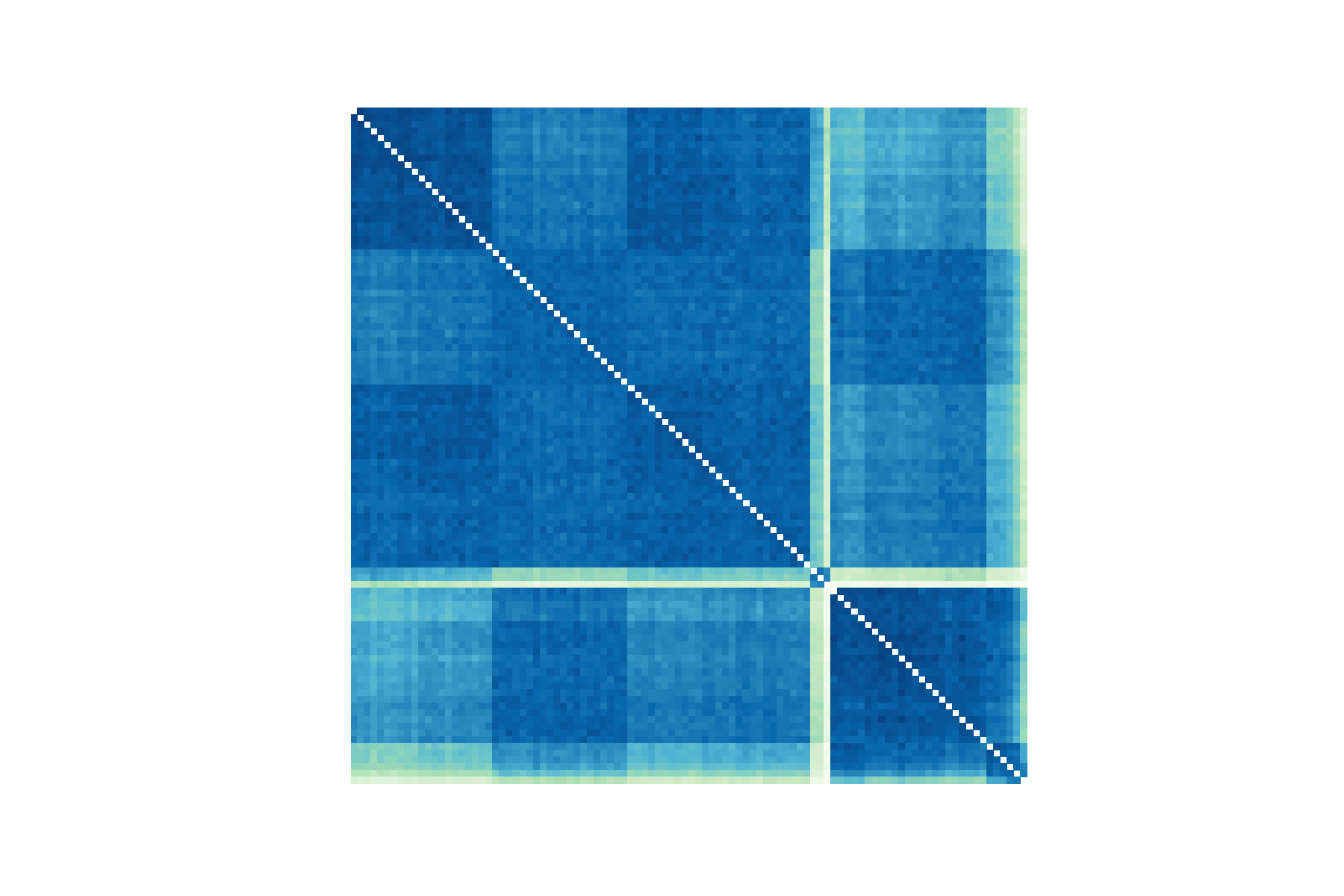}
		\caption{co-clustering}
	\end{subfigure}
\end{figure}

\begin{figure}[H]
	\centering
	\begin{subfigure}{.4\textwidth}
		\centering
		\includegraphics[ width=\linewidth]{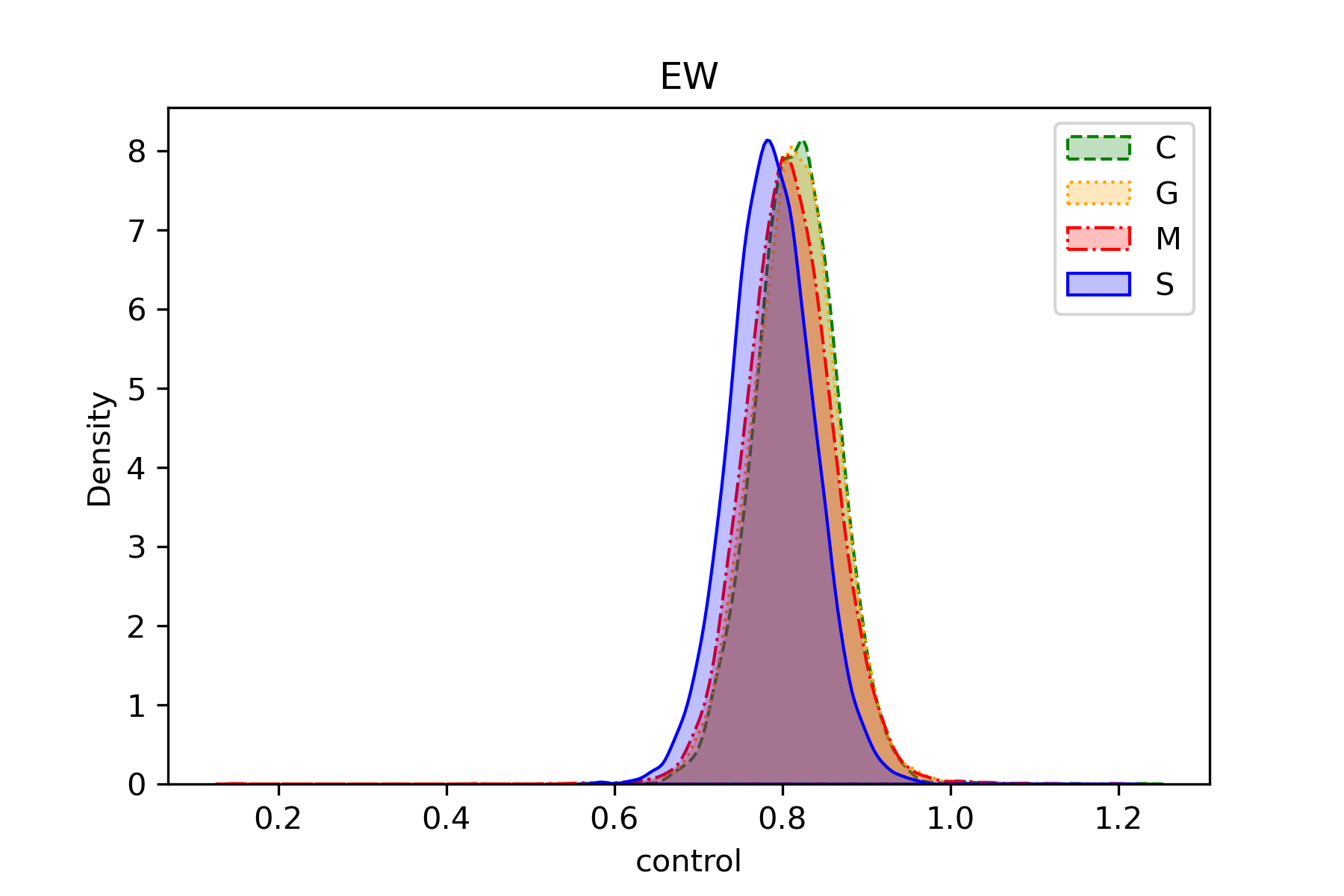}
		\caption{density estimation}
	\end{subfigure}\hspace{0.05\textwidth}%
	\begin{subfigure}{.25\textwidth}
		\centering
		\includegraphics[trim={3cm 1cm 3cm 1cm}, clip,  width=\linewidth]{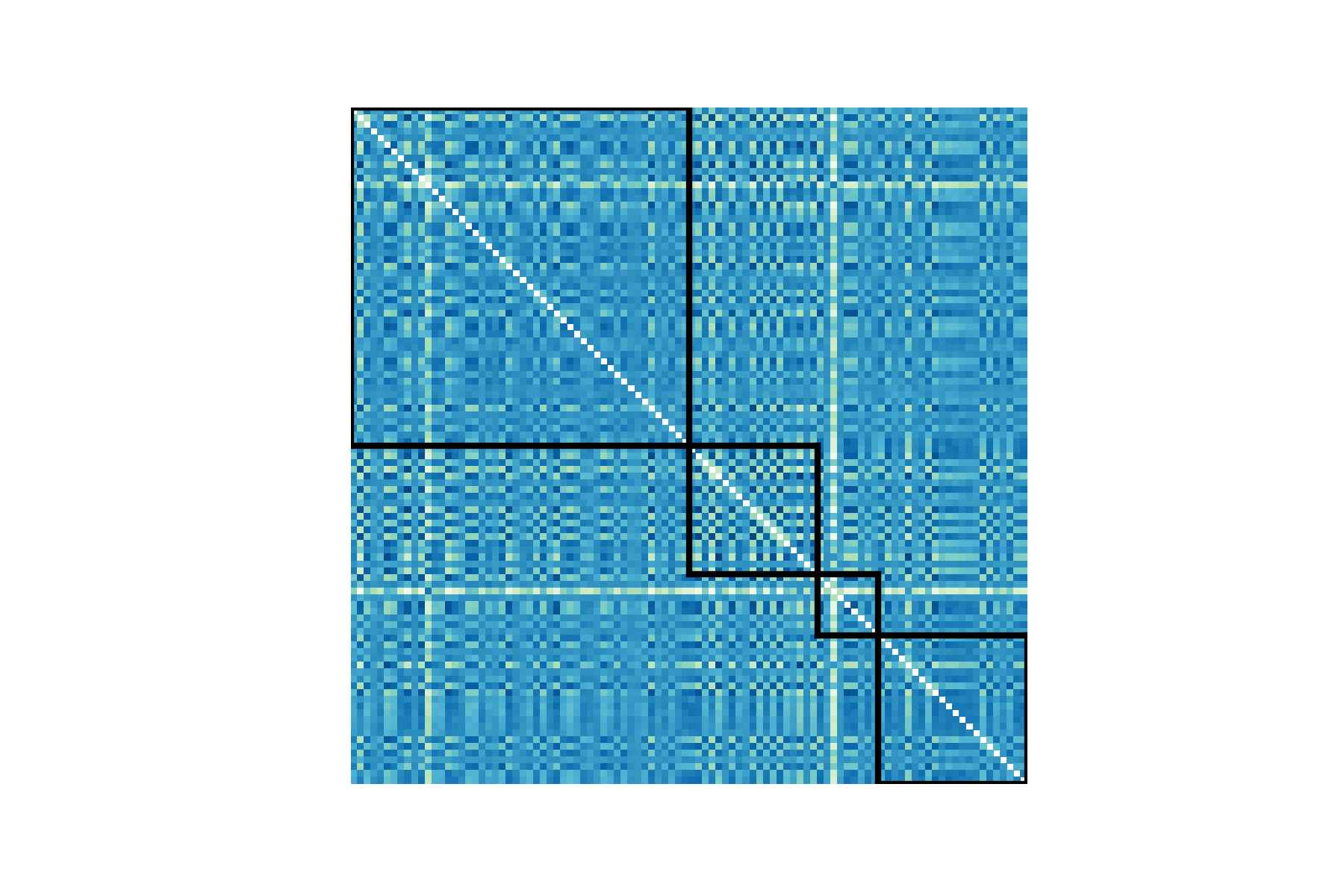}
		\caption{co-clustering}
	\end{subfigure}\hspace{0.05\textwidth}%
	\begin{subfigure}{.25\textwidth}
		\centering
		\includegraphics[trim={3cm 1cm 3cm 1cm}, clip,  width=\linewidth]{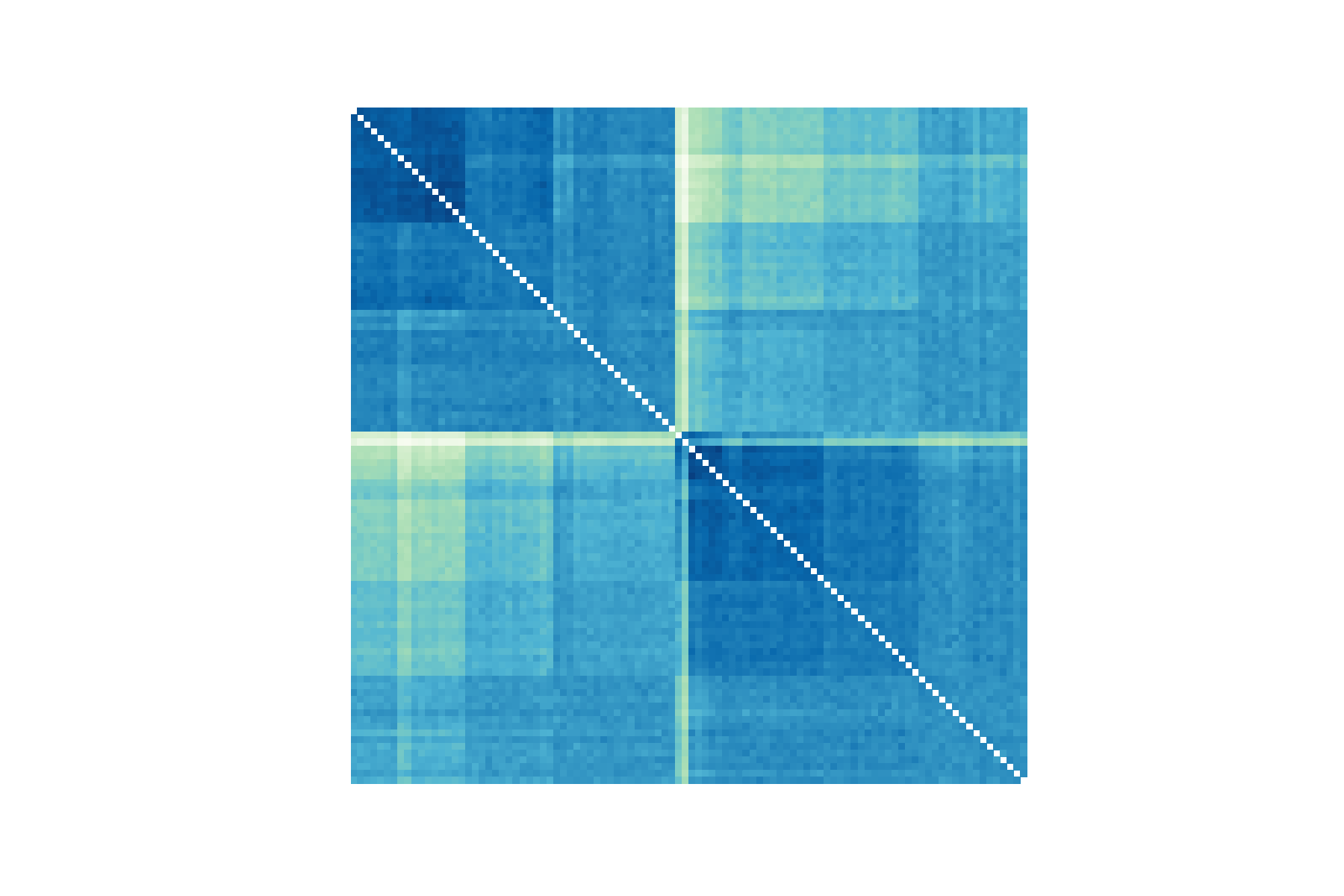}
		\caption{co-clustering}
	\end{subfigure}
\end{figure}

\begin{figure}[H]
	\centering
	\begin{subfigure}{.4\textwidth}
		\centering
		\includegraphics[width=\linewidth]{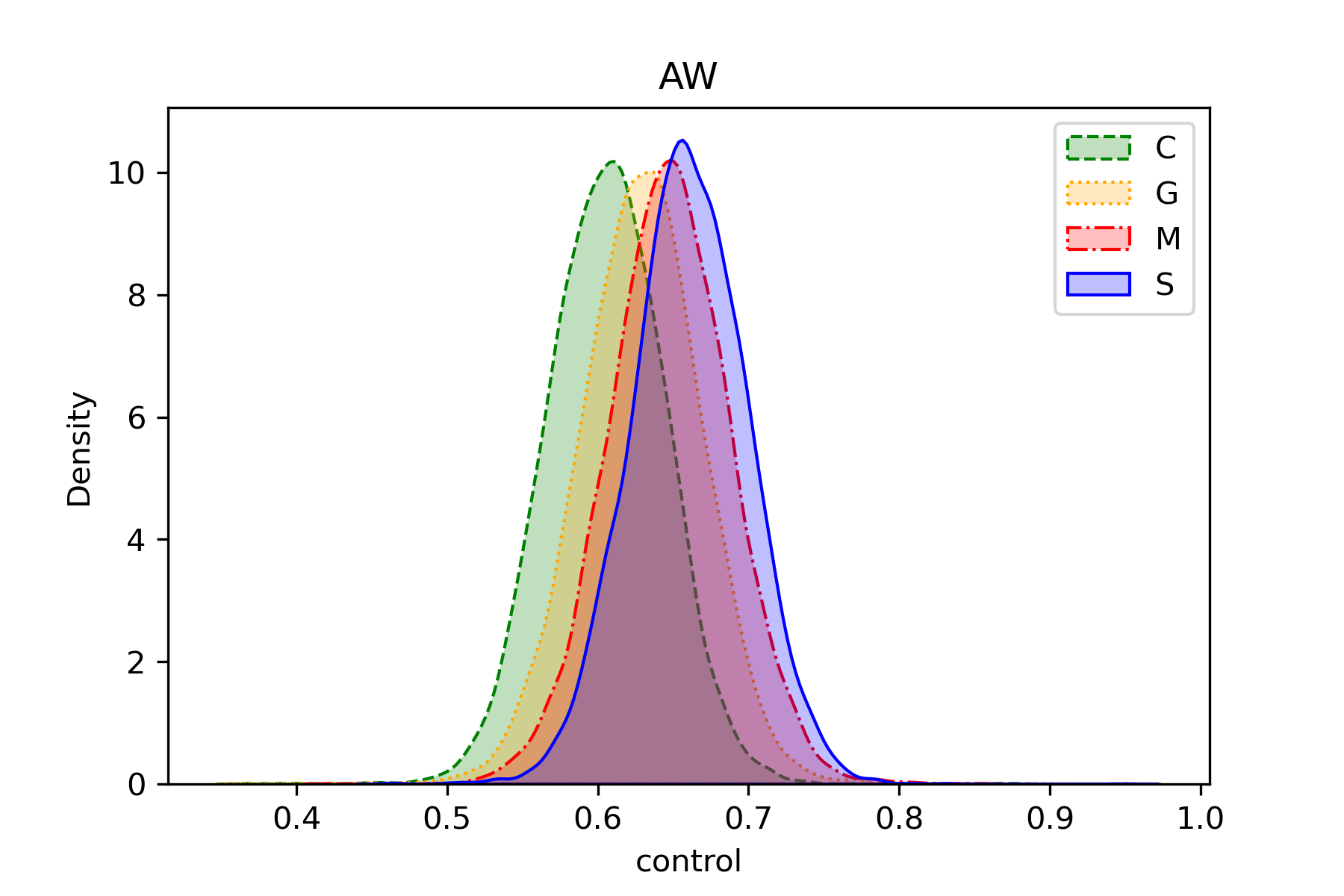}
		\caption{density estimation}
	\end{subfigure}\hspace{0.05\textwidth}%
	\begin{subfigure}{.25\textwidth}
		\centering
		\includegraphics[trim={3cm 1cm 3cm 1cm}, clip,  width=\linewidth]{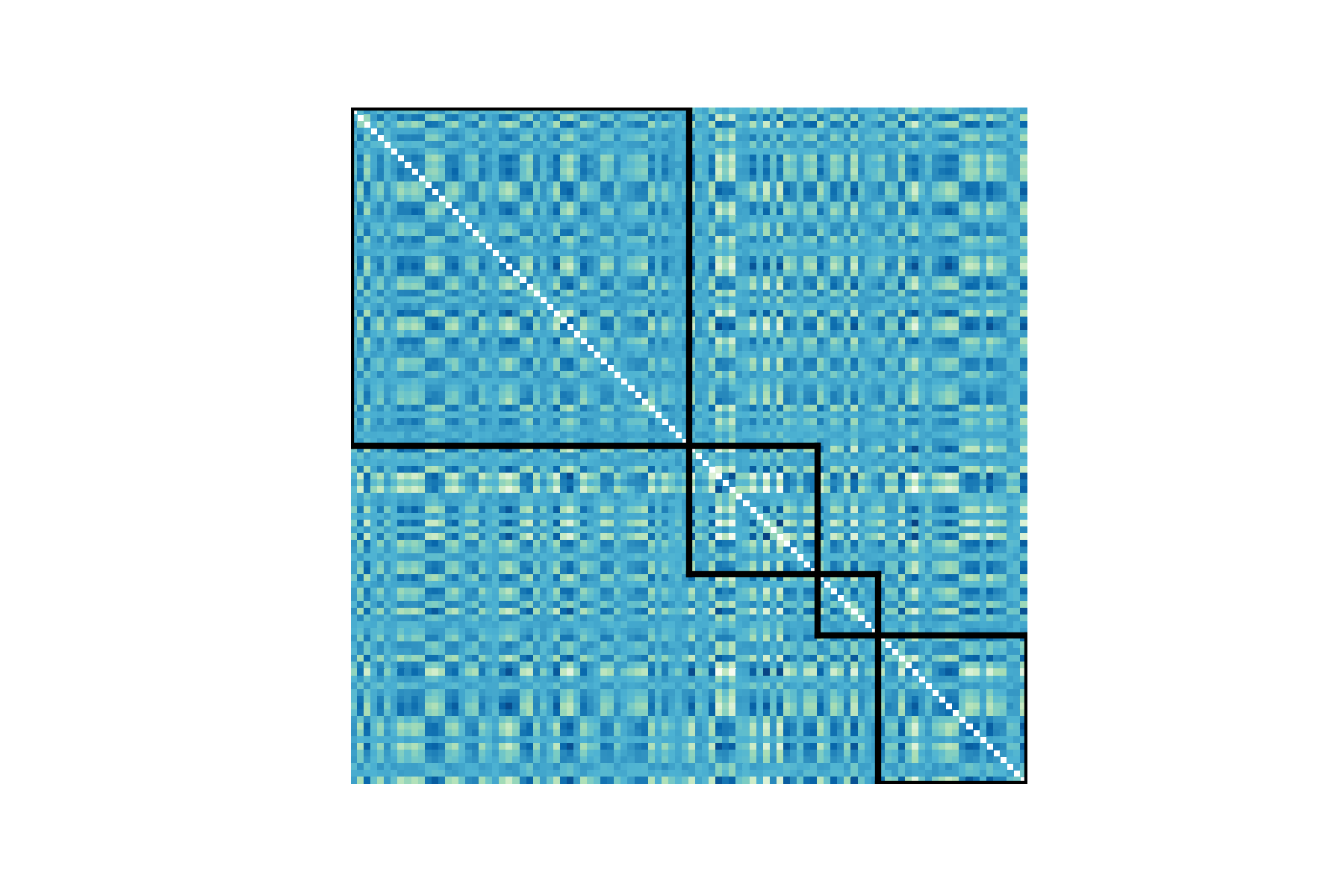}
		\caption{co-clustering}
	\end{subfigure}\hspace{0.05\textwidth}%
	\begin{subfigure}{.25\textwidth}
		\centering
		\includegraphics[trim={3cm 1cm 3cm 1cm}, clip,  width=\linewidth]{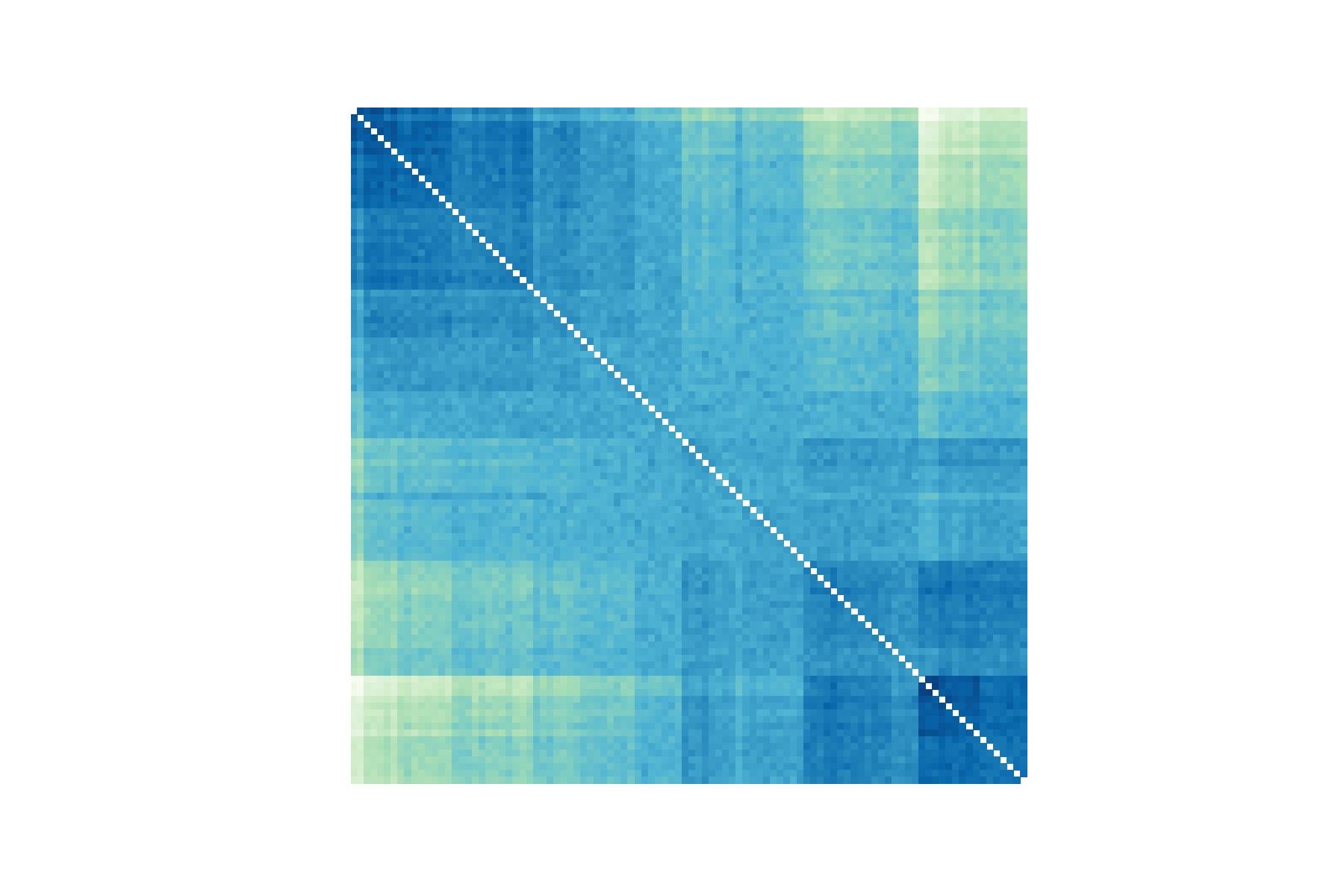}
		\caption{co-clustering}
	\end{subfigure}
\end{figure}

\begin{figure}[H]
	\centering
	\begin{subfigure}{.4\textwidth}
		\centering
		\includegraphics[ width=\linewidth]{figures/EA}
		\caption{density estimation}
	\end{subfigure}\hspace{0.05\textwidth}%
	\begin{subfigure}{.25\textwidth}
		\centering
		\includegraphics[trim={3cm 1cm 3cm 1cm}, clip,  width=\linewidth]{figures/cluster_EA}
		\caption{co-clustering}
	\end{subfigure}\hspace{0.05\textwidth}%
	\begin{subfigure}{.25\textwidth}
		\centering
		\includegraphics[trim={3cm 1cm 3cm 1cm}, clip,  width=\linewidth]{figures/cluster_ord_EA}
		\caption{co-clustering}
	\end{subfigure}
\end{figure}

\begin{figure}[H]
	\centering
	\begin{subfigure}{.45\textwidth}
		\centering
		\includegraphics[width=\linewidth]{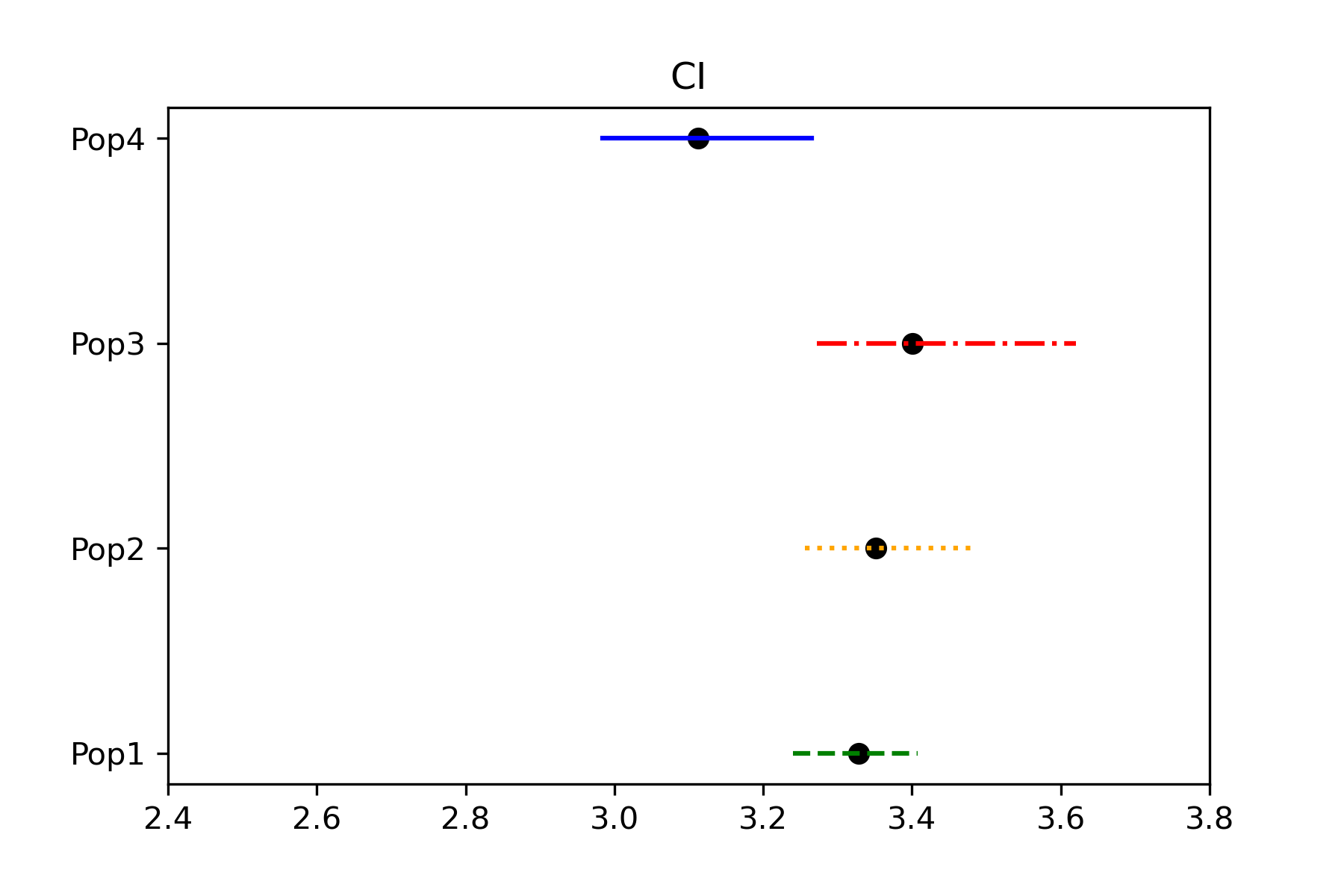}
	\end{subfigure}\hspace{0.05\textwidth}%
	\begin{subfigure}{.45\textwidth}
		\centering
		\includegraphics[width=\linewidth]{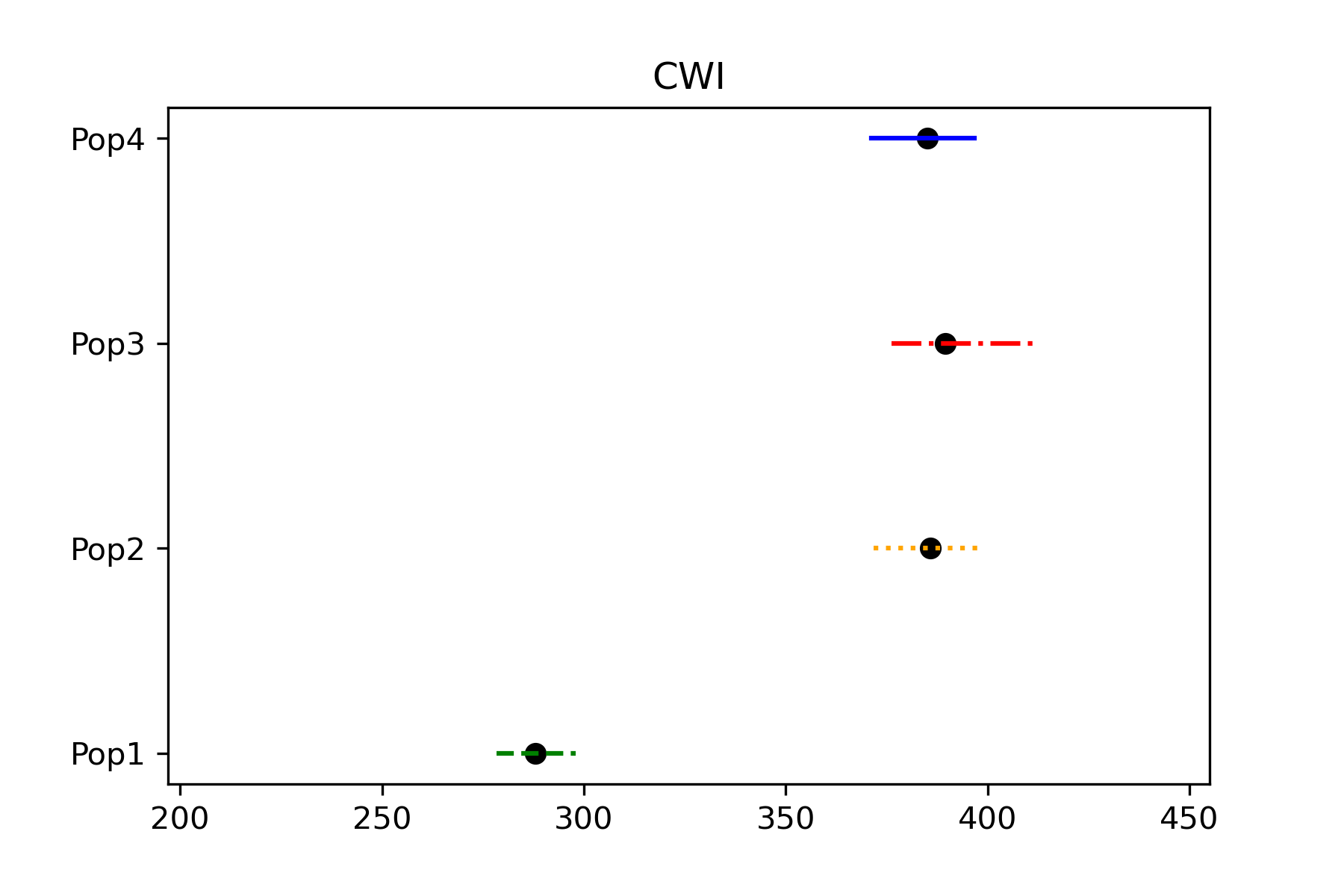}
	\end{subfigure}
\end{figure}

\begin{figure}[H]
	\centering
	\begin{subfigure}{.45\textwidth}
		\centering
		\includegraphics[width=\linewidth]{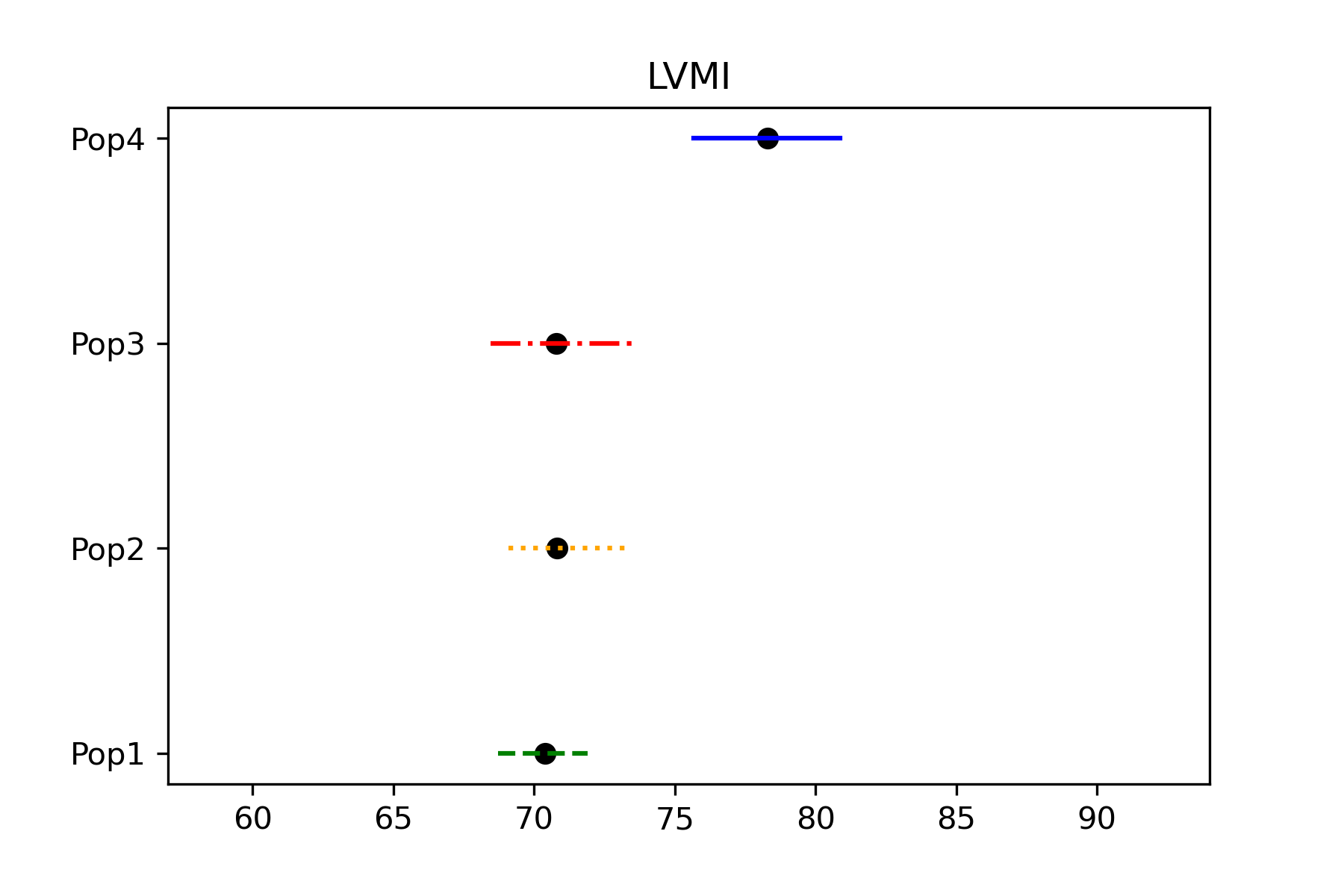}
	\end{subfigure}\hspace{0.05\textwidth}%
	\begin{subfigure}{.45\textwidth}
		\centering
		\includegraphics[width=\linewidth]{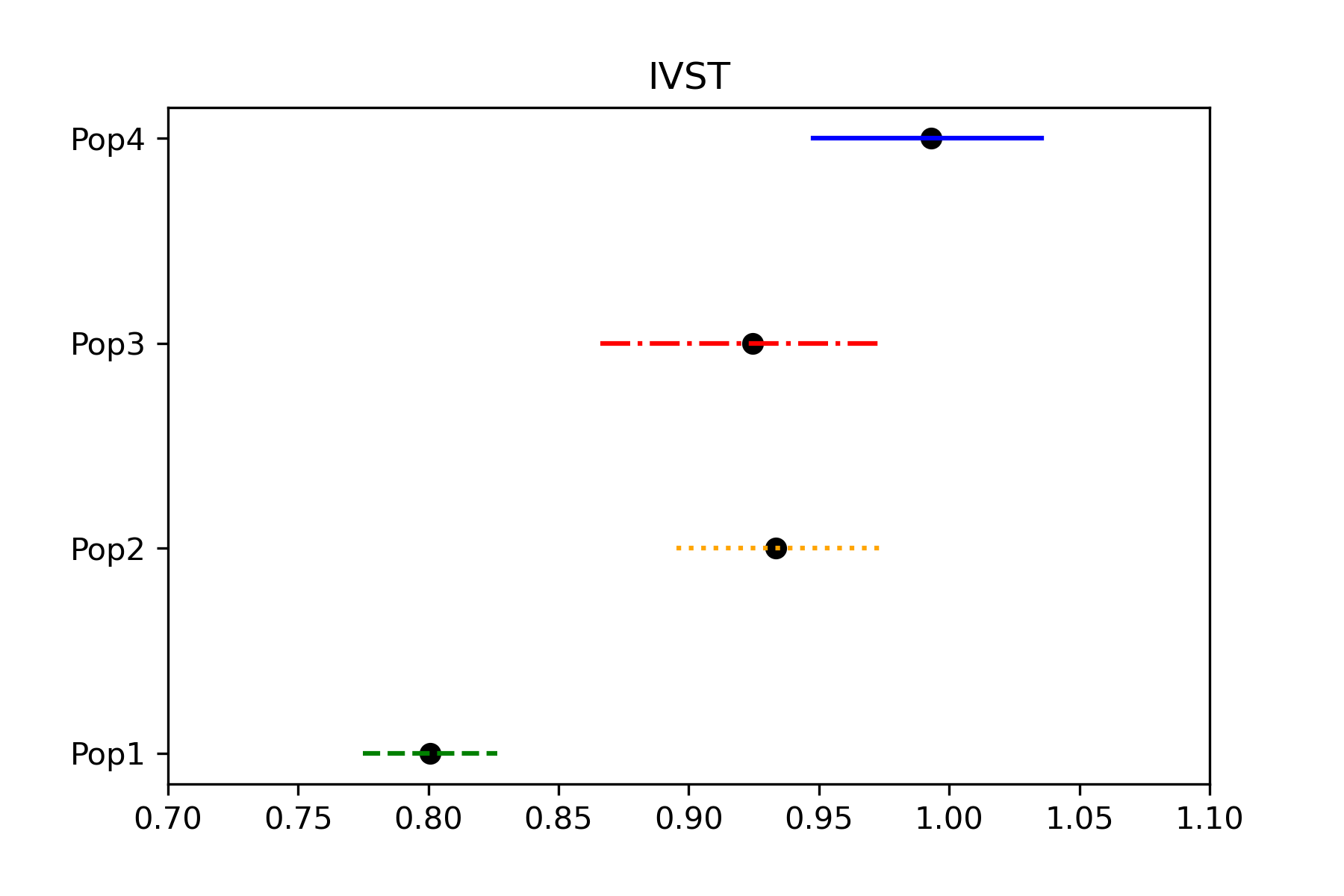}
	\end{subfigure}
\end{figure}

\begin{figure}[H]
	\centering
	\begin{subfigure}{.45\textwidth}
		\centering
		\includegraphics[width=\linewidth]{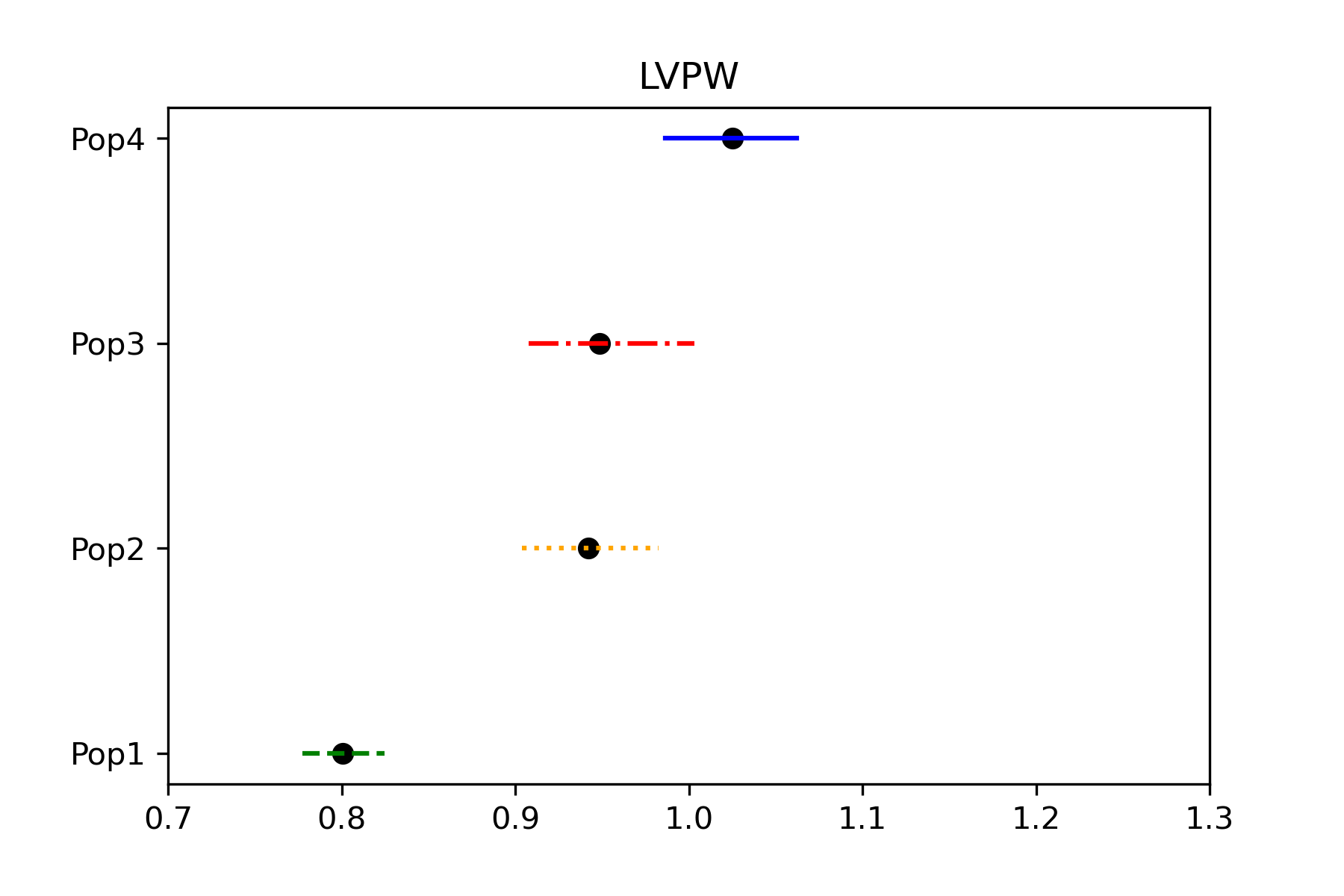}
	\end{subfigure}\hspace{0.05\textwidth}%
	\begin{subfigure}{.45\textwidth}
		\centering
		\includegraphics[width=\linewidth]{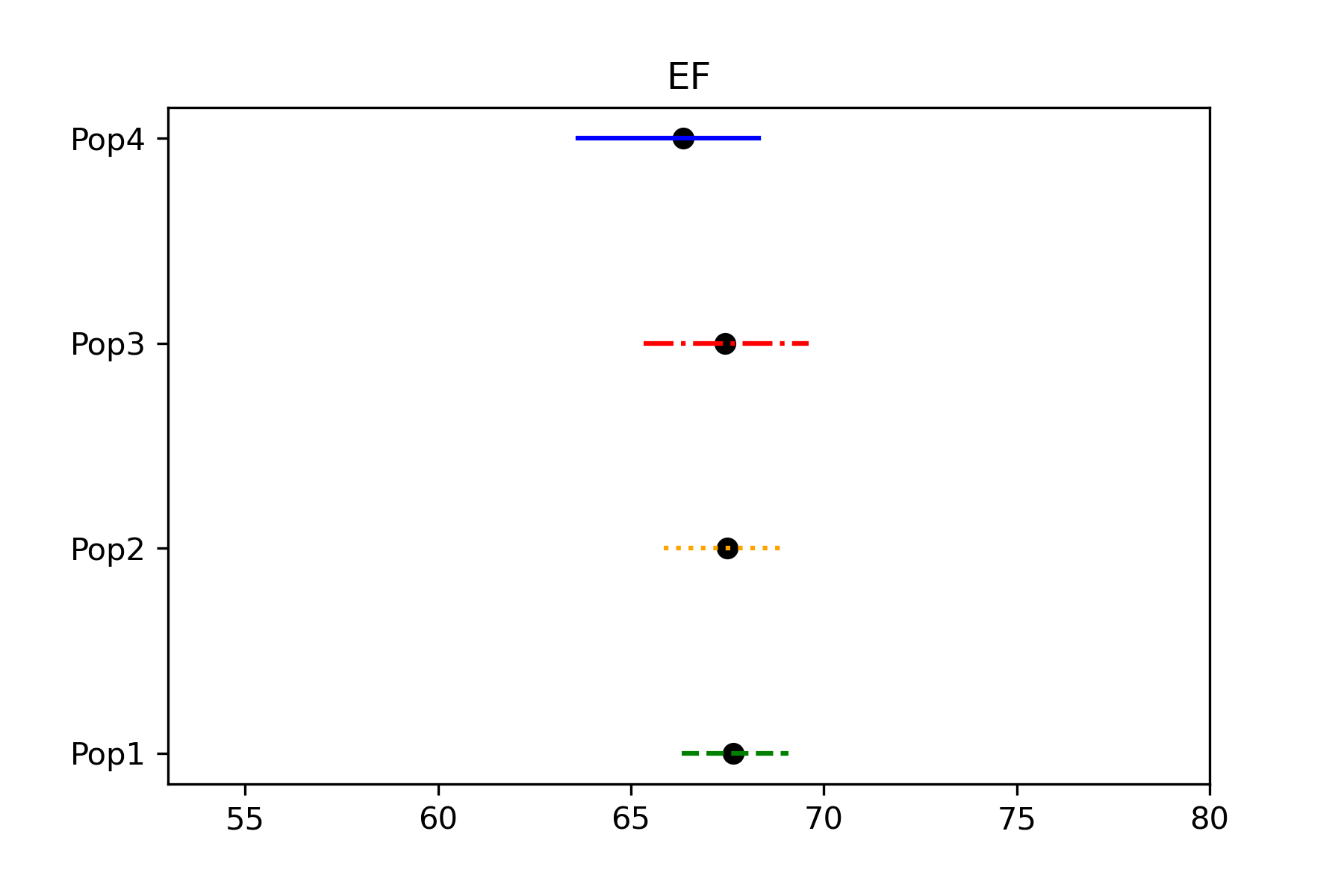}
	\end{subfigure}
\end{figure}

\begin{figure}[H]
	\centering
	\begin{subfigure}{.45\textwidth}
		\centering
		\includegraphics[width=\linewidth]{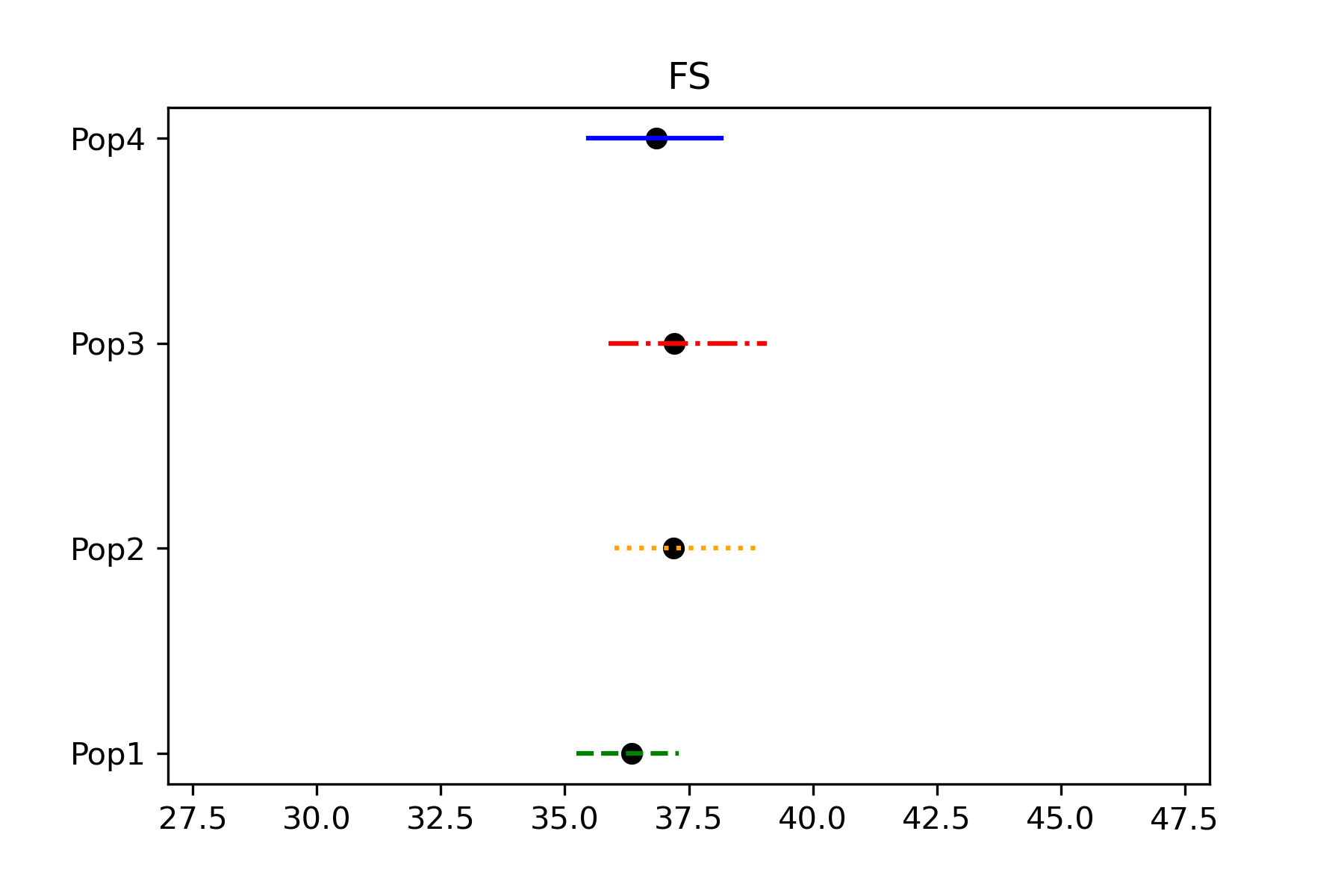}
	\end{subfigure}\hspace{0.05\textwidth}%
	\begin{subfigure}{.45\textwidth}
		\centering
		\includegraphics[width=\linewidth]{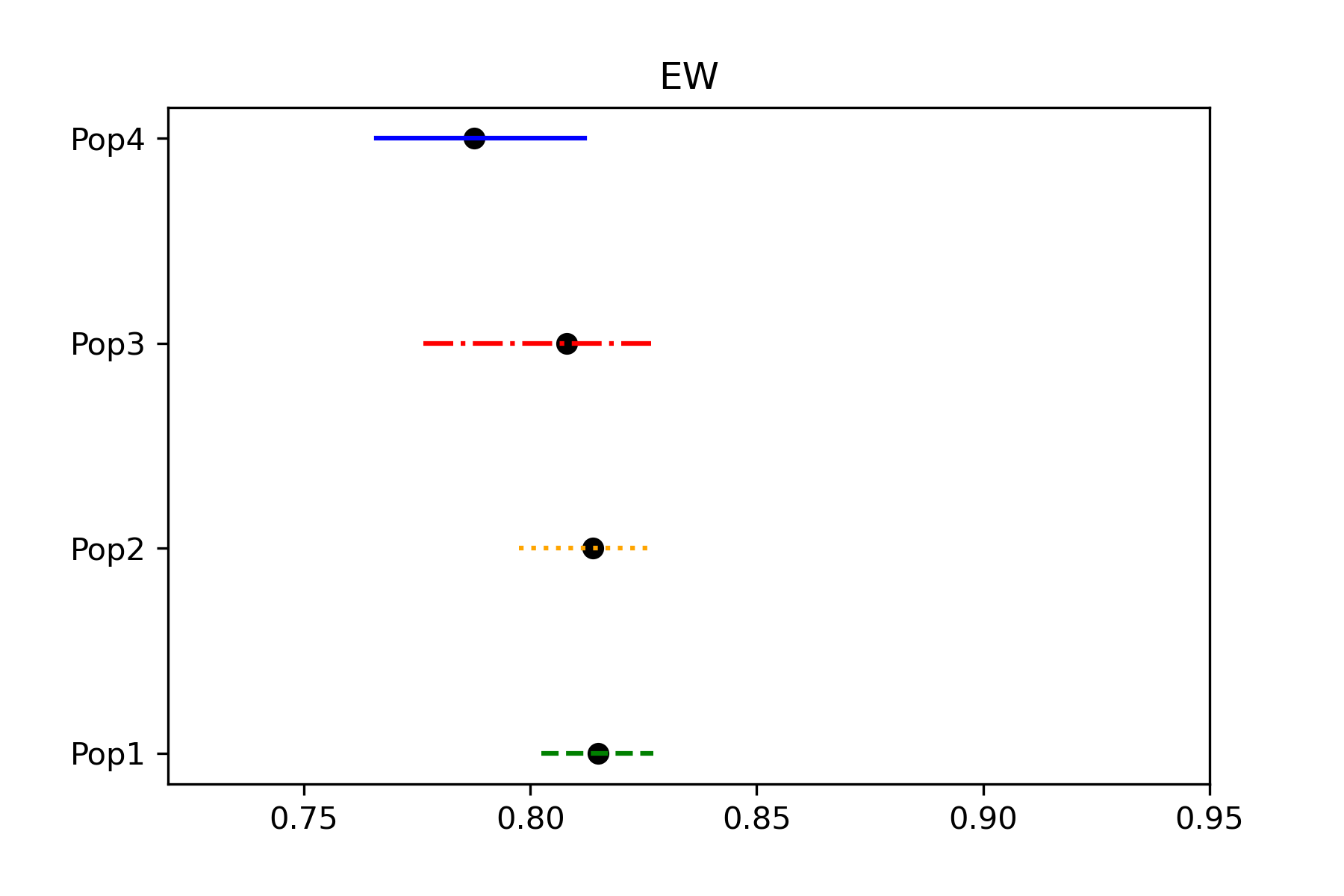}
	\end{subfigure}
\end{figure}

\begin{figure}[H]
	\centering
	\begin{subfigure}{.45\textwidth}
		\centering
		\includegraphics[width=\linewidth]{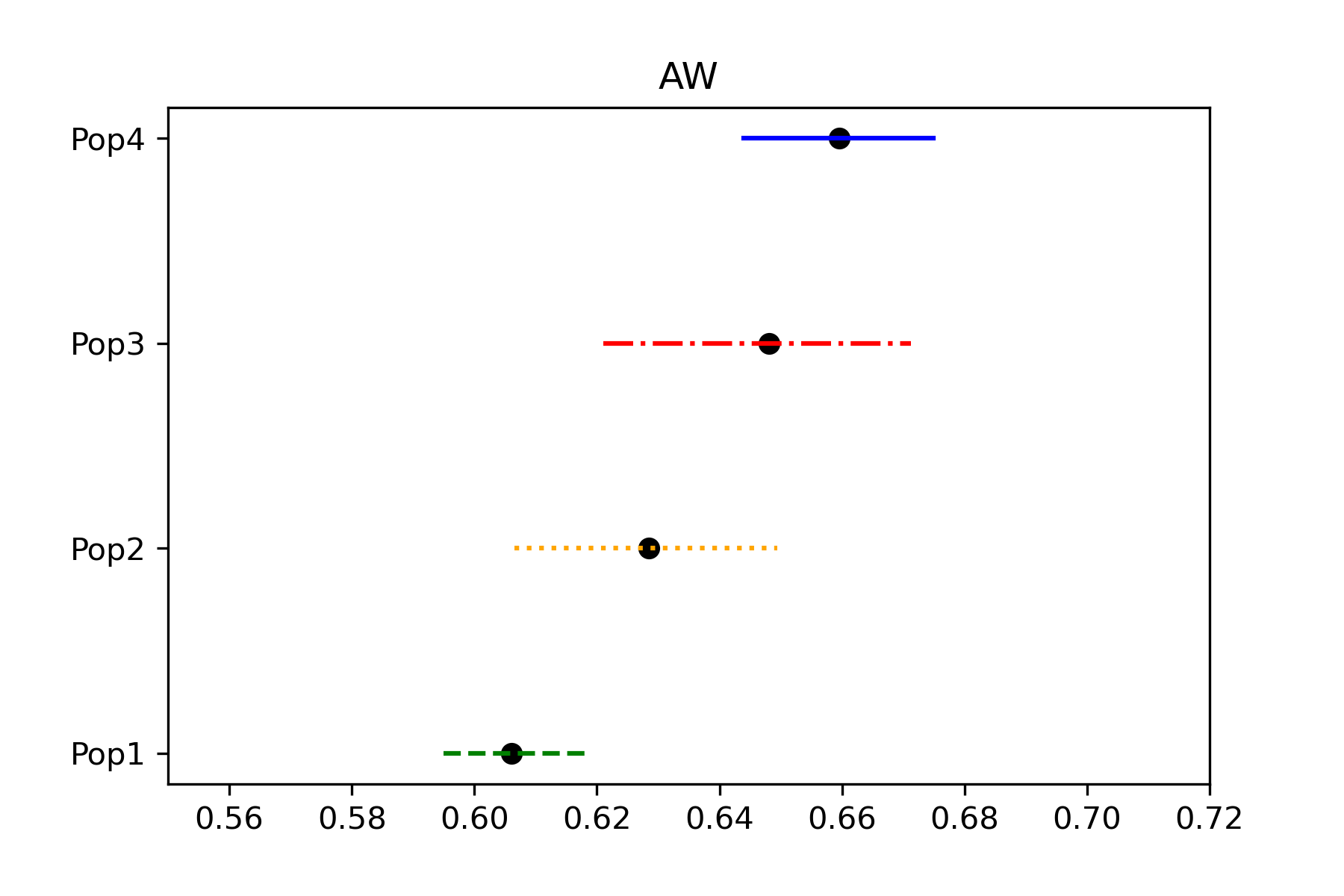}
	\end{subfigure}\hspace{0.05\textwidth}%
	\begin{subfigure}{.45\textwidth}
		\centering
		\includegraphics[width=\linewidth]{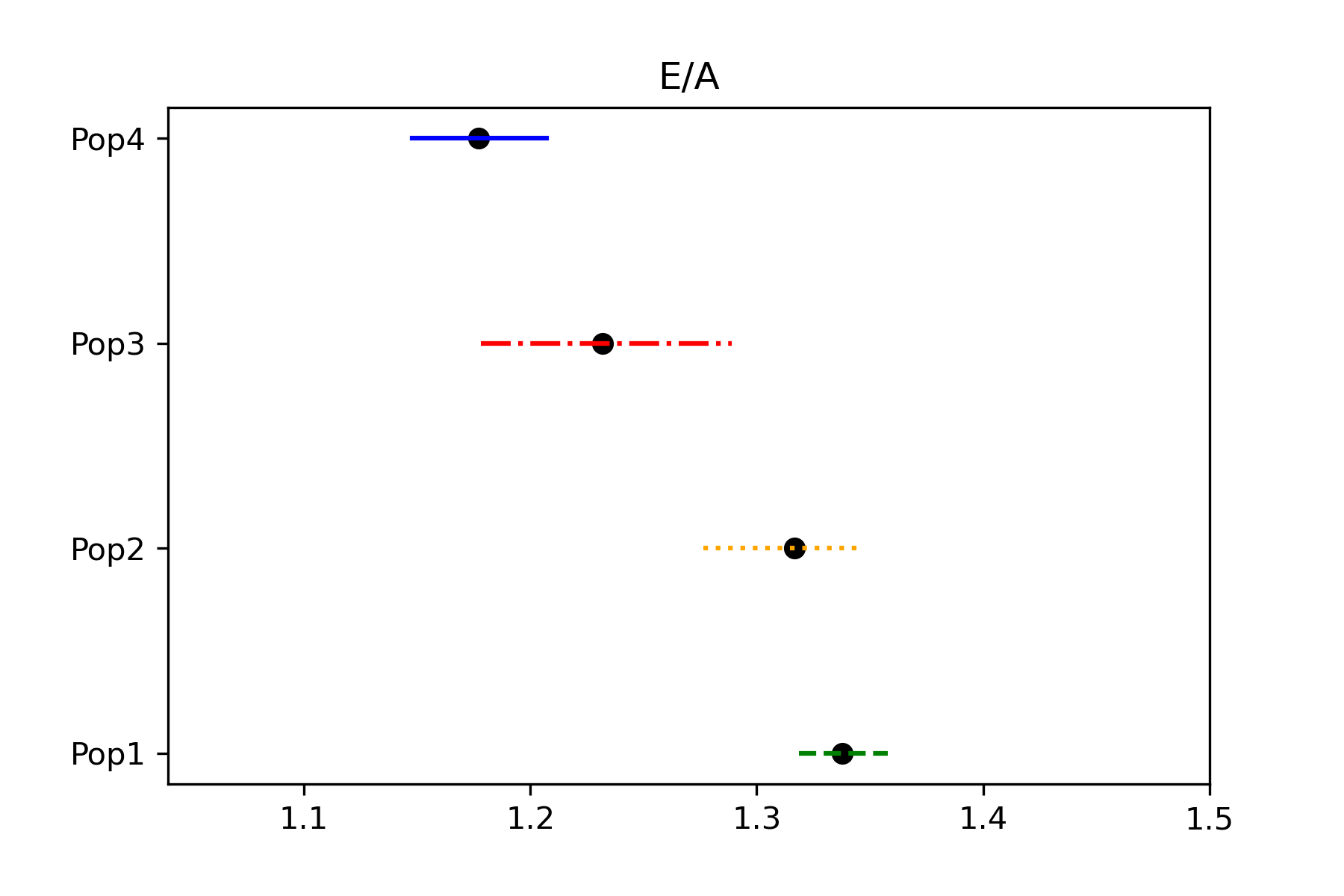}
	\end{subfigure}
\end{figure}

\vspace*{\fill}

\subsection{Prior sensitivity to hyperpriorparameters}
Here we verify the robustness of the model w.r.t.~different specifications of the hyperparameters. We consider two alternative specifications for the hyperparameters, which differ from the one employed in Section 5.2. 
\begin{itemize}
	\item[ ] \textbf{Prior specification 1:} $G_{m}=\text{N}(0,\,1)$; $P_{0,m}\equiv \text{NIG}(\mu=0, \,\tau=0.01,\, \alpha=3,\,\beta=3)$; all concentration parameters have prior equal to $\text{Gamma}(3,\,3)$
	\item[ ] \textbf{Prior specification 2:} $G_{m}=\text{N}(0,\,2)$; $P_{0,m}\equiv \text{NIG}(\mu=0, \,\tau=1,\, \alpha=2,\,\beta=4)$; all concentration parameters have prior equal to $\text{Gamma}(0.1,\,0.1)$
\end{itemize}

The model turns out to be rather robust w.r.t.~the choice of the hyperparameters, leading to the same selected models for all cardiac indexes under all considered specifications. The detailed results are in the following tables and figures, which report the posterior over partitions of locations, the density estimates, and the posterior similarity matrices for the last cardiac index.

\vspace*{\fill}

\newpage

\vspace*{\fill}

\begin{table}[H]
	\caption{Posterior probabilities over partitions of means, using prior specification 1. Maximum a posteriori probabilities are in \textbf{bold}.}
	\label{tab:table4}
	\begin{center}
		\resizebox{\columnwidth}{!}{
			\begin{tabular}{lcccccccccc}
				partitions&CI&CWI&LVMI&IVST&LVPW&EF&FS&EW&AW&E/A\\\hline\hline
				\{\textcolor{aoe}{C},\textcolor{burntorange}{G},\textcolor{bostonuniversityred}{M},\textcolor{burgundy}{S}\}
				&0.018&0.000&0.000&0.000&0.000&\textbf{0.371}&\textbf{0.276}&0.100&0.000&0.000\\
				\{\textcolor{aoe}{C}\}\{\textcolor{burntorange}{G},\textcolor{bostonuniversityred}{M},\textcolor{burgundy}{S}\}
				&0.002&\textbf{0.526}&0.001&0.086&0.015&0.068&0.207&0.025&0.028&0.000\\
				\{\textcolor{aoe}{C},\textcolor{burntorange}{G}\}\{\textcolor{bostonuniversityred}{M},\textcolor{burgundy}{S}\}
				&0.002&0.000&0.000&0.000&0.000&0.038&0.035&0.058&0.072&0.045\\
				\rowcolor{shadecolor}\{\textcolor{aoe}{C},\textcolor{bostonuniversityred}{M},\textcolor{burgundy}{S}\}\{\textcolor{burntorange}{G}\}
				&0.000&0.000&0.000&0.000&0.000&0.000&0.000&0.000&0.000&0.000\\
				\{\textcolor{aoe}{C}\}\{\textcolor{burntorange}{G}\}\{\textcolor{bostonuniversityred}{M},\textcolor{burgundy}{S}\}
				&0.001&0.139&0.000&0.021&0.023&0.025&0.087&0.034&0.244&0.054\\
				\{\textcolor{aoe}{C},\textcolor{burntorange}{G},\textcolor{bostonuniversityred}{M}\}\{\textcolor{burgundy}{S}\}
				&\textbf{0.436}&0.000&\textbf{0.612}&0.000&0.000&0.279&0.04&\textbf{0.499}&0.007&0.001\\
				\rowcolor{shadecolor}\{\textcolor{aoe}{C},\textcolor{burgundy}{S}\}\{\textcolor{burntorange}{G},\textcolor{bostonuniversityred}{M}\}
				&0.000&0.000&0.000&0.000&0.000&0.000&0.000&0.000&0.000&0.000\\
				\{\textcolor{aoe}{C}\}\{\textcolor{burntorange}{G},\textcolor{bostonuniversityred}{M}\}\{\textcolor{burgundy}{S}\}
				&0.157&0.100&0.180&\textbf{0.542}&\textbf{0.678}&0.073&0.172&0.103&0.265&0.026\\
				\rowcolor{shadecolor}\{\textcolor{aoe}{C},\textcolor{bostonuniversityred}{M}\}\{\textcolor{burntorange}{G},\textcolor{burgundy}{S}\}
				&0.000&0.000&0.000&0.000&0.000&0.000&0.000&0.000&0.000&0.000\\
				\rowcolor{shadecolor}\{\textcolor{aoe}{C},\textcolor{burntorange}{G},\textcolor{burgundy}{S}\}\{\textcolor{bostonuniversityred}{M}\}
				&0.000&0.000&0.000&0.000&0.000&0.000&0.000&0.000&0.000&0.000\\
				\rowcolor{shadecolor}\{\textcolor{aoe}{C}\}\{\textcolor{burntorange}{G},\textcolor{burgundy}{S}\}\{\textcolor{bostonuniversityred}{M}\}
				&0.000&0.000&0.000&0.000&0.000&0.000&0.000&0.000&0.000&0.000\\
				\{\textcolor{aoe}{C},\textcolor{burntorange}{G}\}\{\textcolor{bostonuniversityred}{M}\}\{\textcolor{burgundy}{S}\}
				&0.252&0.000&0.092&0.000&0.000&0.081&0.054&0.113&0.087&0.361\\
				\rowcolor{shadecolor}\{\textcolor{aoe}{C},\textcolor{bostonuniversityred}{M}\}\{\textcolor{burntorange}{G}\}\{\textcolor{burgundy}{S}\}
				&0.000&0.000&0.000&0.000&0.000&0.000&0.000&0.000&0.000&0.000\\
				\rowcolor{shadecolor}\{\textcolor{aoe}{C},\textcolor{burgundy}{S}\}\{\textcolor{burntorange}{G}\}\{\textcolor{bostonuniversityred}{M}\}
				&0.000&0.000&0.000&0.000&0.000&0.000&0.000&0.000&0.000&0.000\\
				\{\textcolor{aoe}{C}\}\{\textcolor{burntorange}{G}\}\{\textcolor{bostonuniversityred}{M}\}\{\textcolor{burgundy}{S}\}
				&0.131&0.234&0.116&0.351&0.284&0.066&0.130&0.068&\textbf{0.295}&\textbf{0.513}
		\end{tabular}}
	\end{center}
\end{table} 

\begin{figure}[H]
	\centering
	\begin{subfigure}{.4\textwidth}
		\centering
		\includegraphics[ width=\linewidth]{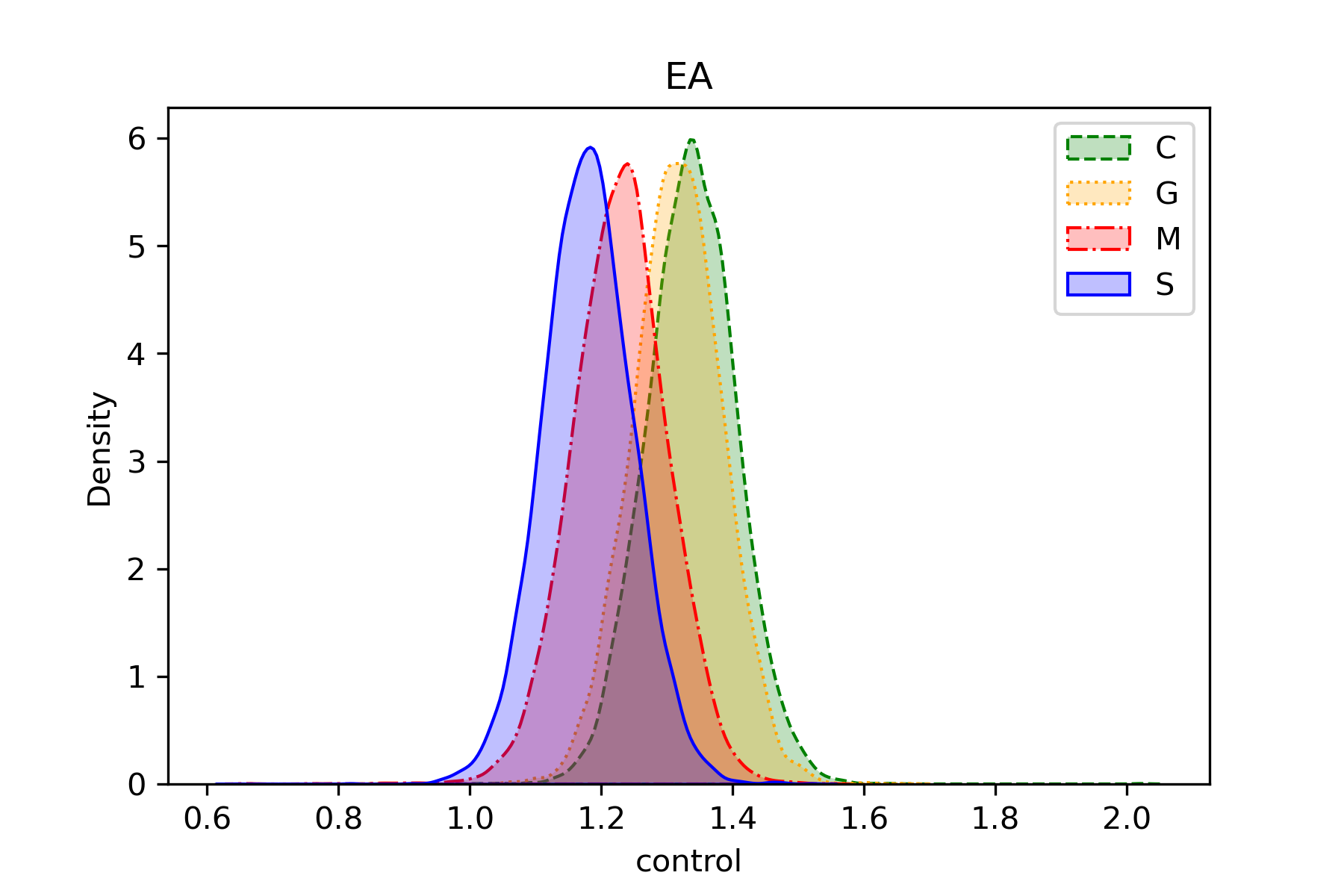}
		\caption{density estimation}
	\end{subfigure}\hspace{0.05\textwidth}%
	\begin{subfigure}{.25\textwidth}
		\centering
		\includegraphics[trim={3cm 1cm 3cm 1cm}, clip,  width=\linewidth]{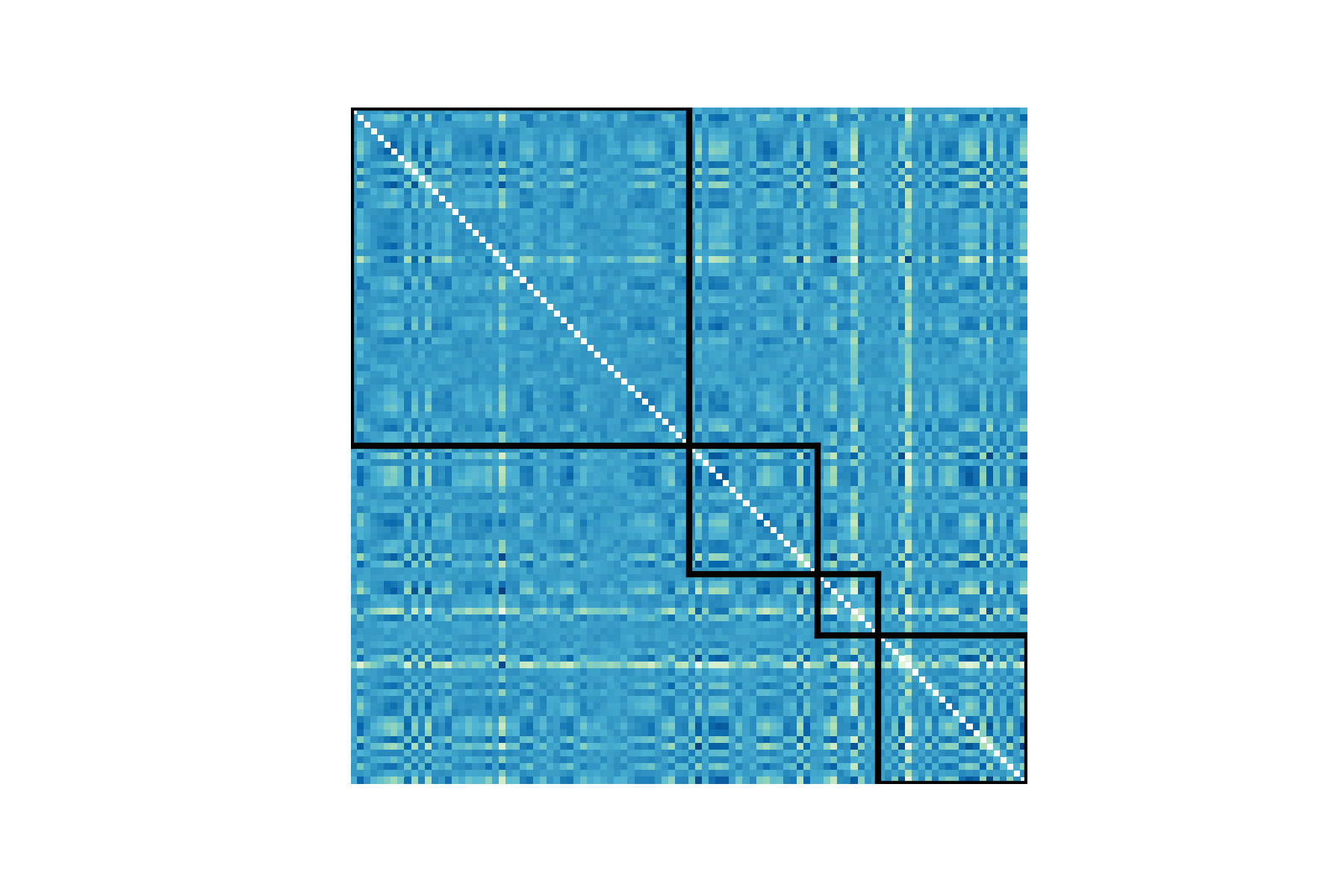}
		\caption{co-clustering}
	\end{subfigure}\hspace{0.05\textwidth}%
	\begin{subfigure}{.25\textwidth}
		\centering
		\includegraphics[trim={3cm 1cm 3cm 1cm}, clip,  width=\linewidth]{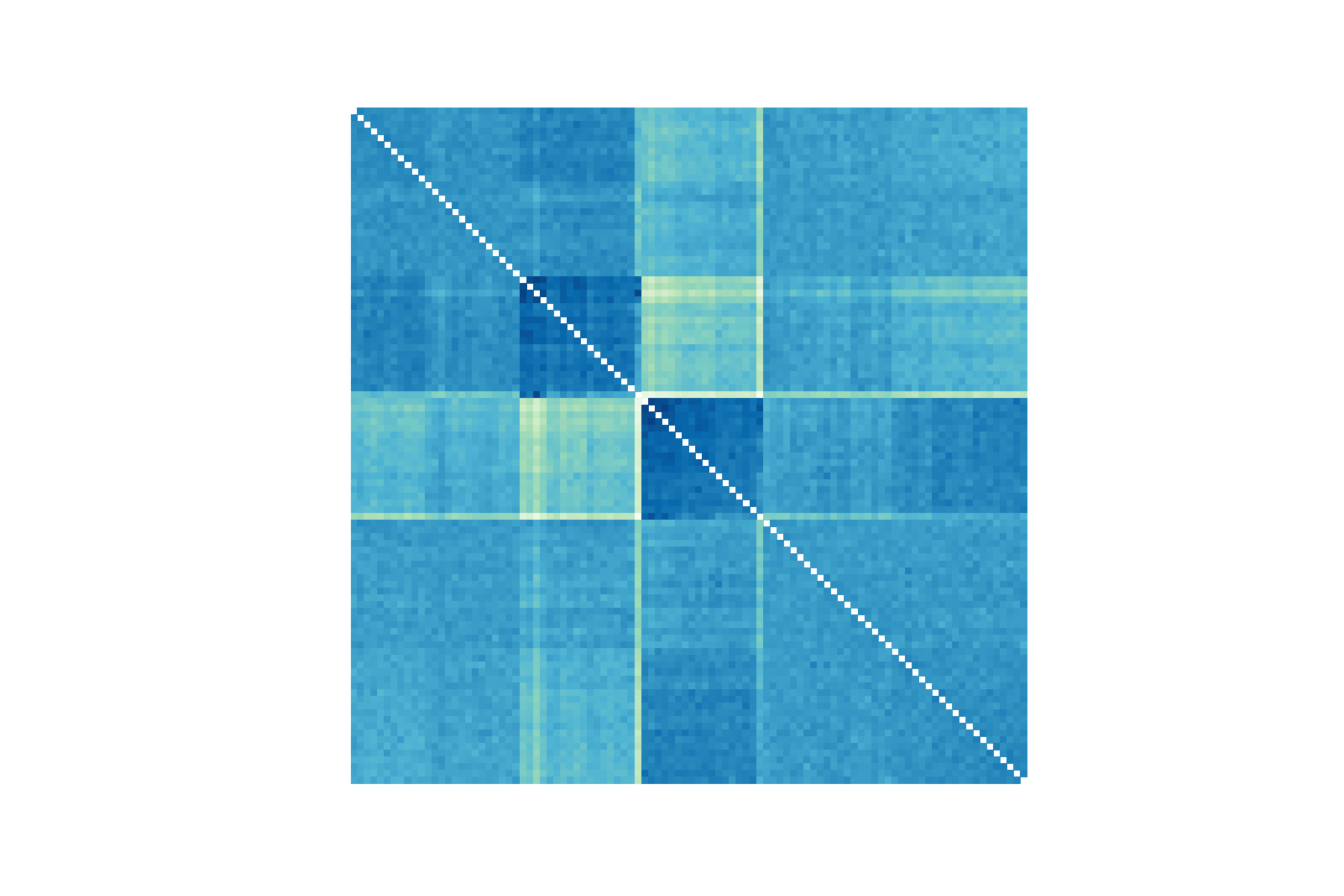}
		\caption{co-clustering}
	\end{subfigure}
\end{figure}
\vspace*{\fill}

\newpage

\vspace*{\fill}

\begin{table}[H]
	\caption{Posterior probabilities over partitions of means, using prior specification 2. Maximum a posteriori probabilities are in \textbf{bold}.}
	\label{tab:table5}
	\begin{center}
		\resizebox{\columnwidth}{!}{
			\begin{tabular}{lcccccccccc}
				partitions&CI&CWI&LVMI&IVST&LVPW&EF&FS&EW&AW&E/A\\\hline\hline
				\{\textcolor{aoe}{C},\textcolor{burntorange}{G},\textcolor{bostonuniversityred}{M},\textcolor{burgundy}{S}\}
				&0.023&0.000&0.000&0.000&0.000&\textbf{0.341}&\textbf{0.281}&0.109&0.000&0.000\\
				\{\textcolor{aoe}{C}\}\{\textcolor{burntorange}{G},\textcolor{bostonuniversityred}{M},\textcolor{burgundy}{S}\}
				&0.002&\textbf{0.484}&0.000&0.097&0.055&0.079&0.199&0.028&0.029&0.000\\
				\{\textcolor{aoe}{C},\textcolor{burntorange}{G}\}\{\textcolor{bostonuniversityred}{M},\textcolor{burgundy}{S}\}
				&0.002&0.000&0.001&0.000&0.000&0.036&0.029&0.042&0.074&0.058\\
				\rowcolor{shadecolor}\{\textcolor{aoe}{C},\textcolor{bostonuniversityred}{M},\textcolor{burgundy}{S}\}\{\textcolor{burntorange}{G}\}
				&0.000&0.000&0.000&0.000&0.000&0.000&0.000&0.000&0.000&0.000\\
				\{\textcolor{aoe}{C}\}\{\textcolor{burntorange}{G}\}\{\textcolor{bostonuniversityred}{M},\textcolor{burgundy}{S}\}
				&0.001&0.134&0.001&0.022&0.028&0.033&0.090&0.033&0.238&0.068\\
				\{\textcolor{aoe}{C},\textcolor{burntorange}{G},\textcolor{bostonuniversityred}{M}\}\{\textcolor{burgundy}{S}\}
				&\textbf{0.408}&0.000&\textbf{0.585}&0.000&0.000&0.254&0.036&\textbf{0.494}&0.014&0.001\\
				\rowcolor{shadecolor}\{\textcolor{aoe}{C},\textcolor{burgundy}{S}\}\{\textcolor{burntorange}{G},\textcolor{bostonuniversityred}{M}\}
				&0.000&0.000&0.000&0.000&0.000&0.000&0.000&0.000&0.000&0.000\\
				\{\textcolor{aoe}{C}\}\{\textcolor{burntorange}{G},\textcolor{bostonuniversityred}{M}\}\{\textcolor{burgundy}{S}\}
				&0.145&0.111&0.184&\textbf{0.530}&\textbf{0.643}&0.077&0.172&0.105&0.254&0.019\\
				\rowcolor{shadecolor}\{\textcolor{aoe}{C},\textcolor{bostonuniversityred}{M}\}\{\textcolor{burntorange}{G},\textcolor{burgundy}{S}\}
				&0.000&0.000&0.000&0.000&0.000&0.000&0.000&0.000&0.000&0.000\\
				\rowcolor{shadecolor}\{\textcolor{aoe}{C},\textcolor{burntorange}{G},\textcolor{burgundy}{S}\}\{\textcolor{bostonuniversityred}{M}\}
				&0.000&0.000&0.000&0.000&0.000&0.000&0.000&0.000&0.000&0.000\\
				\rowcolor{shadecolor}\{\textcolor{aoe}{C}\}\{\textcolor{burntorange}{G},\textcolor{burgundy}{S}\}\{\textcolor{bostonuniversityred}{M}\}
				&0.000&0.000&0.000&0.000&0.000&0.000&0.000&0.000&0.000&0.000\\
				\{\textcolor{aoe}{C},\textcolor{burntorange}{G}\}\{\textcolor{bostonuniversityred}{M}\}\{\textcolor{burgundy}{S}\}
				&0.247&0.000&0.097&0.000&0.000&0.089&0.050&0.106&0.076&0.346\\
				\rowcolor{shadecolor}\{\textcolor{aoe}{C},\textcolor{bostonuniversityred}{M}\}\{\textcolor{burntorange}{G}\}\{\textcolor{burgundy}{S}\}
				&0.000&0.000&0.000&0.000&0.000&0.000&0.000&0.000&0.000&0.000\\
				\rowcolor{shadecolor}\{\textcolor{aoe}{C},\textcolor{burgundy}{S}\}\{\textcolor{burntorange}{G}\}\{\textcolor{bostonuniversityred}{M}\}
				&0.000&0.000&0.000&0.000&0.000&0.000&0.000&0.000&0.000&0.000\\
				\{\textcolor{aoe}{C}\}\{\textcolor{burntorange}{G}\}\{\textcolor{bostonuniversityred}{M}\}\{\textcolor{burgundy}{S}\}
				&0.172&0.270&0.131&0.351&0.274&0.091&0.144&0.084&\textbf{0.315}&\textbf{0.508}
		\end{tabular}}
	\end{center}
\end{table} 

\begin{figure}[H]
	\centering
	\begin{subfigure}{.4\textwidth}
		\centering
		\includegraphics[ width=\linewidth]{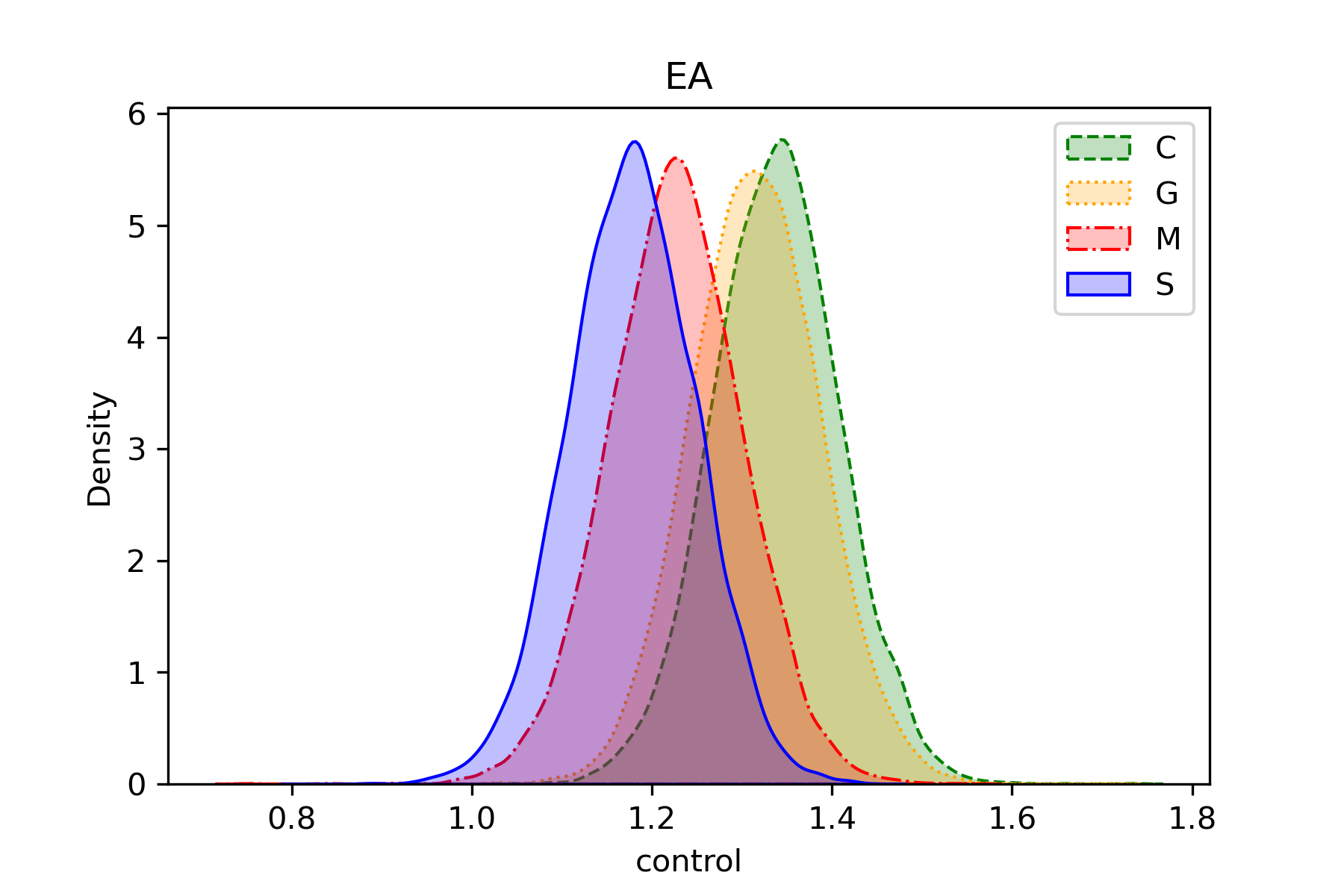}
		\caption{density estimation}
	\end{subfigure}\hspace{0.05\textwidth}%
	\begin{subfigure}{.25\textwidth}
		\centering
		\includegraphics[trim={3cm 1cm 3cm 1cm}, clip,  width=\linewidth]{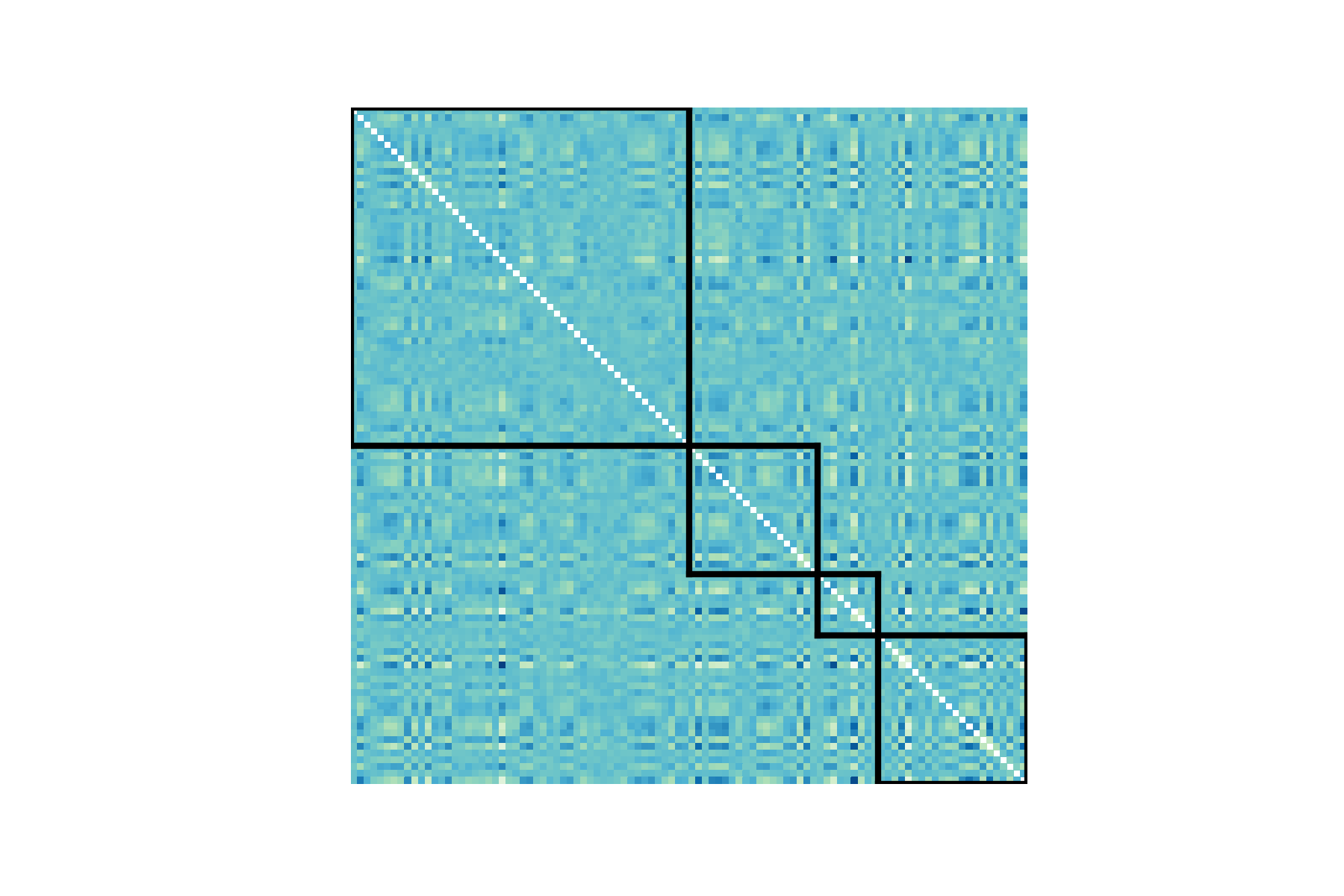}
		\caption{co-clustering}
	\end{subfigure}\hspace{0.05\textwidth}%
	\begin{subfigure}{.25\textwidth}
		\centering
		\includegraphics[trim={3cm 1cm 3cm 1cm}, clip,  width=\linewidth]{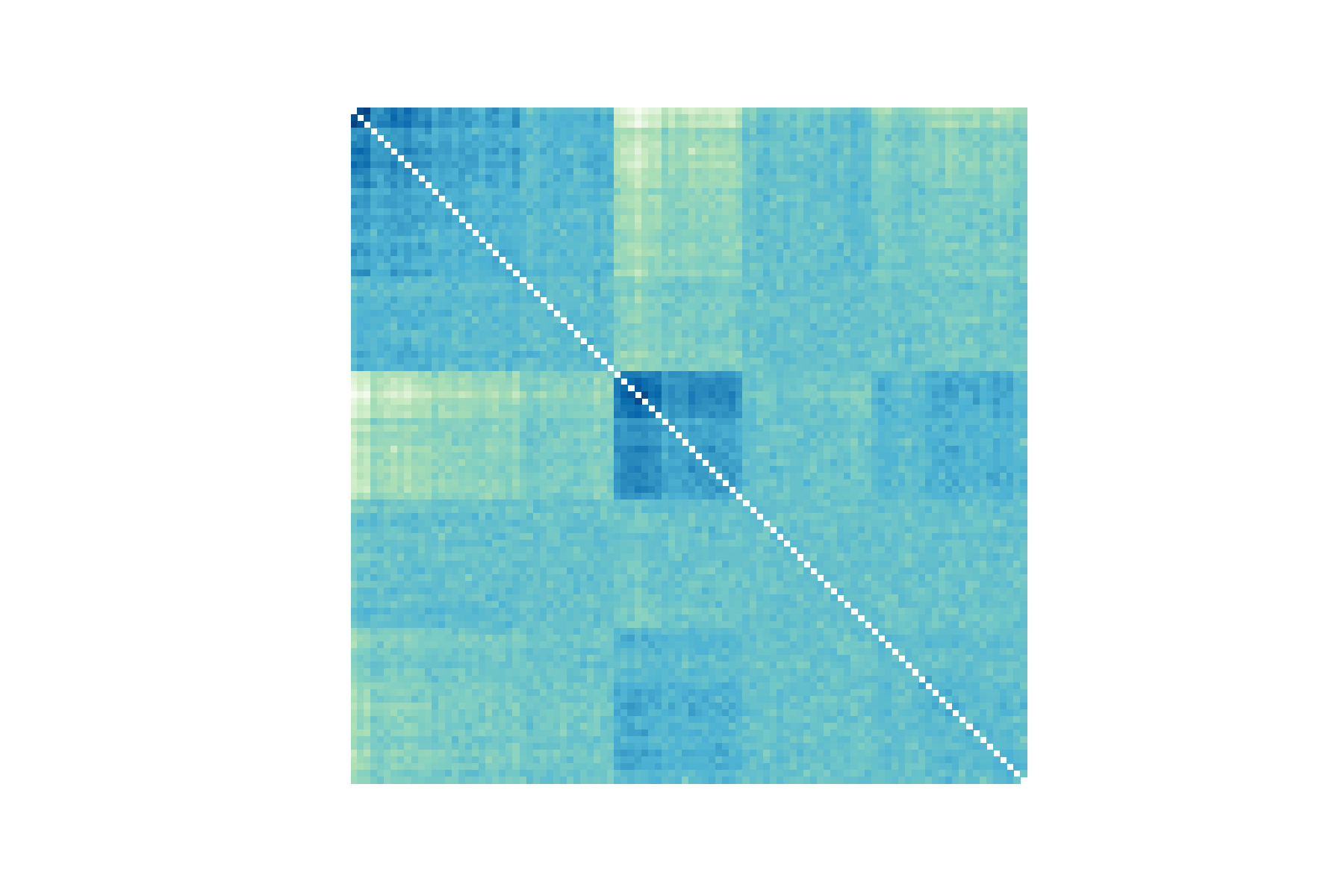}
		\caption{co-clustering}
	\end{subfigure}
\end{figure}
\vspace*{\fill}

\subsection{s-HDP with uniform prior estimates on the Hypertensive Dataset}
Here we report the results on the real dataset, obtained with the s-HDP with independent uniform priors on disease-specific locations, described in Section C.1. 
This prior induces independence between different cardiac indexes and no borrowing of information (i.e.\ penalization for multiplicity) is applied. Moreover, compared to the priors used in Section 5 of the main paper, here the prior associates higher probability to finer partitions and, thus, does not apply a Ockham's-razor penalty, resulting in a different MAP for the EF.

\vspace*{\fill}
\begin{table}[H]
	\caption{Posterior probabilities over partitions obtained through independent uniform priors. Maximum a posteriori probabilities are in \textbf{bold}.}
	\label{tab:table6}
	\begin{center}
		\resizebox{\columnwidth}{!}{
			\begin{tabular}{lcccccccccc}
				partitions&CI&CWI&LVMI&IVST&LVPW&EF&FS&EW&AW&E/A\\\hline\hline
				\{\textcolor{aoe}{C},\textcolor{burntorange}{G},\textcolor{bostonuniversityred}{M},\textcolor{burgundy}{S}\}
				&0.009&0.000&0.000&0.000&0.000&0.248&\textbf{0.216}&0.047&0.000&0.000\\
				\{\textcolor{aoe}{C}\}\{\textcolor{burntorange}{G},\textcolor{bostonuniversityred}{M},\textcolor{burgundy}{S}\}
				&0.002&\textbf{0.568}&0.001&0.084&0.014&0.078&0.205&0.027&0.039&0.000\\
				\{\textcolor{aoe}{C},\textcolor{burntorange}{G}\}\{\textcolor{bostonuniversityred}{M},\textcolor{burgundy}{S}\}
				&0.003&0.000&0.002&0.000&0.000&0.082&0.079&0.160&0.102&0.055\\
				\rowcolor{shadecolor}\{\textcolor{aoe}{C},\textcolor{bostonuniversityred}{M},\textcolor{burgundy}{S}\}\{\textcolor{burntorange}{G}\}
				&0.000&0.000&0.000&0.000&0.000&0.000&0.000&0.000&0.000&0.000\\
				\{\textcolor{aoe}{C}\}\{\textcolor{burntorange}{G}\}\{\textcolor{bostonuniversityred}{M},\textcolor{burgundy}{S}\}
				&0.001&0.143&0.001&0.024&0.029&0.027&0.087&0.041&0.262&0.064\\
				\{\textcolor{aoe}{C},\textcolor{burntorange}{G},\textcolor{bostonuniversityred}{M}\}\{\textcolor{burgundy}{S}\}
				&\textbf{0.376}&0.000&\textbf{0.555}&0.000&0.000&\textbf{0.324}&0.060&\textbf{0.422}&0.005&0.002\\
				\rowcolor{shadecolor}\{\textcolor{aoe}{C},\textcolor{burgundy}{S}\}\{\textcolor{burntorange}{G},\textcolor{bostonuniversityred}{M}\}
				&0.000&0.000&0.000&0.000&0.000&0.000&0.000&0.000&0.000&0.000\\
				\{\textcolor{aoe}{C}\}\{\textcolor{burntorange}{G},\textcolor{bostonuniversityred}{M}\}\{\textcolor{burgundy}{S}\}
				&0.157&0.115&0.188&\textbf{0.614}&\textbf{0.730}&0.078&0.189&0.096&\textbf{0.304}&0.045\\
				\rowcolor{shadecolor}\{\textcolor{aoe}{C},\textcolor{bostonuniversityred}{M}\}\{\textcolor{burntorange}{G},\textcolor{burgundy}{S}\}
				&0.000&0.000&0.000&0.000&0.000&0.000&0.000&0.000&0.000&0.000\\
				\rowcolor{shadecolor}\{\textcolor{aoe}{C},\textcolor{burntorange}{G},\textcolor{burgundy}{S}\}\{\textcolor{bostonuniversityred}{M}\}
				&0.000&0.000&0.000&0.000&0.000&0.000&0.000&0.000&0.000&0.000\\
				\rowcolor{shadecolor}\{\textcolor{aoe}{C}\}\{\textcolor{burntorange}{G},\textcolor{burgundy}{S}\}\{\textcolor{bostonuniversityred}{M}\}
				&0.000&0.000&0.000&0.000&0.000&0.000&0.000&0.000&0.000&0.000\\
				\{\textcolor{aoe}{C},\textcolor{burntorange}{G}\}\{\textcolor{bostonuniversityred}{M}\}\{\textcolor{burgundy}{S}\}
				&0.353&0.000&0.173&0.000&0.000&0.125&0.088&0.162&0.087&0.378\\
				\rowcolor{shadecolor}\{\textcolor{aoe}{C},\textcolor{bostonuniversityred}{M}\}\{\textcolor{burntorange}{G}\}\{\textcolor{burgundy}{S}\}
				&0.000&0.000&0.000&0.000&0.000&0.000&0.000&0.000&0.000&0.000\\
				\rowcolor{shadecolor}\{\textcolor{aoe}{C},\textcolor{burgundy}{S}\}\{\textcolor{burntorange}{G}\}\{\textcolor{bostonuniversityred}{M}\}
				&0.000&0.000&0.000&0.000&0.000&0.000&0.000&0.000&0.000&0.000\\
				\{\textcolor{aoe}{C}\}\{\textcolor{burntorange}{G}\}\{\textcolor{bostonuniversityred}{M}\}\{\textcolor{burgundy}{S}\}&0.099&0.174&0.079&0.278&0.227&0.039&0.077&0.045&0.201&\textbf{0.457}\\
				\hline
				$\sum\log_{15} \left(p_i ^{-p_i}\right)$&0.493&0.426&0.432&0.352&0.269&0.664&0.725&0.624&0.603&0.448
		\end{tabular}}
	\end{center}
\end{table}

\begin{figure}[H]
	\centering
	\begin{subfigure}{.4\textwidth}
		\centering
		\includegraphics[ width=\linewidth]{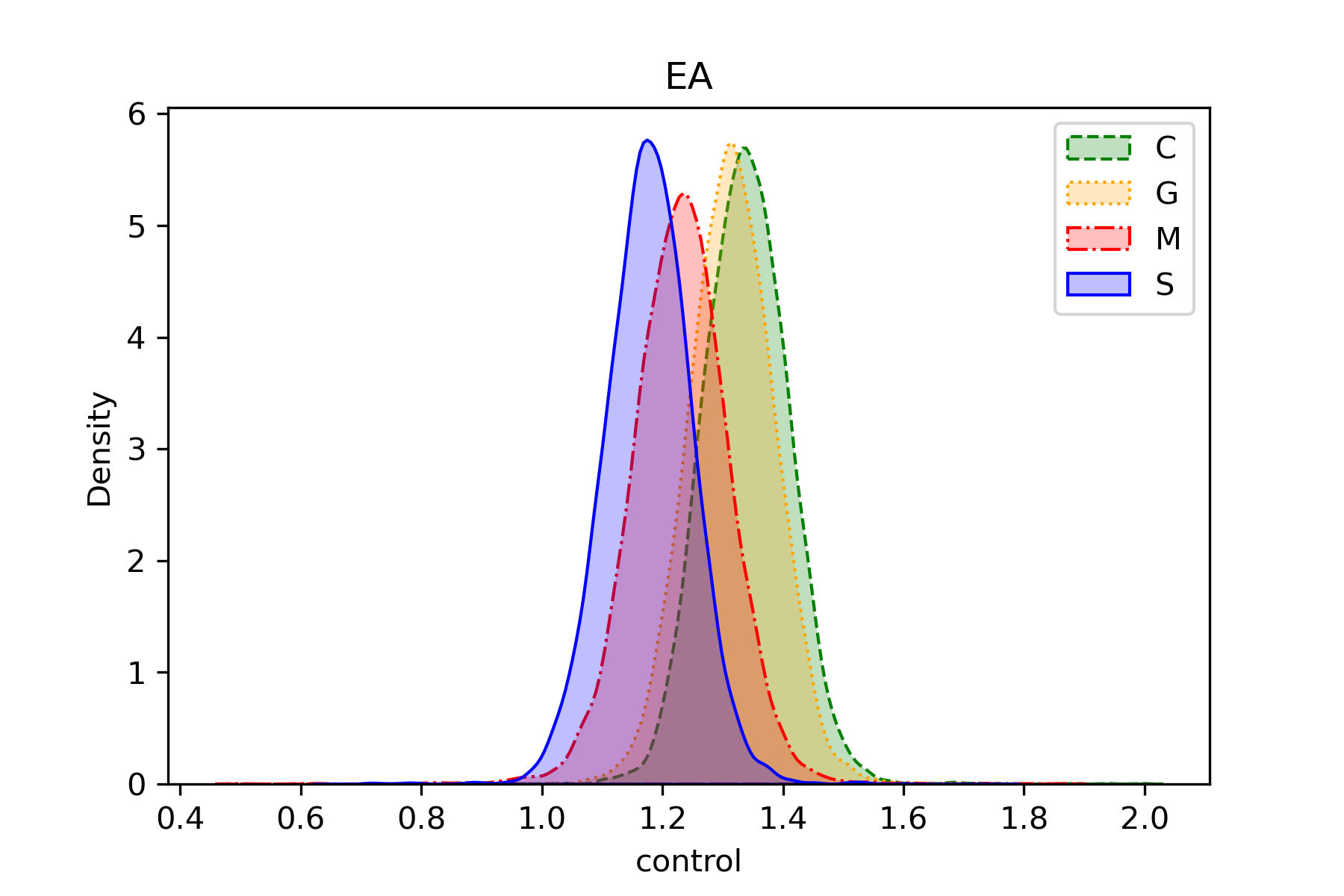}
		\caption{density estimation}
	\end{subfigure}\hspace{0.05\textwidth}%
	\begin{subfigure}{.25\textwidth}
		\centering
		\includegraphics[trim={3cm 1cm 3cm 1cm}, clip,  width=\linewidth]{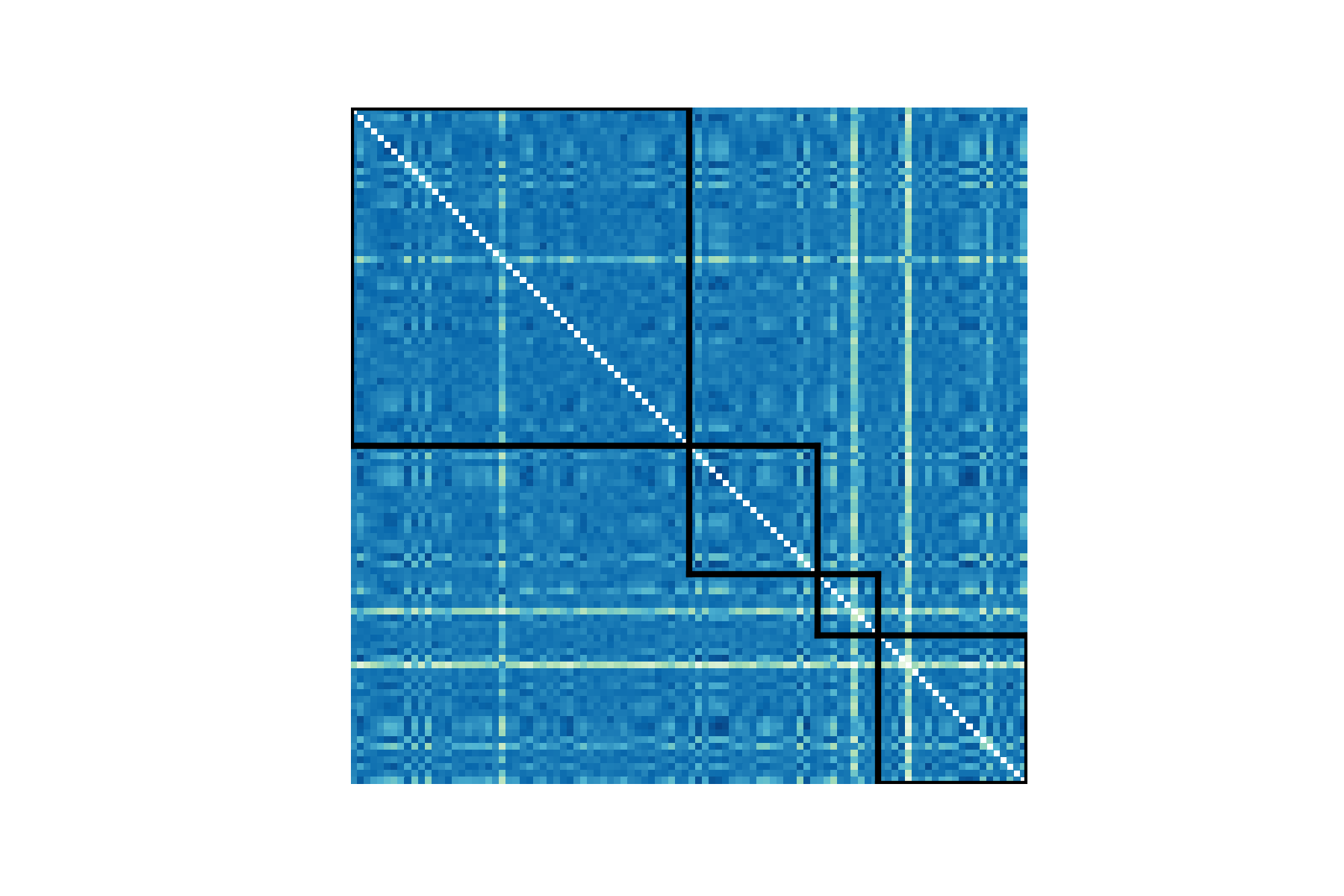}
		\caption{co-clustering}
	\end{subfigure}\hspace{0.05\textwidth}%
	\begin{subfigure}{.25\textwidth}
		\centering
		\includegraphics[trim={3cm 1cm 3cm 1cm}, clip,  width=\linewidth]{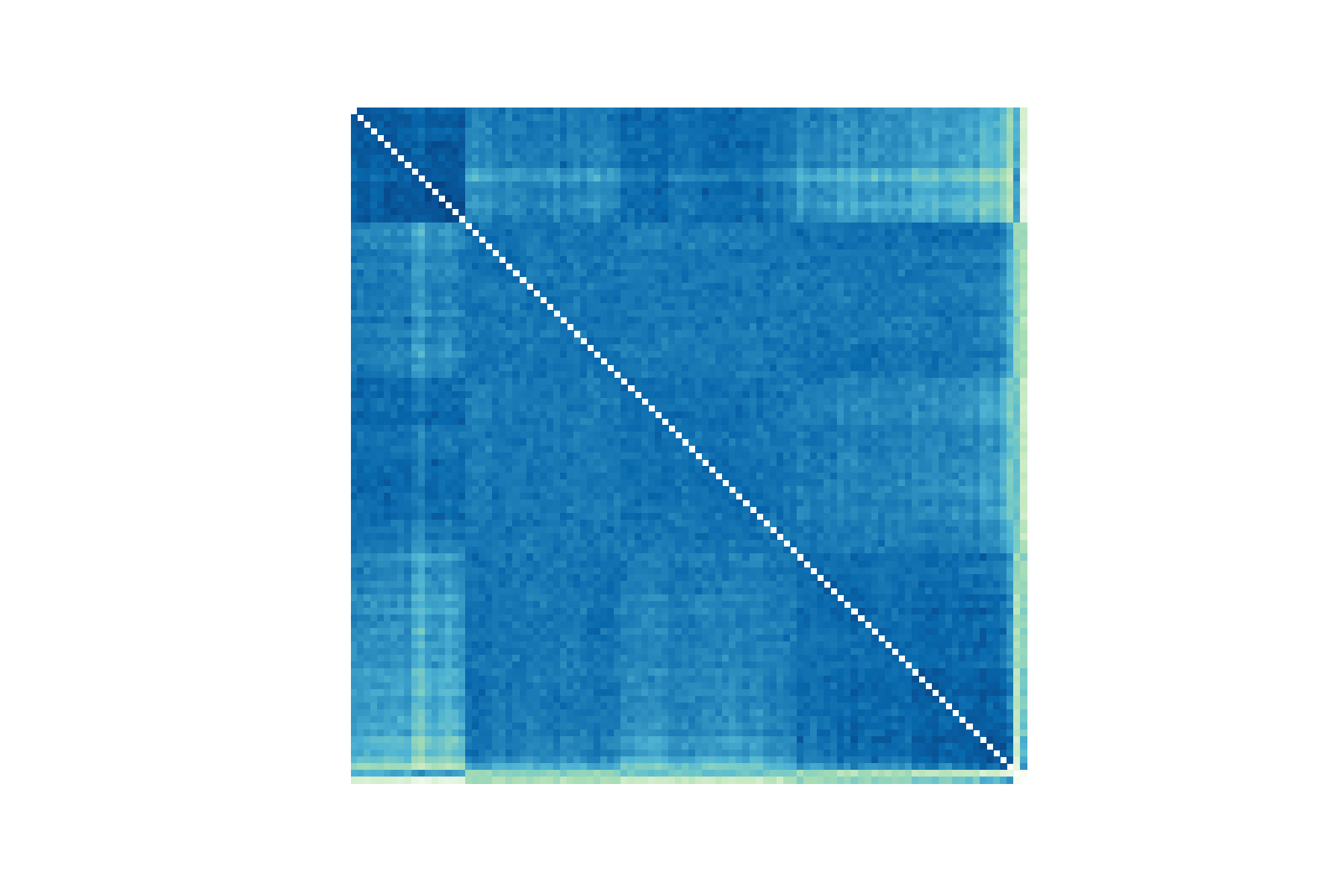}
		\caption{co-clustering}
	\end{subfigure}
\end{figure}
\vspace*{\fill}

\subsection{NDP estimates on the Hypertensive Dataset}
Finally we display the results obtained with ten independent NDPs used on the real dataset. As expected, the NDP tends to identify coarser partitions. Moreover, the independence between cardiac indexes of the NDP approach leads to more concentrated posterior probabilities, because no borrowing of information (i.e.\ penalization for multiplicity) is applied.
\begin{table}[H]
	\caption{Posterior probabilities over partitions obtained through independent NDPs. Maximum a posteriori probabilities are in \textbf{bold}.}
	\begin{center}
		\vspace{-\baselineskip}
		\vspace{-0.5\baselineskip}
		\resizebox{\columnwidth}{!}{
			\begin{tabular}{lcccccccccc}
				partitions&CI&CWI&LVMI&IVST&LVPW&EF&FS&EW&AW&E/A\\\hline\hline
				\{\textcolor{aoe}{C},\textcolor{burntorange}{G},\textcolor{bostonuniversityred}{M},\textcolor{burgundy}{S}\}
				&0.117&0.000&0.000&0.000&0.000&\textbf{0.613}&\textbf{0.394}&0.116&0.000&0.000\\
				\{\textcolor{aoe}{C}\}\{\textcolor{burntorange}{G},\textcolor{bostonuniversityred}{M},\textcolor{burgundy}{S}\}
				&0.004&\textbf{0.999}&0.001&\textbf{0.696}&\textbf{0.663}&0.047&0.099&0.049&0.313&0.000\\
				\{\textcolor{aoe}{C},\textcolor{burntorange}{G}\}\{\textcolor{bostonuniversityred}{M},\textcolor{burgundy}{S}\}
				&0.010&0.000&0.014&0.000&0.001&0.027&0.035&\textbf{0.206}&0.043&\textbf{0.768}\\
				\rowcolor{shadecolor}\{\textcolor{aoe}{C},\textcolor{bostonuniversityred}{M},\textcolor{burgundy}{S}\}\{\textcolor{burntorange}{G}\}
				&0.013&0.000&0.000&0.000&0.000&0.040&0.067&0.051&0.000&0.000\\
				\{\textcolor{aoe}{C}\}\{\textcolor{burntorange}{G}\}\{\textcolor{bostonuniversityred}{M},\textcolor{burgundy}{S}\}
				&0.001&0.000&0.001&0.013&0.163&0.005&0.015&0.088&\textbf{0.468}&0.013\\
				\{\textcolor{aoe}{C},\textcolor{burntorange}{G},\textcolor{bostonuniversityred}{M}\}\{\textcolor{burgundy}{S}\}
				&\textbf{0.552}&0.000&\textbf{0.906}&0.000&0.000&0.103&0.091&0.154&0.002&0.000\\
				\rowcolor{shadecolor}\{\textcolor{aoe}{C},\textcolor{burgundy}{S}\}\{\textcolor{burntorange}{G},\textcolor{bostonuniversityred}{M}\}
				&0.070&0.000&0.000&0.000&0.000&0.025&0.069&0.029&0.000&0.000\\
				\{\textcolor{aoe}{C}\}\{\textcolor{burntorange}{G},\textcolor{bostonuniversityred}{M}\}\{\textcolor{burgundy}{S}\}
				&0.077&0.001&0.010&0.207&0.136&0.010&0.032&0.050&0.093&0.006\\
				\rowcolor{shadecolor}\{\textcolor{aoe}{C},\textcolor{bostonuniversityred}{M}\}\{\textcolor{burntorange}{G},\textcolor{burgundy}{S}\}
				&0.009&0.000&0.003&0.023&0.000&0.035&0.045&0.017&0.001&0.000\\
				\rowcolor{shadecolor}\{\textcolor{aoe}{C},\textcolor{burntorange}{G},\textcolor{burgundy}{S}\}\{\textcolor{bostonuniversityred}{M}\}
				&0.047&0.000&0.000&0.000&0.000&0.068&0.081&0.073&0.000&0.000\\
				\rowcolor{shadecolor}\{\textcolor{aoe}{C}\}\{\textcolor{burntorange}{G},\textcolor{burgundy}{S}\}\{\textcolor{bostonuniversityred}{M}\}
				&0.003&0.001&0.000&0.052&0.027&0.007&0.022&0.012&0.030&0.000\\
				\{\textcolor{aoe}{C},\textcolor{burntorange}{G}\}\{\textcolor{bostonuniversityred}{M}\}\{\textcolor{burgundy}{S}\}
				&0.065&0.000&0.047&0.000&0.000&0.011&0.016&0.071&0.007&0.208\\
				\rowcolor{shadecolor}\{\textcolor{aoe}{C},\textcolor{bostonuniversityred}{M}\}\{\textcolor{burntorange}{G}\}\{\textcolor{burgundy}{S}\}
				&0.025&0.000&0.017&0.004&0.000&0.007&0.017&0.033&0.001&0.000\\
				\rowcolor{shadecolor}\{\textcolor{aoe}{C},\textcolor{burgundy}{S}\}\{\textcolor{burntorange}{G}\}\{\textcolor{bostonuniversityred}{M}\}
				&0.006&0.000&0.000&0.000&0.000&0.006&0.014&0.023&0.000&0.000\\
				\{\textcolor{aoe}{C}\}\{\textcolor{burntorange}{G}\}\{\textcolor{bostonuniversityred}{M}\}\{\textcolor{burgundy}{S}\}&0.007&0.000&0.002&0.007&0.012&0.001&0.007&0.032&0.044&0.006\\
				\hline
				$\sum\log_{15} \left(p_i ^{-p_i}\right)$&0.603&0.016&0.167&0.349&0.368&0.567&0.785&0.898&0.509&0.239
		\end{tabular}}
	\vspace{-\baselineskip}
	\vspace{-\baselineskip}
	\end{center}
\end{table}

\end{document}